\documentclass[prd,aps,superscriptaddress,tightenlines,nofootinbib,floatfix,preprintnumbers,11pt,longbibliography]{revtex4-1}
\usepackage{epsfig}
\usepackage{amsfonts}
\usepackage{amsmath}
\usepackage{amssymb}
\usepackage{dsfont}
\usepackage{hyperref}
\usepackage{xspace}
\usepackage{slashed}
\hypersetup{
    colorlinks=true,       
    linkcolor=blue,          
    citecolor=blue,        
    filecolor=blue,      
    urlcolor=blue           
}
\usepackage{xcolor}
\usepackage{color}
\usepackage{graphicx}  %
\usepackage{bm}  %
\usepackage{graphics}

\usepackage{multirow}

\newcommand{\isovec}[1]{{\vec{#1}}} 
\newcommand{\spacevec}[1]{{\mathbf #1}} 

\newcommand{\simge}{\hspace*{0.2em}\raisebox{0.5ex}{$>$}
     \hspace{-0.8em}\raisebox{-0.3em}{$\sim$}\hspace*{0.2em}}
\newcommand{\simle}{\hspace*{0.2em}\raisebox{0.5ex}{$<$}
     \hspace{-0.8em}\raisebox{-0.3em}{$\sim$}\hspace*{0.2em}}

\newcommand{\al}{\alpha}
\newcommand{\bt}{\beta}

\newcommand{\dt}{\delta}

\newcommand{\la}{\lambda}

\newcommand{\slashpi}{\protect{\slash\hspace{-0.5em}\pi}}

\newcommand{\Nb}{\bar N}

\newcommand{\mpi}{m_{\pi}}

\newcommand{\Or}{\mathcal O}

\newcommand{\vL}{\ensuremath{\mathcal{L}}}

\newcommand{\sq}{^{2}}

\newcommand{\cq}{\mathcal Q}

\newcommand{\MS}{$\overline{\rm MS}$ }

\newcommand{\dslash}[1]{#1 \llap{/\kern-0.5pt}}
\newcommand{\Dslash}[1]{#1 \llap{/\kern+1.5pt}}
\newcommand{\DDslash}[1]{#1 \llap{/\kern+2.3pt}}
\newcommand{\dslashh}[1]{#1 \llap{/\kern+1pt}}

\newcommand{\boldtau}{\mbox{$\isovec \tau$}}

\newcommand{\boldsigma}{\mbox{\boldmath $\sigma$}}

\newcommand{\bea}{\begin{eqnarray}}
\newcommand{\eea}{\end{eqnarray}}
\newcommand{\be}{\begin{equation}}
\newcommand{\ee}{\end{equation}}
\newcommand{\bma}{\begin{pmatrix}}
\newcommand{\ema}{\end{pmatrix}}
\newcommand{\nn}{\nonumber}

\newcommand{\NLDBD}{$0 \nu \beta \beta$}

\newcommand{\nnpp}{$nn \rightarrow p p\, e^- e^-$ }

\begin{document}
\preprint{LA-UR-19-26002, RBRC-1317}

\title{
\vspace*{0.5cm}
\huge
A renormalized approach to
\\
neutrinoless double-beta decay 
\vspace*{.5cm}
}

\vspace*{.5cm}

\author{V. Cirigliano}
\affiliation{Theoretical Division, Los Alamos National Laboratory, Los Alamos, 
NM 87545, USA}

\author{W. Dekens}
\affiliation{Department of Physics, University of California at San Diego,
La Jolla, CA 92093-0319, USA}

\author{J. de Vries}
\affiliation{Amherst Center for Fundamental Interactions, 
Department of Physics, University of Massachusetts Amherst, 
Amherst, MA 01003, USA}
\affiliation{RIKEN BNL Research Center, Brookhaven National Laboratory, 
Upton, NY 11973-5000, USA}

\author{M.~L. Graesser}
\affiliation{Theoretical Division, Los Alamos National Laboratory, Los Alamos, 
NM 87545, USA}

\author{E.~Mereghetti}
\affiliation{Theoretical Division, Los Alamos National Laboratory, Los Alamos, 
NM 87545, USA}

\author{S.~Pastore}
\affiliation{Physics Department, Washington University, St Louis, MO 63130, USA}

\author{M.~Piarulli}
\affiliation{Physics Department, Washington University, St Louis, MO 63130, USA}

\author{U. van Kolck}
\affiliation{Institut de Physique Nucl\'eaire, CNRS/IN2P3, 
Universit\'e Paris-Sud, Universit\'e Paris-Saclay, 91406 Orsay, France}
\affiliation{Department of Physics, University of Arizona, 
Tucson, AZ 85721, USA}

\author{R.~B.~Wiringa}
\affiliation{Physics Division, Argonne National Laboratory, 
Argonne, IL 60439, USA}

\begin{abstract}
\vspace*{.75cm}

The process at the heart of neutrinoless double-beta decay, \nnpp induced 
by a light Majorana neutrino, is investigated in pionless and chiral effective field theory.
We show in various regularization schemes the need to introduce a short-range 
lepton-number-violating operator at leading order, 
confirming earlier findings. We demonstrate that such a short-range operator is 
only needed in spin-singlet $S$-wave transitions, while leading-order 
transitions involving higher partial waves depend solely on long-range 
currents. Calculations are extended to include next-to-leading corrections in 
perturbation theory, 
where to this order no additional undetermined parameters appear.
We establish a connection based on chiral symmetry between neutrinoless 
double-beta decay and nuclear charge-independence breaking 
induced by electromagnetism. Data on the latter confirm the need for a 
leading-order short-range operator,
but do not allow for a full determination of the 
corresponding lepton-number-violating coupling. 
Using a crude estimate of this coupling, we perform {\it ab initio} 
calculations of the matrix elements for neutrinoless double-beta decay 
for $^6$He and $^{12}$Be.
We speculate on the phenomenological impact of the 
leading short-range operator on the basis of these results.

\end{abstract}
\maketitle

\tableofcontents

\newpage

\section{Introduction}

The observation of neutrino oscillations has demonstrated that neutrinos are 
massive particles, with masses constrained by single-beta ($\beta$) decay 
experiments \cite{Aseev:2011dq} and cosmological observations 
\cite{Akrami:2018vks} to be several orders of magnitude smaller than those 
of the charged leptons.
The smallness of the neutrino masses suggests that they have a different origin
with respect to other Standard Model (SM) particles.  
In particular, neutrinos, the only fundamental charge-neutral fermions in the 
SM, could have a Majorana mass, whose small value naturally arises in the 
``see-saw'' mechanism \cite{Minkowski:1977sc,Mohapatra:1979ia,GellMann:1980vs}.
A distinctive signature of the Majorana nature of neutrino masses is the 
violation of lepton number ($L$) by two units ($|\Delta L|=2$)
\cite{Schechter:1981bd}, which would manifest itself in processes such as 
neutrinoless double-beta decay ($0\nu\beta\beta$), nuclear muon-to-positron 
conversion, or rare meson decays such as $K^+ \rightarrow \pi^- e^+ e^+$.
$0\nu\beta\beta$ \cite{Haxton:1985am}
is by far the most sensitive laboratory probe of lepton number
violation (LNV).
Current experimental limits are very stringent \cite{Gando:2012zm,Agostini:2013mzu,Albert:2014awa,Andringa:2015tza,KamLAND-Zen:2016pfg,Elliott:2016ble,Agostini:2017iyd,Aalseth:2017btx, Albert:2017owj,Alduino:2017ehq,Agostini:2018tnm, Azzolini:2018dyb,Anton:2019wmi}, 
e.g. $T^{0\nu}_{1/2}>1.07\times 10^{26}$ yr for ${}^{136}$Xe 
\cite{KamLAND-Zen:2016pfg}, with the next-generation ton-scale experiments 
aiming for improvements by one or two orders of magnitude. 

The interpretation of $0\nu\beta\beta$ experiments and the constraints on 
fundamental LNV parameters,
such as the Majorana masses of left-handed neutrinos, rely on having a 
general theoretical framework 
that provides reliable 
predictions with controlled uncertainties.
Contributions to $0\nu\beta\beta$ can be organized in terms of 
$SU(3)_{\rm c} \times U(1)_{\rm em}$-invariant operators 
\cite{Prezeau:2003xn,Graesser:2016bpz,Cirigliano:2017djv,Cirigliano:2018yza} 
at the scale $\Lambda_\chi \sim 1$ GeV characteristic of QCD nonperturbative 
effects.
The operator of lowest dimension is a Majorana mass term for light, left-handed
neutrinos,
\begin{equation}\label{eq:intro.0}
\mathcal L_{|\Delta L| = 2} = - \frac{m_{\beta\beta}}{2}\, \nu_{eL}^T C \, \nu_{eL} 
+ \ldots,
\end{equation}
where $C = i \gamma_2 \gamma_0$ denotes the charge conjugation matrix
and the effective neutrino mass $m_{\beta\beta} = \sum U_{e i}^2 m_{i}$ combines
the neutrino masses $m_i$ and the elements $U_{ei}$ of the 
Pontecorvo-Maki-Nakagawa-Sato (PMNS) matrix. 
Because of the  $SU(2)_L$ invariance of the SM, $m_i \sim v^2/\Lambda$, 
where $v\simeq 246$ GeV is the vacuum expectation value of the Higgs field
and
$\Lambda$ is the high-energy scale at which LNV arises~\cite{Weinberg:1979sa}. 
The dots in Eq. \eqref{eq:intro.0} denote higher-dimensional LNV operators, 
which
are suppressed by more powers of $v/\Lambda$ and $\Lambda_\chi/v$ 
\cite{Cirigliano:2018yza}. 

In this paper we focus on the $0\nu\beta\beta$ transition operator induced by 
$m_{\beta\beta}$. The quark-level Lagrangian we consider is 
given by  
\begin{equation}
{\cal L}_{\rm eff} = {\cal L}_{\rm QCD}  
- \frac{4 G_F}{\sqrt{2}} V_{ud} \, \bar{u}_L \gamma^\mu d_L \,\bar{e}_L \gamma_\mu
\nu_{eL} 
- \frac{m_{\beta \beta}}{2} \, {\nu}_{eL}^T C\nu_{eL} + {\rm H.c.},   
\label{eq:intro.1}
\end{equation}
where the first term denotes the strong interactions among quarks and gluons,
and the second term represents the weak interactions of up and down 
quarks and leptons, whose strength is determined by the Fermi constant $G_F$ and
the $V_{ud}$ element of the Cabibbo-Kobayashi-Maskawa (CKM) matrix.  
In order to calculate $0\nu\beta\beta$ transitions, the Lagrangian in 
Eq. \eqref{eq:intro.1} needs to be matched onto a theory of hadrons,
\begin{eqnarray}
\mathcal L  &=&  \mathcal L_{\rm strong}(\pi,N,\Delta)   
- \frac{4G_F}{\sqrt{2}} V_{ud} \, \mathcal J_\mu(\pi,N,\Delta) \, 
\bar{e}_L \gamma^\mu \nu_{eL}
\nonumber \\
& &- \frac{1}{2} m_{\beta \beta} \, {\nu}_{eL}^T C \nu_{eL}   
- \frac{4 G_F}{\sqrt{2}} V_{ud} \, \mathcal O(\pi,N,\Delta) \, 
\bar{e} \, \Gamma C \bar{\nu}_{e L }^T
- G_F^2 \, \mathcal O^\prime(\pi,N,\Delta) \, \bar{e} \, \Gamma C \bar e  
+ \textrm{H.c.}, 
\label{eq:intro.2}
\end{eqnarray}
where again the first and second terms represent the strong and 
weak interactions, respectively, while 
the operators in the second line
violate $L$ by two units. 
Here $\mathcal L_{\rm strong}$, $\mathcal J_\mu$, $\mathcal O$, and 
$\mathcal O^\prime$ are 
combinations of pion, nucleon, and Delta isobar fields. 
$\Gamma$ represents the possible Dirac structures of the leptonic bilinear,
and we are suppressing, for simplicity, possible Lorentz indices on 
$\Gamma$, $\mathcal O$, and $\mathcal O^\prime$.
For the short-range operators induced by light Majorana-neutrino exchange, 
$\Gamma = 1$.
For low-energy hadronic and nuclear processes, Eq. \eqref{eq:intro.2}
can be organized using chiral effective field theory ($\chi$EFT) 
\cite{Weinberg:1978kz,Weinberg:1990rz,Weinberg:1991um}
according to 
the scaling of operators in powers of 
the typical momentum in units of the breakdown scale,
\be
\epsilon_\chi= Q/\Lambda_\chi~, 
\qquad   Q \sim m_\pi,
\qquad \Lambda_\chi \sim 4\pi F_\pi,
\label{eq:scales}
\ee
where $m_\pi\simeq 140$ MeV and $F_\pi \simeq 92.2$ MeV are the pion mass 
and decay constant, respectively.
Given that the quark-level Lagrangian breaks $L$, all possible $|\Delta L|=2$ 
operators 
are generated at some order in $G_F$ and $\epsilon_\chi$. 

The Lagrangian in Eq. \eqref{eq:intro.2} can then be used to derive the 
$0\nu\beta\beta$ transition operator, often referred to as the 
``neutrino potential''. 
A leading contribution to the transition operator is induced by the exchange 
of neutrinos between two nucleons, mediated by the single-nucleon vector and 
axial currents. Defining the effective Hamiltonian as
\begin{eqnarray}
H_{\rm eff} =  H_{\rm strong} + 2 G_F^2 V_{ud}^2 \, m_{\beta \beta} \,  
\bar e_{L} C \bar e_{L}^T  \, \sum_{a\neq b} V^{(a,b)}_\nu  \,,
\label{eq:HV}
\end{eqnarray}
the long- and pion-range contributions to the neutrino potential between
two nucleons labeled 1 and 2 are given, at leading order (LO), by
\begin{eqnarray}
V^{(1,2)}_{\nu\, \rm L} = \frac{\tau^{(1)+} \tau^{(2)+} }{\spacevec{q}^2} 
\left[
1- \frac{2g_A^2}{3} \boldsigma^{(1)} \cdot \boldsigma^{(2)} 
\left(1 + \frac{m_\pi^4}{2(\spacevec q^2 + m_\pi^2)^2}\right)  
- \frac{g_A^2 }{3} S^{(12)} \left(1 - \frac{m_\pi^4}{(\spacevec q^2 + m_\pi^2)^2}
\right)
\right],
 \label{eq:Vnu0}
\end{eqnarray}
where $g_A = 1.27$ is the nucleon axial coupling, 
$\spacevec{q}$ is the transferred momentum,
$\tau^{+}$ is the isospin-raising Pauli matrix,
$\boldsigma$ are the Pauli spin matrices,
and
$S^{(12)} = \boldsigma^{(1)} \cdot \boldsigma^{(2)} - 3 \boldsigma^{(1)} \cdot 
\spacevec q \, \boldsigma^{(2)} \cdot \spacevec q/\spacevec q^2$ 
is the spin tensor operator.
We 
use the subscript L to indicate that Eq. \eqref{eq:Vnu0} is a 
long-range potential. In the rest of the paper, we will drop the nucleon labels
in $V_\nu$. Corrections from the momentum dependence of the nucleon vector and axial form 
factors, as well as from weak magnetism, are usually included in the neutrino 
potential, see for example Ref. \cite{Engel:2016xgb,Ejiri:2019ezh}. 
These corrections contribute at next-to-next-to-leading order (N$^{2}$LO) in 
$\chi$EFT.
At this order, there appear many other contributions, 
for instance from pion loops that dress the neutrino exchange and from 
processes involving new hadronic interactions with 
the associated parameters, or ``low-energy constants'' 
(LECs)~\cite{Cirigliano:2017tvr}.

The $0\nu\beta\beta$ transition operator in Eq. \eqref{eq:Vnu0} has a 
Coulomb-like behavior at large $|\spacevec q|$, which induces ultraviolet (UV) 
divergences in LNV scattering amplitudes, such as \nnpp, when both 
the two neutrons in the initial state and the two protons in the final state 
are in the $^1S_0$ channel. Our main goal in this work is to investigate these 
divergences and their consequence: the need for a new short-range 
$0\nu\beta\beta$ operator at LO \cite{Cirigliano:2018hja}.
This situation is analogous to charge-independence breaking (CIB) in 
nucleon-nucleon ($N\!N$) scattering, which receives long-range contributions 
from Coulomb-photon exchange and from the pion-mass splitting in pion exchange.
The consistency of the EFT requires then that, in addition to these long-range 
contributions, one should include 
also short-range CIB $N\!N$
operators. This observation is 
consistent with fits to $N\!N$ scattering data, which, for both chiral 
potentials \cite{Machleidt:2011zz,Piarulli:2014bda,Epelbaum:2014efa,Ekstrom:2015rta,Piarulli:2016vel,Reinert:2017usi} and phenomenological 
potentials such as Argonne $v_{18}$ \cite{Wiringa:1994wb}
and CD-Bonn \cite{Machleidt:2000ge}, require sizable short-range CIB. 
A short-range $0\nu\beta\beta$ operator also appears at LO 
\cite{Cirigliano:2017tvr} in a simpler EFT, pionless EFT ($\slashpi$EFT), 
where all hadronic
degrees of freedom other than the nucleon are integrated out.

In this paper we build upon Refs. \cite{Cirigliano:2017tvr,Cirigliano:2018hja} 
and study the $0\nu\beta\beta$ transition operator 
up to next-to-leading order (NLO) in $\chi$EFT. 
We begin in Sec.~\ref{sec:problem} by illustrating the problem of having 
just a long-range neutrino-exchange transition operator at LO, without going 
into any technical detail. 
The lepton-number-violating operators in the two EFTs, 
pionless EFT and chiral EFT, are constructed
in Sec. \ref{EFTs}.
(Operators with multiple quark-mass insertions are relegated to
App. \ref{mass}.)
In Sec. \ref{LNVLO} we study the scattering amplitude \nnpp at LO,
using different regulators and renormalization schemes.  
(Details about the $\overline{\rm MS}$ scheme are given in 
App. \ref{app:MSbar}.)
In all schemes, and 
independently of the inclusion of pions as dynamical degrees of freedom, the 
matrix element of the neutrino potential $V_{\nu\,\rm L}$ between 
$N\!N$ wavefunctions in the $^1S_0$ state shows logarithmic sensitivity to 
short-distance physics, which is cured by including an LO LNV counterterm.
In Sec. \ref{higherwaves} we study the transition operator in higher partial 
waves, such as $^3P_J$ and $^1D_2$. Weinberg's original power counting 
\cite{Weinberg:1990rz,Weinberg:1991um} 
leads to inconsistencies for $N\!N$ interactions in certain spin-triplet waves 
such as $^3P_0$ 
\cite{Nogga:2005hy,PavonValderrama:2005uj}, which require the promotion of 
contact operators to LO in these waves. Yet, 
we show that, after the strong 
interaction is properly renormalized, LNV matrix elements are well defined, 
and do not require further renormalization.
In Sec. \ref{LNV@NLO} we extend the analysis beyond LO.
We consider only the $^1S_0$ channel, which receives a new contribution from
strong interactions at NLO \cite{Long:2012ve}.
We again study 
the two EFTs, and show that no additional independent LNV counterterms are 
needed at 
this order.
In Sec. \ref{CIB} we discuss the relation between $0\nu\beta\beta$ and CIB 
in $N\!N$ scattering and argue that scattering data show evidence for a CIB 
contact interaction in the $^1S_0$ channel at LO in $\mathcal O(e^2)$,
where $e$ is the proton charge. 
While chiral and isospin symmetry allow us to derive relations between the CIB 
and LNV contact interactions, we show that unfortunately scattering data 
are not enough to unambiguously determine the latter. 
In Sec. \ref{pheno} we explore the implications of the LO short-range 
contribution to the neutrino potential on the $0\nu\beta\beta$ nuclear matrix 
elements in light nuclei, whose wavefunctions can be computed 
$\textit{ab initio}$, and we conclude in Sec. \ref{conclusion}.

\section{The problems of the leading-order neutrino potential}
\label{sec:problem}

\begin{figure}
\includegraphics[width=\textwidth]{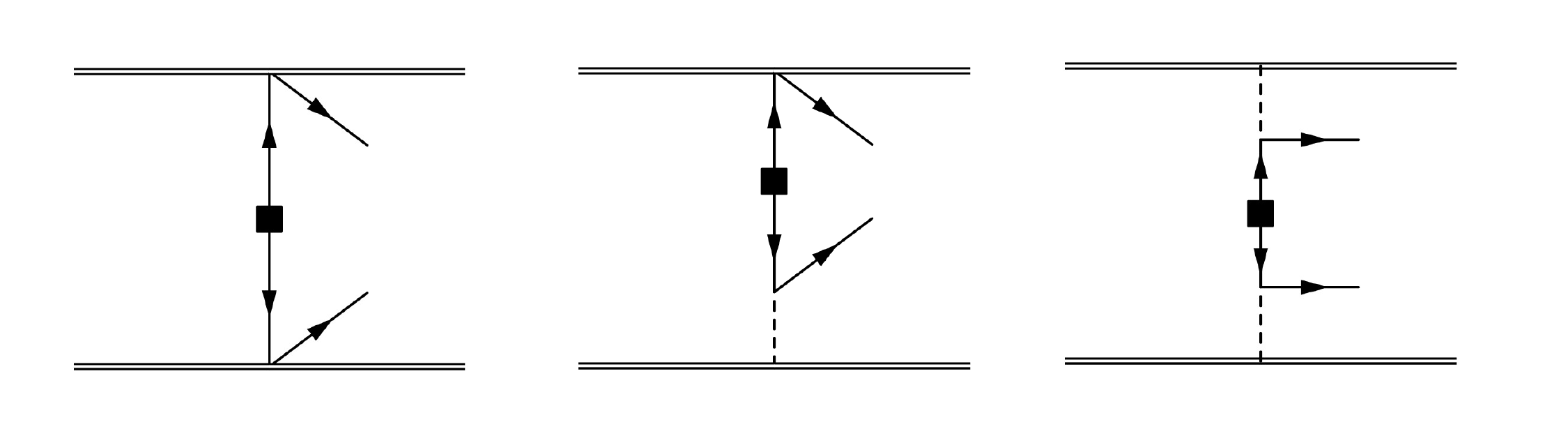}
\caption{Long-range contributions to the neutrino potential. Double and dashed 
lines denote, respectively, nucleons and pions. Single lines denote electrons 
and neutrinos, and a square an insertion of $m_{\beta\beta}$.}
\label{treelevel}
\end{figure}

The main theoretical problem is to connect the Majorana mass term in 
Eq.~\eqref{eq:intro.0} to the experimental $0\nu\beta\beta$
rate for various nuclear isotopes. 
Traditionally,  this connection is made by considering the exchange of a 
neutrino between two neutrons in a nucleus. An insertion of the neutrino 
Majorana mass on the neutrino propagator is required to account for the 
violation of lepton number by two units. At tree level, the process \nnpp 
can happen either via a direct neutrino exchange between neutrons or via 
intermediate pions which then decay into a neutrino and electron. 
The relevant diagrams are shown in Fig.~\ref{treelevel} and lead to the 
so-called neutrino transition operator or neutrino potential in 
Eq.~\eqref{eq:Vnu0}.

To obtain the $0\nu\beta\beta$ nuclear matrix element this transition operator 
is inserted between all pairs of neutrons in a nucleus using advanced nuclear 
many-body methods \cite{Engel:2016xgb}. Typically such calculations apply a 
closure approximation to take into account the effects of intermediate nuclear 
excited states, effectively shifting 
$\spacevec{q}^{-2}\rightarrow |\spacevec q|^{-1}(|\spacevec q|+\bar E)^{-1}$ 
in terms of the closure 
energy $\bar E = \mathcal O(\textrm{MeV})$. Such corrections can be shown 
to occur at higher order if the neutrino transition operator is derived with 
the 
$\chi$EFT power-counting rules \cite{Cirigliano:2017tvr}. We will not 
consider it here and set $\bar E=0$ for simplicity. Our concerns in this 
section involve large values of $|\spacevec q|$ and therefore 
are not affected by the closure approximation.

For a theoretical study of the neutrino transition operator it is convenient 
to perform a {\it Gedanken} experiment involving two neutrons in the ${}^1S_0$ 
state, the simplest nuclear system where the operator can act. Higher partial 
waves will be studied 
in a later section. The transition operator can be straightforwardly 
projected onto the ${}^1S_0 \rightarrow {}^1S_0$ channel, where it takes a 
simpler form
\begin{equation}
V_{\nu\, \rm L}^{^1S_0}(\spacevec q) = \frac{\tau^{(1)+} \tau^{(2)+} }{\spacevec{q}^2} 
\left[
1+ 2 g_A^2  + \frac{ g_A^2 m_\pi^4}{ (\spacevec q^2 + m_\pi^2)^2}\right]\,.
\label{eq:Vnu1}
\end{equation}
The transition operator is clearly Coulomb-like, scaling as $\spacevec q^{-2}$, 
and therefore typically expected to drop off sufficiently fast for large 
$|\spacevec q|$ (or short distances $|\spacevec r|$) to give rise to finite 
nuclear matrix elements.
As we demonstrate in this section, and study in significant detail below, this 
expectation turns out to be false.

Before going into a more detailed analysis we wish to explicitly demonstrate 
the problem here. We want to calculate the amplitude~\footnote{
The amplitude $\mathcal{A}_{fi}$ is related to the S-matrix element by  
$S_{fi} = i  (2 \pi)^4 \, \delta^{(4)}(p_f - p_i) \, {\mathcal A}_{fi}$.}
\begin{equation}
\mathcal A_\nu(E,E') = 
-\langle \Psi_{pp}(E') | V_{\nu\, \rm L}^{{}^1S_0} | \Psi_{nn}(E) \rangle ,
\label{Anu}
\end{equation}
for the process \nnpp where both initial $|\Psi_{nn}(E)\rangle$ and 
final $|\Psi_{pp}(E')\rangle$ states are in the ${}^1S_0$ channel. 
We denote by $E= \spacevec p^2/m_n$ and $E' = \spacevec p'^2/m_p$ the 
center-of-mass energies of the incoming neutrons and outgoing protons
of masses $m_n$ and $m_p$, respectively, and by
$\spacevec p$ and $\spacevec p'$ the corresponding relative momenta. 
Without loss of generality, in this section we assume the outgoing electrons 
to be at rest such that 
\begin{equation}
E' = E + 2(m_n-m_p - m_e), \qquad |\spacevec p^\prime |  = 
\sqrt{\spacevec p^2 + 2 m_N (m_n - m_p - m_e)},
\label{kin}
\end{equation}
with $m_e$ the electron mass and $2 m_N = m_n + m_p$.
When working at the kinematic point \eqref{kin}, we will drop, for simplicity, 
the second argument in $\mathcal A_\nu$.

The initial- and final-state wavefunctions are obtained by solving 
the Lippmann-Schwinger or Schr{\"o}dinger equation involving the strong $N\!N$
potential. Of the latter there exist many variants but 
most include the long-ranged one-pion exchange 
and short-range pieces,
which are described by the exchange of heavier mesons and/or by arbitrary
short-range functions (phenomenological potentials), 
or else by $N\!N$
contact interactions ($\chi$EFT potentials). Our arguments are best 
illustrated by use of the 
LO $\chi$EFT 
potential in the ${}^1S_0$ channel, which consists of only two terms,
\begin{equation}
V_{N\!N}^{{}^1S_0}  = 
C + V_{\pi}^{{}^1S_0}(\spacevec q),  
\label{1S0LOpot1}
\end{equation}
where 
\begin{equation}
V_{\pi}^{{}^1S_0}(\spacevec q) = - \frac{g_A^2 }{4F_\pi^2}
\frac{m_\pi^2}{\spacevec q^2 + m_\pi^2}
\label{1S0LOpot2}
\end{equation}
is the Yukawa potential written in terms of the transferred 
momentum $\spacevec q = \spacevec p - \spacevec p^\prime$,
and $
C$ is a contact interaction that
accounts for short-range physics from pion exchange and other QCD effects.
The latter is needed for renormalization and 
to generate the observed, shallow $^1S_0$ virtual state. 
It is expected at LO \cite{Weinberg:1990rz,Weinberg:1991um} 
on the basis of the naive dimensional analysis (NDA)
\cite{Manohar:1983md}, and
discussed in detail in Sec. \ref{chiEFT}.

\begin{figure}
\includegraphics[width=.45\textwidth]{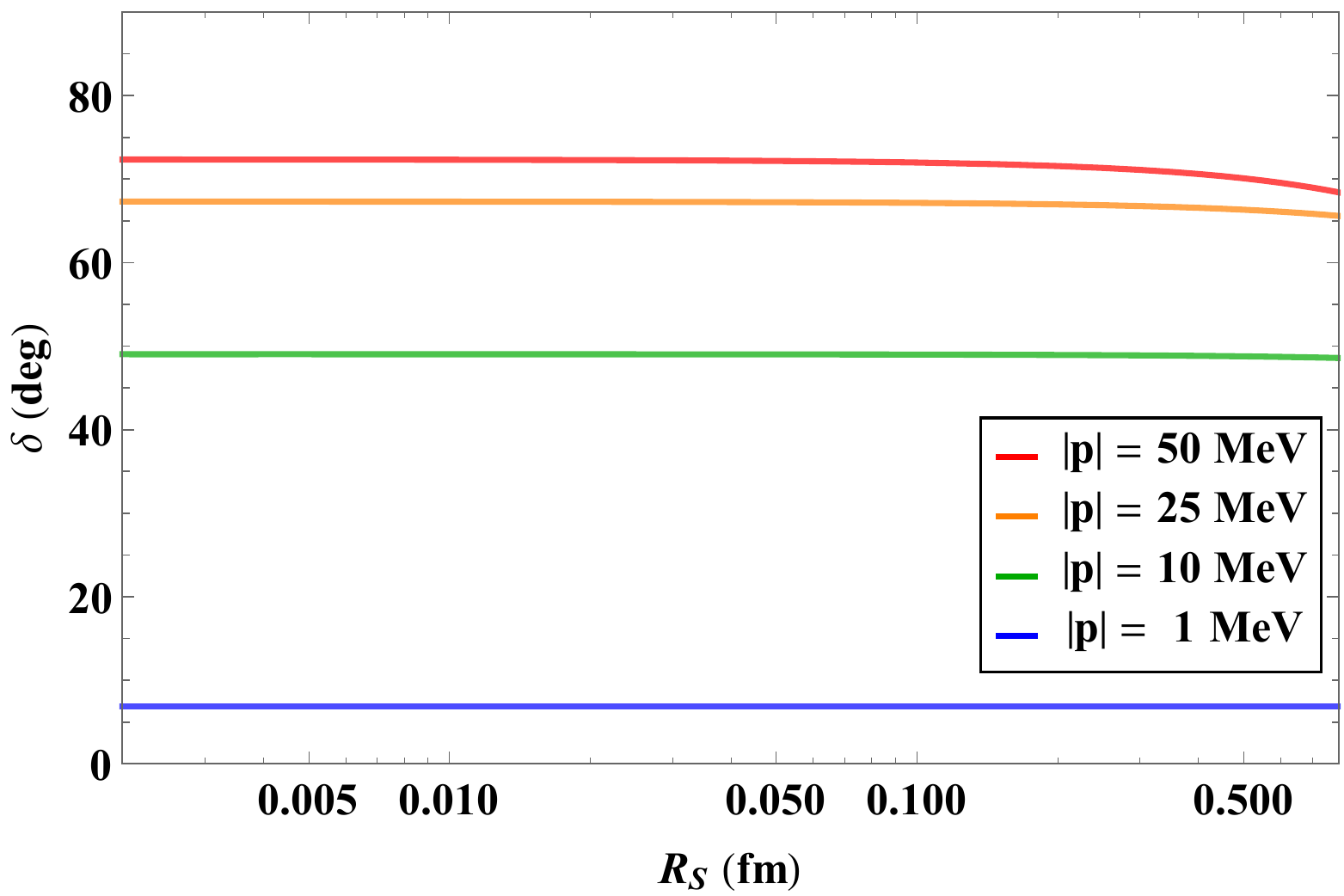}
\hfill
\includegraphics[width=.45\textwidth]{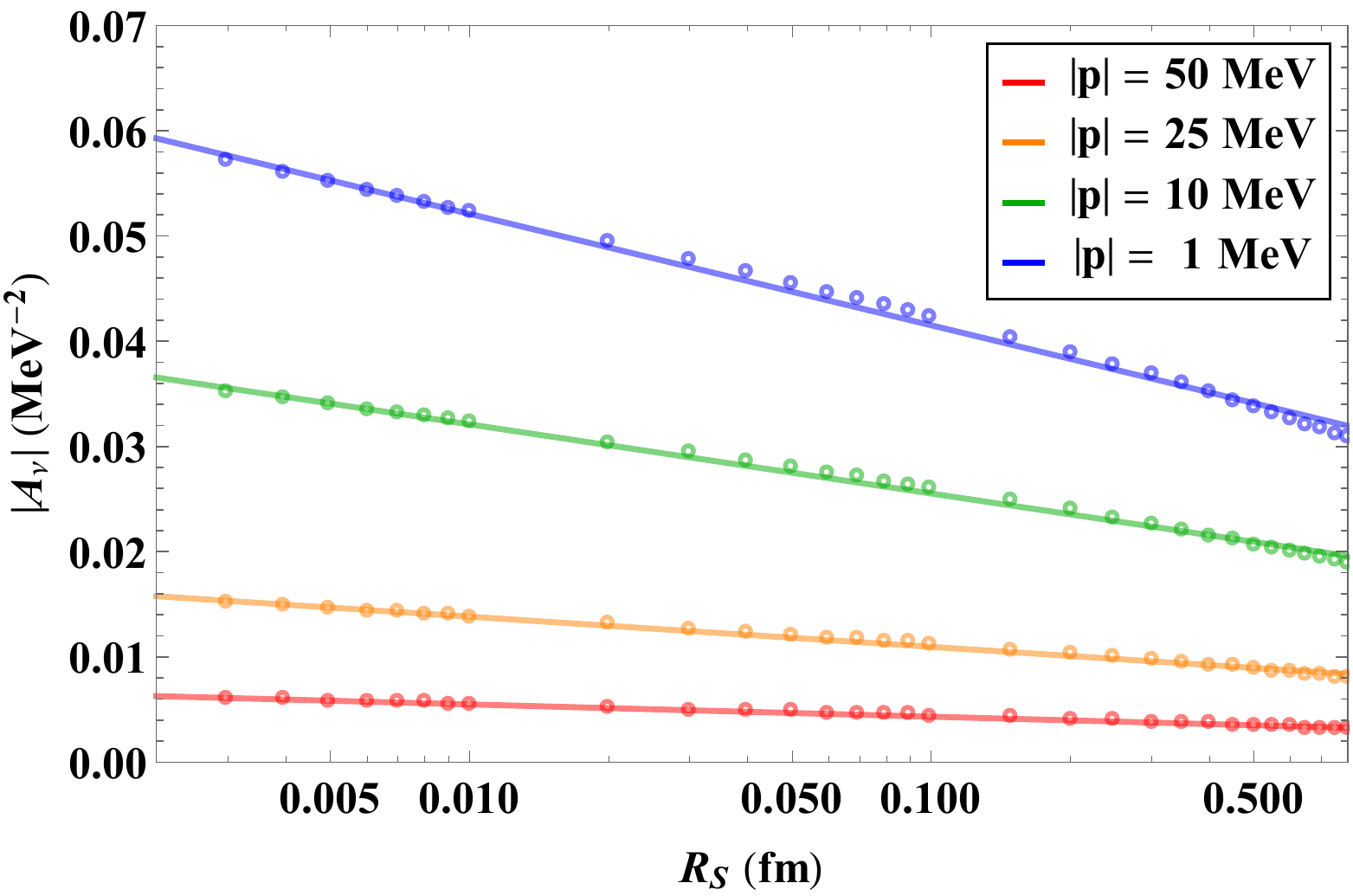}	
\caption{The left panel shows the phase shifts in the $^1S_0$ channel for 
several momenta $|\spacevec p|$
as functions of the cutoff, $R_S$. The right panel shows the 
$0\nu\bt\bt$ amplitude of Eq.\ \eqref{Anu}.
The dots result from explicitly evaluating 
Eq. \eqref{Anu}, while the straight lines are due to a fit of the form 
$\mathcal A_\nu = a+b\ln R_S$. 
}
\label{fig:sec2plots}
\end{figure}

To obtain the $N\!N$
wavefunctions, $V_{N\!N}^{^1S_0}$ must be iterated to all orders, which we do by 
numerically solving the Lippmann-Schwinger or Schr{\"o}dinger equation. 
Because of the short-range pieces in the potential, the involved integrals are 
in general divergent and require regularization. In this section we will use a 
coordinate-space cutoff $R_S$ but other regulators will be discussed 
throughout this paper. For each choice of $R_S$, the short-range $N\!N$
LEC $
C(R_S)$ is fitted to the observed ${}^1S_0$ scattering length.
Since $
C(R_S)$ is not an observable,
its cutoff dependence 
is of no concern. We can then 
calculate the ${}^1S_0$ phase shifts at other energies as a function of $R_S$,
and observe that these observables have well-defined values for 
small $R_S$: the cutoff dependence cannot be seen in the left panel of 
Fig.~\ref{fig:sec2plots}
for $R_S 
\simle 0.2$ fm.
That is, the $N\!N$
interaction is properly renormalized.
These results are in agreement with Ref. \cite{Beane:2001bc}.

Having obtained the wavefunctions $\Psi_{nn}$ and $\Psi_{pp}$, all that is left is
to evaluate the amplitude in Eq.~\eqref{Anu}. This expression only depends on 
the values of the LECs $g_A$ and $
C(R_S)$, and the effective neutrino 
mass $m_{\beta\beta}$. $
C(R_S)$ has been linked to the $N\!N$
scattering length and is thus known for each value of $R_S$. As 
$|\mathcal A_\nu(E)|^2$ is an observable (although experimentally it will be 
hard to measure!) it should not depend on the value of the regulator.
The value of $\mathcal A_\nu(E)$ for various energies as a function of the 
regulator $R_S$ is shown in the right panel of  Fig.~\ref{fig:sec2plots}. 
The amplitude is clearly not regulator independent and the dependence 
at small $R_S$ can be fitted with a 
$\ln R_S$ function. 
We will derive the form of this $R_S$ dependence in later sections.
At larger $R_S$, power corrections in $R_S$ induce a 
deviation from this simple behavior.

The consequences of the dependence of $\mathcal A_\nu(E)$ on $R_S$ are severe. 
Such dependence implies that $m_{\beta\beta}$ cannot be directly obtained from a 
measurement of the \nnpp transition (or alternatively, $m_{\beta\beta}$ cannot be 
limited from an experimental upper bound on the transition rate) as the matrix 
element linking the measurement to $m_{\beta\beta}$ depends on the unphysical 
parameter $R_S$. While we have studied a very simple state of just two 
nucleons, the arguments and conclusions do not depend on it: the same 
$R_S$ dependence occurs in nuclear transitions as long as the corresponding
nuclei are described in 
$\chi$EFT. In practice it might be 
difficult to observe this regulator dependence due to the nature of many-body 
calculations where regulators are often fixed or can only be varied in a small 
range. In this light, {\it ab initio} calculations on lighter nuclei can 
provide an important intermediate step.

Of course, in an EFT setting an observable that depends on the regulator simply
indicates that there exists a counterterm with a corresponding LEC that absorbs
the divergence. In the context of $0\nu\beta\beta$ the counterterm is provided 
by a short-range \nnpp interaction that adds a term to the neutrino transition 
operator 
\begin{equation}
V_{\nu\,{\rm L}}
\rightarrow 
V_{\nu\, {\rm L}}
- 2 g^{\rm NN}_\nu \, \tau^{(1) +} \tau^{(2) +}
\equiv V_{\nu\, \rm L} + V_{\nu\, \rm S} ,
\label{LOnupot}
\end{equation}
where $g^{\rm NN}_\nu$ is the corresponding LEC. In 
Weinberg's power counting \cite{Weinberg:1990rz,Weinberg:1991um} 
such an interaction 
appears at N$^2$LO.
Renormalization, however, requires 
it at LO.
This is in agreement with what was already anticipated on general grounds in 
Ref.~\cite{Valderrama:2014vra} for other weak currents acting in the $^1S_0$ 
channel. The LEC $g^{\rm NN}_\nu(R_S)$ 
depends on the regulator $R_S$ in such a way
as to make $\mathcal A_\nu(E)$ regulator independent.  

We stress that $g^{\rm NN}_\nu$ corresponds to a genuine new contribution due to 
high-momentum neutrino exchange involving inter-nucleon distances 
$R \simle \Lambda_\chi^{-1}$. 
It is an intrinsically two-nucleon effect beyond that of the radii of
weak form factors, which also lead to a short-range neutrino
potential but can be determined in principle from one-nucleon 
processes---one-nucleon form-factor radii are N$^2$LO contributions
unaffected by two-nucleon physics.
In contrast, two-nucleon weak currents, also a higher-order effect, generate
a neutrino potential involving three nucleons\footnote{Two-nucleon weak currents also induce loop corrections to two-body $0\nu\beta\beta$ transition operators \cite{Wang:2018htk}. These corrections are UV divergent, and the divergence is absorbed by N$^3$LO corrections to $g_\nu^{\rm NN}$ \cite{Wang:2018htk}.}.
Despite being a two-nucleon quantity,
$g^{\rm NN}_\nu$ cannot be described by a modification of the 
$N\!N$
potential itself which, in this example, is already correctly renormalized. 
Likewise, $g^{\rm NN}_\nu$ is not part of the so-called
``short-range correlations''~\cite{Miller:1975hu,Haxton:1985am,Simkovic:2009pp,Engel:2011ss,Benhar:2014cka},
if the latter are intended to describe nucleon correlations missed 
in approaches built on independent-particle states. 
As we have seen, $g^{\rm NN}_\nu$ is needed even when 
we use fully correlated wavefunctions, which are exact solutions of 
the Schr\"odinger equation. In {\it ab initio} calculations,
such as those described in Sec. \ref{pheno}, 
the input should be the $^1S_0$ neutrino potential \eqref{LOnupot}, not 
Eq. \eqref{eq:Vnu1}.
In many-body calculations where an {\it ab initio} approach is not 
possible, the $^1S_0$ neutrino potential should still be Eq. \eqref{LOnupot},
on top of the correlations necessary to 
produce good wavefunctions starting from an independent-nucleon basis.
Now, short-range correlations can be viewed as a modification of
the neutrino potential---see e.g. the discussion in 
Ref. \cite{Benhar:2014cka}. 
Missing correlations at distances $R\simge \Lambda_\chi^{-1}$
can thus be mocked up by a $g^{\rm NN}_\nu(R)$.
However, the converse is in principle not true: a 
$g^{\rm NN}_\nu(R_S\simle \Lambda_\chi^{-1})$ cannot be replaced by 
correlations at distances where the nucleon can be considered a well-defined
entity.
The situation is analogous to $\beta$ decay,
where two-nucleon weak currents and short-range correlations
are both present even if each can be viewed as an ``in-medium quenching''
of $g_A$---see Refs. \cite{Pastore:2017uwc,Gysbers:2019uyb} 
for recent discussions.

$g^{\rm NN}_\nu$ accounts for neutrino exchange 
between quarks taking place at the characteristic QCD scale,
which is needed for the very definition of the neutrino potential
between nucleons and requires input from QCD. 
Indeed, while it is fairly easy to obtain part of $g^{\rm NN}_\nu (R_S)$ by 
demanding that $\mathcal A_\nu(E)$ 
be regulator independent, the finite 
contribution of $g^{\rm NN}_\nu$ to the amplitude cannot be so obtained.
The only way to get the total value of $g^{\rm NN}_\nu(R_S)$ is 
to fit to data---similar to how we obtained $
C(R_S)$ by fitting 
the $N\!N$
scattering length.
Fitting to  
LNV data is for obvious reasons impossible at present, and even if there were 
data it would be undesirable: we want to use a nonzero $0\nu\beta\beta$ rate 
to infer the value of the neutrino Majorana mass. 
Fortunately there are ways out. We will argue in Sec.~\ref{CIB} that the 
problems associated to light Majorana-neutrino exchange also affect another 
well-known long-range potential, the Coulomb potential. In that case, the 
corresponding counterterm can be fitted to data on electromagnetic 
isospin-violating processes. Chiral symmetry relates the electromagnetic 
counterterms to $g^{\rm NN}_\nu(R_S)$, but at present this is insufficient to 
fully determine it. Nevertheless, this approach explicitly demonstrates the 
necessity of including $g^{\rm NN}_\nu$ at LO.
An alternative is to match 
this counterterm to 
results from a direct nonperturbative QCD calculation, something which is 
imaginable with lattice QCD (LQCD) methods, both for light Majorana exchange
\cite{Shanahan:2017bgi,Tiburzi:2017iux,Feng:2018pdq,Cirigliano:2019jig}
and TeV-scale LNV mechanisms \cite{Nicholson:2018mwc,Monge-Camacho:2019nby}.

\section{Effective field theories}
\label{EFTs}

In this section we describe the EFTs that we employ to discuss LNV in the 
two-nucleon sector. In Sec.~\ref{piless} we introduce 
pionless EFT, an EFT 
without explicit pionic degrees of freedom that allows us to 
derive more explicit expressions for the \nnpp amplitude than in 
chiral EFT.
In Sec.~\ref{chiEFT} we restore pions, and discuss some of the problems 
associated with them.
Finally, in Sec. ~\ref{LNVLag} we describe long- and short-range LNV 
operators at leading 
orders. 

\subsection{Pionless EFT}
\label{piless}

Few-body systems characterized by momentum scales $p$ much smaller than the 
pion mass can be described in pionless EFT ($\slashpi$EFT) 
\cite{Bedaque:1997qi,Bedaque:1998mb,Chen:1999tn,Kong:1999sf,
Bedaque:1999ve}---for a review, see Ref. \cite{Bedaque:2002mn}.
$\slashpi$EFT has been shown to converge very well in the 
two- \cite{Chen:1999tn,Kong:1999sf} and 
three- \cite{Vanasse:2013sda,Konig:2015aka}
nucleon sectors, and works 
within LO error bars for nuclei as large as  $^{40}$Ca 
\cite{Platter:2004zs,Stetcu:2006ey,Contessi:2017rww,Bansal:2017pwn}.    
While it is 
unclear whether its regime of validity 
extends to experimentally relevant $0\nu\beta\beta$ emitters, LNV amplitudes in 
$\slashpi$EFT have a simple form, which allows 
analytical insight on the structure of
the $0\nu\beta\beta$ transition operator \cite{Cirigliano:2017tvr}. 
Lowest-order interactions contribute only to $N\!N$ $S$ waves.
Since the $^1S_0$ channel is the most important for $0\nu\beta\beta$, 
we will see that many conclusions 
drawn in $\slashpi$EFT continue to hold in $\chi$EFT.
Furthermore,  $\slashpi$EFT will be useful in light of a possible matching to 
LQCD calculations of $0\nu\beta\beta$ matrix elements performed at 
heavy pion masses.
A similar matching between LQCD and $\slashpi$EFT for strong and electroweak 
processes has been carried out in Refs.~\cite{Barnea:2013uqa,Beane:2015yha,Kirscher:2015yda,Savage:2016kon,Contessi:2017rww,Shanahan:2017bgi,Tiburzi:2017iux,Bansal:2017pwn}.

The
strong-interaction Lagrangian in $\slashpi$EFT is made out of all interactions 
among nucleons---the relevant low-energy degrees of freedom in this 
case---constrained only by the symmetries of QCD.
While an infinite set of such interactions exist, they can be ordered in a 
power-counting scheme.
At LO in the two-nucleon $^1S_0$ channel, 
\begin{equation}
\mathcal L^{(0)}_{\slashpi} = 
\bar N \left(i \partial_0- \frac{\boldsymbol{\nabla}^2}{2 m_N} \right)N 
- C
\left(N^T \isovec{P}_{^1S_0} N \right)^\dagger \cdot  
\left(N^T {\isovec P}_{^1S_0} N\right),
\label{CTchiral}
\end{equation}
where the nucleon isospin doublet is represented by $N= (p\; n)^T$ 
and the projector is
\begin{equation}
P^a_{^1S_0} = \frac{1}{\sqrt{8}} \tau_2 \tau^a \sigma_2. 
\label{proj1S0}
\end{equation}
Here $\tau^a$ are the Pauli matrices in isospin space, where a vector
is denoted by an arrow. 
The four-nucleon interaction scales as $C = \mathcal O(4\pi/(m_N \aleph))$ 
where $\aleph$ is a 
fine-tuned scale much smaller 
than the breakdown scale $\Lambda_{\slashpi}\sim m_\pi$,
in order to produce a low-energy pole in the $N\!N$ $^1S_0$ amplitude. 
At momenta $Q\sim \aleph$, the LO amplitude consists of a resummation
of $C$ interactions, and 
coincides with that of the effective-range expansion
truncated at the level of the scattering length.
$C$ can thus be determined from matching to the $N\!N$ $^1S_0$
scattering length $a = -23.714$ fm or to the position of the virtual state
in the complex momentum plane, which agree within the 
relative LO error $\sim r_0/a$ set by
the effective range $r_0 = 2.7$ fm.
The latter arises from the NLO Lagrangian
\begin{equation}
\mathcal L^{(1)}_{\slashpi} = +\frac{C_2}{8} 
\left( N^T  \isovec{P}_{^1S_0} \overleftrightarrow{\boldsymbol \nabla}^2 N \right)   \cdot
\left( N^T \isovec{P}_{^1S_0} N\right)^{\dagger} + \textrm{H.c.}\,,
\label{eq:C2}
\end{equation}
where $\overleftrightarrow{\boldsymbol \nabla} = \overrightarrow{\boldsymbol \nabla}- \overleftarrow{\boldsymbol \nabla}$
and $C_2$ scales as $C_2=\mathcal O(4\pi/(m_N \aleph^2 \Lambda_{\slashpi}))$.

To ensure regulator independence of the scattering amplitude, 
the LECs $C$ and $C_2$ must obey renormalization-group equations (RGEs).
For example, in the power divergence subtraction (PDS) 
scheme \cite{Kaplan:1998tg,Kaplan:1998we},
\begin{equation}
\frac{d}{d\ln \mu}
C = \frac{\mu m_N}{4\pi} C^2, \qquad 
\frac{d}{d\ln \mu}
\left(\frac{C_2}{C^2}\right) = 0,
\end{equation}
where $\mu$ is the renormalization scale.
Solving the RGEs determines
\begin{equation}
C_{} = \frac{4\pi}{m_N} \frac{1}{1/a - \mu}\,, 
\qquad 
C_2 = \frac{2 \pi}{m_N} \frac{r_0}{\left(1/a - \mu\right)^2}\,.
\label{C0C2}
\end{equation}
Similar RGEs hold in schemes that employ 
momentum cutoffs with the replacement 
$\mu \rightarrow c \Lambda$, where $c$ a scheme-dependent constant.
With such regulators one sees explicitly that the amplitude calculated
from the Lagrangian \eqref{CTchiral} contains
a residual cutoff dependence, which contributes $\propto k^2/\Lambda$
to the effective-range expansion, with $k$ the on-shell 
momentum. This indicates that in the absence of further fine tuning 
$C_2$ enters at NLO and $r_0=\mathcal O(\Lambda_{\slashpi}^{-1})$,
consistent with its numerical value.
Renormalization beyond LO can only be achieved if subleading 
corrections such as
$C_2$ are treated in perturbation theory \cite{Beane:1997pk,vanKolck:1998bw}.

At higher orders 
a four-derivative operator appears  
whose coefficient, $C_4$, is fixed at N$^2$LO
and determined by the shape parameter at N$^3$LO. 
Except for interactions in the $^3S_1$ channel analogous to those 
above, all other two-nucleon interactions contribute at N$^2$LO or higher,
including interactions in other isospin-triplet channels relevant to 
$0\nu\beta\beta$ such as $^3P_0$.
The power counting is reviewed in Ref. \cite{Bedaque:2002mn}.
While our work focuses on the two-body sector, it is interesting that 
three-body forces appear already at LO in 
$\slashpi$EFT \cite{Bedaque:1998kg,Bedaque:1998km,Bedaque:1999ve}.
To our knowledge, the possible implications for the $0\nu\beta\beta$ 
transition operators have not been studied.

\subsection{Chiral EFT}
\label{chiEFT}

The low-energy EFT of QCD that incorporates pions explicitly 
is often called chiral EFT, 
a generalization of chiral perturbation
theory ($\chi$PT) \cite{Weinberg:1978kz} to systems with more than one 
nucleon \cite{Weinberg:1990rz,Weinberg:1991um}.
Pions play an important role as they emerge as pseudo-Goldstone bosons of the 
spontaneously broken, approximate chiral symmetry of QCD. This symmetry would 
be exact were it not for the small quark masses and electromagnetic charges. 
Contrary to $\slashpi$EFT, in whose regime it is badly broken, 
approximate chiral symmetry is implemented in the $\chi$EFT Lagrangian:
all interactions either 
conserve chiral symmetry or break 
it in the same way as the chiral-breaking sources at the quark level. 
In addition to the nucleon contact interactions of $\slashpi$EFT, there are also
pion interactions with nucleons and pions themselves.
Because the Delta isobar is heavier than the nucleon by only about 300 MeV,
it should also be included in order not to limit the range of validity of
the theory too stringently \cite{Pandharipande:2005sx}.
However, Delta isobars appear only in loops in the nuclear potential
at orders higher than our discussion of renormalization below 
\cite{Ordonez:1993tn,Ordonez:1995rz}, and will not be explicitly displayed.
``Chiral potentials'' obtained from $\chi$EFT---for a review, see Ref. 
\cite{Machleidt:2011zz}---have been extensively used as input to modern 
{\it ab initio} methods. 
It is hoped that $\chi$EFT converges for the nuclei employed in searches
for $0\nu\beta\beta$.

For processes with at most one nucleon, $\chi$EFT can be treated in 
perturbation theory ($\chi$PT) in a systematic 
expansion in the small 
ratio $\epsilon_\chi$, Eq. \eqref{eq:scales} \cite{Weinberg:1978kz}.
However, as in $\slashpi$EFT, the existence of nuclei
requires a resummation
of a class of diagrams. 
In Weinberg's original papers \cite{Weinberg:1990rz,Weinberg:1991um},
it was recognized that the nonperturbative nature of $N\!N$
interactions is 
due to an infrared enhancement in the propagation of 
nucleons, leading to the presence of the large nucleon mass $m_N$ in
the numerator of integrals
arising from loops with only nucleons in intermediate states---a pinch 
singularity
when $m_N\to \infty$.
Weinberg then proposed to calculate nuclear 
amplitudes in two steps. 
In the first step one calculates a nuclear potential from diagrams that do not 
contain pinch singularities. Such diagrams are expected to follow the standard 
$\chi$PT power-counting rules, as long as nucleon contact interactions obey 
NDA \cite{Manohar:1983md}. In a second step, the 
truncated nuclear potential is iterated 
to all orders by solving the Schr{\"o}dinger equation. 
Most work on nuclear physics has followed this prescription. 

While the potential in $\slashpi$EFT consists of only contact interactions---all
loops contain pinch singularities---in $\chi$EFT the potential
contains also pion exchange.
In Weinberg's original prescription \cite{Weinberg:1990rz,Weinberg:1991um}, 
static one-pion exchange (OPE) appears at LO in the potential,
\begin{equation}
V_{\pi}(\spacevec q) = -\frac{g_A^2}{12 F_\pi^2} \, 
\boldtau^{(1)} \cdot \boldtau^{(2)}
\left[\left(1-\frac{m_\pi^2}{\spacevec q^{\,2}+m_\pi^2}\right)
\boldsigma^{(1)} \cdot \boldsigma^{(2)}
-\frac{\spacevec q^2}{\spacevec q^{\,2}+m_\pi^2}  S^{(12)}
\right] ,
\label{strong}
\end{equation}
which is treated nonperturbatively together with 
contact interactions that arise from dynamics of shorter range than 
$m_\pi^{-1}$.
The size of the contact LECs was assumed to be given by NDA, so at LO
only two non-derivative, chiral-symmetric contact interactions were supposed 
to appear, one in $^1S_0$, the other in $^3S_1$.
The question of renormalizability of the $N\!N$ amplitude was left unanswered.
Initial numerical evidence \cite{Ordonez:1993tn,Ordonez:1995rz} suggested
no problems. Unfortunately,
it has been known from the mid-90s that Weinberg's prescription leads
to amplitudes that depend sensitively on the regularization procedure.
Two types of problems have been identified:

\begin{enumerate}

\item In the $^1S_0$ channel, the LO potential reduces to 
Eqs. \eqref{1S0LOpot1} and
\eqref{1S0LOpot2}. According to NDA, $
C$ consists of a contribution
from pions plus the undetermined LEC $C_0$
of a chiral-symmetric contact interaction.
The contact interaction is singular and must be renormalized.
As we have seen in Sec. \ref{sec:problem}, allowing $
C$ to be
cutoff dependent is sufficient for renormalization at a fixed pion mass. 
However, 
Ref. \cite{Kaplan:1996xu} showed that the 
cutoff dependence contains an $m_\pi^2$-dependent logarithmic divergence
that originates in
the interference between the contact and Yukawa interactions.
The presence of additional chiral-symmetry-breaking interactions is
thus required for renormalization, even though such interactions appear at 
higher orders in Weinberg's power counting. 
In a 
cutoff scheme, an operator with LEC
$D_2m_\pi^2$ is sufficient to produce an $N\!N$ amplitude that approaches a
constant as the cutoff is increased \cite{Beane:2001bc}, so that
\begin{equation}
C =  C_0 + D_2 \, m_\pi^2 + \frac{g_A^2}{4 F^2_\pi}.
\label{somedefs}
\end{equation}

\item In each attractive triplet wave where OPE is iterated,
its $-r^{-3}$ singularity in coordinate
space requires a chiral-symmetric contact interaction for renormalization
\cite{Nogga:2005hy,PavonValderrama:2005uj}.
While in the $^3S_1$-$^3D_1$ coupled channels such an interaction is
already predicted by NDA, in other waves it only appears at higher orders 
in Weinberg's power counting.
It is at present unclear in which waves OPE must be iterated.
A semi-analytical argument \cite{Birse:2005um} implies that $D$ waves and 
higher can be treated perturbatively, while Refs. \cite{Wu:2018lai,Kaplan:2019znu} suggest 
even $^3P_2$-$^3F_2$ is perturbative. 
Unfortunately, treating pion exchange perturbatively 
\cite{Kaplan:1998tg,Kaplan:1998we} does not work in the low triplet waves
\cite{Fleming:1999ee,Kaplan:2019znu}.
\end{enumerate}
\noindent
In summary, for the isospin-triplet channels relevant for $0\nu\beta\beta$,
the LO strong-interaction Lagrangian is
\begin{eqnarray}
\mathcal L^{(0)}_{\chi} &=& 
\frac{1}{2} \partial_\mu \isovec \pi \cdot \partial^\mu \isovec \pi  
-\frac{1}{2}\mpi^2\isovec \pi^2 
+ \bar N \left(i \partial_0- \frac{\boldsymbol{\nabla}^2}{2 m_N} \right)N 
-\frac{g_A}{2F_\pi}\boldsymbol{\nabla}\isovec \pi \cdot \Nb \boldtau\boldsigma N
\nonumber\\ 
&& - \left( C_0 + m_\pi^2 D_2 \right) 
\left(N^T \isovec P_{^1S_0} N \right)^\dagger \cdot 
\left(N^T \isovec P_{^1S_0} N\right)
- C_{^3P_0} \left(N^T \isovec P_{^3P_0}  N \right)^\dagger \cdot 
\left(N^T \isovec P_{^3P_0} N\right)
\nonumber\\ 
&& + \dots,
\label{LOchiralLAG} 
\end{eqnarray}
where $\isovec \pi$ stands for the pion isospin triplet,
the projector $P_{^1S_0}$ is defined in Eq. \eqref{proj1S0},
the projector on the $^3P_0$ channel
is 
\begin{equation}
\isovec P_{^3P_0} = - \frac{i}{\sqrt{8}}  \sigma_2 \boldsigma \cdot  
\overleftrightarrow{\boldsymbol \nabla}  \tau_2 \boldtau, 
\end{equation}
and the dots denote terms with additional pion fields that are not relevant 
for our purposes. (Note, however, that these terms differentiate
between $m_\pi^2 D_2$ and $C_0$, so in higher orders or in processes with 
external pions these LECs do not always appear in the combination 
$C_0 + m_\pi^2 D_2$.)
In $\chi$PT, as well as in the nuclear potential,
one can demote the nucleon recoil term to NLO.
The remaining terms on the first line of Eq. \eqref{LOchiralLAG} give rise to the static OPE 
potential \eqref{strong}.
The LECs $C_0$ and $m_\pi^2 D_2$ contribute to Eq. \eqref{somedefs},
while $C_{^3P_0}$ ensures the renormalization of the $^3P_0$ wave at LO.
The scaling $C_0 =\mathcal O(4\pi/(m_N Q))$ is the same as in NDA and 
$\slashpi$EFT, but the LECs $D_2\sim C_{^3P_0}=\mathcal O(4\pi/(m_N Q^3))$
are enhanced with respect to NDA by $\epsilon_\chi^{-2}$.

By consideration of the corrections in the $^1S_0$ channel similar to that
done in the previous section, one finds \cite{Long:2012ve}
that a nonvanishing NLO correction---that is, one order down in the expansion
parameter $\epsilon_\chi$---
exists despite being expected only two orders down the expansion in
Weinberg's power counting. The NLO strong-interaction Lagrangian 
can thus be written just as in $\slashpi$EFT as
\begin{equation}
\mathcal L^{(1)}_{\chi} =
+ \frac{1}{8} C_2 
\left( N^T \isovec P_{^1S_0} \overleftrightarrow{\boldsymbol{\nabla}}^2 N \right)\cdot  
\left( N^T \isovec P_{^1S_0} N\right)^{\dagger} + \textrm{H.c.},
\label{NLOchiralLAG} 
\end{equation}
with $C_2=\mathcal O(4\pi/(m_N Q^2 \Lambda_\chi))$. 
It leads to an NLO correction to the LO potential in Eq. \eqref{1S0LOpot2}, given by
\begin{equation}
V_{N\!N}^{^1S_0 (1)}(\spacevec p, \spacevec p^\prime) = C^{(1)} 
+ C_2 \, \frac{\spacevec p^2 + \spacevec p^{\prime\, 2}}{2},
\label{eq:1}
\end{equation}
where $C^{(1)}=\mathcal O(4\pi/(m_N\Lambda_\chi))$ denotes a subleading component 
of the nonderivative operator defined in Eq.~\eqref{LOchiralLAG}. 
Again as in $\slashpi$EFT, this potential and other subleading interactions can
only be renormalized in perturbation theory, in stark contrast to Weinberg's
prescription.

While the renormalization issues with Weinberg's prescription have been
extensively documented, there has been relatively little work done
in applying properly renormalized $\chi$EFT to nuclear physics.
It has in fact been argued that ``chiral potentials'' derived and treated
according to Weinberg's prescription give rise to better phenomenology,
as long as the cutoff is chosen somewhat, but not too far, below the
breakdown scale \cite{Epelbaum:2006pt}.
The drawbacks of Weinberg's prescription have limited impact on our
conclusions below about the renormalization of the $0\nu\beta\beta$ amplitude.
Our main results concern $^1S_0$ transitions at LO, as discussed 
in Sec. \ref{LNVLO}. 
The enhancement of $D_2$ has implications on the chiral properties
of the contact in Eq. \eqref{LOnupot}, 
but does not affect its existence in the first place.
In Sec. \ref{higherwaves} we show that the presence of counterterms in
attractive triplet channels has no additional implications for the 
renormalization of the $0\nu\beta\beta$ amplitude.
The effects of NLO corrections will be considered in Sec. \ref{LNV@NLO}.

\subsection{Lepton-number-violating operators}
\label{LNVLag}

The quark-level Lagrangian that is relevant to $0\nu\beta\beta$ transitions 
induced by a light Majorana neutrino is given in Eq. \eqref{eq:intro.1},
and its matching onto $\chi$EFT is sketched in Eq. \eqref{eq:intro.2}.
The first ingredient required to derive the neutrino potential is the weak 
current $\mathcal J^\mu(\pi,N)$. $\mathcal J^\mu$ 
has vector and axial components, and it is dominated by one-body contributions.
Writing  
\begin{equation}
\mathcal J^\mu = \frac{1}{2}\bar N \tau^+ \left[J_V^\mu+ J_A^\mu \right] N + \ldots,
\end{equation}
where $\ldots$ denote two- and higher-body contributions, 
the expressions of $J_V$ and $J_A$ through NLO in the chiral expansion are 
\bea \label{eq:currents}
J^\mu_V  &=& g_V(\spacevec q^2) 
\left( v^\mu + \frac{p^\mu + p^{\prime \mu}}{2m_N} \right)
+ i g_M(\spacevec q^2)\,\epsilon^{\mu \nu \alpha \beta}\,
\frac{v_\alpha S_\beta q_\nu}{m_N}\,, 
\nn\\
J^\mu_A  &=& - 2g_A(\spacevec q^2)  
\left(S^\mu  - \frac{S \cdot (p + p^\prime)}{2 m_N}\, v^\mu
+\frac{S \cdot q}{\spacevec q^2 + m_\pi^2}\,  q^\mu \right)\,.
\eea
Here $p$ and $p'$ stand for the momentum of the incoming neutron and outgoing 
proton, respectively, $q^\mu=(q^0,\, \spacevec q) = p^\mu-p^{\prime \mu}$,
and $v^\mu$ and $S^\mu$ are respectively the nucleon velocity and spin
($v^\mu = (1,\,\spacevec 0)$ and $S^\mu = (0,\,\boldsigma/2)$ in the nucleon 
rest frame). 
Furthermore, $ \epsilon^{\mu \nu \alpha \beta}$ is the totally antisymmetric 
tensor, with $\epsilon^{0123}=+1$. At LO, the 
vector, axial, and magnetic form factors are given by
\bea \label{eq:FF}
g_V(\spacevec q^2) &=& g_V = 1\, ,
\qquad g_A(\spacevec q^2) = g_A \simeq 1.27\,,
\qquad g_M(\spacevec q^2) = 1+\kappa_1\simeq 4.7\,,
\eea
where $\kappa_1\simeq 3.7$ is the nucleon isovector anomalous magnetic moment.
In the literature, the momentum dependence of the vector and axial form 
factors, and the contribution of weak magnetism to the neutrino potential 
are usually included---see for example Ref. \cite{Engel:2016xgb}. 
Since these are N$^2$LO effects, we will neglect them in most of the paper.

Equation \eqref{eq:currents} can be used to derive the long-range component of 
the $0\nu\beta\beta$ transition operator given in Eq. \eqref{eq:Vnu0}.
The expression in $\slashpi$EFT can be obtained by taking the 
$m_\pi \rightarrow \infty$ limit in Eq. \eqref{eq:Vnu0}. In this limit, the 
tensor component of $V_\nu$ vanishes.
The most singular part of $V_{\nu}$, which we denote by $\tilde V$, has a 
$1/\spacevec q^2$ behavior. 
The projections on the waves discussed in this paper are
\begin{equation}
\tilde{V}_{\nu\, {\rm L}}^{^1S_0}(\spacevec q) = \tilde{V}_{\nu\, {\rm L}}^{^1D_2}(\spacevec q) 
= \tau^{(1)+} \tau^{(2)+}  \,  \frac{1+2g_A^2}{\spacevec{q}^2}, 
\qquad
\tilde{V}_{\nu\, \rm L}^{^3P_J}(\spacevec q) = \tau^{(1)+} \tau^{(2)+}  
\,  \frac{1}{\spacevec{q}^2} \left(1 - \frac{2}{3} g_A^2 - \frac{g_A^2}{3}  S^{(12)} |_{^3P_J} \right),
\label{potsing}
\end{equation}
where the tensor operator in Eq. \eqref{potsing} is meant to be projected in the appropriate $P$ wave, as discussed in more detail in Sec. \ref{higherwaves}.
In $\slashpi$EFT Eq. \eqref{potsing} reduces to 
\begin{equation}
\tilde{V}_{\nu\, \rm L}^{^1S_0}(\spacevec q) = \tilde{V}_{\nu\, \rm L}^{^1D_2}(\spacevec q) 
= \tau^{(1)+} \tau^{(2)+}  \,  \frac{1+3g_A^2}{\spacevec{q}^2}, 
\qquad
\tilde{V}_{\nu\, \rm L}^{^3P_J}(\spacevec q) = \tau^{(1)+} \tau^{(2)+}  
\,  \frac{1 - g_A^2}{\spacevec{q}^2}.
\end{equation}
\noindent
In coordinate space,
the long-range neutrino potential defined in Eq.~\eqref{eq:Vnu0} is
\begin{equation}
V_{\nu\, \rm L}=\tau^{(1)+}\tau^{(2)+}  \left( V_{F}(r) 
- g_A^2 \, V_{GT}(r)\, {\bm \sigma}^{(1)}\cdot{\bm \sigma}^{(2)}  
- g_A^2 \, V_{T}(r) \, S^{(12)}\right) \, ,
\label{Vsaori}
\end{equation}
where the tensor operator
$S^{(12)} (\hat{r}) \equiv 3\boldsigma^{(1)}\cdot\hat{\spacevec r} 
\boldsigma^{(2)}\cdot\hat{ \spacevec r} - \boldsigma^{(1)}\cdot\boldsigma^{(2)}$
and the radial functions 
\begin{eqnarray}
V_F(r) &=& \frac{1}{4\pi r}, \qquad V_{GT}(r) = \frac{1}{4\pi r} 
\left[ 1 - \frac{e^{- m_\pi r}}{6} \left( 2+m_\pi r\right) \right],
\nonumber\\
V_T(r) &=& \frac{1}{2\pi r (m_\pi r)^2} 
\left[1 - e^{-m_\pi r}
\left(1 + m_\pi r + \frac{5}{12} (m_\pi r)^2 + \frac{1}{12} (m_\pi r)^3 \right) 
\right].
\label{FGTT}
\end{eqnarray}
In various channels considered below,
\begin{eqnarray}
V_{\nu\, {\rm L}}^{^1S_0}(r) &=& V_{\nu\, \rm L}^{^1D_2}(r) = \tau^{(1)+} \tau^{(2)+}  
\left(V_F(r) + 3 g_A^2 \, V_{GT}(r)\right),
\label{proSr}\\
V_{\nu\, {\rm L}}^{^3P_J}(r) &=&
\tau^{(1)+} \tau^{(2)+}\left(V_F(r) - g_A^2 \, V_{GT}(r) + a_J g_A^2 \, V_T(r)\right),
\label{proPr}
\end{eqnarray}
where $a_0 = 4$, $a_1 = -2$ and $a_2 = 2/5$.

Of the remaining terms in Eq. \eqref{eq:intro.2}, 
the operators schematically denoted by $\mathcal O$ induce 
LNV corrections to $\beta$ decay processes and long-range contributions
to $0\nu\beta\beta$. $\mathcal O$ is 
produced at tree level
by $SU(2)_L \times U(1)_Y$-invariant LNV operators of dimension seven 
and higher \cite{Cirigliano:2017djv,Cirigliano:2018yza}.
If one considers only LNV induced by a neutrino Majorana mass, however, 
these operators are suppressed by electroweak loops with respect to the 
leading contributions, and we will neglect them here. 

The operators denoted by $\mathcal O^\prime$ represent local LNV interactions 
between nucleons, pions and electrons, which are induced either by operators 
of dimension nine and higher~\cite{Prezeau:2003xn,Graesser:2016bpz,Cirigliano:2017djv,Cirigliano:2018yza}   
or  by the exchange of 
hard Majorana neutrinos  \cite{Cirigliano:2017tvr}.
The latter operators have the same transformation properties as the product of two 
weak currents, and their construction is detailed in Sec. \ref{CIB}.
For $0\nu\beta\beta$, the most important interaction is
\begin{equation}
\mathcal L_{|\Delta L| =2}^{NN}=  - \left(2\sqrt{2}\, G_F  V_{ud} \right)^2 
m_{\beta \beta} \,  \bar e_L C\bar e_L^T \, g_{\nu}^{\rm NN}  
\left[(N^T   P_{^1S_0}^{\,+}   N) (N^T  P_{^1S_0}^{\,-} N)^{\dagger}
\right]+ \textrm{H.c.}
+\ldots \,,
\label{gnudef}
\end{equation}
where interactions with additional pion fields required by chiral symmetry
are not written explicitly.
Here the projectors  $P_{^1S_0}^{\,\pm}$
are defined in terms of those in Eq. \eqref{proj1S0} as 
$P_{^1S_0}^{\,\pm} = (P_{^1S_0}^{\,1} \pm i P_{^1S_0}^{\,2})/2$.
In Weinberg's power counting, $g_{\nu}^{\rm NN} = \mathcal O( (4\pi F_\pi)^{-2})$ 
would contribute to the neutrino potential at N$^2$LO. 
As we argued in Sec. \ref{sec:problem} and will discuss in more detail in 
Sec. \ref{LNVLO}, the logarithmic dependence of LNV scattering amplitudes 
on the regulator induced by light-neutrino exchange
requires $g_{\nu}^{\rm NN}$ to be promoted to LO, 
$g_{\nu}^{\rm NN} = \mathcal O(F_\pi^{-2})$ instead. 
In addition, Ref. \cite{Kaplan:1996xu} demonstrated that an $m_\pi^2$ 
expansion might not be appropriate for four-nucleon operators. 
As we will explicitly show below, the counterterm needed for $0\nu\beta\beta$ 
inherits in $\chi$EFT some dependence on the quark mass. 
We 
need to construct $|\Delta L| =2$ operators with one and two 
insertions of the quark masses, which we discuss in detail in 
App. \ref{mass}. 
In the limit of equal up and down quark masses, $m_u = m_d$, considering 
insertions of the common quark mass 
leads to operators in the form of Eq. \eqref{gnudef},
but differing in the pion interactions lumped into the ellipsis.
The coupling of the four-nucleon operator 
is replaced by 
\begin{equation}
g_{\nu}^{\rm NN}  = \sum_n g_{n} m_\pi^{2n},
\label{LNVmass}
\end{equation}
where the coefficients $g_n$ scale as $\Lambda^{-2n-2}$, 
with $n=0, 1, \ldots$ an integer. 
NDA suggests $\Lambda \sim \Lambda_\chi$, implying that the mass dependence is 
suppressed.
We will however see that renormalization requires 
$g_\nu^{\rm NN} \propto 
C^2$, implying that, at least for $g^{}_{1}$ and 
$g_2$, the scale $\Lambda$ should be $\Lambda \sim F_\pi$.
In addition to the operator in Eq. \eqref{LNVmass}, additional mass-dependent 
operators can be constructed, but they contain at least two pion fields as 
described in 
App. \ref{mass}.

Beyond LO, additional contact interactions 
contribute to $0\nu\beta\beta$. 
In Sec. \ref{LNV@NLO} we will consider the derivative operator
\begin{eqnarray}
\mathcal L_{|\Delta L| =2}^{NN} &=&  \left(2\sqrt{2} G_F  V_{ud} \right)^2 
m_{\beta \beta} \,  \bar e_L C\bar e_L^T \, \frac{g_{2\, \nu}^{\rm NN} }{8}  
\nonumber \\ 
& & \times
\left[ (N^T  \overleftrightarrow{\boldsymbol\nabla}^{2} P_{^1S_0}^{\,+}   N) 
(N^T  P_{^1S_0}^{\,-} N)^{\dagger}  
+ (N^T P_{^1S_0}^{\,+} N)(N^T\overleftrightarrow{\boldsymbol\nabla}^{2} 
P_{^1S_0}^{\,-}N)^{\dagger}
\right] 
+ \textrm{H.c.},
\label{g2nudef}
\end{eqnarray}
which also acts between two $^1S_0$ waves 
\footnote{
The two-nucleon part of the operator 
is, up to an isospin factor, related to a linear combination of 
four-nucleon operators  $4C_1 + C_2 -12 C_3 -3 C_4 -4 C_6 - C_7$ 
defined in Ref.~\cite{Epelbaum:2004fk}.}, 
and we discuss the power counting for its LEC $g_{2\, \nu}^{\rm NN}$.

The contact interactions in Eqs. \eqref{gnudef} and \eqref{g2nudef} give 
short-range contributions to the $0\nu\beta\beta$ transition operator. 
Factoring out $G_F$, $m_{\beta\beta}$ and the lepton fields as in 
Eq. \eqref{eq:HV}, the short-range potential in the $^1S_0$ channel is 
\begin{equation}\label{Vshort}
V_{\nu\, \rm S}(\spacevec p, \spacevec p^\prime) = -2  \tau^{(1) +} \, \tau^{(2) +} 
\left(g_{\nu}^{\rm NN}+g_{2\nu}^{\rm NN}\,\frac{\spacevec p^2+\spacevec p^{\prime\, 2}}{2}
\right).
\end{equation}
It turns out that the short-distance operators induced by the exchange 
of hard neutrinos are related by isospin symmetry to isospin-two 
operators induced by the exchange of hard photons.
In Sec. \ref{CIB} we will discuss 
this relation in detail, and explore 
its implications for $0\nu\beta\beta$.

\section{The LNV scattering amplitude at leading order}
\label{LNVLO}

In this section we study the \nnpp
scattering amplitude
at LO in the $^1S_0$ channel, and show how the need for a short-range component 
of the neutrino potential arises in $\slashpi$EFT and $\chi$EFT. 
The section is based on the results of 
Refs. \cite{Cirigliano:2017tvr,Cirigliano:2018hja},
which we discuss in greater detail. 
We start by examining the amplitude in 
$\slashpi$EFT in Sec. \ref{LOpiless}. 
This allows us to derive an analytic expression for the amplitude.
In $\chi$EFT, the iteration of the pion-exchange Yukawa 
potential makes it impossible to provide a simple closed expression for the 
\nnpp
scattering amplitude, but one can still 
identify the leading divergent behavior in dimensional regularization
as we show in Sec. \ref{pifuldimreg}.
However, dimensional regularization is rarely used in few-body calculations. 
In Sec. \ref{cutoffLO} we therefore perform the same analysis with 
cutoff schemes that are widely used in the literature.

\begin{figure}
\includegraphics[width=\textwidth]{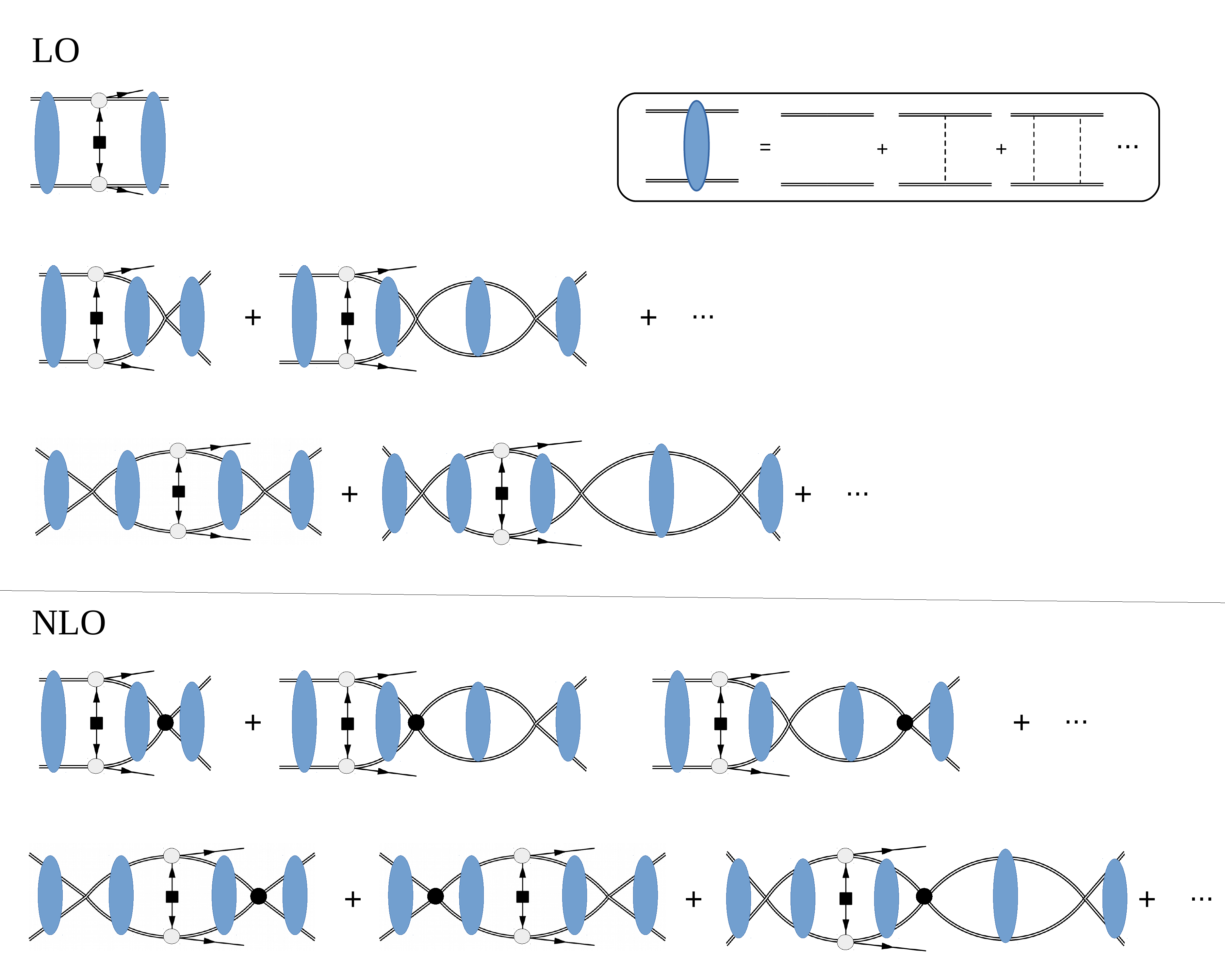}
\caption{Diagrammatic representation of LO and NLO contributions to 
\nnpp induced by long-range neutrino exchange. Double, dashed, and plain lines 
denote nucleons, pions, and leptons, respectively. 
Gray circles denote the nucleon axial and vector currents, and 
the black square an insertion of $m_{\beta \beta}$. 
The blue ellipse represents iteration of $V_\pi$,
while an unmarked contact interaction stands for $
C$.
In the NLO diagrams, the black circle denotes an insertion of $C_2$.
Diagrams analogous to those in the second and fourth rows,
but with contact interactions to the left of neutrino exchange, are not
shown.
The diagrams for $\slashpi$EFT are obtained by neglecting pion exchange in 
the blue ellipse.
}\label{Fig1}
\end{figure}

The LO 
contributions to \nnpp
from the exchange of a light neutrino are shown in the top panel of 
Fig. \ref{Fig1}. 
The blue ellipse denotes the iteration of the Yukawa potential 
$V_\pi(\spacevec q)$, while the contact interaction comes from 
the LEC $
C$. 
For the diagrams in the second and third row of 
Fig. \ref{Fig1}, one has to include an infinite number of bubbles, 
dressed with iterations of the Yukawa potential. 
The diagrams for $\slashpi$EFT can be obtained from those in Fig. \ref{Fig1} 
by neglecting the pion exchange potential.
Without loss of generality for our arguments, we  use  the kinematics 
of Eq. \eqref{kin}, with the electrons emitted at zero momentum.
For incoming neutrons with $|\spacevec p | = 1$ MeV, the relative momentum 
of the outgoing protons is $|\spacevec p^\prime| \simeq 38$ MeV.

Following Refs.~\cite{Kaplan:1996xu,Long:2012ve},
from the free Hamiltonian $H_0$ and the pion potential $V_\pi$
we introduce~\footnote{Note the following 
useful relations: 
$\chi^-_\spacevec{p} (\spacevec r)^* = \chi^+_\spacevec{p} (- \spacevec r)$
and 
$ G^+_E(\spacevec 0,\spacevec r) =  [ G^{-}_E(\spacevec r,\spacevec 0)]^* $.} 
the retarded ($+$) and advanced ($-$) propagators 
\begin{eqnarray}
\hat{G}^\pm_E = \frac{1}{E - H_0 - V_\pi \pm i \varepsilon}, 
\qquad  
G^\pm_{E}(\spacevec r, \spacevec r^\prime) 
= \int \frac{d^3 \spacevec k }{(2\pi)^3} 
\int \frac{d^3 \spacevec k^\prime}{(2\pi)^3}
\, e^{i \spacevec k \cdot r} e^{-i \spacevec k^\prime \cdot \spacevec r^\prime}
\langle \spacevec k | \hat{G}^\pm_E | \spacevec k^\prime \rangle,
\end{eqnarray}
and the Yukawa 
``in'' ($+$) and ``out''($-$) wavefunctions
\begin{eqnarray}
\chi^\pm_{\spacevec p}(\spacevec r) &=& \int \frac{d^3 \spacevec k}{(2\pi)^3} 
\, e^{i \spacevec k \cdot \spacevec r} \langle \spacevec k | (1 +  \hat{G}^\pm_E V_\pi) | 
\spacevec p \rangle.  
\label{chiGE}
\end{eqnarray}
Reference~\cite{Kaplan:1996xu} shows that the bubble diagrams in 
Fig.~\ref{Fig1} are related to 
$ G^+_E(\spacevec 0,\spacevec 0) =[ G^{-}_E(\spacevec 0,\spacevec 0)]^*$, 
while the triangles dressed by Yukawas are related to 
$\chi^+_{\spacevec p}(\spacevec 0)$ 
and 
$\chi^-_{\spacevec p^\prime}(\spacevec 0) ^*= \chi^+_{\spacevec p^\prime}  (\spacevec 0)$ 
(see Fig. 5 in Ref. \cite{Kaplan:1996xu}).
It is also convenient to introduce 
\begin{eqnarray}
K_{E} &=& \frac{
C}{1 - 
C G^+_{E}(\spacevec 0, \spacevec 0)} .
\label{KE}
\end{eqnarray}
The divergence in $G^+_{E}(\spacevec 0, \spacevec 0)$ is absorbed by  
$
C^{-1}$, so that $K_E$ 
is well defined and scale/scheme independent~\cite{Kaplan:1996xu}.

The chains of bubbles in the second and third rows of Fig.~\ref{Fig1} can be 
resummed, and at LO 
the amplitude can be expressed as
\begin{eqnarray}
 \mathcal A^{\rm LO}_{\nu} 
&=& \mathcal A_A  + 
\chi^+_{\spacevec p^\prime}(\spacevec 0) \, K_{E^\prime} \, \mathcal A_B         
+ \mathcal{\bar A}_B \, K_E \, \chi^+_{\spacevec p}(\spacevec 0) 
+
\chi^+_{\spacevec p^\prime}(\spacevec 0) \, K_{E^\prime}  \, \mathcal A_C  \, 
K_E \, \chi^+_{\spacevec p}(\spacevec 0), 
\label{eq:amplitude1}
\end{eqnarray}
where $\mathcal A_A$, $\mathcal A_B$, 
and $\mathcal A_C$ denote the first diagram in the first, second, and third row 
of Fig.~\ref{Fig1}, respectively 
(without the wavefunctions at $\spacevec 0$, in the case of $B$ and $C$),
while $\mathcal {\bar A}_B$ stands for the analog of the second row where
contact interactions come before neutrino exchange.

\begin{figure}
\includegraphics[width=\textwidth]{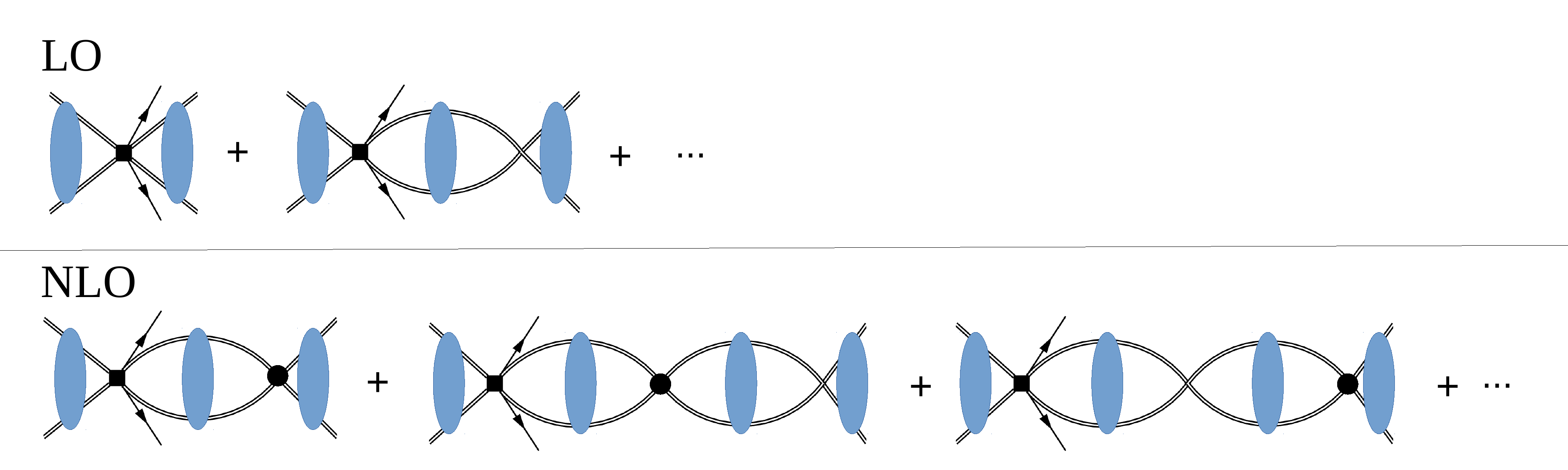}
\caption{Diagrammatic representation of LO and NLO contributions to 
\nnpp induced by the short-range operator $g_\nu^{\rm NN}$.  
The notation is as in Fig. \ref{Fig1}.
The diagrams for $\slashpi$EFT are obtained by neglecting pion exchange 
in the blue ellipse.
}\label{Fig2}
\end{figure}

The contribution of the operator $g_{\nu}^{\rm NN}$, defined in Eq. \eqref{gnudef},
is shown in the first row of Fig. \ref{Fig2}. 
It is easy to sum these diagrams, which modify the amplitude 
into
\begin{eqnarray}
\mathcal A^{\rm LO}_{\nu} 
&=& \mathcal A_A  
+ \chi^+_{\spacevec p^\prime}(\spacevec 0) \, K_{E^\prime}  \, \mathcal A_B         
+ \mathcal {\bar A}_B \, K_E \, \chi^+_{\spacevec p}(\spacevec 0) 
+
\chi^+_{\spacevec p^\prime}(\spacevec 0) \, K_{E^\prime}  
\left(\mathcal A_C + \frac{2 g_\nu^{\rm NN}}{
C^2} \right)
K_E  \, \chi^+_{\spacevec p}(\spacevec 0).
\label{eq:amplitude2}
\end{eqnarray}
Since $g_\nu^{\rm NN}$ appears together with $
C^{-2}$, it proves convenient 
to define the dimensionless parameter
\begin{equation}
\tilde g_\nu^{\rm NN}=\left(\frac{4\pi}{m_N 
C}\right)^2 g_{\nu}^{\rm NN}. 
\label{gtilde}
\end{equation}

The amplitude has the same structure in $\slashpi$EFT and $\chi$EFT. 
In $\slashpi$EFT, 
$\mathcal A_A$, $\mathcal A_B$, $\bar{\mathcal A}_B$ and $\mathcal A_C$ 
contain a single diagram, which can be analytically computed in dimensional 
regularization.
In $\chi$EFT, on the other hand, they still contain an infinite series 
of diagrams.
We now examine the two theories in more detail.

\subsection{Pionless EFT}
\label{LOpiless}

In $\slashpi$EFT the Yukawa wavefunction $\chi_{\spacevec p}^{\pm}$ reduces to a 
plane wave, with
$\chi_{\spacevec p}(0) = 1$.
${G}_E^{+}(\spacevec r,\spacevec r^\prime)$  is the free nucleon propagator, 
\begin{equation}\label{I0}
\left({G}_{E}^{+}(\spacevec 0,\spacevec 0) \right)_{\slashpi} = -I_0(\spacevec p)  
= \int \frac{d^{d-1} k}{(2\pi)^{d-1}} \,
\frac{m_N}{\spacevec p^2 - \spacevec k^2+i\varepsilon}  
=  -\frac{m_N}{4\pi} \left(\mu + i |\spacevec p|\right)
\end{equation}
in $d$ spacetime dimensions, 
where in the last equality we have used the PDS scheme~\cite{Kaplan:1998tg}.
$K_E$ reduces to the full strong scattering amplitude,
\begin{equation}
\left( K_{E} \right)_{\slashpi} = -T^{(0)}_{^1S_0} 
= \frac{1}{
C^{-1} + I_0(\spacevec p)}
=  \frac{4\pi}{m_N} \frac{1}{1/a + i |\spacevec p|}. 
\end{equation}
Equation \eqref{eq:amplitude2} then becomes
\begin{eqnarray}
\mathcal A_\nu^{\rm LO}  &=& 
\mathcal A_{A}
- \mathcal A_B(\spacevec p^{2},\spacevec p^{\prime\, 2})\, 
T^{(0)}_{^1S_0}(\spacevec p^{\prime\, 2})
- T^{(0)}_{^1S_0}(\spacevec p^2)\, 
\bar{\mathcal{A}}_B(\spacevec p^2, \spacevec p^{\prime\, 2})
\nonumber\\ 
&&+ T^{(0)}_{^1S_0}(\spacevec p^2) 
\left(\mathcal A_C(\spacevec p^2, \spacevec p^{\prime\, 2}) +  
\frac{2 g_{\nu}^{\rm NN}}{C^2}\right)  T^{(0)}_{^1S_0}(\spacevec p^{\prime\, 2})  
\label{eq:piless1}.
\end{eqnarray}
Here $\mathcal  A_{A}$ is the projection of the neutrino potential in the 
$^1S_0$ channel,
\begin{equation}
\mathcal  A_{A} = -\frac{1+3 g_A^2}{2} \int d\cos\theta  
\, \frac{1}{ (\spacevec p - \spacevec p^\prime)^2 },
\label{AApionless}
\end{equation}
with $\theta$ the angle between $\spacevec p$ and $\spacevec p^\prime$, 
while $\mathcal A_B$ and $\mathcal A_C$ reduce to 
one- and two-loop integrals,
\bea
{\mathcal A}_B(\spacevec p^{\prime 2},\spacevec p^2)&=& 
\bar{\mathcal A}_B(\spacevec p^2, \spacevec p^{\prime\, 2}) 
= - m_N\int \frac{d^{d-1}k}{(2\pi)^{d-1}} \,
\frac{1}{\spacevec p^2-\spacevec k\sq +i\varepsilon}
\,\frac{1+3g_A\sq}{(\spacevec k-\spacevec p')^2}, 
\label{ABpionless}
\\
\mathcal A_C(\spacevec p \sq, \spacevec p^{\prime\, 2})&=&
-m_N^2\int \frac{d^{d-1}k}{(2\pi)^{d-1}} \int \frac{d^{d-1}q}{(2\pi)^{d-1}} \,
\frac{1}{\spacevec p^2 -\spacevec k^2+i\varepsilon}
\, \frac{1+3g_A\sq}{(\spacevec k -\spacevec q)\sq} 
\,\frac{1}{\spacevec p^{\prime\, 2}-\spacevec q\sq +i\varepsilon}.
\label{ACpionless}
\eea
$\mathcal{A}_B$ is UV finite, and for $|\spacevec p^\prime| > |\spacevec p|$ it 
is given by
\begin{equation}
\bar{\mathcal A}_B(\spacevec p^2, \spacevec p^{\prime\, 2}) =  
\frac{m_N}{4\pi} \, \frac{1 + 3 g_A^2}{2} \,
\frac{i}{|\spacevec p^\prime|} 
\ln\frac{|\spacevec p| + |\spacevec p^\prime|}
{|\spacevec p| - |\spacevec p^\prime| + i 0^+} .
\end{equation}
On the other hand, $\mathcal A_C$ is logarithmically divergent,
\begin{equation}
{\mathcal A}_C(\spacevec p^2, \spacevec p^{\prime\, 2})  = 
- \left(\frac{m_N}{4\pi}\right)^2 \frac{1 + 3 g_A^2}{2}
\left(
 \frac{1}{4-d} - \gamma + 
\ln4\pi 
+ 2L_{\spacevec p,\spacevec p^\prime}(\mu) 
\right)~,
\label{I2piless}
\end{equation}
where $\gamma$ is the Euler-Mascheroni constant and, in the PDS scheme\footnote{
We notice that the sign of the 
$i 0^+$ prescription in the argument of the logarithm given in Ref. \cite{Cirigliano:2018hja} 
is incorrect.},  
\begin{equation}
L_{\spacevec p,\, \spacevec p^\prime}(\mu) =\frac{1}{2} 
\left(
\ln\frac{\mu^2}{-4(|\spacevec p|+|\spacevec p^\prime|)^2 -i 0^+}+1\right).
\label{log}
\end{equation}

Equations \eqref{eq:piless1} and \eqref{I2piless} clearly show that the 
scattering amplitude $\mathcal A_{\nu}$ is UV divergent unless 
$g_{\nu}^{\rm NN}$ appears at LO. Moreover, Eq.~\eqref{eq:piless1} allows one to 
derive the RGE for $g_\nu^{\rm NN}$ or equivalently for $\tilde g_{\nu}^{\rm NN}$ (see Eq. \eqref{gtilde}) 
\begin{equation}
\frac{d}{d\ln\mu}
\tilde g_{\nu}^{\rm NN} = \frac{1 + 3 g_A^2}{2}\equiv \beta.
\label{rge:piless}
\end{equation}
The solution is
\begin{equation}
\tilde g_{\nu}^{\rm NN}(\mu) = \beta \,
\ln(\mu/\mu_0) + \tilde g_{\nu}^{\rm NN}(\mu_0),
\label{rge:pilesssolution}
\end{equation}
with an initial condition $\tilde g_{\nu}^{\rm NN}(\mu_0)$ at some scale $\mu_0$.
There might be a scale $\mu_0$ for which 
$\tilde g_{\nu}^{\rm NN}(\mu_0) \ll 1$ rather than $\mathcal O(1)$.
However, at a comparable scale $\mu_0 + \delta \mu_0$ with 
$\delta \mu_0\sim \mu_0$,
$\tilde g_{\nu}^{\rm NN}(\mu_0+ \delta \mu_0) ={\mathcal O}(\beta)$.
Thus, it is natural to assume  
that $g_\nu^{\rm NN} = \mathcal O(1/\aleph^{2})$
\cite{Cirigliano:2017tvr}
as expected from the fact that $g_{\nu}^{\rm NN}$ connects two S waves 
\cite{Bedaque:2002mn}.

Similar expressions 
can be obtained 
in cutoff schemes, if loop diagrams are regulated in such a way that multiple 
loops with the insertion of contact interactions factorize into a product of 
one-loop diagrams, thus allowing the bubbles to be resummed.
This happens for ``separable'' regulators in the form of a product
of a function of the incoming momentum ($\spacevec p$) and 
a function of the outgoing momentum ($\spacevec p^\prime$).
One example, which will be used in the following sections, is when
contact interactions such as $
C$ and $C_2$ are replaced,
\begin{equation}
 C \rightarrow   
f(|\spacevec p^{\prime}|/\Lambda)
\, C \,  
f(|\spacevec p|/\Lambda)
\label{Lambda1}
\end{equation}
where $f$ is a function such that $f(0)=1$ and $f(\infty)=0$, and
$\Lambda$ is the cutoff parameter. 
For separable regulators,
$\mathcal A_\nu^{\rm LO}$ still has the expression given in 
Eq. \eqref{eq:piless1}. 
It exhibits a logarithmic dependence on the cutoff $\Lambda$,
and the same argument for the presence of $\tilde{g}_\nu^{\rm NN}$ at LO
goes through.

The non-separable regulator we will consider here involves the transferred momentum which
produces a regularization of the three-dimensional delta function in
coordinate space,
\begin{equation}
\delta^{(3)}(\spacevec r) \rightarrow \delta^{(3)}_{R_S}(\spacevec r) \,,
\label{Lambda2}
\end{equation}
where $R_S$ is a cutoff parameter such that
$\lim_{R_S\to 0} \delta^{(3)}_{R_S}(\spacevec r)= \delta^{(3)}(\spacevec r)$.
For non-separable regulators,
one has to resort to numerical calculations
even in the two-nucleon sector of $\slashpi$EFT. 
In numerical calculations, where 
negative powers of $\Lambda$ cannot be isolated and
simply dropped, the cutoff parameter should be taken beyond the 
EFT breakdown scale so that cutoff artifacts are no larger than the effects
of higher-order LECs.  We note that while we have only sketched the calculation of 
LNV amplitudes in $\slashpi$EFT with cutoff regulators, 
they can be obtained with the numerical procedure of Sec. \ref{cutoffLO}, by setting
 $g_A\to 0$ in the strong potential.

\subsection{Chiral EFT with dimensional regularization}
\label{pifuldimreg}

In $\chi$EFT, $\mathcal A_A$, $\mathcal A_B$,  $\bar{\mathcal A}_B$, 
and $\mathcal A_C$ contain an infinite sum of diagrams.
In order to study the renormalization of the neutrino potential, let us 
discuss the divergence structure of ${\mathcal A}_\nu$.
We note that:
\begin{itemize}
\item All the diagrams in $\mathcal A_A$ are finite. The tree level is 
obviously finite, as Eq. \eqref{AApionless} in $\slashpi$EFT.
Each iteration of the Yukawa interaction brings in 
a factor of $d^3 \spacevec k/(\spacevec k^2)^2$,
where one $\spacevec k^{-2}$ comes from the pion propagator, the other from the 
two-nucleon propagators after integrating over $k_0$. So, every Yukawa insertion
improves the convergence.

\item All the diagrams in $\mathcal A_B$ and $\bar{\mathcal A}_B$ are finite
as well. 
The first loop is 
similar to the result in $\slashpi$EFT, Eq. \eqref{ABpionless},
which is finite. 
Again, insertions of the Yukawa interaction improve the convergence.

\item The first two-loop diagram in $\mathcal A_C$ is logarithmically divergent.
The divergence arises from insertion of the most singular component of the 
neutrino potential, namely $\tilde V_{\nu}^{^1S_0}$ defined in Eq. \eqref{potsing}.
This is analogous to Eq. \eqref{ACpionless} in $\slashpi$EFT.
The two-loop diagram with an insertion of $V_\nu - \tilde{V}_\nu$ and 
higher-loop diagrams with one or more Yukawa insertions are convergent. 
\end{itemize}

We thus focus on 
$\mathcal A_C \equiv  \mathcal{A}_C^{\rm sing} +  \delta \mathcal A_C$.
The singular two-loop diagram  $\mathcal{A}_C^{\rm sing}$ is the same as 
in $\slashpi$EFT, with $1 + 3 g_A^2 \rightarrow 1 + 2 g_A^2$
due to the pion contribution to the induced pseudoscalar form factor.
The renormalized amplitude in the PDS and $\overline{\rm MS}$ schemes  
is obtained by the 
replacement
\be
\mathcal A_C + \frac{2 g_\nu^{\rm NN}}{
C^2} \to 
\left( \frac{m_N }{4\pi} \right)^2
\left[2\tilde{g}_{\nu}^{\rm NN}(\mu) - 
\left(1 + 2 g_A^2\right) L_{\spacevec p,\, \spacevec p^\prime}(\mu)  
\right]+ \delta  \mathcal A_C
\label{eq:amplitude2b}
\ee
in Eq.~\eqref{eq:amplitude2}.
Instead of Eq. \eqref{rge:piless}, the renormalized coupling obeys the RGE
\be
\frac{d}{d\ln\mu} 
\tilde{g}_\nu^{\rm NN} =  \frac{1 + 2 g_A^2}{2}~.
\label{gnurunningchiral}
\ee
The above argument shows that, as in $\slashpi$EFT, the counterterm 
$g_\nu^{\rm NN}= \mathcal O(1/Q^{2})$ 
must be included at LO in Eq. \eqref{eq:amplitude2}. 

The finite part of the coupling can be obtained in principle by matching the 
S-matrix element in Eqs.~\eqref{eq:amplitude2} and 
\eqref{eq:amplitude2b} 
to a LQCD calculation, performed at the same kinematic point. 
In order to carry out this program, one needs a non-perturbative calculation 
of the S-matrix element in $\chi$EFT, 
which amounts to a resummation of the  infinite number of Feynman diagrams 
building up to  $\mathcal A_A$, $\mathcal A_B$ and $\delta \mathcal A_C$.   
This is equivalent to solving the  Schr\"odinger equation~\cite{Kaplan:1996xu},
as we recall below.

One can re-express the amplitudes in Fig. \ref{Fig1} as 
\begin{eqnarray}\label{eq:2}
\mathcal A_A &=&  - \int d^3 \spacevec r\, \chi^-_{\spacevec p^\prime}(\spacevec r)^* 
\, V_{\nu\, \rm L}^{^1S_0}(\spacevec r) \, \chi^+_{\spacevec p}(\spacevec r) ,
\\
\mathcal A_B + \mathcal {\bar A}_B  &=& -  \int d^3 \spacevec r 
\left({G}^-_{E^\prime}(\spacevec r, \spacevec 0)^* \, V_{\nu\, \rm L}^{^1S_0}(\spacevec r) 
\, \chi^+_{\spacevec p}(\spacevec r)  +  
\chi^-_{\spacevec p^\prime}(\spacevec r)^* \, V_{\nu\, \rm L}^{^1S_0}(\spacevec r) \,
{G}^+_E(\spacevec r, \spacevec 0)\right) ,
\\
\mathcal A_C &= &-\int d^3\spacevec r\,{G}^-_{E^\prime}(\spacevec r, \spacevec 0)^*
\, V_{\nu\, \rm L}^{^1S_0}(\spacevec r) \, {G}^+_E(\spacevec r, \spacevec 0).
\end{eqnarray}
The three sets of diagrams combine to give 
\begin{eqnarray}
\mathcal A_\nu = - \int d^3 \spacevec r \, \psi^-_{\spacevec p^\prime}(\spacevec r)^* 
\,  V_{\nu\, \rm L}^{^1S_0}(\spacevec r) \, \psi^+_{\spacevec p}(\spacevec r)
\label{eq:anutot}
\end{eqnarray}
in terms of the 
solutions 
\begin{equation}
\psi^\pm_\spacevec p(\spacevec r) = \chi^\pm_\spacevec p(r) 
+ \chi^\pm_\spacevec p(\spacevec 0)\, K_E\, { G}^\pm_E(\spacevec r, 0)
\label{eq:3}
\end{equation}
of the Schr\"odinger equation with the potential in 
Eq. \eqref{1S0LOpot1}.
The expression \eqref{eq:anutot} simply represents first-order perturbation 
theory in the very weak $\Delta L=2$ operator $V_{\nu\, \rm L}^{^1S_0}$  
acting on the wavefunctions of the 
LO strong potential \eqref{1S0LOpot1}.

In this coordinate-space picture the UV convergence or divergence of the 
amplitudes can be simply recovered from the $r \to 0$ behavior. 
For  $r\rightarrow 0$, the long-range neutrino potential goes as $1/r$, 
while the Yukawa wavefunction $\chi_{\spacevec p}^\pm(r)$ tends to a constant.  
This confirms that $\mathcal A_A$ is finite.  
On the other hand, for the propagator $G_E^\pm(\spacevec r, \spacevec 0)$ one has
\begin{equation}
G^\pm_E(\spacevec r, \spacevec 0) \rightarrow \frac{m_N}{4\pi r}  + \ldots 
\end{equation}
$\mathcal A_B$  and $\bar{\mathcal A}_B$ are still finite, 
but $\mathcal A_C$ is logarithmically divergent. 
The singular component $\mathcal{A}_C^{\rm sing}$  is obtained by using 
the free Green's functions, namely 
\be
\mathcal{A}_C^{\rm sing} = - \int d^3 \spacevec r\, 
{G}^{(0)-}_{E^\prime}(\spacevec r, \spacevec 0)^* \, 
\tilde{V}_{\nu\, \rm L}^{^1S_0}(\spacevec r) \, {G}^{(0)+}_E(\spacevec r, \spacevec 0)~,
\qquad 
{G}^{(0)\pm}_E(\spacevec r, \spacevec 0) = - \frac{m_N}{4 \pi r} \, e^{\pm i p r}~.
\ee
Defining 
$\delta G_E^\pm (\spacevec{r}) \equiv {G}^{\pm}_E(\spacevec r, \spacevec 0) 
-  {G}^{(0)\pm}_E(\spacevec r, \spacevec 0)$, 
the finite part can be expressed as 
\bea
\delta \mathcal A_C  &=&
- \int d^3 \spacevec r\, {G}^{(0)-}_{E^\prime}(\spacevec r)^* \, 
V_{\nu\, \rm L}^{^1S_0}(\spacevec r) \, {\delta G}^+_E(\spacevec r)
- \int d^3 \spacevec r\, {\delta G}^-_{E^\prime}(\spacevec r)^* \,  
V_{\nu\, \rm L}^{^1S_0}(\spacevec r) \, {G}^{(0)+}_E(\spacevec r)
 \\
&&- \int d^3 \spacevec r\, {\delta G}^-_{E^\prime}(\spacevec r)^* \,
V_{\nu\, \rm L}^{^1S_0}(\spacevec r) \, {\delta G}^+_E(\spacevec r)
- \int d^3 \spacevec r\, {G}^{(0)-}_{E^\prime}(\spacevec r)^*  
\left( V_{\nu\, \rm L}^{^1S_0}(\spacevec r) - \tilde{V}_{\nu\, \rm L}^{^1S_0}(\spacevec r) 
\right) 
{G}^{(0)+}_E(\spacevec r)~.\nn
\eea

As discussed above, renormalization requires that we consider also the diagrams
of Fig. \ref{Fig2}, which lead to Eq. \eqref{eq:anutot} with the replacement
in Eq. \eqref{LOnupot}.
For given $E$ and $m_\pi$ (and corresponding phase shifts),  
$\chi^\pm_{\spacevec p} (\spacevec r)$ and $G_E^\pm (\spacevec r,0)$ can be obtained
in a straightforward way by numerically solving the Schr\"odinger 
equation~\cite{Kaplan:1996xu}, see Appendix \ref{app:MSbar},
so that $\mathcal A_{A,B}$ and $\delta \mathcal A_C$ can be readily computed 
numerically. One can  then use our representation of the amplitude 
in Eq.~\eqref{eq:amplitude2} to match to future LQCD calculations and  
extract the short-range coupling $\tilde{g}_\nu^{\rm NN}$. 

\subsection{Chiral EFT with cutoff regularization}
\label{cutoffLO}

The analysis of the $\Delta L =2$ \nnpp scattering amplitude in the PDS and 
$\overline{\rm MS}$ schemes is theoretically clean, and it unambiguously shows 
the need for enhanced short-range LNV operators. Furthermore, it can be easily 
matched to future LQCD calculations. 
Such an analysis, however, would yield a value of $\tilde{g}_\nu^{\rm NN}$ 
in a regularization scheme that is distinct from what is used in many-body 
nuclear calculations.  
We therefore repeat the analysis utilizing different regulators for the 
short-range part of the internucleon potential.
These regulators are not only appropriate for use in other channels
(see Sec. \ref{higherwaves}) and heavier nuclei (see Sec. \ref{pheno}),
but also the corresponding calculations can be matched to LQCD (see, 
for example, Refs.~\cite{Barnea:2013uqa,Kirscher:2015yda,Contessi:2017rww}).

We extend the analysis of 
$\chi$EFT in Sec. \ref{pifuldimreg}
by introducing two additional schemes, which effectively work as 
momentum cutoffs.
The first scheme is a non-separable regulator of the type \eqref{Lambda2} with
\begin{equation}
\delta^{(3)}_{R_S}(\spacevec r) =  
\frac{1}{(\sqrt{\pi} R_S)^3} \exp\left(- \frac{r^2}{R_S^2}\right)\,,
\label{eqRS}
\end{equation}
where $r=|\spacevec r|$.
This 
was used, for example, in the definition of the chiral potential in 
Refs. \cite{Ordonez:1993tn,Ordonez:1995rz,Piarulli:2016vel}. 
The wavefunctions 
$\psi^\pm_{\spacevec p}(\spacevec r)$ 
are now solutions of the 
Schr\"odinger equation with the delta function in the strong potential 
regulated using Eq. \eqref{eqRS}, and therefore depend on the cutoff $R_S$.
With the short-range LNV interaction,
the amplitude \eqref{eq:anutot} becomes
\begin{equation}
\mathcal A_\nu = - \int d^3 \spacevec r \,  
\psi^{-\, *}_{\spacevec p^\prime}(\spacevec r)  
\left(V_{\nu\,\rm L}(\spacevec r) - 2 g^{\rm NN}_\nu \delta^{(3)}_{R_S}(\spacevec r) 
\right)
\psi^+_{\spacevec p}(\spacevec r).
\label{ALNV}
\end{equation}

The second scheme is analogous to the cutoff scheme introduced in 
Eq. \eqref{Lambda1} in $\slashpi$EFT and is applied to a momentum-space 
solution of the Lippmann-Schwinger (LS) equation. 
The LS equation for the T matrix can be written in short-hand notation as 
\begin{equation}
T = V + V G_0 T, 
\qquad
G_0 = (E - \spacevec p^{\,2}/m_N + i \varepsilon)^{-1},
\label{LSshorthand}
\end{equation} 
where 
integration is implied. 
In more detail, in the ${}^1S_0$ channel
\begin{equation}
T_{{}^1S_0} (p^\prime, p,E) = V_{{}^1S_0} (p^\prime, p) +\int_0^\infty 
dp^{\prime \prime}\, V_{{}^1S_0} (p^\prime, p^{\prime \prime}) \left(\frac{p^{\prime \prime\,2}}
{E- p^{\prime \prime\,2}/m_N + i\varepsilon}\right) T_{{}^1S_0}(p^{\prime \prime}, p, E)\,,
\label{LS}
\end{equation}
in terms of the partial-wave projection 
\begin{eqnarray}
V_{{}^1S_0}(p^\prime, p) &=& \frac{1}{(2\pi)^3} 
\langle {}^1S_0,\,p'\,|
V_{N\!N}^{{}^1S_0}(\spacevec q) 
| {}^1S_0,\,p \rangle 
\nonumber\\
&=& \frac{1}{(2\pi)^3}\left[
C
- \frac{\pi g_A^2 m_\pi^2}{2F_\pi^2}\, 
\int_{-1}^1 dx\,\frac{1}{p^2+p'^2-2pp'x+m_\pi^2}\right]
\label{Vnu1Somomspace}
\end{eqnarray}
of the potential $V_{N\!N}^{{}^1S_0}(\spacevec q)$ given in Eq.~\eqref{1S0LOpot1}.
Here, and in what follows, we denoted $p = |\spacevec p|$ and $p^\prime = |\spacevec p^\prime|$. 
The on-shell 
T matrix is linked to the S matrix and the phase shifts via
\begin{equation}
S_{{}^1\!S_0}(E) = e^{2i \, \delta_{{}^1\!S_0}(E)} 
= 1 - i \pi m_N q_0\, T_{{}^1S_0}(q_0 , q_0,E), 
\label{SLO}
\end{equation}
where $q_0 = \sqrt{m_N E}$ is the relative momentum of the interacting nucleons
in the center-of-mass frame. The momentum integral in the LS equation 
is divergent and we regulate the potential via a separable regulator of the
form \eqref{Lambda1},
\begin{equation}
V_{{}^1S_0}(p^\prime, p) \rightarrow 
\mathrm{exp}\left[-\left(\frac{p^{\prime\,2}}{\Lambda^2}\right)^n\right]\, 
V_{{}^1S_0}(p^\prime, p)\, 
\mathrm{exp}\left[-\left(\frac{p^{2}}{\Lambda^2} \right)^n\right]\,,
\label{Lambda1b}
\end{equation} 
in terms of a momentum cutoff $\Lambda$. For this paper we choose $n=2$.
The LS equation is solved numerically for different values of $\Lambda$. 
For details of the numerical solution, see e.g. the appendix of 
Ref.~\cite{deVries:2013fxa}. 

\begin{figure}
\includegraphics[width=0.75\textwidth]{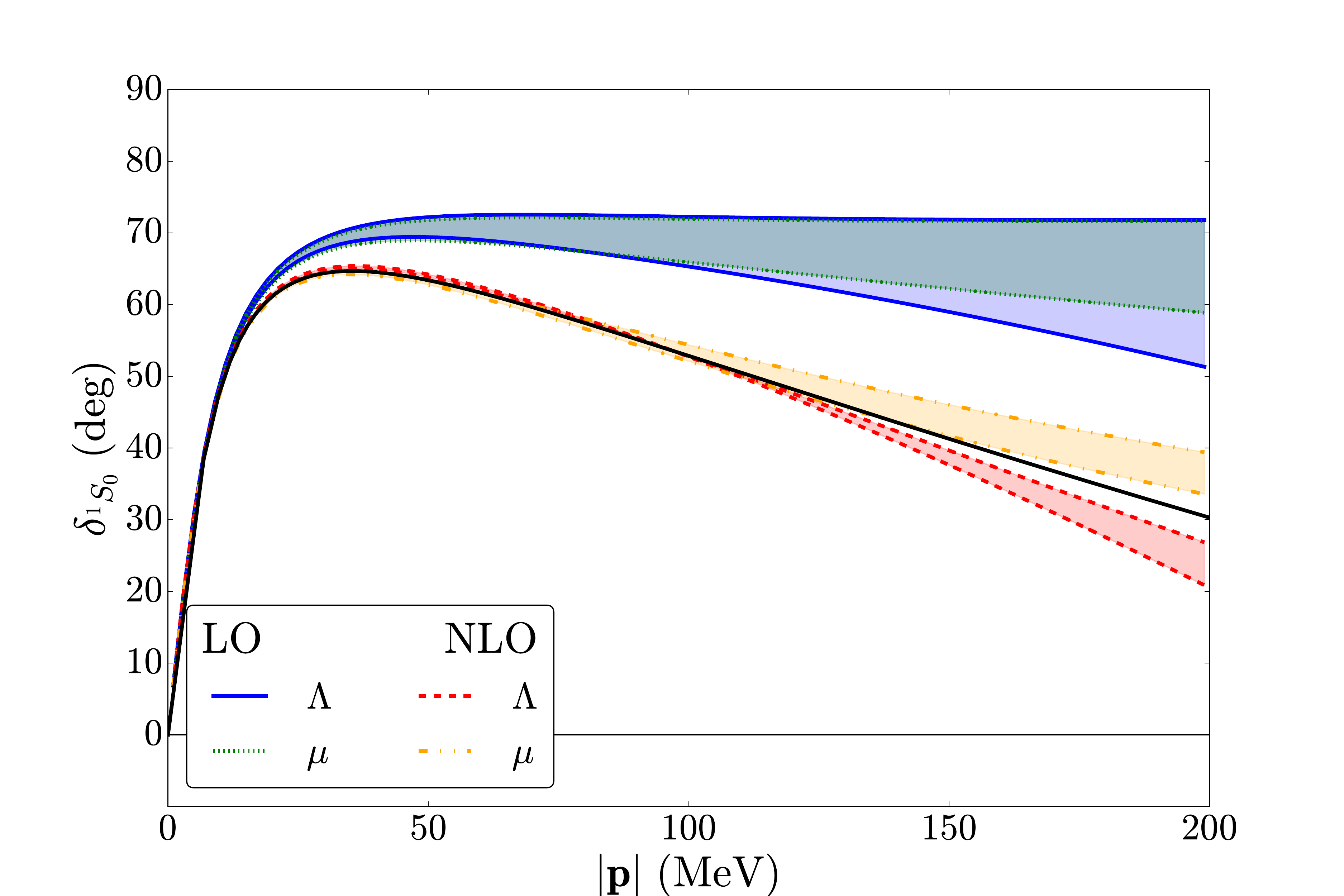}
\caption{Phase shifts for $np$ scattering in the  $^1S_0$ channel, 
computed in $\chi$EFT with a momentum space ($\Lambda$) cutoff and
in dimensional regularization ($\mu$), as function
of the center-of-mass momentum $|\spacevec p|$. 
The solid blue and dashed red lines denote the momentum-cutoff results at 
LO and NLO, respectively.
The bands are obtained by varying $\Lambda$ between $2$ and $20$ fm$^{-1}$.
The dotted green and dash-dotted orange lines are the LO and NLO results 
in dimensional regularization.
The bands in the \MS\ scheme are obtained by varying the 
regulator of the intermediate scheme, as discussed in App. \ref{app:MSbar}, 
between $1/\lambda=0.05$ fm and $1/\lambda =0.7$ fm.
$
C$ is fit to the scattering length, and $C_2$ to the phase shift 
at $|\spacevec p| = 30$ MeV.
The black line shows the Nijmegen partial-wave analysis \cite{Stoks:1993tb}.
}\label{PS1S0}
\end{figure}

In both schemes, we determine $
C$ by fitting to the scattering length 
in the $^1S_0$ channel for a given value of the regulator,
as described in Sec. \ref{sec:problem}. 
$\chi$EFT at LO reproduces the phase shifts in the $^1S_0$ channel
only up to moderate values of the nucleon center-of-mass momentum 
\cite{Kaplan:1996xu}. For the  present discussion, however, what is more 
important is the regulator dependence of the phase shifts. 
As shown in Fig.~\ref{PS1S0}, the regulator dependence in the momentum-space 
scheme is small, and similar results hold in the $R_S$ scheme. 
Results for dimensional regularization are given as well.
NLO corrections to the phase shifts, which improve the agreement with data, 
are discussed in Sec.~\ref{LNV@NLO}.  

After this renormalization exercise we have a consistent description of the 
$N\!N$
system in the ${}^1S_0$ channel. We can now turn to the calculation of the 
\nnpp amplitude. 
In coordinate space, the LNV scattering amplitude is obtained by evaluating 
Eq. \eqref{eq:anutot}.
In the momentum-space scheme, we use its analog,
\begin{eqnarray}
\mathcal A_\nu = - 2 \pi^2 \,\,
{}^{-}\langle {}^1S_0,\,p'\,|
V_\nu^{{}^1S_0} (p^\prime,p) 
| {}^1S_0,\,p \rangle^+ 
\label{Anumomspace}
\end{eqnarray}
where 
\begin{equation}
V^{{}^1S_0}_{\nu}(p^\prime, p) 
= \int_{-1}^1 \frac{dx}{(2\pi)^2}\,\frac{1}{p^2 + p^{\prime\,2}-2pp'x}
\left[1+2 g_A^2 
+ \frac{g_A^2 m_\pi^2}{\left(p^2 + p^{\prime\,2}-2pp'x+m_\pi^2\right)^2}
\right]
- \frac{g_{\nu}^{\rm NN}}{\pi^2} 
\label{Vnumomspace}
\end{equation}
is the partial-wave-projected neutrino potential
and the $\pm$ superscripts indicates that we sandwich $V_\nu$ between 
scattered wavefunctions. 
We then calculate $\mathcal A_\nu $ via the explicit expression
\begin{eqnarray}
\mathcal A_\nu &=& -2 \pi^2 \bigg\{ V^{{}^1S_0}_{\nu}(p^\prime, p)  
\nonumber\\
&&+ \int dp'' \left[ V^{{}^1S_0}_{\nu}(p^\prime, p'')
 \frac{m_Np^{\prime\prime\,2}}{p^2-p^{\prime\prime\,2}+i\varepsilon} T_{{}^1S_0} (p'', p,E) 
+ T_{{}^1S_0} (p^\prime, p'',E') 
\frac{m_Np^{\prime\prime\,2}}{p^{\prime\,2}-p^{\prime\prime\,2}+i\varepsilon}  
V^{{}^1S_0}_{\nu} (p'', p) \right]  
\nonumber\\ 
&&+\int dp'' \int dp'''\,  T_{{}^1S_0} (p^\prime, p'',E') 
\frac{m_Np^{\prime\prime\,2}}{p^{\prime\,2}-p^{\prime\prime\,2}+i\varepsilon}  
V^{{}^1S_0}_{\nu} (p'', p''') 
\frac{m_Np^{\prime\prime\prime\,2}}{p^2-p^{\prime\prime\prime\,2}+i\varepsilon} 
T_{{}^1S_0} (p''', p,E)
 \bigg\}
\label{Anumom}
\end{eqnarray}
or, in short-hand notation,
\begin{equation}
\mathcal A_\nu =  -2 \pi^2 
\left(V_\nu  +V_\nu G_0 T + T G_0 V_\nu  + T G_0 V_\nu G_0 T \right)\,.
\label{AnuLO}
\end{equation}
The solution
is graphically depicted in Figs.~\ref{Fig1} and \ref{Fig2}. 

\begin{figure}
\includegraphics[width=0.75\textwidth]{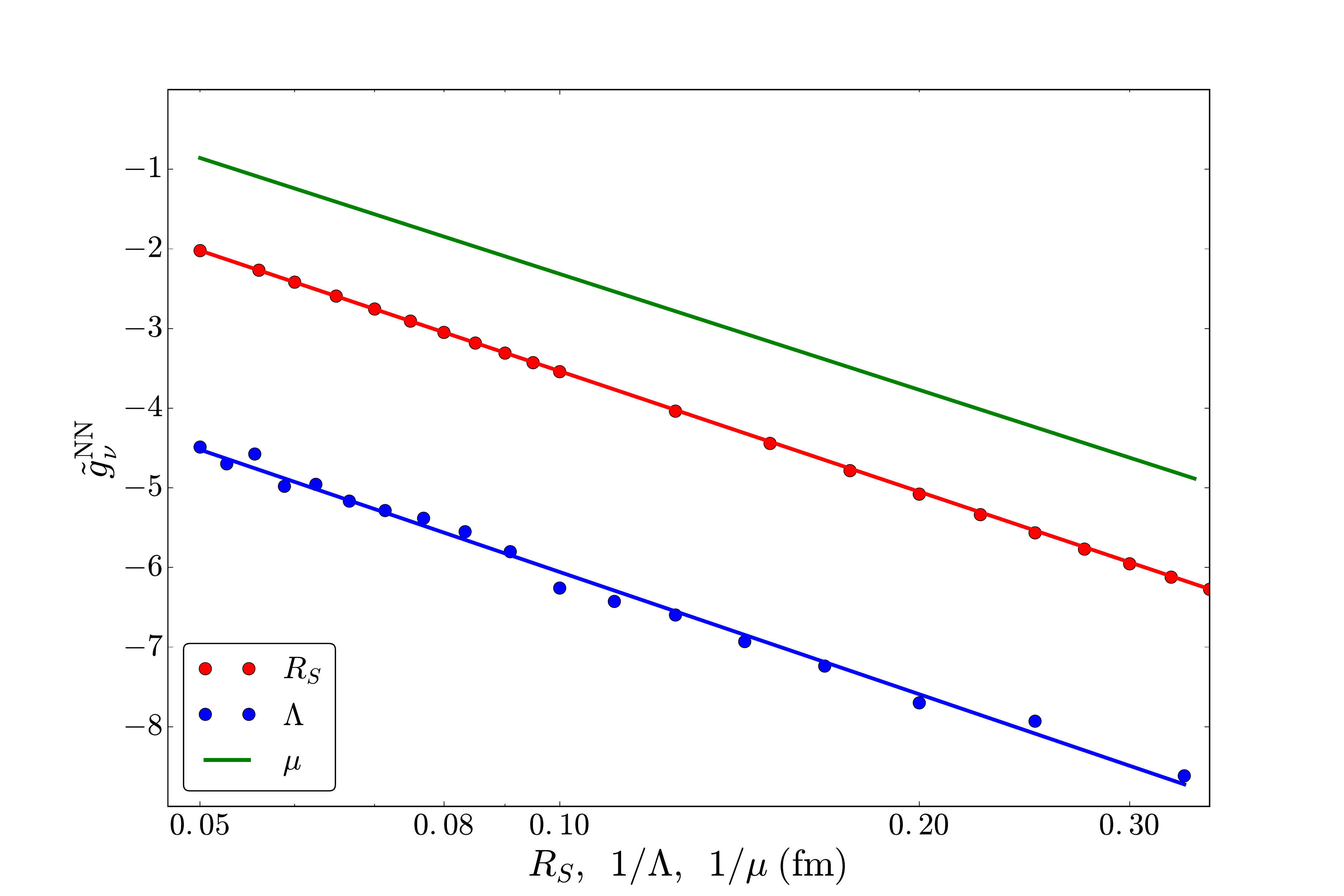}
\caption{Dimensionless counterterm $\tilde g_{\nu}^{\rm NN}$ 
in $\chi$EFT as a function of the 
coordinate- and momentum-space cutoffs $R_S$ (red) and $\Lambda$ (blue), 
and of the dimensional-regularization scale $\mu$ (green). 
Red and blue points denote the results of numerical calculations,
while the corresponding lines are logarithmic fits, as explained in the text.
The counterterm is determined by imposing the (arbitrary) condition  
$ \mathcal A_\nu \exp{(-i (\delta_{^1S_0}(E) + \delta_{^1S_0}(E^\prime)))}  = -0.05$ MeV$^{-2}$ at $|\spacevec p| = 1$ MeV.
}\label{LNV_gnu}
\end{figure}

As illustrated in the right panel of Fig. \ref{fig:sec2plots}, the amplitude 
$\mathcal A_\nu$ computed only with the long-range neutrino-exchange potential
is cutoff dependent. The cutoff dependence is cured by introducing 
$g_{\nu}^{\rm NN}$ at LO. 
Figure \ref{LNV_gnu} shows the values of the dimensionless coupling 
$\tilde g_{\nu}^{\rm NN}$, defined in Eq. \eqref{gtilde},
as a function of $R_S$, $\Lambda$, and the dimensional-regularization 
scale $\mu$.
Because of the lack of data on $\Delta L=2$ processes, $\tilde g_{\nu}^{\rm NN}$ 
was determined here by requiring that the scattering amplitude at 
$|\spacevec p| = 1$ MeV be equal to an arbitrarily chosen value,
\begin{equation}
\mathcal A_\nu(|\spacevec p|=1 \, {\rm MeV},\, |\spacevec p^\prime|=38 \, {\rm MeV}  ) e^{-i (\delta_{^1S_0}(E) + \delta_{^1S_0}(E^\prime))} = -0.05\, {\rm MeV}^{-2}.
\end{equation}
The values of $\tilde g_{\nu}^{\rm NN}$ obtained numerically with the $\Lambda$ 
and $R_S$ regulators are fitted with
\begin{eqnarray}
\tilde g_{\nu}^{\rm NN}(\Lambda) &=& - 12.0  - 2.2 \, 
\ln (m_\pi/\Lambda)\,, 
\qquad   
\tilde g_{\nu}^{\rm NN}(R_S) = - 9.4 - 2.2 \, 
\ln (m_\pi R_S )\, ,   
\nn\\
\tilde g_\nu^{\rm NN}(\mu) &=& - 7.9 - 2.1 \, 
\ln (m_\pi/\mu)\,.
\label{fits}
\end{eqnarray}
The coefficients of the logarithms in Eq. \eqref{fits} 
are close to each other, and close to the dimensional-regularization 
expectation $(1 +2 g_A^2)/2 \simeq 2.1$. While intriguing, there is no proof 
that the coefficient of the logarithm should be universal, 
and counterexamples exist 
in the literature \cite{Beane:2001bc}\footnote{As discussed in 
Sec. \ref{chiEFT}, the pion-exchange potential 
in the $^1S_0$ channel induces a divergence in the strong scattering amplitude 
proportional to  $m_\pi^2 
\ln \Lambda$  \cite{Kaplan:1996xu}, which is absorbed
by promoting $D_2$ to LO. The coefficient of the logarithm can be computed 
analytically in the scheme defined in Ref. \cite{Beane:2001bc}, and differs 
from the dimensional regularization value of Ref. \cite{Kaplan:1996xu}.}.

\begin{figure}
\includegraphics[width=0.75\textwidth]{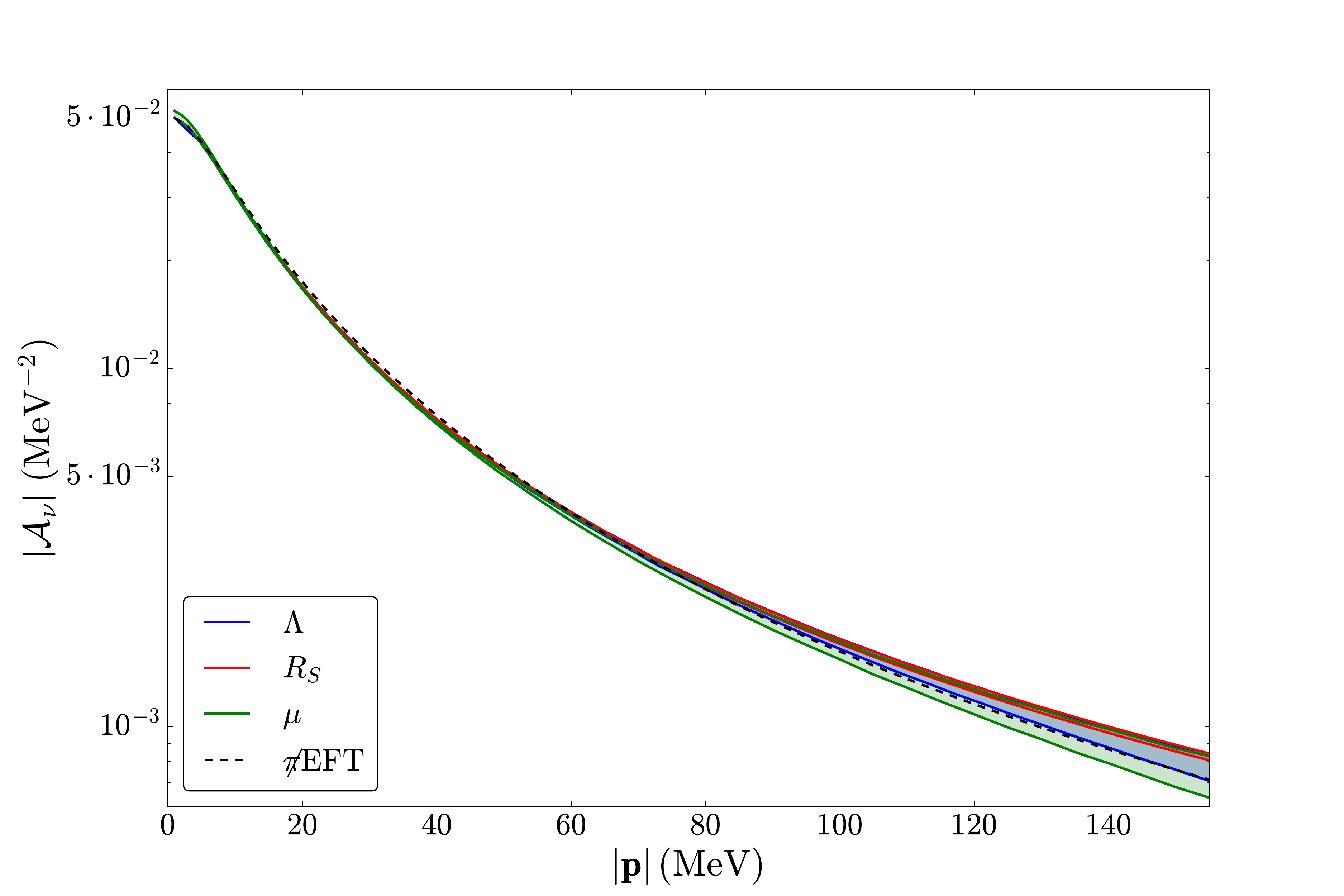}
\caption{Magnitude of the LNV scattering amplitude $\mathcal A_\nu$ 
as a function of the neutron center-of-mass momentum $|\spacevec p|$
at LO.
The red, blue, and green lines represent the results in $\chi$EFT
with coordinate-space, momentum-space, and dimensional regularization,
respectively, using $g_\nu^{\rm NN}$ from Fig. \ref{LNV_gnu}.
The bands indicate residual regulator dependence, and are obtained by 
varying $R_S$ between $0.05$ and $0.7$ fm,
$\Lambda$ between $2$ and $20$ fm$^{-1}$,
and $1/\lambda$ between $0.05$ fm and $0.7$ fm.
The dashed black line is the result in $\slashpi$EFT
with dimensional regularization. 
Since the $\slashpi$EFT amplitude can be made exactly $\mu$-independent at this order,
there is no band associated with the $\mu$ variations.
}\label{LNV_LO}
\end{figure}

In Fig. \ref{LNV_LO} we show the renormalized $\mathcal A_\nu$ as a function of
$\spacevec p$
in $\chi$EFT with momentum-
and coordinate-space 
cutoffs, and in dimensional regularization.
The 
cutoff bands are obtained by varying $\Lambda$ between $0.4$ and $2$ GeV, and
$R_S$ between $0.05$ and $0.7$ fm. The 
dimensional-regularization band, by varying the regulator of the 
intermediate scheme introduced in App. \ref{app:MSbar} between 
$1/\lambda=0.05$ fm and $1/\lambda =0.7$ fm. 
Also shown is the outcome in $\slashpi$EFT with dimensional regularization. In $\slashpi$EFT
the LO amplitude can be made $\mu$-independent, so that no band from scale variation appears. Of course,
there is an uncertainty from missing higher-order corrections.
All results are in excellent agreement. The regulator dependence is negligible 
at small momenta, and is small even at $|\spacevec p| =150$ MeV:
about $15\%$ in the momentum-space scheme and smaller in the other 
schemes. This dependence is 
significantly reduced if we vary $\Lambda$
between $0.6$ and $2$ GeV, indicating that 
$0.4$ GeV might be too low compared to the breakdown scale. 

Note that treating pion exchange perturbatively 
\cite{Kaplan:1998tg,Kaplan:1998we}, which might
be sufficient at low energies \cite{Fleming:1999ee,Kaplan:2019znu}, 
does not avoid the presence of $g_{\nu}^{\rm NN}$ at LO, 
since in this case LO in the strong sector is identical to $\slashpi$EFT.
The conclusion is that after inclusion of $g_{\nu}^{\rm NN}$ the \nnpp amplitude 
is properly renormalized over the whole EFT momentum range.

\section{Neutrino potential in higher partial waves}
\label{higherwaves}

In the previous sections we have demonstrated in various schemes the need to 
introduce an LO short-range counterterm for the \nnpp process for 
${}^1S_0 \rightarrow {}^1S_0$ transitions. We now investigate whether this 
problem also occurs for transitions involving higher partial waves. 
In $\slashpi$EFT nucleons do not interact in these waves until higher orders,
but in $\chi$EFT, as we discussed in Sec. \ref{chiEFT}, there are
renormalization issues already in the strong sector.
We limit ourselves to two $P$-wave transitions 
${}^3P_{0,1} \rightarrow {}^3P_{0,1}$,
which allow us to examine the effects from the singular $N\!N$ tensor force
generated by OPE while avoiding
complications involved in the  ${}^3P_2$-$^3F_2$ coupled channel.
In the ${}^3P_1$ channel this
force is repulsive and one might expect no UV problems. However, 
in the ${}^3P_0$ channel it is attractive and similar problems as in 
the ${}^1S_0$ channel might occur. We also study the ${}^1D_2$ channel as 
representative of a singlet channel (where the OPE tensor force vanishes)
with higher angular momentum $j$. 
In this section we stick to 
a single scheme, where we regulate the LS equation with a momentum cutoff.

We begin by describing the strong force in the $P$ and $D$ waves. 
We perform a partial-wave decomposition of 
the OPE potential \eqref{strong} and obtain
\begin{eqnarray}
\langle (l's')j'\,p'|
V_\pi
| (ls)j\,p\rangle &=& 
2\pi \sum_f \hat f^{\, 3/2}\,(1\,1\,f\,;000)
\sum_{\lambda_1+\lambda_2 = f} \sqrt{\frac{\hat f!}{\hat \lambda_1! \hat \lambda_2!}}
\; (p)^{\lambda_2} (-p^\prime)^{\lambda_1}
\nonumber\\
&&\times\sum_k(-1)^k \hat k^{\, 3/2}\,g_k^f(p,p') 
\bma k&k&0 \\ 
\lambda_1&\lambda_2&f\\ 
l'&l &f  \ema
\sqrt{\hat \lambda_1 \hat \lambda_2}\;
(k\,\lambda_1\,l' ; 000)\, (k\,\lambda_2\,l ; 000)
\nonumber\\
&&\times 6 (-1)^l \sqrt{\hat s \hat {s^\prime}\hat j}
\bma l^\prime&\l&f \\ 
s^\prime&s&f\\ 
j'&j&0  \ema 
\bma 1/2 &1/2&1 \\ 
1/2&1/2&1\\ 
s^\prime&s&f \ema
\times (4t-3)\delta^{tt'}\,,
\end{eqnarray}
where $t$ $(t')$ is the total initial (final) isospin and $\hat z \equiv 2z+1$.
We introduced the function
\begin{equation}
g^f_k(p,p') = \int_{-1}^{1} dx\, P_k(x) \, V(q(x)) \, q^{2-f}(x)\,, 
\end{equation} 
in terms of the Legendre polynomials $P_k(x)$,
$q^2(x) \equiv p^2 +p^{\prime\,2}-2 p p^\prime x$, and the function
\begin{equation}
V(q) = -\frac{g_A^2}{4 F_\pi^2}\, \frac{1}{q^2 + m_\pi^2}\,.
\end{equation}
 
We solve the LS equation \eqref{LSshorthand}
for the potentials
 \begin{equation}
V_{\{{}^3P_{0},{}^3P_{1},{}^1D_{2}\}}(p^\prime, p) = 
\frac{1}{(2\pi)^3} \langle \{{}^3P_{0},{}^3P_{1},{}^1D_{2}\},\,p'|
V_\pi
| \{{}^3P_{0},{}^3P_{1},{}^1D_{2}\}, p\rangle \,,
\end{equation}
where
\begin{eqnarray}
\langle {}^3P_{0},\,p'|
V_\pi
| {}^3P_{0},\,p \rangle &=& \frac{2\pi}{3}
\left[ g^0_1(p,p') -4\left(p^2+p^{\prime\,2}\right) g^2_1(p,p') 
+ \frac{4}{3} \, p p' \left(g^2_2(p,p') + 5 g^2_0(p,p')\right)\right]\,,
\nonumber\\
\langle {}^3P_{1},\,p'|
V_\pi
| {}^3P_{1},\,p \rangle &=& \frac{2\pi}{3}
\left[ g^0_1(p,p') +2\left(p^2+p^{\prime\,2}\right)g^2_1(p,p') 
- \frac{2}{3} \, p p' \left(g^2_2(p,p') + 5 g^2_0(p,p')\right)\right]\,,
\nonumber\\
\langle {}^1D_{2},\,p'|
V_\pi
| {}^1D_{2},\,p \rangle &=& - 2\pi \,g^0_2(p,p')\,,
\end{eqnarray}
and extract the phase shifts from the solution of the T matrix. 

\begin{figure}
\includegraphics[width=1\textwidth]{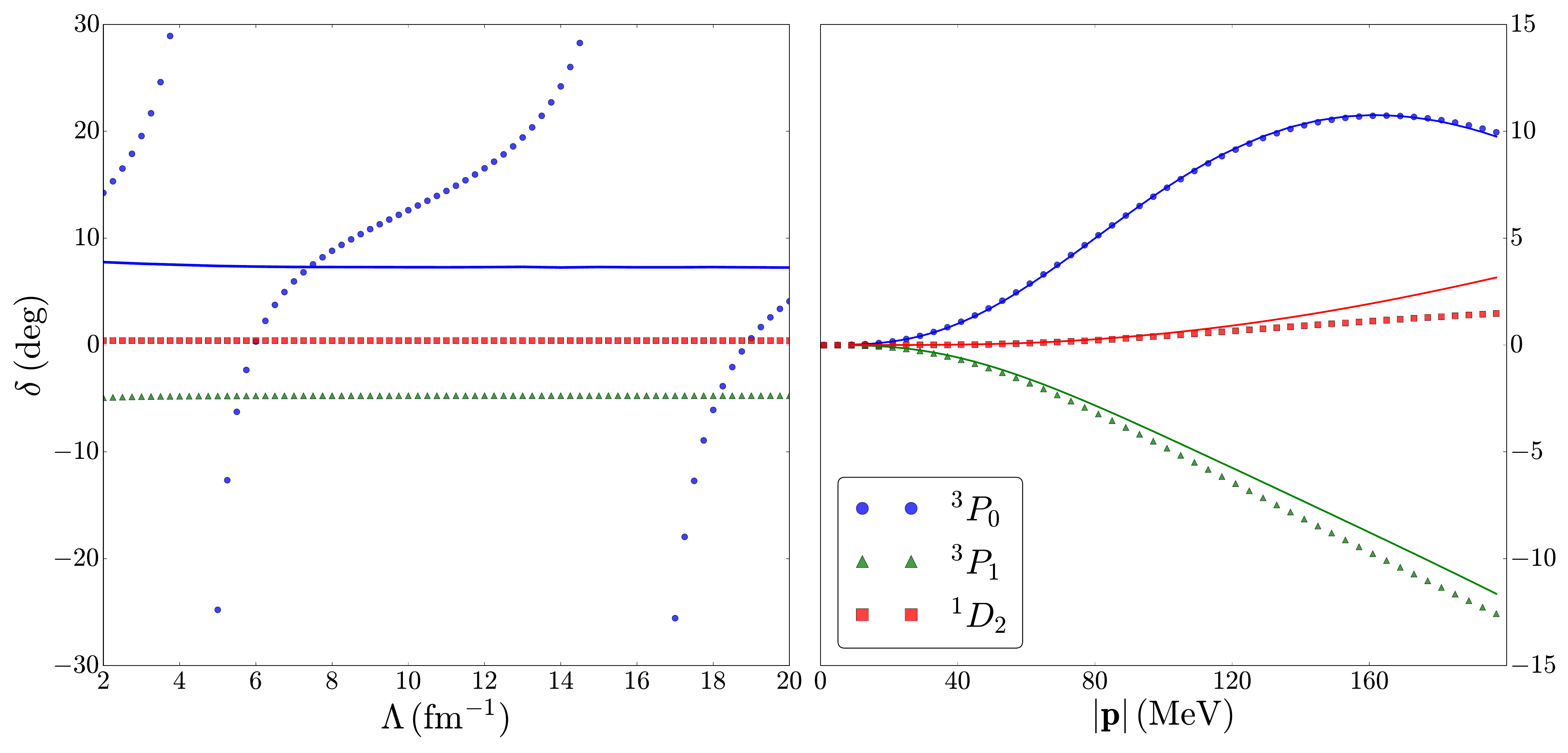}
\caption{Left panel: Phase shifts in the $^3P_{0,1}$ and $^1D_2$ channels as 
a function of the  momentum-space cutoff $\Lambda$. 
The green triangles, red squares, and blue circles denote respectively
the ${}^3P_1$, $^1D_2$, and ${}^3P_0$ phase shifts from the OPE potential.
The blue line is the ${}^3P_0$ phase shift with an additional counterterm.
Right panel: the $^1D_2$ (red squares)
$^3P_{1}$ (green triangles),
and renormalized $^3P_{0}$ (blue circles)
phase shifts as functions of the relative momentum of the nucleon pair 
$|\spacevec p|$,
compared to the Nijmegen partial-wave analysis (solid lines) 
\cite{Stoks:1993tb}.}
\label{PhaseP}
\end{figure}

As was found in Ref.~\cite{Nogga:2005hy}, the pure OPE potential leads to 
cutoff-independent phase shifts in the (repulsive) ${}^3P_1$ and 
(mildly attractive) ${}^1D_2$ channels, but not in the 
(attractive) ${}^3P_0$ channel. 
This behavior is illustrated in the left-panel of Fig.~\ref{PhaseP}, where the  
${}^1D_2$ and ${}^3P_1$ phase shifts at $p=100$ MeV are flat,
but the ${}^3P_0$ phase shift 
shows a limit-cycle-like behavior as a function of $\Lambda$.  
Following Ref.~\cite{Nogga:2005hy},
we promote a counterterm to LO in the ${}^3P_0$ channel---the coupling 
$C_{^3 P_0}$ in Eq.~\eqref{LOchiralLAG}---and
fit it to the phase shift at a centre-of-mass energy $E_{CM}=25$ MeV. 
The resulting phase shifts are essentially cutoff independent as depicted 
in the left panel of Fig.~\ref{PhaseP}. The phase shifts as a function of 
the relative momentum of the nucleons are depicted in the right panel of 
Fig.~\ref{PhaseP} and compared to the Nijmegen partial-wave analysis
\cite{Stoks:1993tb}. After promoting the ${}^3P_0$ counterterm, the phase 
shifts in all three channels are well described at LO in $\chi$EFT. 

\begin{figure}
\includegraphics[width=\textwidth]{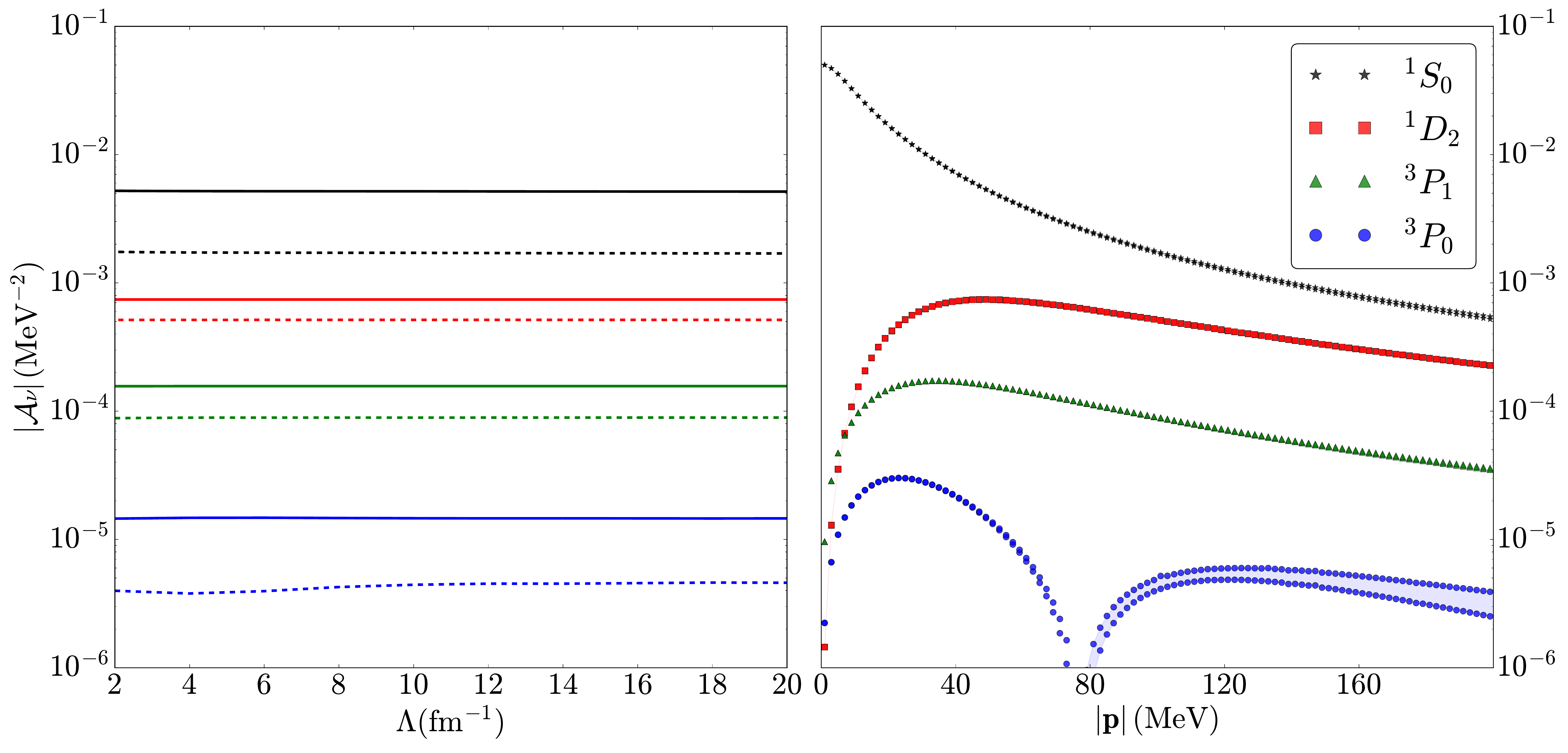}
\caption{ Left panel: absolute value of neutrino amplitude $\mathcal A_\nu$ for
the $^1S_0$ (black), $^1D_2$ (red), ${}^3P_1$ (green), and $^3P_{0}$ (blue)
channels as a function of momentum-space cutoff $\Lambda$. 
Solid (dashed) lines correspond to $|\spacevec p|=50$ MeV  
($|\spacevec p|=100$ MeV). 
Right panel: absolute value of neutrino amplitude $\mathcal A_\nu$ in the 
$^1S_0$ (black stars), ${}^1D_2$ (red squares), ${}^3P_1$ (green triangles), 
and ${}^3P_0$ (blue dots) as functions of 
the neutron center-of-mass momentum $|\spacevec p|$.
The bands (all except blue are nearly invisible)
represent cutoff variation in the range $2$ to $20$ fm$^{-1}$.}
\label{AP}
\end{figure}

Having renormalized the strong interaction in the $P$-wave channels, we now 
turn to the \nnpp amplitude. 
We calculate Eq.~\eqref{AnuLO}
for ${}^3P_{0,1} \rightarrow {}^3P_{0,1}$ and $^1D_2 \rightarrow ^1D_2$ 
transitions. We only consider the long-range neutrino potential and do not 
include additional short-range LNV counterterms. We observe in the left panel 
of Fig.~\ref{AP} that the resulting amplitudes are cutoff independent for the 
repulsive ${}^3P_1$ channel and the attractive ${}^3P_0$ channel,
as well as the mildly attractive ${}^1D_2$ channel. 
Despite the attractive singular nature of the strong $N\!N$
interaction in the ${}^3P_0$ channel, the neutrino amplitude is UV finite. 
We conclude we do not need to promote additional counterterms to LO. 
In the right panel of Fig.~\ref{AP} we plot the neutrino amplitude as a 
function of the neutron
momentum and observe that the $P$-wave amplitudes are small compared to the 
$S$-wave amplitude. The $D$-wave amplitude becomes 
relatively important at higher values of $|\spacevec p|$,
where the ${}^1S_0$ contribution has decreased significantly.

\section{The LNV scattering amplitude at next-to-leading order}
\label{LNV@NLO}

In this section we study NLO corrections to the LNV amplitude 
$nn \rightarrow p p\, e^- e^-$.
The main motivation to go to subleading order is the poor agreement between 
the observed phase shifts in the $^1S_0$ channel and the LO $\chi$EFT 
predictions, shown in Fig. \ref{PS1S0}. The agreement improves by including 
the contribution of the NLO operator $C_2$, and we want to study its impact 
on the \nnpp amplitude. In particular, we 
address the question whether a single counterterm $g_{\nu}^{\rm NN}$ 
is sufficient to renormalize the $\Delta L = 2$ scattering amplitude 
up to NLO. 

At NLO, $\slashpi$EFT and $\chi$EFT contain a single momentum-dependent 
contact interaction in the $^1S_0$ channel, the $C_2$ defined in 
Eqs. \eqref{eq:C2} and \eqref{NLOchiralLAG}.
All other corrections in the singlet channels are expected to be of 
higher order.
To study the strong and LNV scattering amplitudes in a generic scheme,
it is convenient to split the non-derivative  contact interactions 
$C$ and $g_{\nu}^{\rm NN}$ into LO and NLO pieces,
\begin{equation}
C = C^{(0)} + C^{(1)}, 
\qquad  
g_{\nu}^{\rm NN} = g_{\nu}^{\rm NN \,(0)} + g_{\nu}^{\rm NN \, (1)}, 
\end{equation}
with 
\begin{equation}
C^{(0)}  = \mathcal O\left(\frac{4 \pi}{m_N Q}\right),~
C^{(1)}  = \mathcal O\left(\frac{4 \pi}{m_N \Lambda}\right),
\qquad 
g_{\nu}^{\rm NN \, (0)} =  \mathcal O\left(\frac{1}{Q^2}\right),~
g_{\nu}^{\rm NN \, (1)} =  \mathcal O\left(\frac{1}{Q \Lambda}\right).
\end{equation}
Here $\Lambda = \Lambda_{\slashpi}$ in $\slashpi$EFT
and $\Lambda = \Lambda_\chi$ in $\chi$EFT, 
while $Q$ denotes the soft scale,
that is, $Q\sim \aleph$ in $\slashpi$EFT and $Q\sim m_\pi$ in $\chi$EFT.
This splitting does not lead to new LECs; it simply ensures that the LO 
fitting conditions are not affected by NLO corrections.
$C^{(1)}$ and $g_{\nu}^{\rm NN \, (1)}$  absorb power divergences induced by 
$C_2=O(4 \pi/(m_N Q^2 \Lambda))$ 
that appear both in $\slashpi$EFT and $\chi$EFT when using a cutoff scheme. 
In $\chi$EFT,
$C^{(1)}$ absorbs divergences induced by the pion-exchange potential. 
To simplify the notation, we will 
continue to drop the superscript $(0)$ from the LO counterterms.

The diagrams entering the LNV scattering amplitude at NLO are shown in the 
lower panels of Figs. \ref{Fig1} and \ref{Fig2}. 
In the notation of Sec.~\ref{LOpiless},
the NLO scattering amplitude 
takes the form 
\begin{eqnarray}
\mathcal A^{\rm NLO}_{\nu} 
&=& \mathcal A_A 
+ \chi^+_{\spacevec p^\prime}(\spacevec 0) \left( K_{E^\prime} + K^{(1)}_{E^\prime}\right)
 \mathcal A_B         
+ \mathcal {\bar A}_B \left(K_E + K_E^{(1)}\right) \chi^+_{\spacevec p}(\spacevec 0) 
\nonumber \\ 
& & +
\chi^+_{\spacevec p^\prime}(\spacevec 0) \left(   K_{E^\prime} + K^{(1)}_{E^\prime} \right)  
\left(\mathcal A_C + \frac{2 g_\nu^{\rm NN}}{
C^2} \right) 
\left(K_E + K_E^{(1)} \right) \chi^+_{\spacevec p}(\spacevec 0) 
\nonumber \\
& & + \chi^+_{\spacevec p^\prime}(\spacevec 0) \, K_{E^\prime} \, \mathcal A^{(1)}_B
+ \mathcal {\bar A}^{(1)}_B  \, K_E \, \chi^+_{\spacevec p}(\spacevec 0)
+ \chi^+_{\spacevec p^\prime}(\spacevec 0) \, K_{E^\prime} \, \mathcal A^{(1)}_C 
\, K_E \, \chi^+_{\spacevec p}(\spacevec 0)
\, ,  
\label{eq:amplitudeNLO}
\end{eqnarray}
where the superscript ${(1)}$ denotes NLO corrections, and terms quadratic 
in $K_{E,\, E^\prime}^{(1)}$ should be discarded. 
The first two lines in Eq. \eqref{eq:amplitudeNLO} subsume NLO corrections to 
the strong scattering amplitude, and yield a finite, regulator-independent 
result once the strong amplitude is renormalized. 
However, regulator dependence might appear in the remaining terms,
shown in the third line.

To address this possible regulator dependence, we will also include the 
derivative operator $g_{2\nu}^{\rm NN}$, defined in Eq. \eqref{g2nudef}.
This operator only involves the $S$ wave, so when it is inserted into a 
bubble chain it will not cause any mixing 
with other partial waves. We therefore expect \cite{Bedaque:2002mn}
$g_{2\nu}^{\rm NN}$ to be proportional to $
C^2$
in the same way as $g_{\nu}^{\rm NN}$
\cite{Cirigliano:2017tvr},
and we define the rescaled coupling 
\begin{equation}
\tilde g^{\rm NN}_{2 \nu}= \left(\frac{4\pi}{m_N
C}\right)^2 g_{2 \nu}^{\rm NN}
\label{g2rescaled}
\end{equation}
in analogy to Eq. \eqref{gtilde}.
On the basis of the NDA~\cite{Bedaque:2002mn}, we expect 
the ratio of couplings to scale as
\begin{equation}
\frac{g_{2\nu}^{\rm NN}}{g_{\nu}^{\rm NN}} = 
\mathcal O\left(\frac{1}{\Lambda^2}\right),
\label{expectation}
\end{equation}
implying that $g_{2\nu}^{\rm NN}$ contributes at N$^2$LO. 
We will 
examine whether the renormalization of the full neutrino potential 
fulfills 
this 
expectation.

In $\slashpi$EFT, the diagrams
can be analytically resummed in any scheme where additional loops arising from 
insertions of contact interactions factorize. One example is the momentum 
scheme introduced in Eq. \eqref{Lambda1}.
In $\chi$EFT, because of the iteration of the pion-exchange potential, loop 
diagrams in general do not factorize, and we require a numerical solution.
In dimensional regularization, however, the 
structure of the diagrams is simple enough that it is possible to give 
analytical expressions, which closely resemble those of $\slashpi$EFT.
We start by discussing the amplitude at NLO in $\slashpi$EFT
in Sec. \ref{pilessNLO}.
and extend the discussion to $\chi$EFT with 
dimensional and cutoff 
regularization in 
Secs. \ref{chiralNLOdimreg} and \ref{pifullNLO}, respectively.

\subsection{Pionless EFT}
\label{pilessNLO}

In $\slashpi$EFT, $K_E + K_E^{(1)}$ correspond to the full $^1S_0$ NLO scattering
amplitude,
\begin{equation}
\left( K_E + K^{(1)}_{E} \right)_{\slashpi} 
=  \frac{1}{C^{-1} + I_0(\spacevec p)} 
\left[ 1 + 
\left(\frac{C_2}{ C^2} \spacevec p^2 + \frac{C_2}{C} \delta I_0 
+ \frac{C^{(1)}}{C^2} \right) \frac{1}{ C^{-1} + I_0(\spacevec p)} 
\right],
\label{eq:strong}
\end{equation}
where $I_0$ is defined in Eq. \eqref{I0}
and $\delta I_0$ is an integral that vanishes in dimensional regularization
but is non-zero when using a momentum cutoff,
\begin{equation}
\delta I_0 = - m_N \int \frac{d^{3} k}{(2\pi)^{3}} 
\exp\left[-2 \frac{(\spacevec k^2)^2}{\Lambda^4}\right]  
\propto  m_N \Lambda^3\,.
\end{equation}
Its precise value is not important 
because the choice
\begin{equation}
C^{(1)}= - C_2 \, C \,\delta I_0
\label{C(1)pionless}
\end{equation}
exactly cancels 
its contribution. 
With this choice, the scattering length is not affected by NLO corrections,
while $C_2$ is fixed by the effective range. 
Using Eq. \eqref{C0C2} we obtain
\begin{equation}
\left( K_E + K^{(1)}_{E} \right)_{\slashpi} =   \frac{4 \pi}{m_N} 
\frac{1}{1/a + i |\spacevec p|} 
\left( 1 + \frac{r_0 \spacevec p^2}{2}\frac{1}{1/a + i |\spacevec p|}\right)\,,
\label{eq:strong1}
\end{equation}
which is regulator independent. Thanks to $\tilde g_{\nu}^{\rm NN}$,
which compensates for the regulator dependence of ${\cal A}_C$, 
the first two lines of
Eq. \eqref{eq:amplitudeNLO} are indeed independent of the regularization
scheme. 

The remaining corrections to the LNV amplitude 
in Eq. \eqref{eq:amplitudeNLO} are given by
\begin{eqnarray}
\mathcal A_B^{(1)} &=& \frac{C_2}{2 C}\,  I_1(\spacevec p^{\prime\, 2})\,, 
\qquad 
\bar{\mathcal A}^{(1)}_B =  \frac{C_2}{2 C}\, I_1(\spacevec p^2) \, , 
\label{AB(1)pilesscutoff}
\\
\mathcal A_C^{(1)} &=&  
\left(-\frac{4g_{\nu}^{\rm NN}}{C^2}\frac{C_2}{C}+\frac{2 g_{2\nu}^{\rm NN}}{C^2}\right)
\frac{\spacevec p^2 + \spacevec p^{\prime\,2}}{2} 
+\frac{2g_{\nu}^{\rm NN \, (1)}}{C^2} 
+2 \left(C_2 \frac{g_\nu^{\rm NN}}{C^2}  
+\frac{g_{2\nu}^{\rm NN}}{C}\right) \delta I_0    
-\frac{C_2}{C} I_2(\spacevec p^2, \spacevec p^{\prime\, 2}),  \quad 
\label{AC(1)pilesscutoff}
\end{eqnarray}
where we used Eq. \eqref{C(1)pionless}
and defined the cutoff-regularized integrals
\begin{eqnarray}
I_1 (\spacevec p^2) &=&  m_N \int \frac{d^3 k}{(2\pi)^3 } 
\frac{1+3g_A^2}{(\spacevec p - \spacevec k)^2} 
\, \exp \left[- \left(\frac{\spacevec k^2}{\Lambda^2} \right)^2\right]\,,  
\nonumber \\
I_2 (\spacevec p^2, \spacevec p^{\prime\, 2}) &=& - \frac{m_N^2}{2} 
\int \frac{d^3 k_1}{(2\pi)^3 } \int \frac{d^3 k_2}{(2\pi)^3 } 
\frac{1 + 3 g_A^2}{(\spacevec k_1 - \spacevec k_2)^2} 
\frac{1}{\spacevec p^2 - \spacevec k_2^2} 
\, \exp \left[- \left(\frac{\spacevec k_1^2}{\Lambda^2}\right)^2\right] 
\, \exp \left[- \left(\frac{\spacevec k_2^2}{\Lambda^2}\right)^2\right]  
\nonumber \\ 
& & + (\spacevec p \rightarrow \spacevec p^\prime)\,. 
\end{eqnarray}

In dimensional regularization these integrals vanish. 
There is, as a consequence, no scale dependence 
other than in the $\spacevec p^2 + \spacevec p^{\prime\,2}$ term
of $\mathcal A_C^{(1)}$, and we can take 
\begin{equation}
g_{\nu}^{\rm NN \, (1)}=0 \,.
\end{equation}
Since in the absence of the derivative counterterm $g_{2\nu}^{\rm NN}$
the $\spacevec p^2 + \spacevec p^{\prime\,2}$ term in
Eq. \eqref{AC(1)pilesscutoff}
is $\mu$-dependent, 
the two-derivative operator in Eq. \eqref{g2nudef} is 
required to appear at NLO.
It obeys the RGE 
\begin{equation}
\frac{d}{d\ln\mu}
\left( \tilde g^{\rm NN}_{2\nu} 
-2\eta \, C  \,\tilde g_{\nu}^{\rm NN} \right) = 0\,,
\label{eq:rgg2}
\end{equation}
where we used Eq. \eqref{C0C2} and introduced the dimensionless combination 
\begin{equation}
\eta = \frac{m_N r_0}{8\pi}\,.
\label{eta}
\end{equation}
The solution of this RGE is 
\begin{equation}
\tilde g^{\rm NN}_{2\nu} (\mu) = 2 \eta\, C(\mu) \, \tilde g_{\nu}^{\rm NN}(\mu) 
+ \tilde g^0_{2\nu},
\label{solNLO}
\end{equation}
where $\tilde g^0_{2\nu}$ is an integration constant,
and Eqs. \eqref{AB(1)pilesscutoff} and \eqref{AC(1)pilesscutoff} reduce to 
\begin{eqnarray}
\mathcal A_B^{(1)} &=& 0 \,, 
\qquad 
\bar{\mathcal A}^{(1)}_B = 0 \,, 
\label{AB(1)dimreg}
\\
\mathcal A_C^{(1)} &=&  
\mathcal A^{(1)}_{C} = \left(\frac{m_N}{4\pi}\right)^2 2 \tilde g^0_{2\nu} \,
\frac{\spacevec p^2 + \spacevec p^{\prime\,2}}{2} \,, 
\label{AC(1)dimreg}
\end{eqnarray}
which are independent of the renormalization scale.
NDA rules modified to take into account 
S-wave enhancements~\cite{Bedaque:2002mn} imply that
\begin{equation}
 \tilde g^0_{2\nu} = \mathcal O\left(\frac{1}{\Lambda^2_{\slashpi}}\right)\,, 
\label{natural}
\end{equation}
so that Eq. \eqref{AC(1)dimreg} is actually an N${}^2$LO correction. 
So we find that in dimensional regularization with PDS scheme  the coupling $g_{2\nu}^{\rm NN}$  
involves an NLO piece fixed in terms of LO quantities by Eq.~\eqref{eq:rgg2} 
and an N$^2$LO piece parameterized by the constant $ \tilde g^0_{2\nu}$,  whose scaling 
is determined by NDA.    
This non-homogeneous scaling of $g_{2\nu}^{\rm NN}$ is analogous to the one of the four-derivative operator $C_4$
in the strong-interaction Lagrangian, which enters the $N\!N$
scattering amplitude with a fixed coefficient at N$^2$LO, while a new LEC
related to the shape parameter appears at
N$^3$LO \cite{Kaplan:1998tg,vanKolck:1998bw}.
As with an infinite number of other LECs, we cannot {\it a priori}
exclude an enhancement of $ \tilde g^0_{2\nu}$ over NDA,
which could make it NLO or even LO,
but currently we lack evidence for it. Equations \eqref{solNLO} and \eqref{natural} imply that the only NLO corrections to $\mathcal A_\nu$  come in through the strong scattering amplitude $T_{^1S_0}$.



We will now see that this argument is corroborated by a different choice
of regularization.
With a momentum cutoff, 
$I_1$ contains a momentum-independent linear divergence, 
while $I_2$ 
has a logarithmic divergence proportional to the energies
in addition to a momentum-independent quadratic divergence:
\begin{eqnarray}
&&I_1 (\spacevec p^2) \propto m_N \Lambda \,,
\\
&&I_2 (\spacevec p^2, \spacevec p^{\prime\, 2}) \propto m_N^2 
\left( \Lambda^2  + \kappa \frac{\spacevec p^2 + \spacevec p^{\prime\, 2}}{2} 
\ln\frac{\Lambda^2}{\Lambda_0^2}\right)\,,
\end{eqnarray}
where $\kappa$ is a dimensionless constant and $\Lambda_0$ is a constant with dimensions of momentum.
We can thus write
\begin{eqnarray}
&&I_1 (\spacevec p^2) = I_1(0) 
+ {\mathcal O}\left(\frac{m_N \spacevec p^2}{\Lambda}\right)\,,
\\
&&I_2 (\spacevec p^2, \spacevec p^{\prime\, 2}) =
I_2 (0, 0)
+ \frac{\spacevec p^2+\spacevec p'^2}{2}
\left[\left(\frac{\partial}{\partial \spacevec p^2}
+\frac{\partial}{\partial \spacevec p'^2}\right)
I_2 (\spacevec p^2, \spacevec p^{\prime\, 2})\right]_{\spacevec p^2 = \spacevec p^{\prime\, 2} = 0}
+ {\mathcal O}
\left(\frac{m_N^2}{\Lambda^2}
\left(\frac{\spacevec p^2 + \spacevec p^{\prime\, 2}}{2} \right)^2\right) 
\,. \nn\\
\end{eqnarray}
In a cutoff scheme, Eq. \eqref{C0C2} holds with $\mu\to c\Lambda$,
the value of $c$ depending on the choice of regulating function.
With 
$C \propto (m_N \Lambda)^{-1}$ and $C_2 \propto (m_N \Lambda^2)^{-1}$, 
we see that $\mathcal A_B^{(1)}$ is finite as $\Lambda \rightarrow \infty$.
The regulator dependence in the momentum-independent terms of 
$\mathcal A_C^{(1)}$,
\begin{equation}
\frac{2g_{\nu}^{\rm NN \, (1)}}{C^2} 
+2 \left(C_2 \frac{g_\nu^{\rm NN}}{C^2}  
+\frac{g_{2\nu}^{\rm NN}}{C}\right) \delta I_0    
-\frac{C_2}{C} I_2(0,0)\,,
\end{equation}
can be absorbed by a shift in the NLO LEC $g_{\nu}^{\rm NN \,(1)}$.
The terms proportional to $\spacevec p^2 + \spacevec p^{\prime\,2}$ 
converge as the cutoff is sent to infinity, albeit slowly,
\begin{equation}
\frac{\spacevec p^2 + \spacevec p^{\prime\,2}}{2} 
\left\{ -\frac{4C_2}{C} \frac{g_{\nu}^{\rm NN}}{C^2} 
- \frac{C_2}{C} \left[
\left( \frac{\partial}{\partial \spacevec p^2} 
+ \frac{\partial}{\partial \spacevec p^{\prime 2}} \right) 
I_2(\spacevec p^2, \spacevec p^{\prime\, 2}) \right]_{\spacevec p^2 = \spacevec p^{\prime\, 2}=0} \right\}
\propto
\frac{\spacevec p^2 + \spacevec p^{\prime\,2}}{2} 
\frac
{\ln \Lambda }{\Lambda} 
\,.
\label{limit}
\end{equation}
Since higher powers of $\spacevec p^2 + \spacevec p^{\prime\,2}$
converge as well, we conclude that 
in a cutoff scheme there is also no need to include an independent
parameter at NLO.
$g_{2\nu}^{\rm NN}$ can be included at this order with a fixed coefficient 
(as in dimensional regularization) 
as part of an ``improved action'' where cutoff artifacts
scale more favorably as $1/\Lambda$ 
(instead of 
$\ln(\Lambda)/\Lambda$) and is thus of the same
size as corrections that scale with the inverse of the breakdown scale.

In conclusion, the NLO analysis of $\mathcal A_\nu$ in both dimensional and 
cutoff regularizations shows that there appears no new independent LEC in the
NLO neutrino potential. In dimensional regularization, $g_{2\nu}^{\rm NN}$ must 
be introduced to guarantee that $\mathcal A_\nu$ is scale independent, but its 
value is fixed by Eq. \eqref{solNLO} in terms of $g_{\nu}^{\rm NN}$, 
the $^1S_0$ scattering length, and the ${}^1S_0$ effective range. 
In a cutoff scheme, Eq. \eqref{limit} guarantees that for large $\Lambda$ the 
amplitude is correctly renormalized, after momentum-independent 
power divergences are absorbed by a redefinition of the 
LEC $g_{\nu}^{\rm NN}$.
However, $g_{2\nu}^{\rm NN}$ with a cutoff dependence 
fixed by the same parameters as in dimensional regularization
ensures that the error from $\Lambda$ at the breakdown scale is not
unusually large.

\subsection{Chiral EFT with dimensional regularization}
\label{chiralNLOdimreg}

In $\chi$EFT, the NLO correction to the strong scattering amplitude encoded in
\begin{equation}
K_E^{(1)} = K^2_E  \left[\frac{C^{(1)}}{
C^2} 
+ \frac{C_2}{
C^2} \left(\spacevec p^{2} - m_N V_\pi(0) \right) \right]
\end{equation}
contains the additional contribution from the
dimensionally regulated pion potential in coordinate space evaluated at the origin,
\begin{eqnarray}
V_\pi(0) &=& -\frac{g_A^2}{4 F_\pi^2} 
\int \frac{d^{d-1} k}{(2\pi)^{d-1}} \frac{m_\pi^2}{\spacevec k^2 + m_\pi^2} 
= - \frac{g^2_A m_\pi^2}{16 \pi  F_\pi^2}  \left(\mu - m_\pi \right)\,.
\label{Vpiat0}
\end{eqnarray}
The $\mu$ dependence 
signals that the integral is linearly divergent in the PDS scheme. 
The $\mu$ independence of the strong scattering amplitude implies
\begin{equation}
\frac{d}{d\ln\mu}
\left( \frac{C_2}{
C^2} \right) = 0\,,
\end{equation}
but $C_2$ no longer has the simple expression in terms of the effective range 
given in Eq. \eqref{C0C2} due to explicit pion-exchange contributions. Since
 $V_\pi(0)$ does not depend on the nucleon momenta, we can choose 
\begin{equation}
C^{(1)}= m_N \, C_2 \, V_\pi(0)
\label{C(1)}
\end{equation} 
to cancel  the linearly divergent terms. This choice ensures that NLO 
corrections do not change the scattering length.

As in $\slashpi$EFT, the renormalization of the strong scattering amplitude 
implies that the first two lines in Eq. \eqref{eq:amplitudeNLO} are scale 
independent. 
The functions $\mathcal A^{(1)}_B$ and $\mathcal A^{(1)}_C$ are now given by
\begin{eqnarray}
\mathcal A_B^{(1)}  &=& \bar{ \mathcal A}_B^{(1)} = 0 \,,
\\
\mathcal A_C^{(1)}  &=&  
\left( -\frac{4 C_2}{
C}\, \frac{g_{\nu}^{\rm NN}}{
C^2}  
+ \frac{2 g_{2\nu}^{\rm NN}}{
C^2} \right) 
\frac{\spacevec p^2 + \spacevec p^{\prime 2}}{2}
+ \frac{2g^{\rm NN (1)}_{\nu}}{
C^2}    
- 2 m_N V_\pi(0) \, \frac{g_{2\nu}^{\rm NN}}{
C^2}
+ \frac{C_2}{
C^2} \, m_N V_{\nu\, \rm L}(0)\,,
\label{eq:nlo}
\end{eqnarray}
where 
\begin{equation}
V_{\nu\, \rm L}(0) = \int \frac{d^{d-1} k}{(2\pi)^{d-1}} V^{^1S_0}_{\nu\, \rm L}(\spacevec k) 
=  \frac{g_A^2 m_\pi}{8\pi}\,
\label{someintegral}
\end{equation}
is the dimensionally regulated neutrino-exchange potential evaluated 
at the origin.
$V_\nu(0)$ is finite in $\overline{\rm MS}$ and PDS, but would be linearly 
divergent in a cutoff scheme.

The subleading, momentum-independent $g^{\rm NN\, (1)}_{\nu}$ can be chosen to 
cancel the last three terms in Eq. \eqref{eq:nlo}. This choice implies that 
once $g_{\nu}^{\rm NN}$ is fitted to reproduce $\mathcal A_\nu$ at 
$\spacevec p = 0$, its value is not affected by NLO corrections. 
Finally, the momentum dependent piece leads to the same RGE as in $\slashpi$EFT,
\begin{equation}
\frac{d}{d\ln\mu}
\left( \tilde g^{\rm NN}_{2\nu}  
- \frac{2C_2}{
C} \tilde g_{\nu}^{\rm NN} \right) = 0\,.
\end{equation}
We conclude that also in $\chi$EFT $g_{2\nu}^{\rm NN}$ is completely determined
at NLO by $g_{\nu}^{\rm NN}$ (in terms of $C_2$ and $
C$), and new 
independent parameters appear only at N${}^2$LO or higher.

\subsection{Chiral EFT with cutoff regularization}
\label{pifullNLO}

Depending on the subtraction scheme, certain positive powers of a momentum 
cutoff have no analog in dimensional regularization. 
As a consequence, the need for a LEC at a given order might not be apparent 
in this regularization scheme, while it is in a cutoff scheme. 
We now check that the conclusion
reached about $g_{2\nu}^{\rm NN}$ in $\chi$EFT
does not depend on dimensional regularization.
We 
repeat the analysis of Sec. \ref{chiralNLOdimreg}
for the cutoff schemes introduced in Sec. \ref{cutoffLO}. 

In coordinate space, the amplitude at NLO is obtained by computing the integral
\eqref{ALNV}
where now 
$\psi^{+}_{\spacevec p}(\spacevec r)= \psi^{+ \, (0)}_{\spacevec p}(\spacevec r)  
+  \psi^{+\, (1)}_{\spacevec p}(\spacevec r)$,
with $\psi^{+ \, (0)}_{\spacevec p}(\spacevec r)$ the LO wavefunction
and 
\begin{equation}
\psi^{+\, (1)}_{\spacevec p}(\spacevec r) = \frac{1}{E - H + i \varepsilon} 
\left[- \frac{C_2}{2} 
\left(\overleftarrow{\boldsymbol \nabla}^2  \delta^{(3)}_{R_S}(\spacevec r)
+  \delta^{(3)}_{R_S}(\spacevec r) \overrightarrow{\boldsymbol \nabla}^2 \right) 
+ C^{(1)} \delta^{(3)}_{R_S}(\spacevec r) \right] 
\psi^{+\, (0) }_{\spacevec p}(\spacevec r) 
\end{equation}
the NLO correction, where $H$ is the LO Hamiltonian.
To work consistently at NLO, we expand Eq. \eqref{ALNV} and neglect terms 
quadratic in $\psi^{\pm (1)}_{\spacevec p, \spacevec p^\prime}$.
As in the momentum-space treatment below, $C_2$ and $C^{(1)}$ induce 
power-divergent corrections in the amplitude,
which can be absorbed by introducing $g_{\nu}^{\rm NN \,(1)}$ in perturbation
theory.

We also consider a momentum cutoff, where we start by solving 
the LS equation \eqref{LSshorthand}. Schematically, in first order in the NLO
strong-interaction potential \eqref{eq:1},
\begin{equation}
T^{(1)} = V^{(1)} + V^{(1)} G_0 T + T G_0 V^{(1)} + T G_0 V^{(1)} G_0 T\,,
\end{equation}
where $T$ denotes the LO T matrix.
This NLO correction to the T matrix induces a correction 
\begin{equation}
S_{{}^1\!S_0}^{(1)}(E) = - i \pi m_N q_0\,T_{{}^1\!S_0}^{(1)}(q_0,q_0,E)
\label{SNLO}
\end{equation}
in the S matrix in the ${}^1\!S_0$ channel.
We introduce the NLO phase shifts as 
\begin{equation}
e^{2i (\delta_{{}^1\!S_0}(E) + \delta_{{}^1\!S_0}^{(1)}(E))} = S_{{}^1\!S_0}(E) + S_{{}^1\!S_0}^{(1)}(E) 
\qquad \rightarrow \qquad 
\delta_{{}^1\!S_0}^{(1)}(E) = \frac{1}{2i} \,
\frac{S_{{}^1\!S_0}^{(1)}(E) }{S_{{}^1\!S_0}(E)}
\,,
\end{equation}
where $S_{{}^1\!S_0}(E)$ is the LO S matrix given by Eq. \eqref{SLO}.

We now fit $C_2$ and $C^{(1)}$ by demanding that the 
scattering length, which was already correctly described at LO, be unaffected 
and, simultaneously, by fitting the ${}^1S_0$ phase shift at 
$|\spacevec p| = 30$ MeV. 
More details of this procedure can be found in Ref. \cite{Long:2012ve}. 
The resulting $np$ phase shifts with momentum and dimensional
regularizations are shown in Fig. \ref{PS1S0}. 
Compared to LO, significantly better agreement with 
the Nijmegen partial-wave analysis \cite{Stoks:1993tb}
is obtained, but
there is plenty of room for further improvement at higher orders.
Results for the coordinate-space regulator are similar.

Having obtained the NLO T matrix, $T^{(1)}$, the calculation of the NLO 
neutrino amplitude is straightforward. 
Expanding Eq.~\eqref{AnuLO} to first order
\begin{eqnarray}
\mathcal A^{(1)}_{\nu} &=&  -2 \pi^2 
\left(V_\nu G_0 T^{(1)} + T^{(1)} G_0 V_\nu + T^{(1)} G_0 V_\nu G_0 T   
+ T G_0 V_\nu G_0 T^{(1)} 
\right.\nonumber\\
&&\left.
+V^{(1)}_\nu + V^{(1)}_\nu G_0 T + T G_0 V^{(1)}_\nu  + T G_0 V^{(1)}_\nu G_0 T \right)
\,,
\label{A(1)nu}
\end{eqnarray}
one sees that there are two types of corrections.
The first type comes from the perturbative insertion of the NLO $T$ matrix. 
The contributions from $C_2$ and $C^{(1)}$ to $T^{(1)}$ induce power-divergent 
corrections to the amplitude. These can be absorbed by introducing 
$g_{\nu}^{\rm NN \,(1)}$, the momentum-independent NLO counterterm that corresponds 
to the NLO neutrino potential $V^{(1)}_\nu$. 
This piece then gives a second type of correction to the NLO amplitude.

The NLO counterterm $g_{\nu}^{\rm NN \,(1)}$ is fitted by demanding that 
$\mathcal A^{(1)}_\nu(|\spacevec p| =1\,\mathrm{MeV}) = 0$, 
such that the (arbitrary) LO fit condition at this energy 
employed in Sec.~\ref{LNVLO} is not affected.
We stress that $g_{\nu}^{\rm NN \,(1)}$ does not correspond to a new LEC but simply 
to a perturbative shift in the LO LEC.
Only the sum $g_{\nu}^{\rm NN} + g_{\nu}^{\rm NN \,(1)}$ is relevant. 
In practice, 
$g_{\nu}^{\rm NN} + g_{\nu}^{\rm NN\, (1)}$, is quite 
different from $g_{\nu}^{\rm NN}$:
even in the limited range of cutoffs commonly used in the literature, 
$R_S \sim 0.5 - 0.7$ fm, they differ
by a factor of 2.
While such variation is not unexpected in cutoff schemes, and has no effect 
on the observable $\mathcal A_\nu$, it highlights the importance of using 
consistent nuclear interactions in the extraction of $g_{\nu}^{\rm NN}$ and 
the calculation of $0\nu\beta\beta$ nuclear matrix elements. 

\begin{figure}
\includegraphics[width=\textwidth]{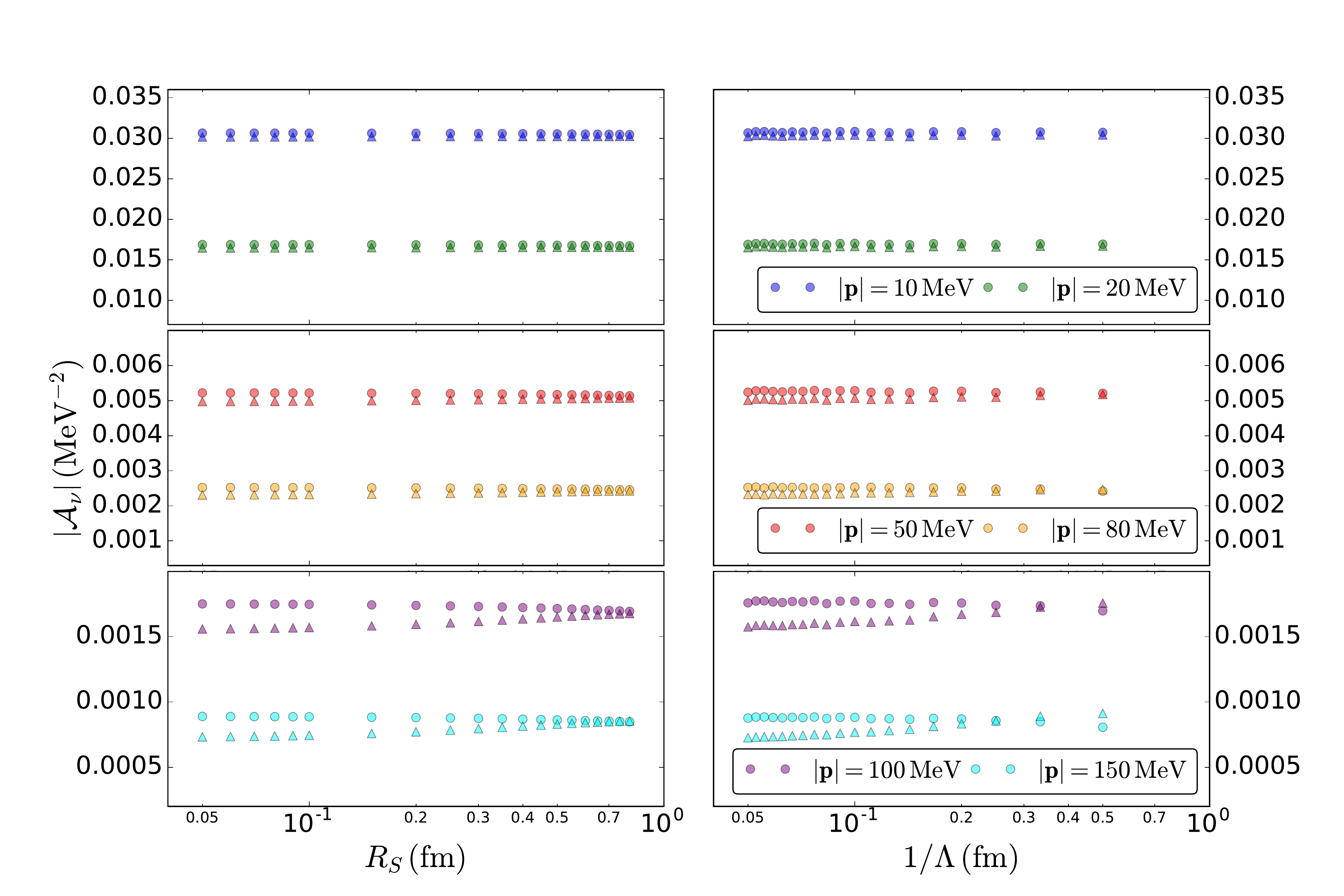}
\caption{Magnitude of the LNV matrix element $\mathcal A_{\nu}$
at various values of the neutron center-of-mass momentum:
$|\spacevec p| = 10$, $20$, $50$, $80$, $100$, and $150$ MeV.
Left and right panels show $\mathcal A_{\nu}$ as function of
coordinate- and momentum-space regulators, respectively. 
Circles and triangles denote results at, respectively, LO and NLO.
For the purpose of illustration, the 
LNV counterterm 
is determined  by imposing the (arbitrary) condition  
$\left| \mathcal A^{}_\nu \right| = 0.05$ MeV$^{-2}$ at $|\spacevec p| = 1$ MeV.
}\label{ALNV_RS_1}
\end{figure}

The magnitude of the resulting $\Delta L=2$ scattering amplitude,
\begin{equation}
|\mathcal A_\nu| = | \mathcal A_\nu^{(0)}| 
+ \frac{1}{| \mathcal A_\nu^{(0)}|} 
\textrm{Re} \left(\mathcal A_\nu^{(0) *} \mathcal A_\nu^{(1)}\right) 
+ \dots\,,
\label{ampNLO}
\end{equation}
is shown in Fig. \ref{ALNV_RS_1} up to NLO 
for six values of $|\spacevec p|$ 
(namely  $10,\,20,\, 50,\, 80,\, 100,\, 150$ MeV). 
The left panels correspond to the coordinate-space regulator and 
the right panels to the momentum-space regulator. 
Both schemes agree 
very well. 
LO results are the same as in Fig. \ref{LNV_LO} and given for comparison.
NLO corrections to the amplitude are small and, more importantly, 
cutoff independent for sufficiently large cutoff. 
There is no numerical evidence for the need of an NLO counterterm. 
This observation is in agreement with the analysis in $\slashpi$EFT 
with a hard cutoff, which showed that $g^{\rm NN}_{2 \nu}$ is 
not needed for convergence as the cutoff increases.
It is also consistent with the analysis in dimensional regularization in 
both $\slashpi$EFT and $\chi$EFT, where it was concluded that 
no new 
LNV parameters appear until N${}^2$LO.

Of course, in the absence of data one cannot be sure $g^{\rm NN}_{2 \nu}$
is not numerically large because of some fine tuning at small
distances. Our arguments only show that there is no 
renormalization-group reason for it to be enhanced with respect to the estimate
\eqref{expectation}.
In cutoff-regulated $\chi$EFT too, 
$g^{\rm NN}_{2 \nu}$ could be included to 
accelerate convergence, but given the numerical nature of the calculation
it could only be determined after the slow-converging results are obtained
for one nucleus.
Such an improved action would only be useful as input for
calculations on a different nucleus.

\section{The connection to charge-independence breaking
}
\label{CIB}

The analysis of Sec. \ref{LNVLO} shows that matrix elements of the long-range 
neutrino potential $V_{\nu\, {\rm L}}$, defined in Eq. \eqref{eq:Vnu0}, are 
ultraviolet divergent. 
The amplitude can be made independent of the UV regulator only by including 
at LO a short-range neutrino potential parametrized by $g_\nu^{\rm NN}$.
While one can determine the 
dependence of $g^{\rm NN}_\nu$ on the renormalization scale $\mu$ or on the 
cutoff $\Lambda$ or $R_S$, 
knowledge of the finite piece of the LEC is necessary to make predictions 
for the $0\nu\beta\beta$ half-lives in terms of the effective neutrino 
Majorana mass $m_{\beta \beta}$.  
The argument in Sec. \ref{LNV@NLO} then shows that $g^{\rm NN}_\nu$ is the only 
LNV input needed up to NLO.
It can
in principle be extracted by matching the scattering amplitude for \nnpp
in $\chi$EFT to LQCD. Such LQCD calculations are extremely 
challenging \cite{Cirigliano:2019jig}, but are beginning to be investigated. 
For instance, Ref.~\cite{Feng:2018pdq} calculated the LEC associated 
to an N$^2$LO LNV pion-electron coupling.
In the absence of LQCD results,
we discuss here how the size of LNV LECs, including $g_{\nu}^{\rm NN}$, 
can be estimated by studying 
their relation to analogous counterterms that are needed to describe 
isospin-breaking effects.

\subsection{The $I=2$ electromagnetic Lagrangian}
\label{CIBLag}

In Sec. \ref{LNVLag} we derived the long-range neutrino potential, and 
discussed the form of short-range operators mediated by hard-neutrino exchange. 
We now explore the formal relation between LNV interactions and 
electromagnetic charge-independence breaking (CIB).

The starting point is the quark-level electromagnetic and weak Lagrangian
\begin{equation}\label{quark}
\mathcal L = \bar q_L \gamma^\mu \left(l_\mu + \hat{l}_\mu \right) q_L  
+ \bar q_R \gamma^\mu \left(r_\mu + \hat{r}_\mu\right) q_R  
\,,
\end{equation}
where $q$ denotes the quark doublet 
$q = (u\; d)^T$ and we defined
\begin{eqnarray}
l_\mu = \frac{e}{2} A_\mu \, \tau^3 
- 2 \sqrt{2}\, G_F \left[V_{ud} \, \bar e_L \gamma_\mu  \nu_L \, \tau^+  + {\rm H.c.}\right]\,,  
& & \qquad \hat{l}_\mu = \frac{e}{6} A_\mu\,, 
\\
r_\mu = \frac{e}{2} A_\mu \, \tau^3\,,  
& & \qquad \hat{r}_\mu = \frac{e}{6} A_\mu\,.
\end{eqnarray}
We neglect weak neutral-current interactions that are not relevant to the 
present discussion.
This Lagrangian gives rise to long-distance effects through couplings of 
photons and leptons to nucleons and pions. 
It induces the following one-body isovector  amplitude 
\begin{equation}
\mathcal A = \bar N  \left[\frac{l_\mu+r_\mu}{2}J_V^\mu+\frac{l_\mu-r_\mu}{2} J_A^\mu \right] N ,
\end{equation}
where  $J_V^\mu$ and $J_A^\mu$ are the vector and axial currents of  Eq.\ \eqref{eq:currents}.
In addition, short-range operators are generated by the insertion of two 
currents 
connected by the exchange of hard photons (in the electromagnetic case) 
or neutrinos (in the case of \NLDBD). For \NLDBD\ this mechanism gives rise 
to the $N\!N$
interactions of Sec. \ref{LNVLag} and additional $\pi N$ and $\pi\pi$ 
interactions, 
while electromagnetism (EM) induces very similar isospin $I=2$ interactions. 
This analogy between the two cases can be made precise by noticing that 
the insertion of two weak currents connected by a neutrino propagator with 
a single insertion of $m_{\beta\beta}$ leads to a massless (up to neutrino-mass 
corrections) boson propagator (in the Feynman gauge). 
The exchange of hard neutrinos therefore leads to identical contributions 
as photon exchange, up to an overall factor~\cite{Cirigliano:2017tvr}: 
hard-neutrino exchange is multiplied by $8G_F^2 V_{ud}^2 m_{\bt\bt}\bar e_L e^c_L$ 
compared to the usual $e^2$ in the EM case.
To elucidate this relation, we first construct the chiral Lagrangian in the 
$\pi\pi$ and $\pi N$ sectors for the LNV and EM cases, before discussing the 
short-range $N\!N$ interactions.

To construct operators that transform like two insertions of the weak and 
electromagnetic currents, we introduce the spurion fields for the left- and right-handed currents
\begin{eqnarray}
\mathcal Q_L  &=& u^\dagger Q_L u, 
\qquad  
\mathcal Q_R  = u Q_R u^\dagger\,,
\end{eqnarray}
where $u^2 = U = \exp(i \boldtau \cdot \isovec \pi/F_\pi)$
incorporates the pion fields.
Under left- and right-handed chiral rotations $L$ and $R$, respectively, 
the meson $u$ and nucleon $N$ fields transform 
as $u \rightarrow L u K^{\dagger} = K u R^\dagger$ and $N \rightarrow K N$,
where $K$ is an $SU(2)$ matrix that depends nonlinearly on the pion field.
(For a review of chiral symmetry, see for example Ref. \cite{Bernard:1995dp}).
The spurions $Q_{L,R}$ transform like currents,
\begin{eqnarray}
Q_L & \rightarrow& L Q_L L^\dagger\,, 
\qquad 
\,\,\, Q_R \rightarrow R Q_R R^\dagger\,, \\
\mathcal Q_L  
&\rightarrow& K \mathcal Q_L K^\dagger\,, 
\qquad 
\mathcal Q_R  
\rightarrow K \mathcal Q_R K^\dagger\,.
\end{eqnarray}
One then writes the most general Lagrangian involving $\mathcal Q_{L,R}$
that is invariant under chiral symmetry.
The way weak currents break the symmetry is recovered by taking
$Q\to Q^{\rm w}$ with
\begin{equation}
Q^{\rm w}_L = \tau^+\,,\qquad Q^{\rm w}_R = 0\,.
\end{equation}
In the 
EM case, $Q\to Q^{\rm em}$ with 
\begin{equation}
Q^{\rm em}_L = Q^{\rm em}_R = \tau_3/2\,.
\end{equation}
Because two insertions of $\cq^{\rm w}_L$ give rise to $I=2$ interactions in the 
\NLDBD\ case, in the EM case we will investigate $I=2$ operators that induce 
CIB interactions. 

In the mesonic sector, the only operator that can be constructed with
two insertions of $Q_{L,R}$ and no derivatives is the $I=2$ interaction
\begin{equation}
\mathcal L^{\pi\pi}_{e^2} = 
 Z e^2 F_\pi^4 \, 
\textrm{Tr} [\mathcal Q^{\rm em}_L \mathcal Q^{\rm em}_R ]\,,
\label{CIBmeson0}
\end{equation}
where, at LO in $\chi$PT, $Z$ is related to the pion-mass (squared) splitting by 
\begin{equation}
Z e^2 F_\pi^2 = \frac{1}{2}\delta m^2_\pi  = \frac{1}{2} \left( m_{\pi^\pm}^2 - m_{\pi^0}^2 \right).
\end{equation}
There is no interaction of the type $Q_{L}^2$ that would lead to $|\Delta L|=2$. 
The first such interaction contains two chiral-covariant derivatives 
of the pion field,
\begin{equation}
u_\mu = 
- u^\dagger \left[ i \partial_\mu +\left(l_\mu+\hat{l}_\mu\right)\right] u 
+ u \left[ i \partial_\mu +\left(r_\mu + \hat{r}_\mu\right)\right] u^\dagger  \,,
\end{equation}
and it is given by \cite{VanKolck:1993ee,Gasser:2002am,Cirigliano:2017tvr}    
\begin{eqnarray}
 \mathcal L^{\pi\pi}_{e^2}  &=&  
-e^2 F_\pi^2 \, \kappa_3 \left[ \textrm{Tr}( \mathcal Q^{\rm em}_L u^\mu) 
\, \textrm{Tr}( \mathcal Q^{\rm em}_L u_\mu )
- \frac{1}{3}\,\textrm{Tr} \left(\mathcal Q^{\rm em }_L\mathcal Q^{\rm em }_L\right) 
\, \textrm{Tr} \left( u^\mu u_\mu\right) +  (L \rightarrow R)\right] \,,
\nonumber\\
\mathcal L^{\pi\pi}_{|\Delta L| = 2}   &= & 
\left(2 \sqrt{2}\, G_F V_{ud}\right)^2  m_{\beta \beta} \, 
\bar e_L C\bar e_L^T \, \frac{}{} \frac{5g_\nu^{\pi \pi}}{3(16\pi)^2} F_\pi^2 
\nonumber \\
&& \times \left[   
\textrm{Tr}(\mathcal Q^{\rm w}_L u^\mu) 
\, \textrm{Tr}( \mathcal Q^{\rm w}_L u_\mu )
- \frac{1}{3}\, \textrm{Tr}\left(\mathcal Q^{\rm w}_L\mathcal Q^{\rm w}_L\right)
\, \textrm{Tr} \left( u^\mu u_\mu\right)
\right]  + {\rm H.c.}\,,
\label{eq:gnupipi} 
\end{eqnarray}
where we used the notation of Ref. \cite{Gasser:2002am} for the EM operator
\footnote{Differently from Ref. \cite{Gasser:2002am}, we subtracted the trace 
part of the $\kappa_3$ operator to isolate the $I=2$ representation. This shift
in the $I=0$ part can be absorbed in a redefinition of the isospin-invariant 
operator $\kappa_1$ defined in Ref. \cite{Gasser:2002am}.}.
$g_{\nu}^{\pi\pi}$ is a LEC of $\mathcal O(1)$, so that the operator in 
Eq. \eqref{eq:gnupipi} contributes to the neutrino potential at N$^2$LO, 
together with the pion-neutrino loops discussed in 
Ref. \cite{Cirigliano:2017tvr}. 
The factors of $e^2$ and $(2 \sqrt{2} G_F V_{ud})^2
m_{\beta \beta} \, \bar e_L C\bar e_L^T$ appear due to two insertions of EM 
and weak currents, respectively. 
This allows us to identify \cite{Cirigliano:2017tvr}
\begin{equation}
g_{\nu}^{\pi\pi}= - \frac{3}{5} \left(16\pi\right)^2 \kappa_3\,. 
\end{equation}
The model estimate of Ref. \cite{Ananthanarayan:2004qk} for $\kappa_3$ gives
$g_{\nu}^{\pi\pi}(\mu = m_\rho) = -7.6$, in agreement with  a recent LQCD 
extraction that found $g_{\nu}^{\pi\pi}$ 
between $-12$ and $-8.5$  \cite{Feng:2018pdq,Feng:private}.

In the single-nucleon sector, 
the lowest-order $|\Delta L|=2$ interaction involves one derivative. Focusing on terms with only $\mathcal Q^{\rm em}_L$ ($\mathcal Q^{\rm w}_L$) or $\mathcal Q^{\rm em}_R$, one can write 
\cite{VanKolck:1993ee,vanKolck:1996rm,Gasser:2002am,Cirigliano:2017tvr}
\footnote{We again subtracted the trace terms compared to the $O_4$ and $O_5$ 
operators in Ref. \cite{Gasser:2002am}, such that the operators in 
Eq. \eqref{eq:CT1em} 
have $I=2$. These redefinitions would be absorbed by shifting the couplings of 
the $O_1$ and $O_2$ operators of Ref. \cite{Gasser:2002am}.}
\begin{eqnarray} 
\vL_{e^2}^{\pi N}&=& e^2 F_\pi^2 \, \frac{g_4 + g_5}{4} 
\left[{\rm Tr}\left(u_\mu \mathcal Q^{\rm em}_L \right)
\bar N S^\mu \mathcal Q^{\rm em}_L N 
-\frac{1}{3} \,\textrm{Tr}\left(\mathcal Q^{\rm em}_L\,\mathcal Q^{\rm em}_L\right) 
\bar N S^\mu u_\mu N
+ (L \rightarrow R) 
\right] \,, 
\nonumber \\
\vL_{|\Delta L| = 2}^{\pi N}  &=&
\left(2\sqrt{2} G_F V_{ud}\right)^2  m_{\beta \beta} \, \bar e_L C\bar e_L^T
\, \frac{ g_A  g_\nu^{\pi N}}{4 (4\pi)^2} 
\nonumber \\
&&
\times \left[{\rm Tr}\left(u_\mu \mathcal Q^{\rm w}_L \right) 
\bar N S^\mu \mathcal Q^{\rm w}_L N
- \frac{1}{3} \,\textrm{Tr}\left(\mathcal Q^{\rm w }_L\mathcal Q^{\rm w }_L\right) 
\bar N S^\mu u_\mu N \right] + {\rm H.c.} \,,
\label{eq:CT1em}
\end{eqnarray}
where the LEC $g_{\nu}^{\pi N}=\mathcal O(1)$
is related to the EM LEC $g_4+g_5$ by \cite{Cirigliano:2017tvr} 
\begin{equation}
g_\nu^{\pi N}= \left(4\pi F_\pi\right)^2 \, \frac{g_4 + g_5}{g_A}
\equiv -\frac{2}{g_A} \left( \frac{4\pi}{e} \right)^2 \bar{\beta}_{10}\,. 
\label{gnupiN}
\end{equation}
The EM interactions induce CIB in the pion-nucleon couplings, but
at the moment there exist no good estimates besides NDA.
There is only a bound $\bar{\beta}_{10}=5(18)\cdot 10^{-3}$ \cite{vanKolck:1996rm}
extracted from 
the Nijmegen partial-wave analysis \cite{Stoks:1993tb,vanKolck:1997fu} 
of $N\!N$ scattering, which translates to $|g_\nu^{\pi N}|\simle 61$.
This introduces a source of uncertainty at N$^2$LO in the chiral expansion
of the neutrino potential.

We  
now come to the $N\!N$
sector, where the failure of Weinberg's power counting requires 
the $|\Delta L|=2$ contact interaction in Eq. \eqref{gnudef} at LO.
The associated EM operators were constructed in Ref. \cite{Walzl:2000cx}. 
In the 
$m_{u,d} \rightarrow 0$ limit, 
there are only two rank-2 isospin operators with two insertions of $Q_{L,R}$ 
\cite{VanKolck:1993ee,Cirigliano:2017tvr},
\begin{eqnarray}
\mathcal L_{e^2}^{NN} 
&= & \frac{e^2}{4} 
\left\{ \bar N \mathcal Q^{\rm em}_L N \, 
\bar N \!\left({\cal C}_1\mathcal Q^{\rm em}_L
+{\cal C}_2\mathcal Q^{\rm em}_R\right) \! N
- \frac{1}{6} 
\textrm{Tr}\left[\mathcal Q^{\rm em}_L
\!\left({\cal C}_1\mathcal Q^{\rm em}_L+{\cal C}_2\mathcal Q^{\rm em}_R\right)\right]
\bar N \boldtau N \cdot \bar N \boldtau N\right\}
\nonumber\\
&&  + (L \rightarrow R)\,,
\nonumber\\
\mathcal L_{|\Delta L| =2}^{NN}  &=& \left(2\sqrt{2} G_F V_{ud}\right)^2  
m_{\beta \beta} 
\bar e_L C\bar e_L^T  \, \frac{g_\nu^{\rm NN}}{4} 
\left[ \bar N \mathcal Q^{\rm w}_L N \, \bar N \mathcal Q^{\rm w}_L N  
- \frac{1}{6} \textrm{Tr}\left(\mathcal Q^{\rm w}_L\,\mathcal Q^{\rm w}_L\right)
\bar N \boldtau N \cdot \bar N \boldtau N \right] \nonumber \\
&& + {\rm H.c.} \,.
\label{C12def}
\end{eqnarray}
As before, the LECs $g_\nu^{\rm NN}$ and ${\cal C}_1$ are related,
$g_\nu^{\rm NN}={\cal C}_1$.
When expanded in powers of the pion field, the $|\Delta L|=2$ 
Lagrangian generates the contact interaction in Eq. \eqref{gnudef}.
As we have seen, operators related to those in Eq. \eqref{C12def}
but containing insertions of the quark masses are also needed
at LO. The full set of 
such $N\!N$ operators with up to two mass insertions is constructed in 
App. \ref{mass}.
In the isospin limit $m_u=m_d$, we can include quark-mass corrections by 
replacing ${\cal C}_1$ and ${\cal C}_2$ by the combinations
\begin{equation}
g_{\nu}^{\rm NN} = {\cal C}_1 = \sum_n c^{(1)}_{n} m_\pi^{2n}\,,   
\qquad
{\cal C}_2 = \sum_n c^{(2)}_n m_\pi^{2n}\,,
\end{equation}
where $c_n^{(1,2)}$ are the couplings of the EM operators with $n$ mass 
insertions.
The equality $g_{\nu}^{\rm NN} = {\cal C}_1$ relies only on isospin symmetry 
and is not spoiled by insertions of the average quark mass. 

If ${\cal C}_1$ and ${\cal C}_2$ can be fixed separately from CIB processes,
then $g_{\nu}^{\rm NN}$ can be determined independently of LNV data.
In Weinberg's power counting, the LECs ${\cal C}_1$ and ${\cal C}_2$ scale 
as ${\cal C}_{1,2} = \mathcal O( (4\pi F_\pi)^{-2})$. 
In the following subsection we will show that renormalization requires 
${\cal C}_{1,2} = \mathcal O(F_\pi^{-2})$, consistently with the
enhancement of $g_{\nu}^{\rm NN}$.
Unfortunately we cannot fix ${\cal C}_1$ and ${\cal C}_2$ 
separately at present, but we will discuss how CIB in the $N\!N$ system 
can be used to extract ${\cal C}_1 + {\cal C}_2$.

\subsection{CIB
in $N\!N$
scattering}
\label{CIBfits}

We now determine the 
coefficient ${\cal C}_1 + {\cal C}_2$ from $N\!N$
scattering data.
By expanding the pion fields in the operators in Eq. \eqref{C12def}, we see 
that ${\cal C}_1$ and ${\cal C}_2$ only differ at the multipion level. 
Any CIB observable that is not sensitive to multipion contributions therefore 
only constrains the sum ${\cal C}_1 + {\cal C}_2$. 
In particular, $N\!N$
scattering data are not sufficient to determine ${\cal C}_1$, and thus 
$g_{\nu}^{\rm NN}$, separately 
as required for 
$0\nu\beta\beta$. Nevertheless, the analysis of CIB in $N\!N$
scattering does convincingly demonstrate the need for short-range operators 
to absorb divergences of Coulomb-like potentials acting in the $^1S_0$ channel.
It provides a concrete data-driven example of the breakdown of Weinberg's 
power counting. In addition, the extraction of ${\cal C}_1 + {\cal C}_2$ 
will provide an estimate of the importance of the short-range neutrino 
potential by assuming $g_\nu^{\rm NN}\sim ({\cal C}_1+{\cal C}_2)/2$. 

We start our discussion by demonstrating the need for CIB counterterms. 
Charge-independence breaking is evident in the difference between the 
${}^1S_0$ $np$, $pp$, and $nn$ scattering lengths. In the $pp$ channel, 
Coulomb photon exchange is an LO effect at small center-of-mass momenta and 
the Coulomb potential must be iterated to all orders. We can define $a_C$ as 
the $pp$ scattering length after subtraction of the pure Coulomb contribution 
to $pp$ scattering. We will use the 
empirical determination of the scattering lengths \cite{Piarulli:2014bda} 
\begin{equation}
a_{np} = -23.7 \pm 0.02 \,\, \textrm{fm}\,, 
\qquad 
a_{nn} = -18.90 \pm 0.40\,\, \textrm{fm}\,,
\qquad 
a_C = -7.804 \pm 0.005 \,\, \textrm{fm}\,.
\label{eq:CIBscatt}
\end{equation}
While we have subtracted long-range photon-exchange contributions, 
$a_C$ still contains short-range 
contributions from hard-photon exchange.
These two types of contributions can be separated within specific models
by defining a Coulomb-subtracted $pp$ scattering length 
\cite{Jackson:1950zz}, which is 
estimated to be $a_{pp} = -17.3 \pm 0.4$ fm.
From $a_{np}$, $a_{nn}$ and $a_{pp}$ we can construct the combination
\begin{equation}
a_{\rm CIB} = \frac{1}{2}\left(a_{pp} + a_{nn}\right) - a_{np} 
= 5.6 \pm 0.6 \, \textrm{fm},
\end{equation}
which demonstrates that CIB effects are sizable in the ${}^1S_0$ channel 
even after subtracting Coulomb contributions \cite{Miller:1990iz}. 
Since the separation between 
$a_{pp}$ and $a_C$ depends (mildly) on the model of the nuclear force, we will 
not use the Coulomb-subtracted scattering length and instead fit to $a_C$. 

\begin{figure}
	\includegraphics[width=0.75\textwidth]{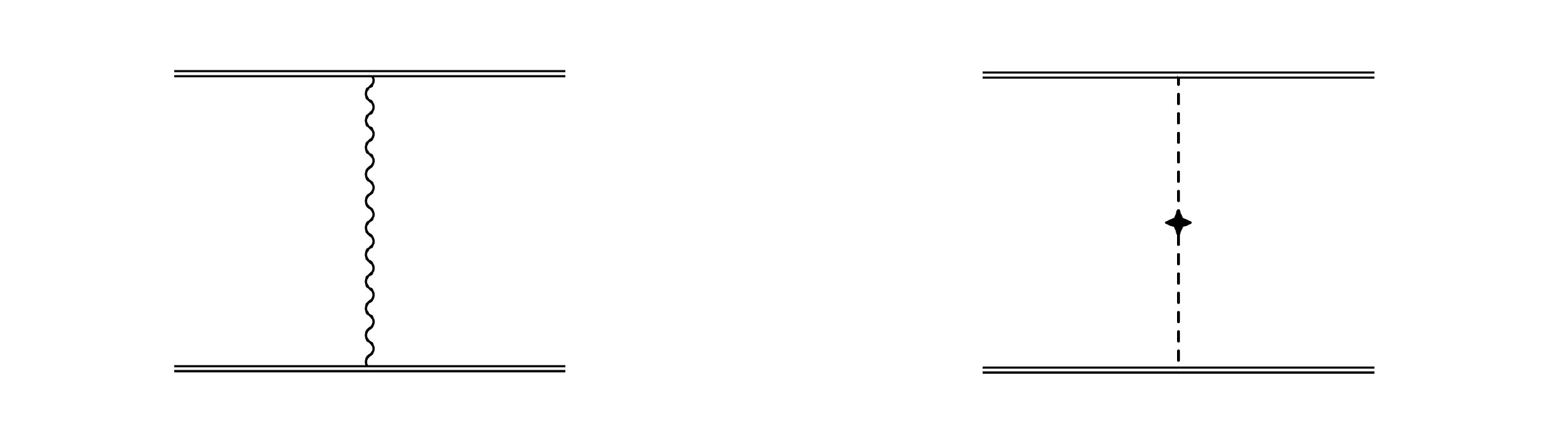}
	\caption{Long-range contributions to the CIB $N\!N$
potential. The wavy line represents a photon and the star denotes an insertion 
of the electromagnetic pion-mass splitting. 
Other 
symbols as in Fig. \ref{treelevel}.}
\label{treelevel2}
\end{figure}

The most important pion-range CIB interaction stems from
the pion-mass splitting, Eq. \eqref{CIBmeson0}.
Together with Coulomb-photon exchange,
it gives rise through 
the diagrams in Fig. \ref{treelevel2} to the 
long-range CIB potential
\cite{VanKolck:1993ee,Friar:1999zr}
\begin{equation}
V_{\rm CIB} =  \frac{e^2}{4} 
\left( \tau^{(1)}_3 \tau^{(2)}_3 - \frac{1}{3}\, \boldtau^{(1)} \cdot \boldtau^{(2)} 
\right) \frac{1}{\spacevec q^2} 
\left[ 1 - 
\frac{g_A^2}{3}  \frac{\delta m^2_\pi}{e^2 F_\pi^2} \,
\left(\boldsigma^{(1)} \cdot \boldsigma^{(2)}-S^{(12)}\right)
\, \left(1-\frac{m_\pi^2}{\spacevec q^2 + m_\pi^2}\right)^2
\right]
\,.
\label{Vcib}
\end{equation}
In the $^1S_0$ channel it reduces to
\begin{equation}
V^{^1S_0}_{\rm CIB} =  \frac{e^2}{4} 
\left(\tau^{(1)}_3\tau^{(2)}_3 -\frac{1}{3}\right) 
\frac{1}{\spacevec q^2} 
\left[ 1 + g_A^2 \frac{\delta m^2_\pi}{e^2 F_\pi^2}    
\left(1-\frac{m_\pi^2}{\spacevec q^2 + m_\pi^2}\right)^2
\right]\,.
\label{Vcib1S0}
\end{equation}
Since by NDA 
$\delta m^2_\pi
={\mathcal O}(e^2F_\pi^2)$, the pion-mass-splitting term 
is expected to contribute sizably to CIB for momenta $Q\sim m_\pi$.

Equations \eqref{Vcib} and \eqref{Vcib1S0} can be directly
compared to, respectively, Eqs. \eqref{eq:Vnu0} and \eqref{eq:Vnu1}.
$V_{\rm CIB}$ has a very similar structure to the long-range neutrino potential
$V_\nu$,
with the difference that Eq. \eqref{eq:Vnu0} contains contributions from the 
couplings of nucleons and pions to the weak axial current. In addition, 
there is no 
analog of the pion-mass-splitting term 
in $V_\nu$.
The Coulombic nature of $V^{^1S_0}_{\rm CIB}$ at short distances implies that 
the same divergence we encountered in the LNV scattering amplitude 
$\mathcal A_\nu$ also affects CIB observables, 
requiring a CIB four-nucleon operator at LO in $e^2$.

The Coulombic nature of $V^{^1S_0}_{\rm CIB}$ also determines its importance relative to the LO strong interactions represented by $V^{^1S_0}_\pi$ in Eq. \eqref{1S0LOpot2}. For momenta $Q\sim m_\pi$, $V^{^1S_0}_{\rm CIB}/V^{^1S_0}_\pi \sim (e F_\pi/m_\pi)^2 \ll 1$ and $V^{^1S_0}_{\rm CIB}$ can be treated in perturbation theory. In this region, the argument of Sec. \ref{LNVLO} for the need of a counterterm goes through essentially unchanged for the CIB amplitude, if we replace the neutrino-exchange diagrams by Coulomb plus pion-mass-splitting OPE. The case of perturbative Coulomb in $\slashpi$EFT has been examined in Ref. \cite{Konig:2015aka}. In contrast, at momenta $Q\simle e F_\pi$ CIB is no longer a small correction since $V^{^1S_0}_{\rm CIB}/V^{^1S_0}_\pi \simge 1$, but pion-mass-splitting OPE is $\simle (Q/m_\pi)^4 \ll 1$ compared to Coulomb. At even smaller momenta, $Q\simle \alpha_{\rm em} m_N \sim e^2 F_\pi$, Coulomb-photon exchange is nonperturbative. The need for a counterterm for nonperturbative Coulomb in $\slashpi$EFT was shown in Ref. \cite{Kong:1999sf}. We generalize this argument now to $\chi$EFT including pion-mass splitting.

In order to interpolate smoothly between the three regions, 
we treat 
the CIB potential nonperturbatively. This is what is done in all 
chiral-potential calculations we are aware of. 
The iteration of the CIB potential does not affect the presence of a 
logarithmic divergence, which is due to diagrams 
where a single photon exchange or a single insertion of the pion-mass splitting
is sandwiched between two short-range operators,
analogous to the diagrams shown in the third row of Fig. \ref{Fig1}. 
The iteration, however, affects the finite pieces of the counterterms
by including corrections suppressed by powers of $e^2 \sim 1/10$.
Since the equality ${\cal C}_1 = g_\nu^{\rm NN}$ is valid at LO in $e^2$, 
we expect the counterterms extracted from $N\!N$ scattering to be a 
good representation of the LNV counterterms up to $10\%$ corrections. 
In summary,  we  replace
the long-range potential $V_\pi$ in Eq. \eqref{chiGE}
by different potentials in the $pp$, $nn$, and $np$ 
channels,  
\begin{equation}
V_{pp}= V_\pi(m_{\pi^0})+\frac{e^2}{4\pi r},
\qquad 
V_{nn}= V_\pi(m_{\pi^0})\,,
\qquad 
V_{np}= 2 V_\pi(m_{\pi^\pm})- V_\pi(m_{\pi^0}),
\label{Vcib2}
\end{equation}
with  $m_{\pi^\pm} = 139.57$ MeV and $m_{\pi^0} = 134.98$ MeV.

\begin{figure}
\includegraphics[width=0.9\textwidth]{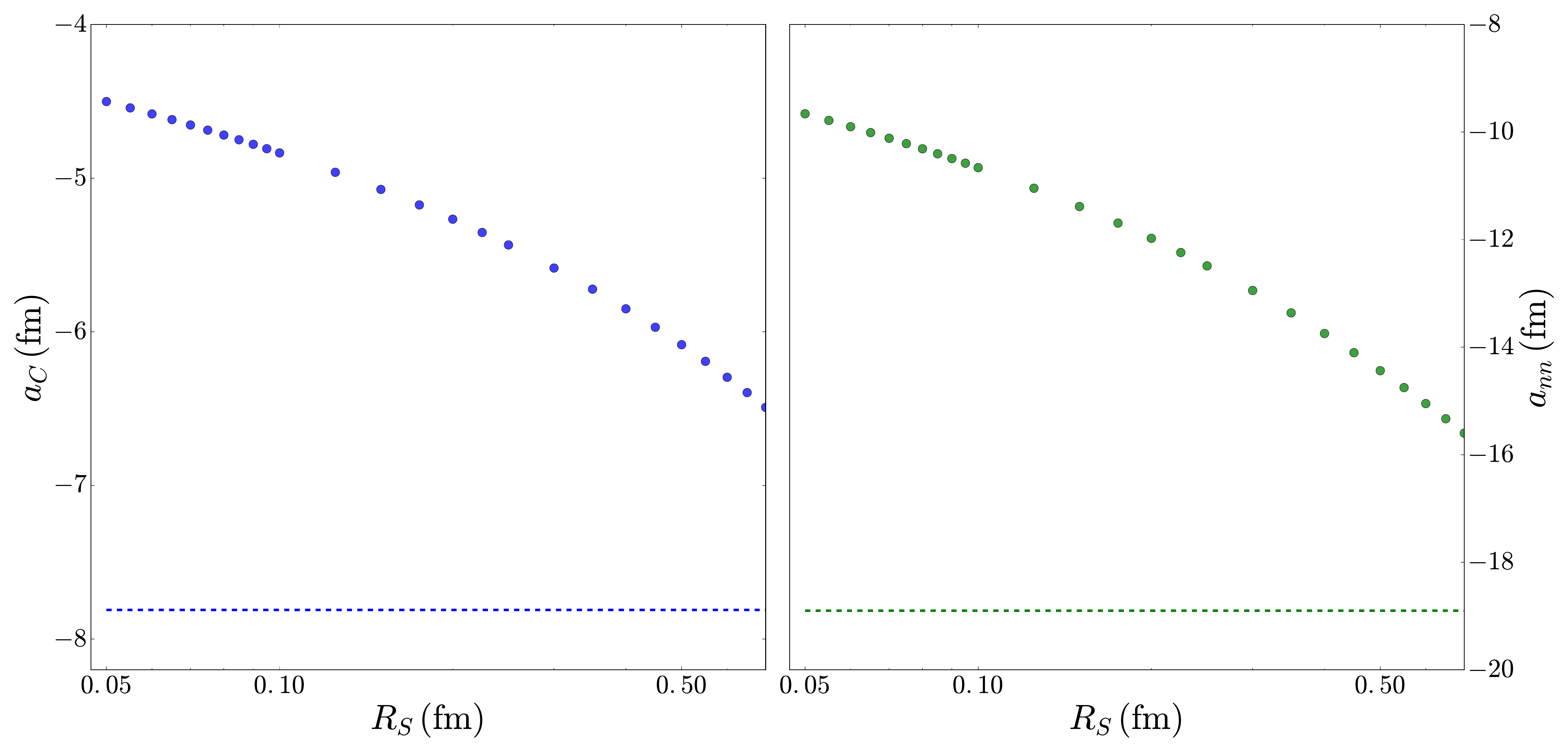}
\caption{
Proton-proton scattering length $a_C$ (left panel) and 
neutron-neutron scattering length $a_{nn}$ (right panel), 
as a function of the coordinate-space cutoff $R_S$. 
The points 
are computed with the long-range potentials $V_{pp}$ and $V_{nn}$ defined in 
Eq. \eqref{Vcib2}, and with a charge-independent short-range potential
with LEC 
$
C = 
C_{np}$ fitted to the $np$ scattering length in the $^1S_0$ channel, $a_{np}$.  
The dashed lines 
indicate the experimental values of $a_C$ and $a_{nn}$. 
}
\label{FigACIB}
\end{figure}

In Weinberg's power counting, the contact interaction 
$
C$ is charge independent at LO. As discussed in Sec. \ref{CIBLag}, 
charge dependence only enters at $\mathcal O(e^2/(4\pi)^2)$,
which is suppressed by $(4\pi)^{-2}$ with respect to the terms in 
Eq. \eqref{Vcib2}.
This implies that once $
C$ is determined in one isospin channel, for example $np$,  
the phase shifts in the remaining channels, $pp$ and $nn$, 
should be independent of the regulator.
We test this prediction 
of Weinberg's power counting in 
Fig. \ref{FigACIB}. We determine $
C$ by fitting to $a_{np}$ in the $np$ channel and define the resulting value 
as $
C_{np}$. We then calculate $a_C$ and $a_{nn}$ using the long-range potentials 
in Eq. \eqref{Vcib2} combined with the short-range interaction with LEC 
$
C_{np}$. Figure \ref{FigACIB} shows that $a_C$ and $a_{nn}$ have a 
strong dependence on the cutoff $R_S$, and for no $R_S$ in the plotted range  
there is agreement between the calculated and the measured values.
As was the case for $0\nu\beta\beta$ decay, our calculations explicitly 
demonstrate that Weinberg's power counting is inadequate for Coulomb-like 
potentials in the ${}^1S_0$ channel.

The regulator dependence can be removed by introducing 
isospin-breaking counterterms
$
C_{np}$, $
C_{nn}$, and $
C_{pp}$. 
This amounts to including short-range charge-symmetry breaking (CSB) as well 
as CIB. According to NDA,
OPE with a CSB 
pion-nucleon coupling \cite{VanKolck:1993ee,vanKolck:1996rm, Friar:2003yv} contributes at the same order as short-range CSB. The value of the CSB pion-nucleon coupling is unknown \cite{vanKolck:1996rm,vanKolck:1997fu}, but its inclusion would not affect the conclusions drawn below that short-range CSB is enhanced over NDA, just as short-range CIB. 
For simplicity we do not include CSB OPE. 
We determine 
$
C_{np}$, $
C_{nn}$, and $
C_{pp}$ 
by reproducing the observed scattering lengths in 
Eq. \eqref{eq:CIBscatt}. 
We then
extract 
the CIB combination
\begin{equation}
\frac{\mathcal C_1 + \mathcal C_2}{2}  
\equiv \left(\frac{m_N
C}{4\pi }\right)^2
\frac{\tilde {\cal C}_1+\tilde {\cal C}_2}{2}
=\frac{1}{e^2}\left(
C_{np}-\frac{
C_{nn}+
C_{pp}}{2}\right)
\,,
\label{eq:CIBc1c2}
\end{equation}
with $
C = (
C_{np} + 
C_{nn} + 
C_{pp})/3$.
At small $R_S$, the dimensionless sum
$\tilde {\cal C}_{1} + \tilde {\cal C}_2$ shows the 
expected logarithmic behavior,
\begin{equation}
\frac{\tilde {\cal C}_1 + \tilde {\cal C}_2}{2}
\simeq 0.4 - 1.95 \, 
\ln( m_\pi R_S) \,.
\label{eq:em2}
\end{equation}
The fit in Eq.\ \eqref{eq:em2} is accurate up to $R_S \sim 0.3$ fm at 
which point power corrections become important.
Note that for the values of $R_S$ commonly used in the literature, 
$0.5 - 0.8$ fm, the numerical value of $F_\pi^2 ({\cal C}_1 + {\cal C}_2)$ 
is in the range $0.15 -0.2$, much larger than the $(4\pi)^{-2}$ 
predicted by Weinberg's power counting.

\begin{figure}
\includegraphics[width=.75\textwidth]{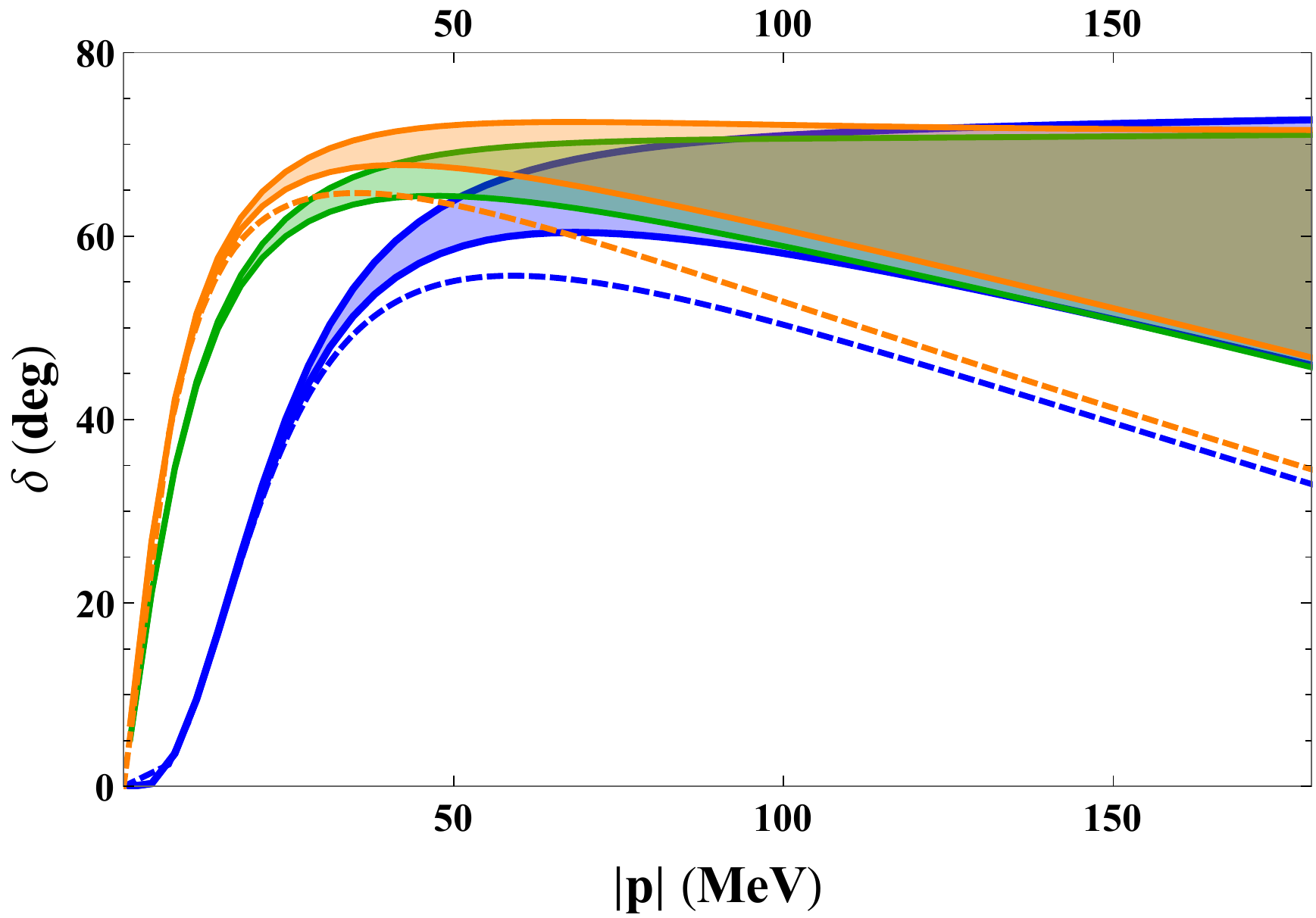}
\caption{Phase shifts 
for the different isospin components in the $^1S_0$ channel as function
of the center-of-mass momentum $|\spacevec p|$.
Orange, blue, and green bands represent an $\overline {\rm MS}$ calculation
of $np$, $pp$, and $nn$ scattering, respectively,
where the regulator of the intermediate scheme, as discussed in 
App. \ref{app:MSbar}, is varied
between $1/\lambda=0.05$ fm and $1/\lambda =0.7$ fm.
The red and blue dashed lines show the Nijmegen partial-wave analysis for $np$ and $pp$ phase shifts, respectively \cite{Stoks:1993tb}. 
}
\label{Fig:CIBPhaseShifts}
\end{figure}

The logarithmic divergence induced by the long-range potential $V_{\rm CIB}$ 
can be seen explicitly using the $\overline{\textrm{MS}}$ scheme. 
The analysis follows that in Sec. \ref{LOpiless}.
The RGEs for $
C_{np}$, $
C_{pp}$ and $
C_{nn}$ 
are modified by the isospin-breaking interactions:
\bea
\frac{d}{d\ln \mu}
C_{pp}^{-1} &=& 
\left(\frac{m_N}{4\pi}\right)^2 \left(e^2-\frac{g_A^2m_{\pi^0}^2}{4 F_\pi^2}\right)
\,,\nn\\
\frac{d}{d\ln \mu}
C_{np}^{-1}  &=&
\left(\frac{m_N}{4\pi}\right)^2 \frac{g_A^2(m_{\pi^0}^2-2m_{\pi^\pm}^2)}{4 F_\pi^2}
\,,\nn\\
\frac{d}{d\ln \mu}
C_{nn}^{-1}  &=& 
-\left(\frac{m_N}{4\pi}\right)^2 \frac{g_A^2m_{\pi^0}^2}{4 F_\pi^2}\,,
\label{eq:CIBcontacts}
\eea
which are solved to reproduce the scattering lengths in 
Eq. \eqref{eq:CIBscatt}. 
The resulting phase shifts are shown in Fig. \ref{Fig:CIBPhaseShifts}. 
Since we iterated the Coulomb potential in the $pp$ channel,
we 
get a good description of the phase shifts at small momentum. 
For simplicity, we also iterated the pion-mass splitting by considering 
the physical pion masses in Eq. \eqref{Vcib2}.
In agreement with the expectation from NDA,
isospin-breaking effects are relatively small at momenta comparable to the 
pion mass.

The RGEs in Eq. \eqref{eq:CIBcontacts} imply 
\begin{eqnarray}
\frac{d}{d\ln\mu} 
\tilde {\cal C}_1 &=&  
\frac{1}{2} \left( 1 + 2 g_A^2\right) 
\simeq 2.1 \,,
\label{DimRegEM0} \\
\frac{d}{d\ln\mu} 
\tilde {\cal C}_2 &=&  
\frac{1}{2} \left( 1 - 2 g_A^2 + 2 g_A^2 \,\frac{\delta m^2_\pi}{e^2F_\pi^2}
\right) 
\simeq 1.5\,.
\label{DimRegEM}
\end{eqnarray}
Equation \eqref{DimRegEM0} agrees with Eq. \eqref{gnurunningchiral}, as
it should.
Using the fit values for $
C_{nn,np,pp}$, we 
obtain
\begin{equation}
\frac{\tilde {\cal C}_1+\tilde {\cal C}_2}{2}
\simeq 2.5-1.8 \, \ln(m_\pi/\mu).
\label{CIBregfit}
\end{equation}
As for the \nnpp case, Eq. \eqref{fits}, the coefficient of the logarithms in 
$\overline{\textrm{MS}}$ and 
$R_S$ schemes agree at the 10\% level.
The coefficients of the logarithms in $(\tilde{\cal C}_1 + \tilde{\cal C}_2)/2$ 
and $\tilde g_{\nu}^{\rm NN}$
are numerically similar, while $\tilde{\cal  C}_1 - \tilde{\cal C}_2$ runs 
more slowly. This appears to be consequence of the numerical accident
$\delta m^2_\pi \approx 2 e^2F_\pi^2$, 
for which we are not aware of an underlying physical reason. 

Finally, we comment on the possibility of using a similar analysis for the two-derivative short-range $0\nu\beta\beta$ operator. In Sect.~\ref{LNV@NLO} we argued, based on a combination of renormalization-group arguments and NDA, that we expect a new LEC only to enter at N${}^2$LO in both pionless and chiral EFT. Focusing for simplicity on pionless EFT, we are concerned whether we can confirm the size of the LEC $\tilde g^0_{2\nu}$ in Eq.~\eqref{natural}. This LEC is connected to the CIB combination of ${}^1S_0$ effective ranges \cite{Kong:1998sx,Kong:1999sf}  
\begin{equation}
(r_0)_{\rm CIB} = \frac{(r_0)_{pp} + (r_0)_{nn} - 2 (r_0)_{np} }{2} \sim e^2  \frac{m_N}{4\pi} \tilde g^0_{2\nu}\,.
\end{equation}
Using the scaling in Eq.~\eqref{natural}, we obtain ${(r_0)_{\rm CIB}}/{r_0} = \mathcal O(e^2)$. This scaling agrees with NN scattering data that give \cite{Piarulli:2014bda}  
$(r_0)_{\rm CIB}/{r_0} \in [-0.04,0.06]$. An NLO scaling of $\tilde g^0_{2\nu}$
would predict much larger CIB corrections,  ${(r_0)_{\rm CIB}}/{r_0} = \mathcal O(e^2 \Lambda_{\slashpi}/\aleph)$, 
which are not supported by data.

\subsection{Impact on the two-body LNV amplitude}
\label{LNVamptwobody}

We can use the value of ${\cal C}_1+{\cal C}_2$ to estimate the numerical 
impact of the short-range component of the neutrino potential
on $\mathcal A_{\nu}$.  
Since $N\!N$ scattering alone does not allow one to isolate ${\cal C}_1$, 
and thus the contribution to $0\nu\beta\beta$,
we will first assume that 
$\tilde {\cal C}_1(\bar R_S) = \tilde {\cal C}_2(\bar R_S)$ 
at a given scale $\bar R_S$. 
Since ${\cal C}_1$ and ${\cal C}_2$ have different runnings,
the renormalization point at which this choice is made influences the value 
of the amplitude.
We assess this dependence by varying $\bar R_S$ in a 
wide range, between 0.05 and 0.7 fm. 
The corresponding results for $\mathcal A_\nu$ 
are shown in Table \ref{Tab1}. 
With the suffixes L and S we denote the matrix elements of the long- and 
short-range neutrino potentials defined in Eqs. \eqref{eq:Vnu0} and 
\eqref{Vshort} (with $g_{2\nu}^{\rm NN}$ set to zero) respectively. 
We observe for $nn \rightarrow pp $ transitions, where the initial and final 
states have the same total isospin, a reduction of the total amplitude by 
10-30\% due to inclusion of the short-range potential.

\begin{table}
\begin{tabular}{c||c|c|c|}
$\bar R_S$ (fm) & $(\mathcal A_\nu)_{\rm L}$ (MeV$^{-2}$) 
& $(\mathcal A_\nu)_{\rm S}$ (MeV$^{-2}$) & $\mathcal A_\nu$ (MeV$^{-2}$)\\
\hline 
0.05  & 0.046 &  $-0.014$  &  0.032 \\
0.1   & 0.043 &  $-0.012$  &  0.031  \\
0.3   & 0.037 &  $-0.007$  &  0.030 \\  
0.7   & 0.032 &  $-0.00$4  &  0.028\\ 
\end{tabular}
\caption{\nnpp
scattering amplitude $\mathcal A_\nu$, divided by the factor $-\exp(i (\delta_{^1S_0}(E) + \delta_{^1S_0}(E^\prime) ))$, evaluated at $|\spacevec p| =$ 1 MeV and 
$|\spacevec p^\prime| = 38$ MeV for selected values
of the coordinate-space cutoff $\bar R_S$ where 
${\cal C}_1(\bar R_S)$ is assumed equal to ${\cal C}_2(\bar R_S)$.
The suffixes L and S label the matrix elements of the long- and short-range 
components of the neutrino potential.
}
\label{Tab1}
\end{table}

\begin{figure}
\center
\includegraphics[width=0.75\textwidth]{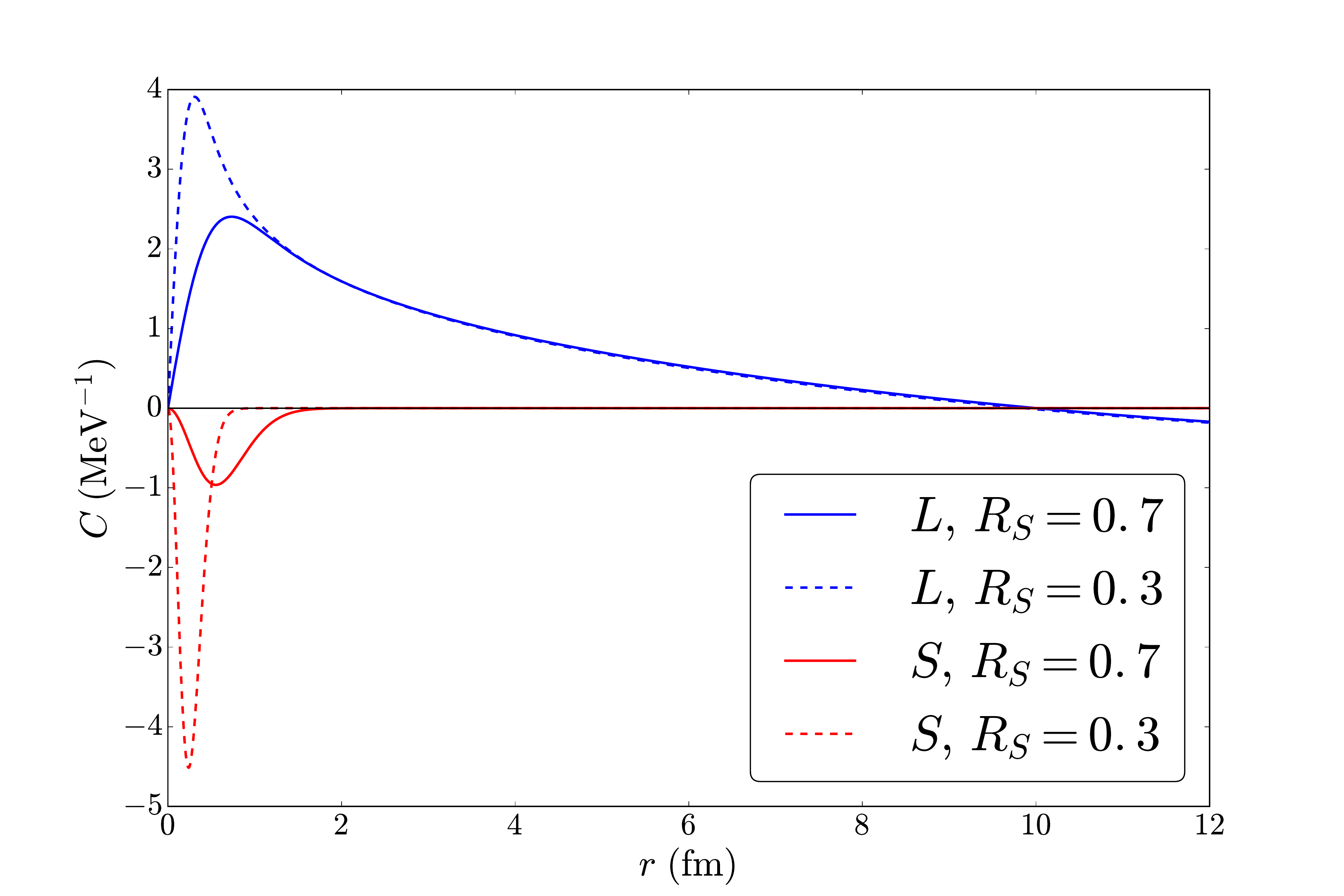}
\caption{Matrix-element densities
for the $nn \to ppe^- e^-$ transition as a function of the radial coordinate.
Curves are shown for the long- (blue) and short-range (red) components 
of the $0\nu\beta\beta$ transition operator
for two choices of the cutoff $R_S$, 0.7 (solid)
and 0.3 (dashed) fm. 
}
\label{C12}
\end{figure}

While significant, the influence of the short-range potential is somewhat 
smaller than the $\Or(1)$ expectation. This smallness can be understood by 
examining the matrix-element density $C(r)$ defined as
\begin{equation}
C_{\rm L,S}(r) =  \int d^3 \spacevec r^\prime \, 
\psi_{\spacevec p^{\prime}}^{-}(\spacevec r^\prime)^* \, 
V_{\nu\, \rm L,\, S}(\spacevec r^\prime) 
\,\delta(r - r^\prime) \, \psi^{+}_{\spacevec p}(\spacevec r^\prime).
\label{eq:meden}
\end{equation}
We see 
in Fig. \ref{C12} that the long-range matrix element $C_{\rm L}(r)$
has support over a wide range of $r$.
In contrast, $C_{\rm S}(r)$ is essentially zero 
for $r \simge 1$ fm. 
Only for the smaller cutoff values does the short-range
component become sufficiently large to partially 
compensate for the smaller range.
Therefore, even if formally LO, the impact of $g_{\nu}^{\rm NN}$ is somewhat 
diluted. We will see in Sec. \ref{pheno} that this is not the case for
transitions in which the nuclear isospin changes by two units, 
which is the case for all nuclei of experimental interest. 

\begin{figure}
\includegraphics[width=0.75\textwidth]{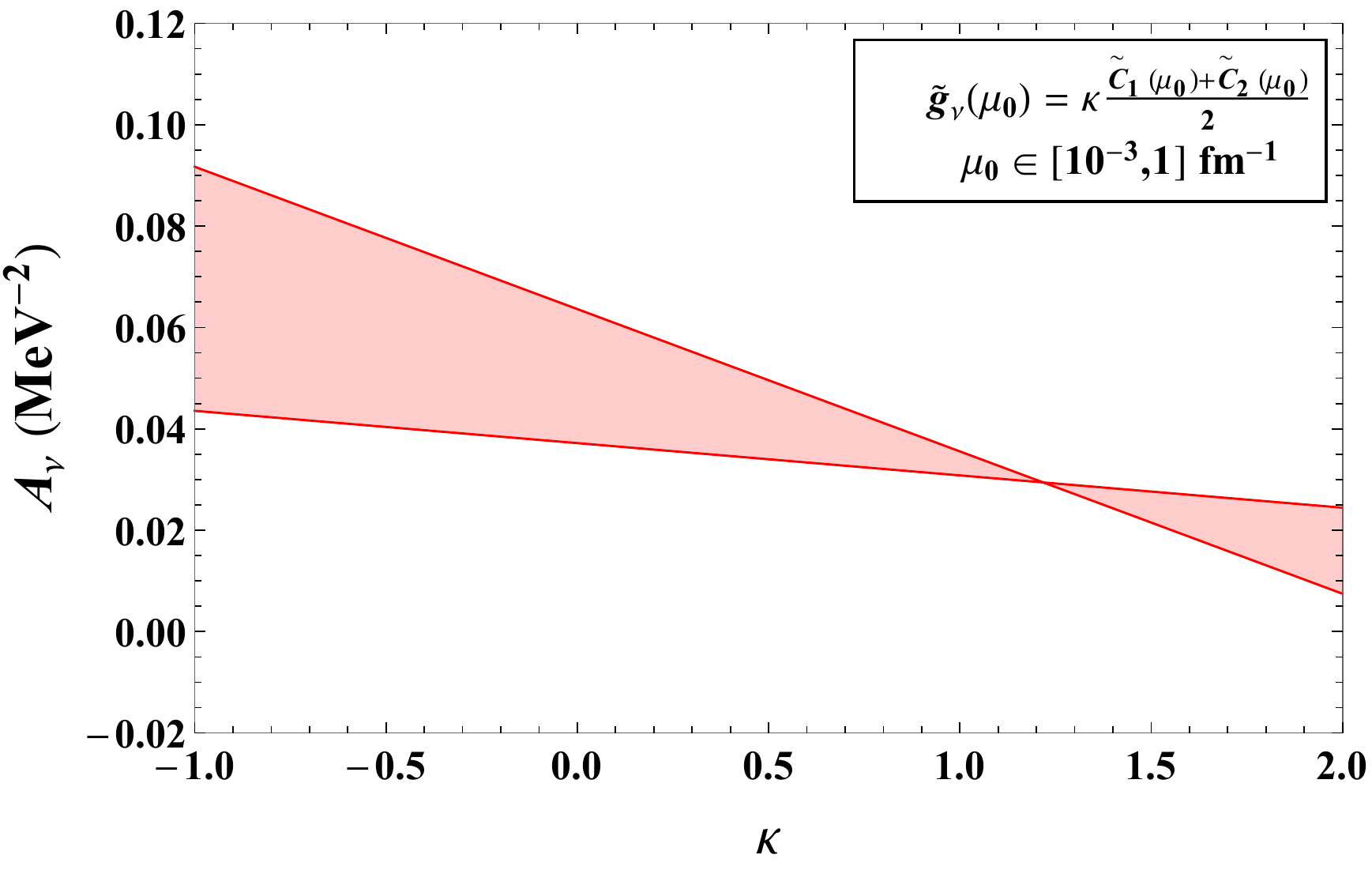}
\caption{Dependence of the sum of the short- and long-range \nnpp
amplitudes at $\spacevec p =1$ MeV, on the parameter $\kappa$ that parametrizes 
the relation between $\tilde g_\nu$ and the CIB counterterms 
$\tilde {\cal C}_1+\tilde {\cal C}_2$ at a given scale $\mu_0$.
The bands are obtained by varying $\mu_0$ between $10^{-3}$ 
and $1$ fm$^{-1}$.}
\label{kappa}
\end{figure}

We stress that the choice ${\cal C}_1 = {\cal C}_2$ was made for 
illustration purposes only, and other choices can be made. 
This arbitrariness leads to an uncontrolled theoretical uncertainty.
We can illustrate the effect of varying the assumption 
${\cal C}_1 = {\cal C}_2$, by considering the more general situation 
${\cal C}_2=\alpha {\cal C}_1$, or, equivalently,
\begin{equation}\label{choice}
\tilde g_\nu^{\rm NN}(\mu_0) = \kappa 
\frac{\tilde {\cal C}_1(\mu_0) + \tilde {\cal C}_2(\mu_0)}{2},
\end{equation}
with $\kappa = 2/(1+\alpha)$ and $\mu_0$ the (arbitrary) scale where 
Eq. \eqref{choice} holds. We show the result of varying $\kappa$ between 
$-1$ and $2$ in Fig. \ref{kappa}. 
The point $\kappa = 0$ corresponds to no counterterm in the neutrino potential,
while $\kappa = 2$ to the situation in which ${\cal C}_2 = 0$
and all CIB arises from ${\cal C}_1$. 
The red bands are obtained by changing the renormalization point $\mu_0$ 
at which the choice in Eq. \eqref{choice} is made.
Figure \ref{kappa} highlights that CIB in $N\!N$ scattering, while providing 
strong evidence for the existence of a counterterm in $0\nu\beta\beta$, 
unfortunately does not allow us to draw robust quantitative conclusions
about its impact in the magnitude of renormalized amplitude. 
It does demonstrate that a better understanding of the short-range 
contributions is crucial, since reasonable $\mathcal O(1)$ choices for $\kappa$ 
lead to variations of the \nnpp amplitude of roughly an order of magnitude. 
This uncertainty must be reduced to reliably extract the effective Majorana 
neutrino mass from $0\nu\beta\beta$ decay experiments.

\section{Phenomenological implications}
\label{pheno}

In the previous sections, we have demonstrated the need to include a 
counterterm to absorb divergences induced by the long-range 
neutrino potential: in coordinate space,
\begin{equation}
V_{\nu\,\rm S}= - 2 g_{\nu}^{\rm NN}\,\tau^{(1)+}\tau^{(2)+} \delta^{(3)}_{R_S}({\bf r})
\ ,
\label{VCTsaori}
\end{equation}
with $\delta^{(3)}_{R_S}({\bf r})$ a regularization of the 
delta function such 
as 
Eq.~\eqref{eqRS}.
The value of $ g_{\nu}^{\rm NN}$ depends on nonperturbative QCD dynamics, and, 
lacking a measurement of the \nnpp
cross section, could be determined by matching to LQCD calculations of 
LNV processes. In Sec. \ref{CIB},
we have established a relation between the contact interactions appearing 
in the LNV and in the EM Lagrangian, 
which leads to 
$g_\nu^{\rm NN}=\mathcal C_1$.
The electromagnetic
counterpart of the LNV contact potential $V_{\nu\,\rm S}$,
\begin{equation}
V_{\rm CIB,\, \rm S}= - \frac{e^2}{6} \frac{{\cal C}_1+{\cal C}_2}{2}  
\, T^{(12)} \, \delta^{(3)}_{R_S}({\bf r}) \ ,
\label{VCIB_pia}
\end{equation}
where the isotensor operator reads 
$T^{(12)}=3\,\tau^{(1)}_3\tau^{(2)}_3-\boldtau^{(1)}\cdot\boldtau^{(2)}$,
is included in all high-quality 
phenomenological~\cite{Machleidt:2000ge,Wiringa:1994wb} and 
chiral potentials~\cite{Machleidt:2011zz,Epelbaum:2014efa,Piarulli:2014bda,Ekstrom:2015rta,Piarulli:2016vel,Reinert:2017usi}, as it was recognized that 
just including the Coulomb interaction and the pion-mass splitting does not 
reproduce the CIB in the $N\!N$ scattering lengths.
In the previous section we have seen that renormalization 
of the amplitude with Coulomb-photon exchange in fact
demands the presence of this short-range interaction.
In this section, we extract the value of $({\cal C}_1+{\cal C}_2)/2$ from the 
phase-shift analysis performed in 
Refs.~\cite{Piarulli:2016vel,Piarulli:2014bda}. 
We then study the impact of the counterterm on $0\nu\beta\beta$ matrix elements 
in light nuclei,
whose wavefunctions are consistently computed with the same chiral potential.

\subsection{CIB in high-quality $N\!N$ potentials}
\label{highqualityCIB}

In Sec. \ref{LNVlightnuclei} we will replace
\begin{equation}
g_{\nu}^{\rm NN} \rightarrow  \frac{{\cal C}_1+{\cal C}_2}{2} \,,
\label{replacechiral}
\end{equation}
with a value determined by the corresponding potential.
The expression in 
Eq. \eqref{VCIB_pia} corresponds to the
short-range charge-dependent (CD) contact potential 
given by the momentum-independent terms in Eq.~(2.7) of 
Ref.~\cite{Piarulli:2014bda}, namely
\begin{equation}
v^{CD}_{12,\, \rm S}= C_0^{\rm IT} \, T^{(12)} \, \delta^{(3)}_{R_S}({\bf r})  \ ,
\end{equation}
from which 
\begin{equation}
\frac{{\cal C}_1+{\cal C}_2}{2} = - \frac{6}{e^2} \, C_0^{\rm IT}\ .
\label{eq:c1c2gnu}
\end{equation}
The values of $C_0^{\rm IT}$ and the corresponding
$({\cal C}_1 + \mathcal C_2)/2$ for two choices of cutoff $R_S$ 
in Eq.~\eqref{eqRS} are reported in Table~\ref{tb:lecs}.
The two interactions NV-I and NV-II are fitted to
$N\!N$ scattering data in the ranges [0-125] and [0-200] MeV, respectively,
of laboratory energies.
Models $a$ and $c$ differ by the choices of the cutoff $R_S$ and $R_L$, where 
the second cutoff regulates, for example, the pion-exchange tensor potential. 
In the models denoted by an asterisk, the three-nucleon
interaction is constrained by the tritium 
$\beta$-decay width and 
trinucleon binding energies~\cite{Baroni:2018fdn}, 
but this choice does not affect $C_{0}^{\rm IT}$.
In addition to $C_{0}^{\rm IT}$, the interactions in 
Refs. \cite{Piarulli:2014bda,Piarulli:2016vel} 
contain four CIB operators with two derivatives, whose effects  
manifest in the dependence of $C_{0}^{\rm IT}$  on the energy range of the fits.

\begin{table}
\begin{tabular}{c | c | c | c | c || c |c | c | c}
{Model} & Ref. & $R_S$ (fm) & $C_0^{\rm IT}$ (fm$^2$) 
& $({\cal C}_1+{\cal C}_2 )/2$ (fm$^2$) &
{Model} & Ref. & $\Lambda$ (MeV) & $({\cal C}_1 + {\cal C}_2 )/2$ (fm$^2$) \\
\hline         
NV-Ia*  & \cite{Piarulli:2016vel} &  0.8  & 0.0158 & $-1.03$  
& Entem-Machleidt &\cite{Machleidt:2011zz}   & 500 & $-0.47$ \\ 
NV-IIa* & \cite{Piarulli:2016vel} &  0.8  & 0.0219 & $-1.44$  
& Entem-Machleidt &\cite{Machleidt:2011zz}   & 600 & $-0.14$ \\
NV-Ic   & \cite{Piarulli:2016vel} &  0.6  & 0.0219 & $-1.44$  
& 
Reinert {\it et al.} & \cite{Reinert:2017usi}& 450 & $-0.67$ \\
NV-IIc  & \cite{Piarulli:2016vel} &  0.6  & 0.0139 & $-0.91$  
& 
Reinert {\it et al.} & \cite{Reinert:2017usi}& 550 & $-1.01$ \\
& & & &							       
& 
NNLO$_{sat}$& \cite{Ekstrom:2015rta}          & 450 & $-0.39$ \\
\end{tabular}
\caption{Values of ${\cal C}_1 + \mathcal C_2$ obtained from the 
CIB contact interactions in various 
chiral potentials 
}  
\label{tb:lecs}
\end{table}

The potentials constructed in Refs. \cite{Machleidt:2011zz,Ekstrom:2015rta} use
the momentum regulator in Eq. \eqref{Lambda1b} with $n=3$ for both 
short-range and long-range potentials.
Reference \cite{Reinert:2017usi} constructed a semilocal potential
in which short-range interactions are regulated also as in 
Eq. \eqref{Lambda1b}, but with $n=1$.
The conversion to
the coefficients 
defined in Refs.\cite{Machleidt:2011zz,Ekstrom:2015rta,Reinert:2017usi} 
is 
\begin{equation}
\frac{\mathcal C_1 + \mathcal C_2}{2} = 
\frac{1}{4\pi e^2}\left(C^{np}_{^1S_0}-\frac{C^{pp}_{^1S_0} +C^{nn}_{^1S_0}}{2}\right)\,.
\end{equation}
The values of $(\mathcal C_{1} + \mathcal C_2)/2$ obtained from the 
$C^{pp}_{^1S_0}$, $C^{nn}_{^1S_0}$ and $C^{np}_{^1S_0}$ of 
Refs. \cite{Machleidt:2011zz,Ekstrom:2015rta,Reinert:2017usi}
for a few choices of $\Lambda$ are also reported in Table~\ref{tb:lecs}. 
While the LECs are not 
observable and depend on the scheme, we notice that the values of   
$F_\pi^2 (\mathcal C_1 + \mathcal C_2)$ in Table \ref{tb:lecs} 
are consistently larger than the prediction of Weinberg's counting.

Phenomenological potentials such as Argonne $v_{18}$ or CD-Bonn also include 
CIB effects of range $\lesssim m_\pi^{-1}$.
In the Argonne $v_{18}$ potential, the short-range component of the CIB 
potential is given in the notation of Ref. \cite{Wiringa:1994wb} by  
\begin{equation}\label{CIB_av18}
v^{cd}_{S1}(r) = -\frac{1}{6}  
\left[v^{c}_{S1,np}(r) - \frac{1}{2}\left(v^c_{S1,pp}(r) + v^c_{S1,nn}(r) \right) 
\right]  T^{(12)}\,.  
\end{equation}
The functions $v^c_{S1, N\!N}$ contain a medium-range component, which models 
two-pion contributions, and a genuine short-range component,
\begin{equation}
v^c_{S1, N\!N}(r) = I^c_{S1} \, T^2_\mu(r) 
+ \left[P^c_{S1,N\!N} + \mu r \, Q^i_{S1,N\!N} + (\mu r)^2\, R^i_{S1,N\!N}\right] 
W(r)\,,
\end{equation}
where $\mu= (2 m_{\pi^\pm} + m_{\pi^0})/3$ denotes the average pion mass. 
The function $T^2_\mu(r)$ is of two-pion-exchange range,  
while $W(r)$ is a Woods-Saxon function with radius $r_0 = 0.5$ fm 
and surface thickness $a=0.2$ fm, representing the short-range core. 
The parameters $I$, $P$ are fit to data in the $^1S_0$ channel, 
while $Q$ and $R$ are determined theoretically, 
as discussed in Ref. \cite{Wiringa:1994wb}.
While the potential in Eq. \eqref{CIB_av18} is not purely short-ranged,
when computing nuclear matrix elements with the Argonne $v_{18}$ wavefunctions 
we will replace
\begin{equation}
g_{\nu}^{\rm NN} \delta^{(3)}_{R_S}(\spacevec r) \rightarrow  
- \frac{6}{e^2} \, v^{cd}_{S1}(r).
\label{replaceAV18}
\end{equation}
This is justified since our long-range neutrino potential does not include 
the two-pion effects mimicked by $T^2_{\mu}(r)$, which were computed in 
$\chi$EFT in Ref. \cite{Cirigliano:2017tvr}.

\subsection{LNV amplitudes in light nuclei}
\label{LNVlightnuclei}

In what follows---since we lack  observables 
to disentangle ${\cal C}_1$ from ${\cal C}_2$---we make the assumption 
that ${\cal C}_1= {\cal C}_2$, in which case
the replacements \eqref{replacechiral} and \eqref{replaceAV18} are
justified.
We stress again that this 
assumption is arbitrary (see Sec.~\ref{CIBfits}), 
but it 
exemplifies the potential impact of short-range physics 
on $0\nu\beta\beta$ matrix elements. 

We studied two transitions corresponding to the cases 
in which the initial and final nucleus have the same isospin, 
$\Delta I=0$, or the nuclear isospin changes by two units, 
$\Delta I=2$. 
The latter is the case for all the experimentally relevant $0\nu\beta\beta$ 
emitters.
We consider $^6$He $\rightarrow$ $^6$Be 
as a $\Delta I=0$
example, and $^{12}$Be $\rightarrow$ $^{12}$C for the  
$\Delta I=2$ case.
In both cases, we provide results obtained from a phenomenological 
potential and a chiral potential.
In the former, nuclear wavefunctions are derived from 
a many-body Hamiltonian with two- and three-body forces corresponding
to the Argonne $v_{18}$~\cite{Wiringa:1994wb} and Illinois-$7$~\cite{IL7} 
potentials. In the figures and in what follows, 
we will denote these calculations with the 
label ``AV18''. Details on the
procedure adopted to construct the Variational Monte Carlo (VMC)
wavefunctions can be found
in Ref.~\cite{Pastore:2017ofx} and references therein.
The second set of calculations is based on nuclear wavefunctions
obtained from chiral two- and three-body forces developed and 
constrained in Refs.~\cite{Piarulli:2016vel,Piarulli:2014bda,Baroni:2018fdn}.
The $A=6$ calculation uses the model NV-IIa*, while the $A=12$ calculation 
is based on the NV-Ia* model. We will refer to this set of calculations with the 
label ``$\chi$EFT''.

\begin{figure}
\includegraphics[width=\textwidth]{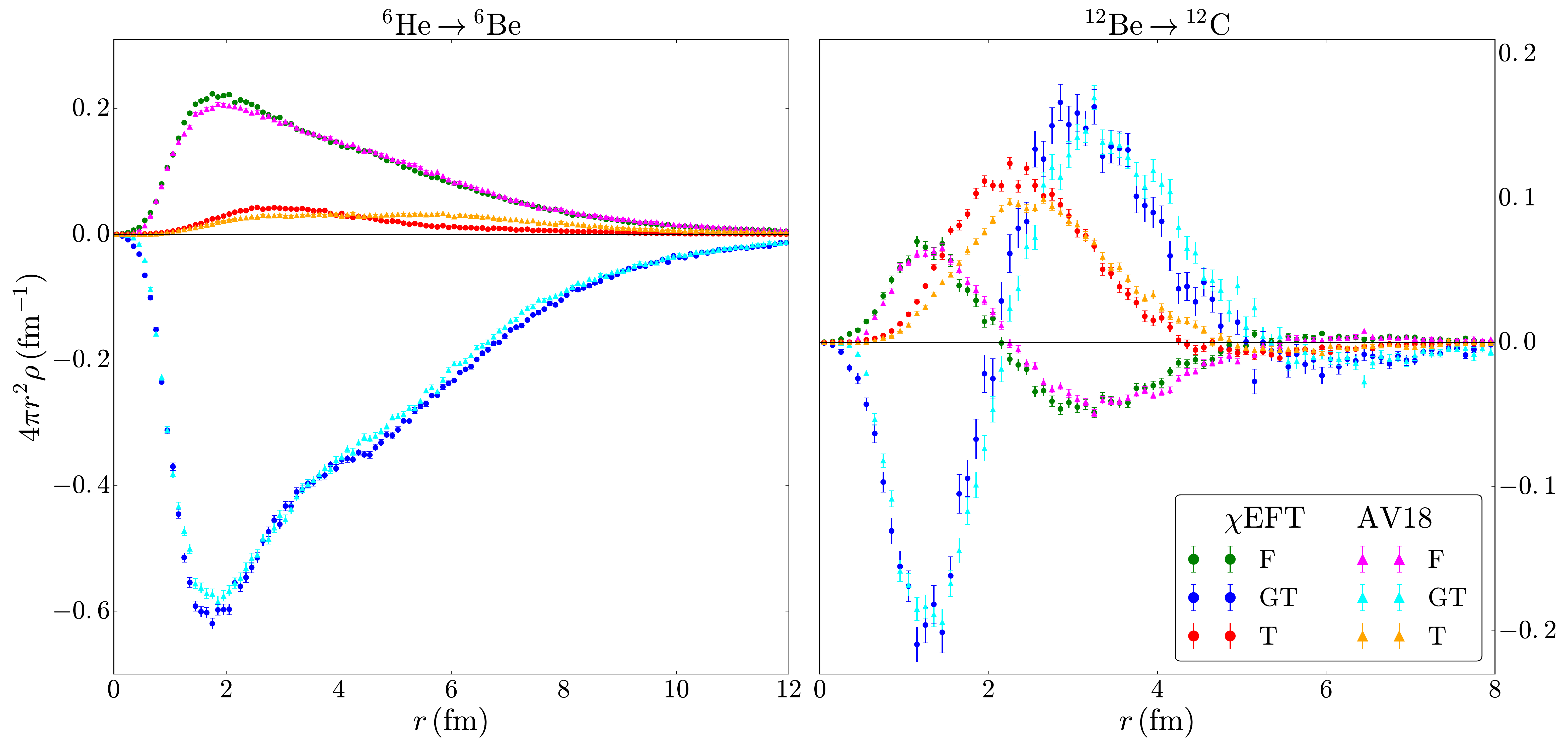}
\caption{VMC calculations of the Fermi (F), Gamow-Teller (GT), and tensor (T) 
densities $\rho(r)$
for 
$^{6}$He$\rightarrow^{6}$Be (left panel)
and $^{12}$Be$\rightarrow^{12}$C (right panel) decays 
with two potentials, labeled AV18 and $\chi$EFT.}
\label{densities}
\end{figure}

In Fig. \ref{densities} we plot the Fermi (F), Gamow-Teller (GT), 
and tensor (T) radial densities $\rho$, defined as
\begin{eqnarray}
4\pi r^2 \rho_{F}(r) &=& \langle \Psi_f | \sum_{a < b} \tau^{(a)+} \tau^{(b)+} \, 
\delta^{}(r_{ab} - r) | \Psi_i \rangle  \, , 
\nonumber\\
4\pi r^2 \rho_{GT}(r) &=& \langle \Psi_f | \sum_{a < b} \tau^{(a)+} \tau^{(b)+} \,
{\bm \sigma}^{(a)}\cdot{\bm \sigma}^{(b)}\,\delta^{}(r_{ab} - r)\,|\Psi_i \rangle 
\, , 
\nonumber \\
4\pi r^2  \rho_{T}(r)  &=& \langle \Psi_f | \sum_{a < b} \tau^{(a)+}  \tau^{(b)+} \, 
S^{(ab)}\,  \delta^{}(r_{ab} - r) \,  | \Psi_i \rangle \, ,  
\label{rhos}
\end{eqnarray}
where $\Psi_{i,f}$ denote the initial- and final-state wavefunctions, 
and $r_{ab}$ is the distance between two nucleons.
Figure \ref{densities} shows an excellent level of agreement between 
the densities computed with the AV18 and $\chi$EFT formulations. 
The 
$\Delta I=2$ F and GT densities exhibit the typical node due to the 
orthogonality of the initial and final wavefunctions, 
and the integrated F density gives the correct, vanishing result.

\begin{figure}
\includegraphics[width=\textwidth]{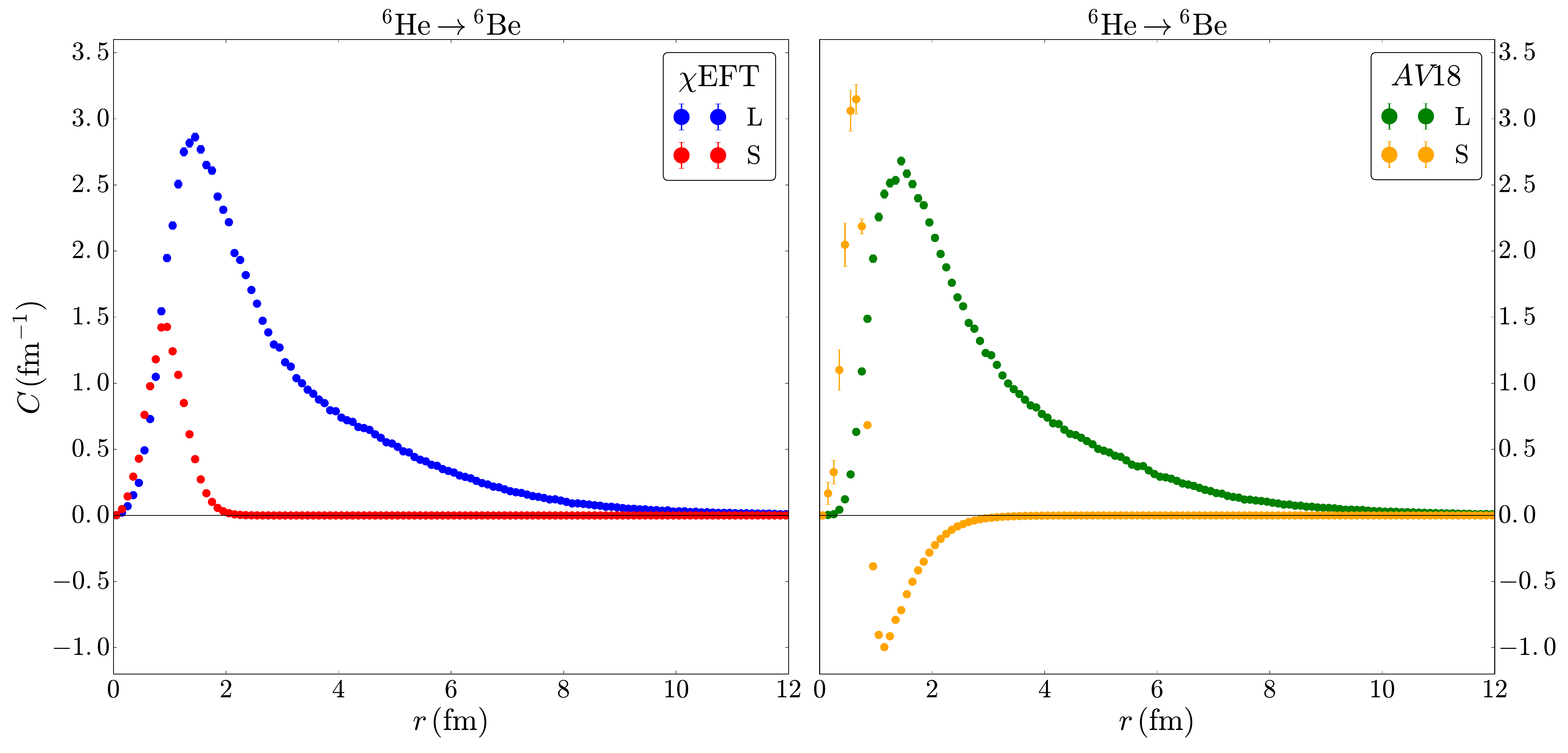}
\includegraphics[width=\textwidth]{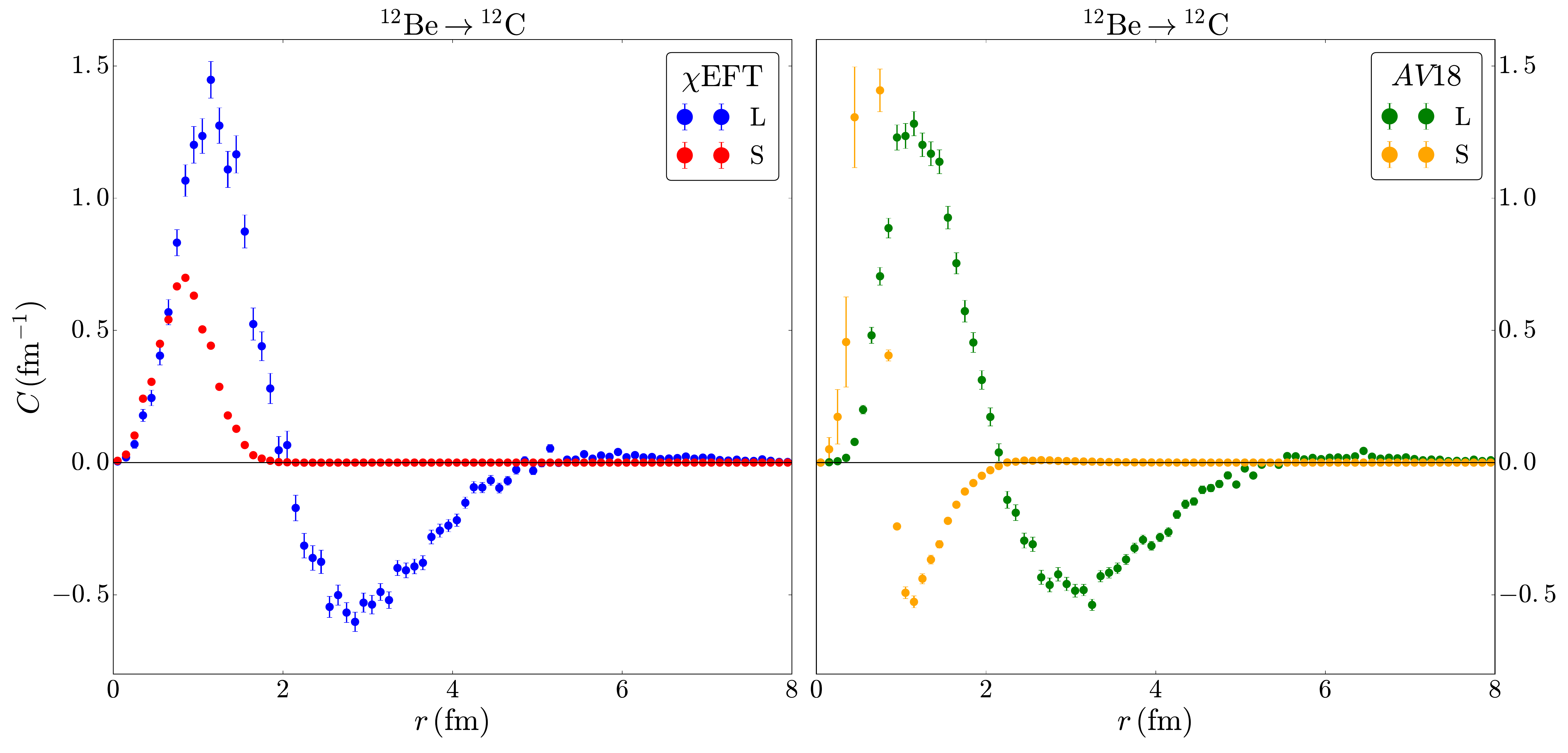}
\caption{VMC calculations of the long- (L) and short- (S) range 
transition densities $C(r)$ 
for 
$^{6}$He$\rightarrow^{6}$Be (upper panel) and 
$^{12}$Be$\rightarrow^{12}$C (bottom panel) decays
with two potentials, labeled $\chi$EFT (left) and AV18 (right).
}\label{fig:ME}
\end{figure}

In order to compare the long- and short-range contributions we define,
similarly to Eq. \eqref{eq:meden}, the transition densities 
\begin{equation}
C_{\rm L,S}(r) = 4\pi R_A \, \langle \Psi_f | \sum_{a, b} V_{\nu \,\rm L,S}(r_{ab}) 
\,\delta^{}(r_{ab} - r) | \Psi_i \rangle \, , \quad  M_{\rm L,S} = \int dr \, C_{\rm L,S}(r) 
\label{eq:cLS}
\end{equation}
where the conventional factor $\propto R_A = 1.2\, A^{1/3}$ fm was
introduced to make the 
$A$-nucleon matrix element dimensionless.
These densities are plotted in Fig.~\ref{fig:ME}.
Integrating $C_{\rm L,S}(r)$ over $r$
yields the values for the matrix elements $M_i$ shown 
in Table \ref{tb:ME},
where we split the long-range neutrino potential in its 
Fermi, Gamow-Teller, and tensor components
\begin{eqnarray}\label{eq:MEexplicit}
M_{i} = 4 \pi R_A \int d r \, V_{i}(r) (4\pi r^2 \rho_{i}(r)), \quad i \in \{ F,\, GT,\, T \}.
\end{eqnarray}
The neutrino potentials $V_{F,\, GT, T}(r)$ are defined in Eq.~\eqref{FGTT}, and $M_L = M_F - g_A^2 (M_{GT} + M_T)$.

\begin{table}
\begin{tabular}{ c |c| c| c| c || c| c}
$A$  &  Model      & $M_F$  &  $M_{GT}$  &   $M_T$  &  $M_{\rm L}$  & $M_{\rm S}$ \\
\hline
$6$  &  AV18       & 1.56   & $-3.66$   & 0.03   &  7.45   &  0.48 \\
     &  $\chi$EFT  & 1.62   & $-3.85$   & 0.03   &  7.82   &  1.15 \\
\hline     
$12$ &  AV18       & 0.198  & $-0.349$  & 0.068  &  0.653  &  0.518 \\
     &  $\chi$EFT  & 0.223  & $-0.394$  & 0.083  &  0.725  &  0.533 \\       
\end{tabular}
\caption{VMC results for the dimensionless matrix elements
of the long- (L) and short- (S) range neutrino-exchange potentials, defined in Eqs. \eqref{eq:cLS} and \eqref{eq:MEexplicit}.
For each row, the total long-distance entry 
$M_{\rm L}$ is obtained via the combination 
$M_F -g_A^2 (M_{GT} + M_{T})$ of its Fermi (F), Gamow-Teller (GT), and tensor (T)
components. 
}
\label{tb:ME}
\end{table}

The matrix elements of the long-range neutrino potentials obtained with the 
AV18 and $\chi$EFT models are in good agreement with each other, and 
with the results of Ref.~\cite{Pastore:2017ofx}, which used the same AV18
model described above for the nuclear Hamiltonian and ``clusterized'' 
wavefunctions obtained by allowing for the formation of clusters in the 
$p$ shell~\cite{Nollett:2000ch}.
In contrast, the profile of the matrix element of the short-range neutrino 
potential is sensitive to the model, which for the $\Delta I = 0$ transition
translates into an uncertainty of a factor 2 in the integrated density.
For $\Delta I = 2$, the integrated density is much less sensitive to the
model.

With the assumption $g_{\nu}^{\rm NN} ={\cal C}_1 = {\cal C}_2$,
the short-range component of the neutrino potential amounts
to only $5-15$\% of the long-range component in the total ${}^6$He$\rightarrow {}^6$Be amplitude.
As for the $nn\rightarrow pp$ transition discussed in Sec. \ref{LNVamptwobody},
the smallness is mostly due to the monotonic long tail of the distribution 
seen in the top panel of Fig. \ref{fig:ME}, which is a feature of $\Delta I=0$ transitions.

In contrast, for the $\Delta I = 2$
transition the orthogonality of the wavefunctions implies a cancellation 
between the long-range contributions from $r \lesssim 2$ fm and 
$r \gtrsim 2$ fm, seen in the lower panel of Fig. \ref{fig:ME}.
In this case, the contribution of
$g_{\nu}^{\rm NN}={\cal C}_1 = {\cal C}_2$ 
is a sizable $75-80$\% of the total long-range contribution.
Although Fig. \ref{fig:ME} appears to show a higher degree of cancellation 
compared to many-body calculations in experimentally relevant nuclei,
the node in the density is a robust feature of 
$\Delta I = 2$ transitions \cite{Simkovic:2007vu,Menendez:2008jp}.
We thus expect the contribution of the short-range operator 
$g_{\nu}^{\rm NN}$ to be non-negligible.

We caution that these results are based on the arbitrary choice 
${\cal C}_1={\cal C}_2$
dictated by the current undetermined value of ${\cal C}_2$. 
Using the more general assumption 
$\mathcal C_2 = \alpha \, \mathcal C_1$ as in Eq. \eqref{choice}
leads to a simple rescaling of the last column of Table \ref{tb:ME}
by $\kappa = 2/(1+\alpha)$. For the $^{12}$Be $\rightarrow$ $^{12}$C transition, 
the short-range component of $V_\nu$ can be reduced to a $20$\% ($-20$\%) 
correction only for large values
for $\alpha \approx 6 $ ($\alpha \approx -8$), 
which would require a sizable deviation from the power-counting expectation 
$\mathcal C_1 \sim \mathcal C_2$.

Standard derivations of the $0\nu\beta\beta$ transition operator include 
short-range effects by introducing the axial, vector, and weak magnetic 
form factors of the nucleon.
We stress that in the analysis of CIB in $N\!N$ scattering the vector 
form factor is included
in both AV18 and $\chi$EFT photon-exchange potentials.
However,
it does not capture the entire short-range dynamics, 
which results in non-zero $C_0^{\rm IT}$ and $v^{cd}_{S1}(r)$.
The contribution of weak magnetism induces corrections to the GT and T 
potentials~\cite{Simkovic:1999re}, which
neglecting the momentum dependence of the magnetic form factor
are~\cite{Pastore:2017ofx}
\begin{equation}
V_{GT,M\!M}(r) = \frac{(1+\kappa_1)^2}{6 g_A^2 m_N^2} \,\delta^{(3)}_{}(\spacevec r)
\,, \qquad
V_{T,M\!M}(r) =\frac{(1+\kappa_1)^2}{16\pi g_A^2 m_N^2} \, \frac{1}{r^3}\,. \\
\end{equation}
$V_{GT,M\!M}$ provides a shift in $g_{\nu}^{\rm NN}$, with coefficient determined 
by the isovector magnetic moment.
The matrix element of $V_{GT,M\!M}$ is however much smaller than the short-range 
component shown in Table \ref{tb:ME}. 
Using the $\chi$EFT wavefunctions we find for example
\begin{equation}\label{MGTMM}
M_{GT,M\!M}(^6{\rm He} \rightarrow\, ^6\mathrm{Be}) = -0.10 \,,
\qquad  
M_{GT,M\!M}(^{12}{\rm Be} \rightarrow\, ^{12}{\rm C}) = -0.060\,,
\end{equation}
where $M_{GT, MM}$ is defined as in Eq. \eqref{eq:MEexplicit}, with $V_{GT} \rightarrow V_{GT, MM}$.
Eq. \eqref{MGTMM} is to be compared to the contributions of $g_{\nu}^{\rm NN}$ in Table~\ref{tb:ME}.
This result is a reflection of the fact that CIB data in $N\!N$ scattering indicate 
the need for an independent local operator, whose coefficient
is large and not determined by couplings in the single-nucleon sector. 
While the extrapolation from CIB to $0\nu\beta\beta$ relies on 
the uncontrolled assumption ${\cal C}_1 = {\cal C}_2$, the results provide 
strong evidence for the importance of short-range dynamics in $0\nu\beta\beta$.

\section{Conclusion}
\label{conclusion}

Neutrinoless double-beta decay is the most sensitive laboratory probe of the 
Majorana nature of neutrinos. 
The limits on the electron-neutrino Majorana mass from current data, or its 
extraction from future observations, rely on calculations of
$0\nu\beta\beta$ nuclear matrix elements.  
The calculation of these transition matrix elements in nuclei such as $^{76}$Ge
or $^{136}$Xe starting from QCD is a daunting task. Nuclear EFTs can help 
bridge this gap by deriving interactions and transition operators that 
capture the symmetries of QCD and providing a theoretically consistent
framework that can be improved order by order.
Nuclear matrix elements of light nuclei, while not directly experimentally 
accessible, play an important role in establishing such a framework. 
The first {\it ab initio} calculations of $0\nu\beta\beta$ matrix elements, 
in which nuclear wavefunctions 
are computed using chiral interactions that are fitted to the properties 
of two- and three-nucleon systems, 
are starting to appear as part of a concerted effort toward 
the reduction of theoretical uncertainties in $0\nu\beta\beta$.
In this paper, we derived the $0\nu\beta\beta$ transition operator 
consistent with these interactions for the case of LNV mediated by 
light Majorana neutrinos.

Our main findings can be summarized as follows:
\begin{itemize}
\item The $0\nu\beta\beta$ transition operator mediated by light 
Majorana neutrinos has both a long-range
and a short-range component at leading order in $\chi$EFT. 
The long-range component can be expressed in terms of the couplings 
of nucleons and pions to the axial and vector weak currents,
while the short-range component is parametrized by a contact operator 
whose coefficient, $g_{\nu}^{\rm NN}$, encodes nontrivial
QCD dynamics and is at the moment unknown.
The need for a short-range component of the $0\nu\beta\beta$ 
transition operator emerges clearly by studying the \nnpp
scattering amplitude in various regularization schemes, as done in Sec. \ref{LNVLO}. 
The matrix element of the long-range neutrino potential, $V_\nu$, 
between two incoming neutrons and two outgoing protons in the 
$^1S_0$ channel depends logarithmically on the  short-range regulator. 
Observables can only be made  regulator-independent by inclusion of a leading-order short-range LNV operator.
Similar sensitivity to UV physics appears in other processes 
involving potentials with Coulombic behavior that act in the $^1S_0$ channel, 
for instance charge-independence breaking.
The analysis of Sec. \ref{CIB} shows that to reproduce the observed 
combination of scattering lengths $a_{nn}+a_{pp}-2 a_{np}$ in the $^1S_0$ channel,
charge-independence-breaking counterterms need to appear at $\mathcal O(e^2)$. 
They are thus enhanced by $(4\pi)^2$, or two powers in the $\chi$EFT power 
counting, with respect to Weinberg's power counting. 
The enhanced contribution of short-range dynamics to charge-independence 
breaking  is observed in both chiral and phenomenological $N\!N$
potentials \cite{Epelbaum:2014efa,Piarulli:2016vel,Wiringa:1994wb,Machleidt:2000ge}.

\item There is no need for the enhancement of short-range LNV operators in 
higher partial waves, such as the $^{3}P_J$ or $^1D_2$.
This can be expected in channels where the pion-exchange tensor force 
is absent or repulsive, like the $^1D_2$ and $^3P_1$ channels.
While the attractive nature of the tensor force requires the promotion of an 
$N\!N$
contact operator to leading order \cite{Nogga:2005hy}, once 
strong interactions are properly renormalized, the matrix element of 
the long-range neutrino potential is cutoff independent
and does not require additional renormalization. We thus expect short-range LNV operators in $P$ and $D$ waves to follow Weinberg's counting.

\item There is no evidence for a short-range momentum-dependent counterterm 
at next-to-leading order in $\chi$EFT. 
The NLO analysis of the scattering amplitude was discussed in 
Sec. \ref{LNV@NLO}, in a variety of schemes. In dimensional regularization,
the scale invariance of the amplitude requires 
inclusion of the derivative operator $g_{2\nu}^{\rm NN}$ at NLO, 
but its coefficient is not independent, and is determined in terms of known 
couplings. The NLO corrections to $\mathcal A_\nu$ then purely stem from 
NLO corrections to the $^1S_0$ strong scattering amplitude. 
In a cutoff scheme, we similarly showed that $\mathcal A_{\nu}$ at NLO 
becomes cutoff independent as the cutoff is removed without
the inclusion of a momentum-dependent counterterm.  The residual cutoff dependence of the NLO \nnpp scattering amplitude exhibits a 
$\ln(\Lambda)/\Lambda$ behavior, which might lead to sizable corrections 
at moderate values of the cutoff, unless $g_{2\nu}^{\rm NN}$ is introduced
with fixed coefficients as in dimensional regularization.
Our analysis indicates that an independent $g_{2\nu}^{\rm NN}$ enters the 
neutrino potential at N$^2$LO, or $\mathcal O(Q^2/\Lambda_\chi^2)$, 
the same order as contributions from nucleon form factors, 
closure corrections, and pion-neutrino loops \cite{Cirigliano:2017tvr}. 
We must say, however, that we cannot completely exclude an independent finite LEC at NLO. While NDA predicts such a term at N${}^2$LO, it should be kept in mind that NDA only provides a guide to what should be included in a calculation. A possible way to verify the presence (or lack thereof) of an independent NLO LEC would by connecting  $g_{2\nu}^{\rm NN}$ to a CIB-breaking combinations of nucleon-nucleon effective ranges, similar to the connection between $g_{\nu}^{\rm NN}$ and CIB scattering lengths (see next bullet). 

\item The determination of the LEC $g_{\nu}^{\rm NN}$ requires an LQCD calculation
of the \nnpp scattering amplitude and its matching to $\slashpi$EFT or 
$\chi$EFT. In the absence of an LQCD calculation, we can get an order 
of magnitude estimate of $g_{\nu}^{\rm NN}$ using symmetry arguments.
Isospin symmetry relates $g_{\nu}^{\rm NN}$ to the
component of the short-range charge-independence-breaking operators that 
transform as the product of two left- or two right-handed currents, 
denoted by ${\cal C}_1$ in Sec. \ref{CIB}. However, the short-range charge-independence-breaking operators also have a left-right component, ${\cal C}_2$, 
and $N\!N$
scattering data cannot completely determine $g_{\nu}^{\rm NN}$.
Using the naturalness assumption ${\cal C}_1 \sim {\cal C}_2$, we showed 
the potential impact of the short-range neutrino potential on $0\nu\beta\beta$ 
matrix elements in light nuclei. In Sec. \ref{pheno} 
we computed the matrix elements for the 
$^6$He $\rightarrow$ $^6$Be ($\Delta I =0$) 
and $^{12}$Be $\rightarrow$ $^{12}$C ($\Delta I =2$) 
transitions using wavefunctions obtained with the $\chi$EFT interactions 
of Ref. \cite{Piarulli:2016vel}, and 
a consistent extraction of ${\cal C}_1 + {\cal C}_2$.  
While its impact on 
$\Delta I =0$ transitions is moderate, $g_{\nu}^{\rm NN}$ can significantly 
affect 
$\Delta I =2$ transitions. 
This observation reinforces the need for a first-principle calculation 
of $g_{\nu}^{\rm NN}$, in particular because relative factors in the relation 
between ${\cal C}_1$ and ${\cal C}_2$ have $\mathcal O(1)$ impact on 
the final 
results.

\item 
We cannot at this point address the relatively large uncertainty 
\cite{Engel:2016xgb} in the calculation of the matrix elements of 
heavier nuclei, which are of experimental interest.
Once this issue is resolved, the extraction of the effective neutrino
mass $m_{\beta\beta}$ could be significantly affected by the short-range
LEC $g_{\nu}^{\rm NN}$. If the effect is similar to that calculated 
in $^{12}$Be $\rightarrow$ $^{12}$C with the assumption	${\cal C}_1 \sim {\cal C}_2$, the nuclear neutrinoless double beta decay amplitude would double.
With another natural assumption, it might instead be halved.
Or it could also be that the small range suppresses its effects
for larger $A$. It would be of great interest to calculate the effects of the 
leading short-range current in heavier nuclei. We suggest to use, as a starting point, the relation 
$g_{\nu}^{\rm NN} \rightarrow  \frac{{\cal C}_1+{\cal C}_2}{2}$ and the values of ${\cal C}_1+{\cal C}_2$ as given in Table~\ref{tb:lecs} corresponding to the strong potential applied to obtain the nuclear wave functions. 

\end{itemize}

The EFT framework presented here can be extended in several  directions.
One of them involves 
the inclusion of next-to-next-to-leading-order corrections to the nuclear 
potential.  
Such terms play an important role in high-quality descriptions of the $N\!N$
database. At this order, the LNV potentials obtain additional corrections 
\cite{Cirigliano:2017tvr} that should be consistently included.  
Three-body LNV operators have been identified as a potential source of 
``$g_A$ quenching'' \cite{Menendez:2011qq,Wang:2018htk}. 
It would be interesting to extend our $\chi$EFT framework to 
three-nucleon processes. 
Our work here has been limited to LNV arising from a 
light-Majorana-neutrino mass term, but in well-motivated scenarios 
of beyond-the-Standard Model physics, $0\nu\beta\beta$ decay rates 
can be dominated by higher-dimensional LNV 
operators \cite{Prezeau:2003xn,Cirigliano:2017djv,Cirigliano:2018yza,Graf:2018ozy}. It was argued in Ref.~\cite{Cirigliano:2018yza} that short-range 
operators must be promoted to leading order for several higher-dimension 
LNV operators, but the impact of higher-order corrections has not been 
investigated so far.
Most importantly, calculations of the leading 
short-range contributions must be carried out for heavier nuclei.

\section*{Acknowledgments}

We acknowledge stimulating discussions with Joe Carlson, Jon Engel,  
Wick Haxton, Martin Hoferichter, Javier Men{\'e}ndez, Rocco Schiavilla, and Andr{\'e} Walker-Loud.
This research was supported in part by the LDRD program at Los Alamos 
National Laboratory (VC, MG, EM), the DOE topical collaboration on 
``Nuclear Theory for Double-Beta Decay and Fundamental Symmetries'' 
(VC, EM), the US DOE, Office of Science, Office of Nuclear Physics, Office of High Energy Physics, under award
numbers DE-FG02-04ER41338 (UvK), DE-SC0009919 (WD), DE-AC52-06NA25396 (VC,MG,EM), and DE-AC02-06CH11357 (RBW), 
the NUCLEI SciDAC and INCITE programs (SP, MP, RBW),  and the European Union Research 
and Innovation program Horizon 2020 under grant agreement No.\ 654002 (UvK).  
JdV is supported by the  RHIC Physics Fellow Program of the RIKEN BNL Research Center.
Computational resources have been provided by
the Argonne Laboratory Computing Resource Center and Argonne Leadership
Computing Facility.

\appendix

\section{LNV and $\Delta I = 2$ operators with multiple mass insertions}
\label{mass}

In this appendix we consider the most important LNV operators with
insertions of the quark mass mentioned in Sec. \ref{CIB}:
one (App. \ref{Onemassinsertion})
and two (App. \ref{Twomassinsertions}) mass insertions.
A summary is given in App. \ref{Massinsertionsum}.

\subsection{One mass insertion}
\label{Onemassinsertion}

The LNV operators involving one quark-mass insertion can be built with the  
elements $Q_{L,R}$ and $M = {\rm diag}\, (m_u,\, m_d)$, as well as 
$u$, $u^\dagger$, $N$, and $\bar N$. We can choose to work with the slightly 
different spurions $\cq_{L,R}$ and $M_{\pm}= u^\dagger M u^\dagger \pm  u M^\dagger u$,
after which all the building blocks transform only under the diagonal subgroup 
(i.e.\ $N\to KN$, $\cq_{L,R}\to K\cq_{L,R} K^\dagger$, and $M_\pm\to KM_\pm K^\dagger$)
apart from $u$ and $u^\dagger$. Thus, whenever an operator includes $u$ and/or 
$u^\dagger$, their indices 
have to be contracted 
with each other, giving rise to factors of $u^\dagger u=uu^\dagger =1$. 
As a result we can forget about the $u$ matrices and use only the spurions 
and nucleon fields.

We are interested in the operators that give rise to $\Delta I=2$ transitions 
in $N\!N$ scattering which can be built from $\cq_X \times \cq_Y\times M_\pm$ 
(where $X,Y\in L,R$). We will therefore need all the $\bar 5$ representations 
that reside in the generic tensor, 
$T^{abc}_{ijk} =  (\cq_X)^a_i(\cq_Y)^b_j(M_\pm)^c_k$, which transforms as 
$T^{abc}_{ijk}\to K_{aa'}K_{bb'}K_{cc'}T^{a'b'c'}_{i'j'k'}(K^\dagger)_{i'i}(K^\dagger)_{j'j}
(K^\dagger)_{k'k}$. 
This tensor can be rewritten as 
$\bar T^{abc\, ijk} =\epsilon^{ii'} \epsilon^{jj'}\epsilon^{kk'} T^{abc}_{i'j'k'}$, 
so that all indices transform in the same way, 
$\bar T^{abc\, ijk}\to K_{aa'}K_{bb'}K_{cc'}\bar T^{a'b'c'\,i'j'k'}K_{ii'}K_{jj'}K_{kk'}$. 
One can then show that the largest dimensional representation, the $\bar 7$, 
is given by $\bar T$ with completely symmetrized indices, the next-largest 
irrep is the one with two antisymmetrized indices (keeping the rest fully 
symmetric), while the second largest has two pairs of antisymmetrized indices, 
{\it etc}. Thus, to find all the $\bar 5$ irreps we need to find all the ways 
in which to contract $\bar T$ with a single $\epsilon^{IJ}$ tensor. There are 5 
independent ways of doing this, in agreement with the decomposition of 
$\bar 2 \otimes\bar 2\otimes\bar 2\otimes\bar 2\otimes\bar 2\otimes\bar 2$.

One can then choose to contract the indices of $\bar T^{abc\, ijk}$ with the 
following 
tensors---after using
$T^{abc\, ijk}\epsilon_{ab} =T^{acb\, ijk}\epsilon_{ab}+T^{cba\, ijk}\epsilon_{ab}$, 
which follows from $p^I \epsilon^{JK}+p^J \epsilon^{KI}+p^K \epsilon^{IJ}=0$, to 
move indices around:
\bea
\epsilon^{ck},\quad \epsilon^{jc},\quad \epsilon^{ic},\quad \epsilon^{ak},
\quad \epsilon^{bk},
\label{eq:contractions}
\eea
where we pick combinations that lead to multiplication of the matrices in $T$ 
(without leaving any explicit $\epsilon^{ij}$s).
In terms of $T$, this leads to the combinations
\bea
&(\cq_X)^a_{i}(\cq_Y)^b_j \,{\rm Tr}\, M_{\pm}\,, 
\quad (\cq_X)^a_{i}(M_\pm \cq_Y)^b_j \,,\quad (\cq_X)^a_{i}( \cq_YM_\pm)^b_j \,,
\nn\\
& (\cq_Y)^a_{i}(M_\pm \cq_X)^b_j \,,\quad (\cq_Y)^a_{i}( \cq_XM_\pm)^b_j \,.
\eea
One then still needs to project the remaining indices onto the $\bar 5$ 
representation, by demanding the upper/lower indices to be symmetric and 
traceless (or simply fully symmetric in the case of $\bar T$). 
An explicit form of this projection is 
\bea
A^a_i B_j^b\bigg|_{\bar 5} &=& \frac{1}{2} A^a_i B^b_j
+ \frac{1}{24}({\rm Tr}\, A\,{\rm Tr}\, B -2 {\rm Tr}\, AB)
(\tau^I)^a_i (\tau^I)^b_j
\nn\\
&&-\frac{1}{4}\,{\rm Tr}\, A(B_i^a\delta^b_j+B_j^b\delta^a_i)
+\frac{1}{8} {\rm Tr}\, A\,{\rm Tr}\, B\dt^a_i \dt^b_j + (A\leftrightarrow B)\,.
\label{eq:project5}
\eea
After projecting, one then has to make sure that the combinations of 
$\cq_{L,R}$ and $M_{+,-}$ have the correct 
properties under charge conjugation ($C$), parity ($P$), and time reversal 
($T$). 

All in all,  this leads to the following operators,
\bea
O_{M1}^{(1)} &=& O_1 \, {\rm Tr}\, M_+\,,
\qquad O_{M1}^{(2)} = O_2 \, {\rm Tr}\, M_+\,,
\qquad O^{(1)}_{M2} = \bar N\cq_L N\, \bar N
\left[ \cq_L,\, M_-\right] N-(L\leftrightarrow R)\,,
\nn\\
O^{(2)}_{M2} &=& \bar N\cq_L N\, \bar N\left[ \cq_R,\, M_-\right] N
-\frac{1}{6} \,{\rm Tr}\left(\cq_L \left[\cq_R,\,M_-\right]\right)
\bar N\tau^IN\, \bar N\tau^IN  -(L\leftrightarrow R)\,,
\label{eq:1massOs}
\eea
where $O_{1,2}$ are the operators without any $M$ insertions,
\bea
O_1 &=&  \bar N\cq_L N\, \bar N \cq_L N
-\frac{1}{6} \,{\rm Tr}\left(\cq_L\cq_L \right)
\bar N\tau^IN\, \bar N\tau^IN+(L\leftrightarrow R)
\,,\nn\\
O_2 &=&  \bar N\cq_L N\, \bar N \cq_R N
-\frac{1}{6} \,{\rm Tr}\left(\cq_L\cq_R \right)
\bar N\tau^IN\, \bar N\tau^IN+(L\leftrightarrow R)\,.
\eea
There are no operators similar to $O^{(1,2)}_{M1}$ with $M_-$ instead of $M_+$ 
since Tr$M_-=0$. In addition, operators of the form of $O^{(1,2)}_{M2}$ involving 
$M_+$ instead of $M_-$ necessarily contain an anticommutator 
$\{M_+,\, \cq_X\}$, which can be rewritten in terms of $O^{(1,2)}_{M1}$
thanks to the fact that the $\cq_X$ are traceless. 

For the $N\!N$ vertices without any pions, only the $O_{M1}^{(1,2)}$ operators 
contribute, simply giving the original operators multiplied by Tr $M$. 
Instead, the $O^{(1,2)}_{M2}$ operators only induce vertices with two additional 
pions.

\subsection{Two mass insertions}
\label{Twomassinsertions}

We can use a similar process to find the $\bar 5$ representations in 
$T^{abcd}_{ijkl} =  (\cq_X)^a_i(\cq_Y)^b_j(M_\pm)^c_k(M_\pm)^d_l$. 
One now has to contract with two epsilon tensors, which can be done in 20 
independent ways. However, after choosing a set of ways to contract the 
indices, not all possibilities contribute, for example due to $P$, $T$, 
or $C$ properties, or because some of our building blocks are traceless. 
The operators involving two insertions of the same $M_\pm$ 
take the schematic form
\bea
&O^{(1,2)}_{M_\pm\sq 1} = {\rm Tr}\, M_{\pm}\sq O_{1,2}\,
\qquad  O^{(1,2)}_{M_+\sq 2} = {\rm Tr}\, \left(M_{+}\right)\sq O_{1,2},\, 
\nn\\
O^{(1)}_{M_\pm\sq 3} :&
\quad A = \cq_L M_\pm \cq_L\,,
\quad B = M_\pm
\qquad +(L\leftrightarrow R)\,,
\nn\\
O^{(2)}_{M_\pm\sq 3} :&
\quad A = \cq_L M_\pm \cq_R\,,
\quad B = M_\pm
\qquad +(L\leftrightarrow R)\,,
\nn\\
O^{(1)}_{M_\pm\sq 4} :&
\quad A = \cq_L\, {\rm Tr}\left(M_\pm \cq_L\right)\,,
\quad B = M_\pm
\qquad +(L\leftrightarrow R)\,,
\nn\\
O^{(2)}_{M_\pm\sq 4} :&
\quad A = \cq_L\, {\rm Tr}\left(M_\pm \cq_R\right)\,,
\quad B = M_\pm
\qquad +(L\leftrightarrow R)\,,
\nn\\
O^{(1)}_{M_\pm\sq 5} :&
\quad A =  M_\pm\cq_L\,,
\quad B = M_\pm\cq_L
\qquad +(L\leftrightarrow R)+(M_{\pm}\leftrightarrow \cq_{L,R})\,,
\nn\\
O^{(2)}_{M_\pm\sq 5} :&
\quad A =  M_\pm\cq_L\,,
\quad B = M_\pm\cq_R
\qquad +(L\leftrightarrow R)+(M_{\pm}\leftrightarrow \cq_{L,R})\,,
\label{eq:2Minsertion1}
\eea
One 
should use the above expressions for $A$ and $B$ to construct the corresponding
operators by projecting them onto the $\bar 5$ representation using 
Eq. \eqref{eq:project5}, and subsequently contracting with 
$\bar N^a \bar N^b N^iN^j$. 
After doing so the $O^{(1,2)}_{M_+\sq 3,4}$ operators do not contribute to 
$(\bar NN)^2$ and $(\pi \bar NN)^2$ vertices, while 
$O^{(1,2)}_{M_+\sq 5}=O^{(1,2)}_{M_+\sq 2}/2$ (at least up to two-pion vertices, 
for $m_u = m_d$).
Instead, the $O^{(1,2)}_{M_-\sq 3,4,5}$ operators only give rise to 
$(\pi \bar NN)^2$ vertices.

The remaining operators are proportional to $M_- M_+$, and take the form
\bea
O_{M_+ M_- 1}^{(1)}:&
\quad A=\cq_L M_+,
\quad B = \cq_L M_-
\qquad -(L\leftrightarrow R)-(M_{\pm}\leftrightarrow \cq_{L,R})\,,
\nn\\
O_{M_+ M_- 1}^{(2)}:&
\quad A=\cq_R M_+,
\quad B = \cq_L M_-
\qquad -(L\leftrightarrow R)-(M_{\pm}\leftrightarrow \cq_{L,R})\,,
\nn\\
O_{M_+ M_- 2}^{(1)}:&
\quad A=\left[M_-,\, \cq_L\right]\,,
\quad B = \cq_L\, {\rm Tr}\, M_+
\qquad -(L\leftrightarrow R)\,,
\nn\\
O_{M_+ M_- 2}^{(2)}:&
\quad A=\left[M_-,\, \cq_R\right]\,,
\quad B = \cq_L\, {\rm Tr}\, M_+
\qquad -(L\leftrightarrow R)\,
\nn\\
O_{M_+ M_- 3}^{(2)}:&
\quad A=M_-\,,
\quad B = \cq_R M_+\cq_L
\qquad -(L\leftrightarrow R)\,.
\label{eq:2Minsertion2}
\eea
After projecting, these operators turn out to be similar to the ones with 
one $M$ insertion, we have
$2O_{M_+ M_- 1}^{(1,2)}=-O_{M_+ M_- 2}^{(1,2)}=O_{M2}^{(1,2)}\,{\rm Tr}\,M_+$ and $2O_{M_+ M_- 3}^{(2)}=O_{M2}^{(2)}\,{\rm Tr}\,M_+$, 
up to two-pion vertices (for $m_u = m_d$). 

To get the operators in Eqs. \eqref{eq:2Minsertion1} and 
\eqref{eq:2Minsertion2} we again used identities like the one above 
Eq.\ \eqref{eq:contractions} to pick combinations that lead to multiplication 
of the matrices in $T$ (to avoid explicit $\epsilon^{ij}$s). 
In addition, we used the fact that $\cq_X$ and $M_-$ are traceless. 
Apart from those assumptions,  Eqs.\ \eqref{eq:2Minsertion1} and 
\eqref{eq:2Minsertion2} provide a complete basis of operators.

\subsection{Mass-insertion summary}
\label{Massinsertionsum}

With the above results in hand, the effective ${\cal C}_{1,2}$ couplings 
defined in Sec. \ref{CIB} become
\bea
{\cal C}_{1,2}^{eff} = {\cal C}_{1,2} 
+ 4\bar m C_{M1}^{(1,2)}
+ 4\bar m\sq\left(2C_{M_+\sq 1}^{(1,2)}+4C_{M_+\sq 2}^{(1,2)}+2C_{M_+\sq 5}^{(1,2)}\right)\,,
\eea
where we set $m_{u,d}=\bar m$. Here the second term should absorb the divergence
proportional to $D_2$, while the third term (in brackets) should do so for 
the $D_2\sq$ term.

The procedure can be extended to a larger number of mass insertions 
straightforwardly but painfully.

\section{The $\overline{\rm MS}$ scheme}
\label{app:MSbar}

Although one can in principle calculate the $N\!N$ amplitudes analytically in 
\MS within $\slashpi$EFT, this is no longer the case in $\chi$EFT. 
Here one needs to numerically evaluate quantities such as 
$\chi_{\spacevec p}^\pm(r)$,  $G_E^\pm(\spacevec r,0)$, and $K_E$
which all involve an arbitrary number of pion exchanges.
To do so we closely follow the method described in Ref. \cite{Kaplan:1996xu}, 
to which we refer for further details. 

The strategy is to first use an intermediate scheme in which one solves the 
Schr\"odinger equation by imposing the boundary conditions at $r=1/\lambda$, 
where $\lambda$ is a regulator. This results in a regular and an irregular 
solution, the latter of which will depend on the regulator, $\lambda$. 
The regular solution obtained in this way allows one to determine 
$\chi_{\spacevec p}^\pm( r)$, while a combination of the regular and irregular 
solutions give rise to  $G_E^\pm(\spacevec r,0)$ and $K_E$.
Because the latter of these quantities is 
regulator- and scheme-independent it can be used to translate from the 
$\lambda$ scheme to the \MS scheme. In particular, we have 
\bea
\frac{1}{\tilde C(\la)}-G_E^\pm(0,0)\big|_{\la} = 
\frac{1}{\tilde C(\mu)_{\overline{\rm MS}}} -G_E^\pm(0,0)\big|_{\overline{\rm MS}}\,.
\eea
In addition, one knows that differences between the $G_E^\pm(0,0)$ in the two 
schemes can only arise from their divergent parts, and only the first two 
diagrams in $G_E^\pm(0,0)$ (i.e. the parts that have zero and one insertion 
of $V_\pi$ after expanding Eq.\ \eqref{chiGE}) lead to divergences. 
This allows one to relate both $\tilde C$ and $G_E^\pm(0,0)$ in \MS to terms 
that can be analytically computed and quantities that can be numerically 
obtained in the $\la$ scheme. 

For example, applying this procedure to $N\!N$ scattering in the $^1S_0$ 
channel leads to 
\begin{equation}
\frac{1}{\tilde C_{}(\mu)}
+\frac{\al_\pi m_N^2}{8\pi}\,\ln 
\frac{\mu\sq}{m_\pi\sq}
\simeq -0.24 \,{\rm fm}^{-2},
\end{equation}
which is in agreement with the results of Ref.\ \cite{Kaplan:1996xu}. 
The same procedure can be used to evaluate the $\Delta L=2$ amplitudes in 
Eqs.\ \eqref{eq:anutot} and \eqref{eq:amplitudeNLO}, as well 
as the $nn$, $pp$, and $np$ amplitudes in the presence of isospin violation 
as discussed in Sec. \ref{CIBfits}.

\bibliography{bibliography}

\begin{thebibliography}{115}%
\makeatletter
\providecommand \@ifxundefined [1]{%
 \@ifx{#1\undefined}
}%
\providecommand \@ifnum [1]{%
 \ifnum #1\expandafter \@firstoftwo
 \else \expandafter \@secondoftwo
 \fi
}%
\providecommand \@ifx [1]{%
 \ifx #1\expandafter \@firstoftwo
 \else \expandafter \@secondoftwo
 \fi
}%
\providecommand \natexlab [1]{#1}%
\providecommand \enquote  [1]{``#1''}%
\providecommand \bibnamefont  [1]{#1}%
\providecommand \bibfnamefont [1]{#1}%
\providecommand \citenamefont [1]{#1}%
\providecommand \href@noop [0]{\@secondoftwo}%
\providecommand \href [0]{\begingroup \@sanitize@url \@href}%
\providecommand \@href[1]{\@@startlink{#1}\@@href}%
\providecommand \@@href[1]{\endgroup#1\@@endlink}%
\providecommand \@sanitize@url [0]{\catcode `\\12\catcode `\$12\catcode
  `\&12\catcode `\#12\catcode `\^12\catcode `\_12\catcode `\%12\relax}%
\providecommand \@@startlink[1]{}%
\providecommand \@@endlink[0]{}%
\providecommand \url  [0]{\begingroup\@sanitize@url \@url }%
\providecommand \@url [1]{\endgroup\@href {#1}{\urlprefix }}%
\providecommand \urlprefix  [0]{URL }%
\providecommand \Eprint [0]{\href }%
\providecommand \doibase [0]{http://dx.doi.org/}%
\providecommand \selectlanguage [0]{\@gobble}%
\providecommand \bibinfo  [0]{\@secondoftwo}%
\providecommand \bibfield  [0]{\@secondoftwo}%
\providecommand \translation [1]{[#1]}%
\providecommand \BibitemOpen [0]{}%
\providecommand \bibitemStop [0]{}%
\providecommand \bibitemNoStop [0]{.\EOS\space}%
\providecommand \EOS [0]{\spacefactor3000\relax}%
\providecommand \BibitemShut  [1]{\csname bibitem#1\endcsname}%
\let\auto@bib@innerbib\@empty
\bibitem [{\citenamefont {Aseev}\ \emph {et~al.}(2011)\citenamefont {Aseev}
  \emph {et~al.}}]{Aseev:2011dq}%
  \BibitemOpen
  \bibfield  {author} {\bibinfo {author} {\bibfnamefont {V.~N.}\ \bibnamefont
  {Aseev}} \emph {et~al.} (\bibinfo {collaboration} {Troitsk}),\ }\bibfield
  {title} {\enquote {\bibinfo {title} {{An upper limit on electron antineutrino
  mass from Troitsk experiment}},}\ }\href {\doibase
  10.1103/PhysRevD.84.112003} {\bibfield  {journal} {\bibinfo  {journal} {Phys.
  Rev.}\ }\textbf {\bibinfo {volume} {D84}},\ \bibinfo {pages} {112003}
  (\bibinfo {year} {2011})},\ \Eprint {http://arxiv.org/abs/1108.5034}
  {arXiv:1108.5034 [hep-ex]} \BibitemShut {NoStop}%
\bibitem [{\citenamefont {Akrami}\ \emph {et~al.}(2018)\citenamefont {Akrami}
  \emph {et~al.}}]{Akrami:2018vks}%
  \BibitemOpen
  \bibfield  {author} {\bibinfo {author} {\bibfnamefont {Y.}~\bibnamefont
  {Akrami}} \emph {et~al.} (\bibinfo {collaboration} {Planck}),\ }\bibfield
  {title} {\enquote {\bibinfo {title} {{Planck 2018 results. I. Overview and
  the cosmological legacy of Planck}},}\ }\href@noop {} {\  (\bibinfo {year}
  {2018})},\ \Eprint {http://arxiv.org/abs/1807.06205} {arXiv:1807.06205
  [astro-ph.CO]} \BibitemShut {NoStop}%
\bibitem [{\citenamefont {Minkowski}(1977)}]{Minkowski:1977sc}%
  \BibitemOpen
  \bibfield  {author} {\bibinfo {author} {\bibfnamefont {Peter}\ \bibnamefont
  {Minkowski}},\ }\bibfield  {title} {\enquote {\bibinfo {title} {{$\mu \to
  e\gamma$ at a Rate of One Out of $10^{9}$ Muon Decays?}}}\ }\href {\doibase
  10.1016/0370-2693(77)90435-X} {\bibfield  {journal} {\bibinfo  {journal}
  {Phys. Lett.}\ }\textbf {\bibinfo {volume} {67B}},\ \bibinfo {pages}
  {421--428} (\bibinfo {year} {1977})}\BibitemShut {NoStop}%
\bibitem [{\citenamefont {Mohapatra}\ and\ \citenamefont
  {Senjanovic}(1980)}]{Mohapatra:1979ia}%
  \BibitemOpen
  \bibfield  {author} {\bibinfo {author} {\bibfnamefont {Rabindra~N.}\
  \bibnamefont {Mohapatra}}\ and\ \bibinfo {author} {\bibfnamefont {Goran}\
  \bibnamefont {Senjanovic}},\ }\bibfield  {title} {\enquote {\bibinfo {title}
  {{Neutrino Mass and Spontaneous Parity Violation}},}\ }\href {\doibase
  10.1103/PhysRevLett.44.912} {\bibfield  {journal} {\bibinfo  {journal} {Phys.
  Rev. Lett.}\ }\textbf {\bibinfo {volume} {44}},\ \bibinfo {pages} {912}
  (\bibinfo {year} {1980})}\BibitemShut {NoStop}%
\bibitem [{\citenamefont {Gell-Mann}\ \emph {et~al.}(1979)\citenamefont
  {Gell-Mann}, \citenamefont {Ramond},\ and\ \citenamefont
  {Slansky}}]{GellMann:1980vs}%
  \BibitemOpen
  \bibfield  {author} {\bibinfo {author} {\bibfnamefont {Murray}\ \bibnamefont
  {Gell-Mann}}, \bibinfo {author} {\bibfnamefont {Pierre}\ \bibnamefont
  {Ramond}}, \ and\ \bibinfo {author} {\bibfnamefont {Richard}\ \bibnamefont
  {Slansky}},\ }\bibfield  {title} {\enquote {\bibinfo {title} {{Complex
  Spinors and Unified Theories}},}\ }\bibfield  {booktitle} {\emph {\bibinfo
  {booktitle} {{Supergravity Workshop Stony Brook, New York, September 27-28,
  1979}}},\ }\href@noop {} {\bibfield  {journal} {\bibinfo  {journal} {Conf.
  Proc.}\ }\textbf {\bibinfo {volume} {C790927}},\ \bibinfo {pages} {315--321}
  (\bibinfo {year} {1979})},\ \Eprint {http://arxiv.org/abs/1306.4669}
  {arXiv:1306.4669 [hep-th]} \BibitemShut {NoStop}%
\bibitem [{\citenamefont {Schechter}\ and\ \citenamefont
  {Valle}(1982)}]{Schechter:1981bd}%
  \BibitemOpen
  \bibfield  {author} {\bibinfo {author} {\bibfnamefont {J.}~\bibnamefont
  {Schechter}}\ and\ \bibinfo {author} {\bibfnamefont {J.~W.~F.}\ \bibnamefont
  {Valle}},\ }\bibfield  {title} {\enquote {\bibinfo {title} {{Neutrinoless
  Double beta Decay in SU(2) x U(1) Theories}},}\ }\href {\doibase
  10.1103/PhysRevD.25.2951} {\bibfield  {journal} {\bibinfo  {journal} {Phys.
  Rev.}\ }\textbf {\bibinfo {volume} {D25}},\ \bibinfo {pages} {2951} (\bibinfo
  {year} {1982})}\BibitemShut {NoStop}%
\bibitem [{\citenamefont {Haxton}\ and\ \citenamefont
  {Stephenson}(1984)}]{Haxton:1985am}%
  \BibitemOpen
  \bibfield  {author} {\bibinfo {author} {\bibfnamefont {W.~C.}\ \bibnamefont
  {Haxton}}\ and\ \bibinfo {author} {\bibfnamefont {G.~J.}\ \bibnamefont
  {Stephenson}},\ }\bibfield  {title} {\enquote {\bibinfo {title} {{Double beta
  Decay}},}\ }\href {\doibase 10.1016/0146-6410(84)90006-1} {\bibfield
  {journal} {\bibinfo  {journal} {Prog. Part. Nucl. Phys.}\ }\textbf {\bibinfo
  {volume} {12}},\ \bibinfo {pages} {409--479} (\bibinfo {year}
  {1984})}\BibitemShut {NoStop}%
\bibitem [{\citenamefont {Gando}\ \emph {et~al.}(2013)\citenamefont {Gando}
  \emph {et~al.}}]{Gando:2012zm}%
  \BibitemOpen
  \bibfield  {author} {\bibinfo {author} {\bibfnamefont {A.}~\bibnamefont
  {Gando}} \emph {et~al.} (\bibinfo {collaboration} {KamLAND-Zen}),\ }\bibfield
   {title} {\enquote {\bibinfo {title} {{Limit on Neutrinoless $\beta\beta$
  Decay of $^{136}$Xe from the First Phase of KamLAND-Zen and Comparison with
  the Positive Claim in $^{76}$Ge}},}\ }\href {\doibase
  10.1103/PhysRevLett.110.062502} {\bibfield  {journal} {\bibinfo  {journal}
  {Phys. Rev. Lett.}\ }\textbf {\bibinfo {volume} {110}},\ \bibinfo {pages}
  {062502} (\bibinfo {year} {2013})},\ \Eprint {http://arxiv.org/abs/1211.3863}
  {arXiv:1211.3863 [hep-ex]} \BibitemShut {NoStop}%
\bibitem [{\citenamefont {Agostini}\ \emph {et~al.}(2013)\citenamefont
  {Agostini} \emph {et~al.}}]{Agostini:2013mzu}%
  \BibitemOpen
  \bibfield  {author} {\bibinfo {author} {\bibfnamefont {M.}~\bibnamefont
  {Agostini}} \emph {et~al.} (\bibinfo {collaboration} {GERDA}),\ }\bibfield
  {title} {\enquote {\bibinfo {title} {{Results on Neutrinoless Double-$\beta$
  Decay of $^{76}$Ge from Phase I of the GERDA Experiment}},}\ }\href {\doibase
  10.1103/PhysRevLett.111.122503} {\bibfield  {journal} {\bibinfo  {journal}
  {Phys. Rev. Lett.}\ }\textbf {\bibinfo {volume} {111}},\ \bibinfo {pages}
  {122503} (\bibinfo {year} {2013})},\ \Eprint {http://arxiv.org/abs/1307.4720}
  {arXiv:1307.4720 [nucl-ex]} \BibitemShut {NoStop}%
\bibitem [{\citenamefont {Albert}\ \emph {et~al.}(2014)\citenamefont {Albert}
  \emph {et~al.}}]{Albert:2014awa}%
  \BibitemOpen
  \bibfield  {author} {\bibinfo {author} {\bibfnamefont {J.~B.}\ \bibnamefont
  {Albert}} \emph {et~al.} (\bibinfo {collaboration} {EXO-200}),\ }\bibfield
  {title} {\enquote {\bibinfo {title} {{Search for Majorana neutrinos with the
  first two years of EXO-200 data}},}\ }\href {\doibase 10.1038/nature13432}
  {\bibfield  {journal} {\bibinfo  {journal} {Nature}\ }\textbf {\bibinfo
  {volume} {510}},\ \bibinfo {pages} {229--234} (\bibinfo {year} {2014})},\
  \Eprint {http://arxiv.org/abs/1402.6956} {arXiv:1402.6956 [nucl-ex]}
  \BibitemShut {NoStop}%
\bibitem [{\citenamefont {Andringa}\ \emph {et~al.}(2016)\citenamefont
  {Andringa} \emph {et~al.}}]{Andringa:2015tza}%
  \BibitemOpen
  \bibfield  {author} {\bibinfo {author} {\bibfnamefont {S.}~\bibnamefont
  {Andringa}} \emph {et~al.} (\bibinfo {collaboration} {SNO+}),\ }\bibfield
  {title} {\enquote {\bibinfo {title} {{Current Status and Future Prospects of
  the SNO+ Experiment}},}\ }\href {\doibase 10.1155/2016/6194250} {\bibfield
  {journal} {\bibinfo  {journal} {Adv. High Energy Phys.}\ }\textbf {\bibinfo
  {volume} {2016}},\ \bibinfo {pages} {6194250} (\bibinfo {year} {2016})},\
  \Eprint {http://arxiv.org/abs/1508.05759} {arXiv:1508.05759
  [physics.ins-det]} \BibitemShut {NoStop}%
\bibitem [{\citenamefont {Gando}\ \emph {et~al.}(2016)\citenamefont {Gando}
  \emph {et~al.}}]{KamLAND-Zen:2016pfg}%
  \BibitemOpen
  \bibfield  {author} {\bibinfo {author} {\bibfnamefont {A.}~\bibnamefont
  {Gando}} \emph {et~al.} (\bibinfo {collaboration} {KamLAND-Zen}),\ }\bibfield
   {title} {\enquote {\bibinfo {title} {{Search for Majorana Neutrinos near the
  Inverted Mass Hierarchy Region with KamLAND-Zen}},}\ }\href {\doibase
  10.1103/PhysRevLett.117.109903, 10.1103/PhysRevLett.117.082503} {\bibfield
  {journal} {\bibinfo  {journal} {Phys. Rev. Lett.}\ }\textbf {\bibinfo
  {volume} {117}},\ \bibinfo {pages} {082503} (\bibinfo {year} {2016})},\
  \bibinfo {note} {[Addendum: Phys. Rev. Lett.117,no.10,109903(2016)]},\
  \Eprint {http://arxiv.org/abs/1605.02889} {arXiv:1605.02889 [hep-ex]}
  \BibitemShut {NoStop}%
\bibitem [{\citenamefont {Elliott}\ \emph {et~al.}(2016)\citenamefont {Elliott}
  \emph {et~al.}}]{Elliott:2016ble}%
  \BibitemOpen
  \bibfield  {author} {\bibinfo {author} {\bibfnamefont {S.~R.}\ \bibnamefont
  {Elliott}} \emph {et~al.},\ }\bibfield  {title} {\enquote {\bibinfo {title}
  {{Initial Results from the MAJORANA DEMONSTRATOR}},}\ \ }(\bibinfo {year}
  {2016})\ \Eprint {http://arxiv.org/abs/1610.01210} {arXiv:1610.01210
  [nucl-ex]} \BibitemShut {NoStop}%
\bibitem [{\citenamefont {Agostini}\ \emph {et~al.}(2017)\citenamefont
  {Agostini} \emph {et~al.}}]{Agostini:2017iyd}%
  \BibitemOpen
  \bibfield  {author} {\bibinfo {author} {\bibfnamefont {M.}~\bibnamefont
  {Agostini}} \emph {et~al.},\ }\bibfield  {title} {\enquote {\bibinfo {title}
  {{Background-free search for neutrinoless double-$\beta$ decay of $^{76}$Ge
  with GERDA}},}\ }\href {\doibase 10.1038/nature21717} {\  (\bibinfo {year}
  {2017}),\ 10.1038/nature21717},\ \bibinfo {note} {[Nature544,47(2017)]},\
  \Eprint {http://arxiv.org/abs/1703.00570} {arXiv:1703.00570 [nucl-ex]}
  \BibitemShut {NoStop}%
\bibitem [{\citenamefont {Aalseth}\ \emph {et~al.}(2018)\citenamefont {Aalseth}
  \emph {et~al.}}]{Aalseth:2017btx}%
  \BibitemOpen
  \bibfield  {author} {\bibinfo {author} {\bibfnamefont {C.~E.}\ \bibnamefont
  {Aalseth}} \emph {et~al.} (\bibinfo {collaboration} {Majorana}),\ }\bibfield
  {title} {\enquote {\bibinfo {title} {{Search for Neutrinoless Double-? Decay
  in $^{76}$Ge with the Majorana Demonstrator}},}\ }\href {\doibase
  10.1103/PhysRevLett.120.132502} {\bibfield  {journal} {\bibinfo  {journal}
  {Phys. Rev. Lett.}\ }\textbf {\bibinfo {volume} {120}},\ \bibinfo {pages}
  {132502} (\bibinfo {year} {2018})},\ \Eprint
  {http://arxiv.org/abs/1710.11608} {arXiv:1710.11608 [nucl-ex]} \BibitemShut
  {NoStop}%
\bibitem [{\citenamefont {Albert}\ \emph {et~al.}(2018)\citenamefont {Albert}
  \emph {et~al.}}]{Albert:2017owj}%
  \BibitemOpen
  \bibfield  {author} {\bibinfo {author} {\bibfnamefont {J.~B.}\ \bibnamefont
  {Albert}} \emph {et~al.} (\bibinfo {collaboration} {EXO}),\ }\bibfield
  {title} {\enquote {\bibinfo {title} {{Search for Neutrinoless Double-Beta
  Decay with the Upgraded EXO-200 Detector}},}\ }\href {\doibase
  10.1103/PhysRevLett.120.072701} {\bibfield  {journal} {\bibinfo  {journal}
  {Phys. Rev. Lett.}\ }\textbf {\bibinfo {volume} {120}},\ \bibinfo {pages}
  {072701} (\bibinfo {year} {2018})},\ \Eprint
  {http://arxiv.org/abs/1707.08707} {arXiv:1707.08707 [hep-ex]} \BibitemShut
  {NoStop}%
\bibitem [{\citenamefont {Alduino}\ \emph {et~al.}(2018)\citenamefont {Alduino}
  \emph {et~al.}}]{Alduino:2017ehq}%
  \BibitemOpen
  \bibfield  {author} {\bibinfo {author} {\bibfnamefont {C.}~\bibnamefont
  {Alduino}} \emph {et~al.} (\bibinfo {collaboration} {CUORE}),\ }\bibfield
  {title} {\enquote {\bibinfo {title} {{First Results from CUORE: A Search for
  Lepton Number Violation via $0\nu\beta\beta$ Decay of $^{130}$Te}},}\ }\href
  {\doibase 10.1103/PhysRevLett.120.132501} {\bibfield  {journal} {\bibinfo
  {journal} {Phys. Rev. Lett.}\ }\textbf {\bibinfo {volume} {120}},\ \bibinfo
  {pages} {132501} (\bibinfo {year} {2018})},\ \Eprint
  {http://arxiv.org/abs/1710.07988} {arXiv:1710.07988 [nucl-ex]} \BibitemShut
  {NoStop}%
\bibitem [{\citenamefont {Agostini}\ \emph {et~al.}(2018)\citenamefont
  {Agostini} \emph {et~al.}}]{Agostini:2018tnm}%
  \BibitemOpen
  \bibfield  {author} {\bibinfo {author} {\bibfnamefont {M.}~\bibnamefont
  {Agostini}} \emph {et~al.} (\bibinfo {collaboration} {GERDA}),\ }\bibfield
  {title} {\enquote {\bibinfo {title} {{Improved Limit on Neutrinoless
  Double-$\beta$ Decay of $^{76}$Ge from GERDA Phase II}},}\ }\href {\doibase
  10.1103/PhysRevLett.120.132503} {\bibfield  {journal} {\bibinfo  {journal}
  {Phys. Rev. Lett.}\ }\textbf {\bibinfo {volume} {120}},\ \bibinfo {pages}
  {132503} (\bibinfo {year} {2018})},\ \Eprint
  {http://arxiv.org/abs/1803.11100} {arXiv:1803.11100 [nucl-ex]} \BibitemShut
  {NoStop}%
\bibitem [{\citenamefont {Azzolini}\ \emph {et~al.}(2018)\citenamefont
  {Azzolini} \emph {et~al.}}]{Azzolini:2018dyb}%
  \BibitemOpen
  \bibfield  {author} {\bibinfo {author} {\bibfnamefont {O.}~\bibnamefont
  {Azzolini}} \emph {et~al.} (\bibinfo {collaboration} {CUPID-0}),\ }\bibfield
  {title} {\enquote {\bibinfo {title} {{First Result on the Neutrinoless
  Double-$\beta$ Decay of $^{82}Se$ with CUPID-0}},}\ }\href {\doibase
  10.1103/PhysRevLett.120.232502} {\bibfield  {journal} {\bibinfo  {journal}
  {Phys. Rev. Lett.}\ }\textbf {\bibinfo {volume} {120}},\ \bibinfo {pages}
  {232502} (\bibinfo {year} {2018})},\ \Eprint
  {http://arxiv.org/abs/1802.07791} {arXiv:1802.07791 [nucl-ex]} \BibitemShut
  {NoStop}%
\bibitem [{\citenamefont {Anton}\ \emph {et~al.}(2019)\citenamefont {Anton}
  \emph {et~al.}}]{Anton:2019wmi}%
  \BibitemOpen
  \bibfield  {author} {\bibinfo {author} {\bibfnamefont {G.}~\bibnamefont
  {Anton}} \emph {et~al.},\ }\bibfield  {title} {\enquote {\bibinfo {title}
  {{Search for Neutrinoless Double-Beta Decay with the Complete EXO-200
  Dataset}},}\ }\href@noop {} {\  (\bibinfo {year} {2019})},\ \Eprint
  {http://arxiv.org/abs/1906.02723} {arXiv:1906.02723 [hep-ex]} \BibitemShut
  {NoStop}%
\bibitem [{\citenamefont {Prezeau}\ \emph {et~al.}(2003)\citenamefont
  {Prezeau}, \citenamefont {Ramsey-Musolf},\ and\ \citenamefont
  {Vogel}}]{Prezeau:2003xn}%
  \BibitemOpen
  \bibfield  {author} {\bibinfo {author} {\bibfnamefont {Gary}\ \bibnamefont
  {Prezeau}}, \bibinfo {author} {\bibfnamefont {M.}~\bibnamefont
  {Ramsey-Musolf}}, \ and\ \bibinfo {author} {\bibfnamefont {Petr}\
  \bibnamefont {Vogel}},\ }\bibfield  {title} {\enquote {\bibinfo {title}
  {{Neutrinoless double beta decay and effective field theory}},}\ }\href
  {\doibase 10.1103/PhysRevD.68.034016} {\bibfield  {journal} {\bibinfo
  {journal} {Phys. Rev.}\ }\textbf {\bibinfo {volume} {D68}},\ \bibinfo {pages}
  {034016} (\bibinfo {year} {2003})},\ \Eprint
  {http://arxiv.org/abs/hep-ph/0303205} {arXiv:hep-ph/0303205 [hep-ph]}
  \BibitemShut {NoStop}%
\bibitem [{\citenamefont {Graesser}(2017)}]{Graesser:2016bpz}%
  \BibitemOpen
  \bibfield  {author} {\bibinfo {author} {\bibfnamefont {Michael~L.}\
  \bibnamefont {Graesser}},\ }\bibfield  {title} {\enquote {\bibinfo {title}
  {{An electroweak basis for neutrinoless double $\beta$ decay}},}\ }\href
  {\doibase 10.1007/JHEP08(2017)099} {\bibfield  {journal} {\bibinfo  {journal}
  {JHEP}\ }\textbf {\bibinfo {volume} {08}},\ \bibinfo {pages} {099} (\bibinfo
  {year} {2017})},\ \Eprint {http://arxiv.org/abs/1606.04549} {arXiv:1606.04549
  [hep-ph]} \BibitemShut {NoStop}%
\bibitem [{\citenamefont {Cirigliano}\ \emph {et~al.}(2017)\citenamefont
  {Cirigliano}, \citenamefont {Dekens}, \citenamefont {de~Vries}, \citenamefont
  {Graesser},\ and\ \citenamefont {Mereghetti}}]{Cirigliano:2017djv}%
  \BibitemOpen
  \bibfield  {author} {\bibinfo {author} {\bibfnamefont {V.}~\bibnamefont
  {Cirigliano}}, \bibinfo {author} {\bibfnamefont {W.}~\bibnamefont {Dekens}},
  \bibinfo {author} {\bibfnamefont {J.}~\bibnamefont {de~Vries}}, \bibinfo
  {author} {\bibfnamefont {M.~L.}\ \bibnamefont {Graesser}}, \ and\ \bibinfo
  {author} {\bibfnamefont {E.}~\bibnamefont {Mereghetti}},\ }\bibfield  {title}
  {\enquote {\bibinfo {title} {{Neutrinoless double beta decay in chiral
  effective field theory: lepton number violation at dimension seven}},}\
  }\href {\doibase 10.1007/JHEP12(2017)082} {\bibfield  {journal} {\bibinfo
  {journal} {JHEP}\ }\textbf {\bibinfo {volume} {12}},\ \bibinfo {pages} {082}
  (\bibinfo {year} {2017})},\ \Eprint {http://arxiv.org/abs/1708.09390}
  {arXiv:1708.09390 [hep-ph]} \BibitemShut {NoStop}%
\bibitem [{\citenamefont {Cirigliano}\ \emph
  {et~al.}(2018{\natexlab{a}})\citenamefont {Cirigliano}, \citenamefont
  {Dekens}, \citenamefont {de~Vries}, \citenamefont {Graesser},\ and\
  \citenamefont {Mereghetti}}]{Cirigliano:2018yza}%
  \BibitemOpen
  \bibfield  {author} {\bibinfo {author} {\bibfnamefont {V.}~\bibnamefont
  {Cirigliano}}, \bibinfo {author} {\bibfnamefont {W.}~\bibnamefont {Dekens}},
  \bibinfo {author} {\bibfnamefont {J.}~\bibnamefont {de~Vries}}, \bibinfo
  {author} {\bibfnamefont {M.~L.}\ \bibnamefont {Graesser}}, \ and\ \bibinfo
  {author} {\bibfnamefont {E.}~\bibnamefont {Mereghetti}},\ }\bibfield  {title}
  {\enquote {\bibinfo {title} {{A neutrinoless double beta decay master formula
  from effective field theory}},}\ }\href {\doibase 10.1007/JHEP12(2018)097}
  {\bibfield  {journal} {\bibinfo  {journal} {JHEP}\ }\textbf {\bibinfo
  {volume} {12}},\ \bibinfo {pages} {097} (\bibinfo {year}
  {2018}{\natexlab{a}})},\ \Eprint {http://arxiv.org/abs/1806.02780}
  {arXiv:1806.02780 [hep-ph]} \BibitemShut {NoStop}%
\bibitem [{\citenamefont {Weinberg}(1979{\natexlab{a}})}]{Weinberg:1979sa}%
  \BibitemOpen
  \bibfield  {author} {\bibinfo {author} {\bibfnamefont {Steven}\ \bibnamefont
  {Weinberg}},\ }\bibfield  {title} {\enquote {\bibinfo {title} {{Baryon and
  Lepton Nonconserving Processes}},}\ }\href {\doibase
  10.1103/PhysRevLett.43.1566} {\bibfield  {journal} {\bibinfo  {journal}
  {Phys. Rev. Lett.}\ }\textbf {\bibinfo {volume} {43}},\ \bibinfo {pages}
  {1566--1570} (\bibinfo {year} {1979}{\natexlab{a}})}\BibitemShut {NoStop}%
\bibitem [{\citenamefont {Weinberg}(1979{\natexlab{b}})}]{Weinberg:1978kz}%
  \BibitemOpen
  \bibfield  {author} {\bibinfo {author} {\bibfnamefont {Steven}\ \bibnamefont
  {Weinberg}},\ }\bibfield  {title} {\enquote {\bibinfo {title}
  {{Phenomenological Lagrangians}},}\ }\href@noop {} {\bibfield  {journal}
  {\bibinfo  {journal} {Physica}\ }\textbf {\bibinfo {volume} {A96}},\ \bibinfo
  {pages} {327} (\bibinfo {year} {1979}{\natexlab{b}})}\BibitemShut {NoStop}%
\bibitem [{\citenamefont {Weinberg}(1990)}]{Weinberg:1990rz}%
  \BibitemOpen
  \bibfield  {author} {\bibinfo {author} {\bibfnamefont {Steven}\ \bibnamefont
  {Weinberg}},\ }\bibfield  {title} {\enquote {\bibinfo {title} {{Nuclear
  forces from chiral Lagrangians}},}\ }\href {\doibase
  10.1016/0370-2693(90)90938-3} {\bibfield  {journal} {\bibinfo  {journal}
  {Phys. Lett.}\ }\textbf {\bibinfo {volume} {B251}},\ \bibinfo {pages}
  {288--292} (\bibinfo {year} {1990})}\BibitemShut {NoStop}%
\bibitem [{\citenamefont {Weinberg}(1991)}]{Weinberg:1991um}%
  \BibitemOpen
  \bibfield  {author} {\bibinfo {author} {\bibfnamefont {Steven}\ \bibnamefont
  {Weinberg}},\ }\bibfield  {title} {\enquote {\bibinfo {title} {{Effective
  chiral Lagrangians for nucleon - pion interactions and nuclear forces}},}\
  }\href {\doibase 10.1016/0550-3213(91)90231-L} {\bibfield  {journal}
  {\bibinfo  {journal} {Nucl. Phys.}\ }\textbf {\bibinfo {volume} {B363}},\
  \bibinfo {pages} {3--18} (\bibinfo {year} {1991})}\BibitemShut {NoStop}%
\bibitem [{\citenamefont {Engel}\ and\ \citenamefont
  {Men\'endez}(2017)}]{Engel:2016xgb}%
  \BibitemOpen
  \bibfield  {author} {\bibinfo {author} {\bibfnamefont {Jonathan}\
  \bibnamefont {Engel}}\ and\ \bibinfo {author} {\bibfnamefont {Javier}\
  \bibnamefont {Men\'endez}},\ }\bibfield  {title} {\enquote {\bibinfo {title}
  {{Status and Future of Nuclear Matrix Elements for Neutrinoless Double-Beta
  Decay: A Review}},}\ }\href {\doibase 10.1088/1361-6633/aa5bc5} {\bibfield
  {journal} {\bibinfo  {journal} {Rept. Prog. Phys.}\ }\textbf {\bibinfo
  {volume} {80}},\ \bibinfo {pages} {046301} (\bibinfo {year} {2017})},\
  \Eprint {http://arxiv.org/abs/1610.06548} {arXiv:1610.06548 [nucl-th]}
  \BibitemShut {NoStop}%
\bibitem [{\citenamefont {Ejiri}\ \emph {et~al.}(2019)\citenamefont {Ejiri},
  \citenamefont {Suhonen},\ and\ \citenamefont {Zuber}}]{Ejiri:2019ezh}%
  \BibitemOpen
  \bibfield  {author} {\bibinfo {author} {\bibfnamefont {H.}~\bibnamefont
  {Ejiri}}, \bibinfo {author} {\bibfnamefont {J.}~\bibnamefont {Suhonen}}, \
  and\ \bibinfo {author} {\bibfnamefont {K.}~\bibnamefont {Zuber}},\ }\bibfield
   {title} {\enquote {\bibinfo {title} {{Neutrino–nuclear responses for
  astro-neutrinos, single beta decays and double beta decays}},}\ }\href
  {\doibase 10.1016/j.physrep.2018.12.001} {\bibfield  {journal} {\bibinfo
  {journal} {Phys. Rept.}\ }\textbf {\bibinfo {volume} {797}},\ \bibinfo
  {pages} {1--102} (\bibinfo {year} {2019})}\BibitemShut {NoStop}%
\bibitem [{\citenamefont {Cirigliano}\ \emph
  {et~al.}(2018{\natexlab{b}})\citenamefont {Cirigliano}, \citenamefont
  {Dekens}, \citenamefont {Mereghetti},\ and\ \citenamefont
  {Walker-Loud}}]{Cirigliano:2017tvr}%
  \BibitemOpen
  \bibfield  {author} {\bibinfo {author} {\bibfnamefont {Vincenzo}\
  \bibnamefont {Cirigliano}}, \bibinfo {author} {\bibfnamefont {Wouter}\
  \bibnamefont {Dekens}}, \bibinfo {author} {\bibfnamefont {Emanuele}\
  \bibnamefont {Mereghetti}}, \ and\ \bibinfo {author} {\bibfnamefont
  {Andr\'e}\ \bibnamefont {Walker-Loud}},\ }\bibfield  {title} {\enquote
  {\bibinfo {title} {{Neutrinoless double-$\beta$ decay in effective field
  theory: The light-Majorana neutrino-exchange mechanism}},}\ }\href {\doibase
  10.1103/PhysRevC.97.065501} {\bibfield  {journal} {\bibinfo  {journal} {Phys.
  Rev.}\ }\textbf {\bibinfo {volume} {C97}},\ \bibinfo {pages} {065501}
  (\bibinfo {year} {2018}{\natexlab{b}})},\ \Eprint
  {http://arxiv.org/abs/1710.01729} {arXiv:1710.01729 [hep-ph]} \BibitemShut
  {NoStop}%
\bibitem [{\citenamefont {Cirigliano}\ \emph
  {et~al.}(2018{\natexlab{c}})\citenamefont {Cirigliano}, \citenamefont
  {Dekens}, \citenamefont {De~Vries}, \citenamefont {Graesser}, \citenamefont
  {Mereghetti}, \citenamefont {Pastore},\ and\ \citenamefont {van
  Kolck}}]{Cirigliano:2018hja}%
  \BibitemOpen
  \bibfield  {author} {\bibinfo {author} {\bibfnamefont {Vincenzo}\
  \bibnamefont {Cirigliano}}, \bibinfo {author} {\bibfnamefont {Wouter}\
  \bibnamefont {Dekens}}, \bibinfo {author} {\bibfnamefont {Jordy}\
  \bibnamefont {De~Vries}}, \bibinfo {author} {\bibfnamefont {Michael~L.}\
  \bibnamefont {Graesser}}, \bibinfo {author} {\bibfnamefont {Emanuele}\
  \bibnamefont {Mereghetti}}, \bibinfo {author} {\bibfnamefont {Saori}\
  \bibnamefont {Pastore}}, \ and\ \bibinfo {author} {\bibfnamefont {Ubirajara}\
  \bibnamefont {van Kolck}},\ }\bibfield  {title} {\enquote {\bibinfo {title}
  {{New Leading Contribution to Neutrinoless Double-Beta Decay}},}\ }\href
  {\doibase 10.1103/PhysRevLett.120.202001} {\bibfield  {journal} {\bibinfo
  {journal} {Phys. Rev. Lett.}\ }\textbf {\bibinfo {volume} {120}},\ \bibinfo
  {pages} {202001} (\bibinfo {year} {2018}{\natexlab{c}})},\ \Eprint
  {http://arxiv.org/abs/1802.10097} {arXiv:1802.10097 [hep-ph]} \BibitemShut
  {NoStop}%
\bibitem [{\citenamefont {Machleidt}\ and\ \citenamefont
  {Entem}(2011)}]{Machleidt:2011zz}%
  \BibitemOpen
  \bibfield  {author} {\bibinfo {author} {\bibfnamefont {R.}~\bibnamefont
  {Machleidt}}\ and\ \bibinfo {author} {\bibfnamefont {D.~R.}\ \bibnamefont
  {Entem}},\ }\bibfield  {title} {\enquote {\bibinfo {title} {{Chiral effective
  field theory and nuclear forces}},}\ }\href {\doibase
  10.1016/j.physrep.2011.02.001} {\bibfield  {journal} {\bibinfo  {journal}
  {Phys. Rept.}\ }\textbf {\bibinfo {volume} {503}},\ \bibinfo {pages} {1--75}
  (\bibinfo {year} {2011})},\ \Eprint {http://arxiv.org/abs/1105.2919}
  {arXiv:1105.2919 [nucl-th]} \BibitemShut {NoStop}%
\bibitem [{\citenamefont {Piarulli}\ \emph {et~al.}(2015)\citenamefont
  {Piarulli}, \citenamefont {Girlanda}, \citenamefont {Schiavilla},
  \citenamefont {Navarro~P\'erez}, \citenamefont {Amaro},\ and\ \citenamefont
  {Ruiz~Arriola}}]{Piarulli:2014bda}%
  \BibitemOpen
  \bibfield  {author} {\bibinfo {author} {\bibfnamefont {M.}~\bibnamefont
  {Piarulli}}, \bibinfo {author} {\bibfnamefont {L.}~\bibnamefont {Girlanda}},
  \bibinfo {author} {\bibfnamefont {R.}~\bibnamefont {Schiavilla}}, \bibinfo
  {author} {\bibfnamefont {R.}~\bibnamefont {Navarro~P\'erez}}, \bibinfo
  {author} {\bibfnamefont {J.~E.}\ \bibnamefont {Amaro}}, \ and\ \bibinfo
  {author} {\bibfnamefont {E.}~\bibnamefont {Ruiz~Arriola}},\ }\bibfield
  {title} {\enquote {\bibinfo {title} {{Minimally nonlocal nucleon-nucleon
  potentials with chiral two-pion exchange including $\Delta$ resonances}},}\
  }\href {\doibase 10.1103/PhysRevC.91.024003} {\bibfield  {journal} {\bibinfo
  {journal} {Phys. Rev.}\ }\textbf {\bibinfo {volume} {C91}},\ \bibinfo {pages}
  {024003} (\bibinfo {year} {2015})},\ \Eprint {http://arxiv.org/abs/1412.6446}
  {arXiv:1412.6446 [nucl-th]} \BibitemShut {NoStop}%
\bibitem [{\citenamefont {Epelbaum}\ \emph {et~al.}(2015)\citenamefont
  {Epelbaum}, \citenamefont {Krebs},\ and\ \citenamefont
  {Mei{\ss}ner}}]{Epelbaum:2014efa}%
  \BibitemOpen
  \bibfield  {author} {\bibinfo {author} {\bibfnamefont {E.}~\bibnamefont
  {Epelbaum}}, \bibinfo {author} {\bibfnamefont {H.}~\bibnamefont {Krebs}}, \
  and\ \bibinfo {author} {\bibfnamefont {U.~G.}\ \bibnamefont {Mei{\ss}ner}},\
  }\bibfield  {title} {\enquote {\bibinfo {title} {{Improved chiral
  nucleon-nucleon potential up to next-to-next-to-next-to-leading order}},}\
  }\href {\doibase 10.1140/epja/i2015-15053-8} {\bibfield  {journal} {\bibinfo
  {journal} {Eur. Phys. J.}\ }\textbf {\bibinfo {volume} {A51}},\ \bibinfo
  {pages} {53} (\bibinfo {year} {2015})},\ \Eprint
  {http://arxiv.org/abs/1412.0142} {arXiv:1412.0142 [nucl-th]} \BibitemShut
  {NoStop}%
\bibitem [{\citenamefont {Ekstr{\"o}m}\ \emph {et~al.}(2015)\citenamefont
  {Ekstr{\"o}m}, \citenamefont {Jansen}, \citenamefont {Wendt}, \citenamefont
  {Hagen}, \citenamefont {Papenbrock}, \citenamefont {Carlsson}, \citenamefont
  {Forss{\'e}n}, \citenamefont {Hjorth-Jensen}, \citenamefont {Navrátil},\
  and\ \citenamefont {Nazarewicz}}]{Ekstrom:2015rta}%
  \BibitemOpen
  \bibfield  {author} {\bibinfo {author} {\bibfnamefont {A.}~\bibnamefont
  {Ekstr{\"o}m}}, \bibinfo {author} {\bibfnamefont {G.~R.}\ \bibnamefont
  {Jansen}}, \bibinfo {author} {\bibfnamefont {K.~A.}\ \bibnamefont {Wendt}},
  \bibinfo {author} {\bibfnamefont {G.}~\bibnamefont {Hagen}}, \bibinfo
  {author} {\bibfnamefont {T.}~\bibnamefont {Papenbrock}}, \bibinfo {author}
  {\bibfnamefont {B.~D.}\ \bibnamefont {Carlsson}}, \bibinfo {author}
  {\bibfnamefont {C.}~\bibnamefont {Forss{\'e}n}}, \bibinfo {author}
  {\bibfnamefont {M.}~\bibnamefont {Hjorth-Jensen}}, \bibinfo {author}
  {\bibfnamefont {P.}~\bibnamefont {Navrátil}}, \ and\ \bibinfo {author}
  {\bibfnamefont {W.}~\bibnamefont {Nazarewicz}},\ }\bibfield  {title}
  {\enquote {\bibinfo {title} {{Accurate nuclear radii and binding energies
  from a chiral interaction}},}\ }\href {\doibase 10.1103/PhysRevC.91.051301}
  {\bibfield  {journal} {\bibinfo  {journal} {Phys. Rev.}\ }\textbf {\bibinfo
  {volume} {C91}},\ \bibinfo {pages} {051301} (\bibinfo {year} {2015})},\
  \Eprint {http://arxiv.org/abs/1502.04682} {arXiv:1502.04682 [nucl-th]}
  \BibitemShut {NoStop}%
\bibitem [{\citenamefont {Piarulli}\ \emph {et~al.}(2016)\citenamefont
  {Piarulli}, \citenamefont {Girlanda}, \citenamefont {Schiavilla},
  \citenamefont {Kievsky}, \citenamefont {Lovato}, \citenamefont {Marcucci},
  \citenamefont {Pieper}, \citenamefont {Viviani},\ and\ \citenamefont
  {Wiringa}}]{Piarulli:2016vel}%
  \BibitemOpen
  \bibfield  {author} {\bibinfo {author} {\bibfnamefont {Maria}\ \bibnamefont
  {Piarulli}}, \bibinfo {author} {\bibfnamefont {Luca}\ \bibnamefont
  {Girlanda}}, \bibinfo {author} {\bibfnamefont {Rocco}\ \bibnamefont
  {Schiavilla}}, \bibinfo {author} {\bibfnamefont {Alejandro}\ \bibnamefont
  {Kievsky}}, \bibinfo {author} {\bibfnamefont {Alessandro}\ \bibnamefont
  {Lovato}}, \bibinfo {author} {\bibfnamefont {Laura~E.}\ \bibnamefont
  {Marcucci}}, \bibinfo {author} {\bibfnamefont {Steven~C.}\ \bibnamefont
  {Pieper}}, \bibinfo {author} {\bibfnamefont {Michele}\ \bibnamefont
  {Viviani}}, \ and\ \bibinfo {author} {\bibfnamefont {Robert~B.}\ \bibnamefont
  {Wiringa}},\ }\bibfield  {title} {\enquote {\bibinfo {title} {{Local chiral
  potentials with $\Delta$-intermediate states and the structure of light
  nuclei}},}\ }\href {\doibase 10.1103/PhysRevC.94.054007} {\bibfield
  {journal} {\bibinfo  {journal} {Phys. Rev.}\ }\textbf {\bibinfo {volume}
  {C94}},\ \bibinfo {pages} {054007} (\bibinfo {year} {2016})},\ \Eprint
  {http://arxiv.org/abs/1606.06335} {arXiv:1606.06335 [nucl-th]} \BibitemShut
  {NoStop}%
\bibitem [{\citenamefont {Reinert}\ \emph {et~al.}(2018)\citenamefont
  {Reinert}, \citenamefont {Krebs},\ and\ \citenamefont
  {Epelbaum}}]{Reinert:2017usi}%
  \BibitemOpen
  \bibfield  {author} {\bibinfo {author} {\bibfnamefont {P.}~\bibnamefont
  {Reinert}}, \bibinfo {author} {\bibfnamefont {H.}~\bibnamefont {Krebs}}, \
  and\ \bibinfo {author} {\bibfnamefont {E.}~\bibnamefont {Epelbaum}},\
  }\bibfield  {title} {\enquote {\bibinfo {title} {{Semilocal momentum-space
  regularized chiral two-nucleon potentials up to fifth order}},}\ }\href
  {\doibase 10.1140/epja/i2018-12516-4} {\bibfield  {journal} {\bibinfo
  {journal} {Eur. Phys. J.}\ }\textbf {\bibinfo {volume} {A54}},\ \bibinfo
  {pages} {86} (\bibinfo {year} {2018})},\ \Eprint
  {http://arxiv.org/abs/1711.08821} {arXiv:1711.08821 [nucl-th]} \BibitemShut
  {NoStop}%
\bibitem [{\citenamefont {Wiringa}\ \emph {et~al.}(1995)\citenamefont
  {Wiringa}, \citenamefont {Stoks},\ and\ \citenamefont
  {Schiavilla}}]{Wiringa:1994wb}%
  \BibitemOpen
  \bibfield  {author} {\bibinfo {author} {\bibfnamefont {R.~B.}\ \bibnamefont
  {Wiringa}}, \bibinfo {author} {\bibfnamefont {V.~G.~J.}\ \bibnamefont
  {Stoks}}, \ and\ \bibinfo {author} {\bibfnamefont {R.}~\bibnamefont
  {Schiavilla}},\ }\bibfield  {title} {\enquote {\bibinfo {title} {{An Accurate
  nucleon-nucleon potential with charge independence breaking}},}\ }\href
  {\doibase 10.1103/PhysRevC.51.38} {\bibfield  {journal} {\bibinfo  {journal}
  {Phys. Rev.}\ }\textbf {\bibinfo {volume} {C51}},\ \bibinfo {pages} {38--51}
  (\bibinfo {year} {1995})},\ \Eprint {http://arxiv.org/abs/nucl-th/9408016}
  {arXiv:nucl-th/9408016 [nucl-th]} \BibitemShut {NoStop}%
\bibitem [{\citenamefont {Machleidt}(2001)}]{Machleidt:2000ge}%
  \BibitemOpen
  \bibfield  {author} {\bibinfo {author} {\bibfnamefont {R.}~\bibnamefont
  {Machleidt}},\ }\bibfield  {title} {\enquote {\bibinfo {title} {{The High
  precision, charge dependent Bonn nucleon-nucleon potential (CD-Bonn)}},}\
  }\href {\doibase 10.1103/PhysRevC.63.024001} {\bibfield  {journal} {\bibinfo
  {journal} {Phys. Rev.}\ }\textbf {\bibinfo {volume} {C63}},\ \bibinfo {pages}
  {024001} (\bibinfo {year} {2001})},\ \Eprint
  {http://arxiv.org/abs/nucl-th/0006014} {arXiv:nucl-th/0006014 [nucl-th]}
  \BibitemShut {NoStop}%
\bibitem [{\citenamefont {Nogga}\ \emph {et~al.}(2005)\citenamefont {Nogga},
  \citenamefont {Timmermans},\ and\ \citenamefont {van Kolck}}]{Nogga:2005hy}%
  \BibitemOpen
  \bibfield  {author} {\bibinfo {author} {\bibfnamefont {A.}~\bibnamefont
  {Nogga}}, \bibinfo {author} {\bibfnamefont {R.~G.~E.}\ \bibnamefont
  {Timmermans}}, \ and\ \bibinfo {author} {\bibfnamefont {U.}~\bibnamefont {van
  Kolck}},\ }\bibfield  {title} {\enquote {\bibinfo {title} {{Renormalization
  of one-pion exchange and power counting}},}\ }\href {\doibase
  10.1103/PhysRevC.72.054006} {\bibfield  {journal} {\bibinfo  {journal} {Phys.
  Rev.}\ }\textbf {\bibinfo {volume} {C72}},\ \bibinfo {pages} {054006}
  (\bibinfo {year} {2005})},\ \Eprint {http://arxiv.org/abs/nucl-th/0506005}
  {arXiv:nucl-th/0506005 [nucl-th]} \BibitemShut {NoStop}%
\bibitem [{\citenamefont {Pav\'on~Valderrama}\ and\ \citenamefont
  {Ruiz~Arriola}(2006)}]{PavonValderrama:2005uj}%
  \BibitemOpen
  \bibfield  {author} {\bibinfo {author} {\bibfnamefont {M.}~\bibnamefont
  {Pav\'on~Valderrama}}\ and\ \bibinfo {author} {\bibfnamefont
  {E.}~\bibnamefont {Ruiz~Arriola}},\ }\bibfield  {title} {\enquote {\bibinfo
  {title} {{Renormalization of NN interaction with chiral two pion exchange
  potential: Non-central phases}},}\ }\href {\doibase
  10.1103/PhysRevC.74.064004, 10.1103/PhysRevC.75.059905} {\bibfield  {journal}
  {\bibinfo  {journal} {Phys. Rev.}\ }\textbf {\bibinfo {volume} {C74}},\
  \bibinfo {pages} {064004} (\bibinfo {year} {2006})},\ \bibinfo {note}
  {[Erratum: Phys. Rev.C75,059905(2007)]},\ \Eprint
  {http://arxiv.org/abs/nucl-th/0507075} {arXiv:nucl-th/0507075 [nucl-th]}
  \BibitemShut {NoStop}%
\bibitem [{\citenamefont {Long}\ and\ \citenamefont
  {Yang}(2012)}]{Long:2012ve}%
  \BibitemOpen
  \bibfield  {author} {\bibinfo {author} {\bibfnamefont {Bingwei}\ \bibnamefont
  {Long}}\ and\ \bibinfo {author} {\bibfnamefont {C.-J.}\ \bibnamefont
  {Yang}},\ }\bibfield  {title} {\enquote {\bibinfo {title} {{Short-range
  nuclear forces in singlet channels}},}\ }\href {\doibase
  10.1103/PhysRevC.86.024001} {\bibfield  {journal} {\bibinfo  {journal} {Phys.
  Rev.}\ }\textbf {\bibinfo {volume} {C86}},\ \bibinfo {pages} {024001}
  (\bibinfo {year} {2012})},\ \Eprint {http://arxiv.org/abs/1202.4053}
  {arXiv:1202.4053 [nucl-th]} \BibitemShut {NoStop}%
\bibitem [{\citenamefont {Manohar}\ and\ \citenamefont
  {Georgi}(1984)}]{Manohar:1983md}%
  \BibitemOpen
  \bibfield  {author} {\bibinfo {author} {\bibfnamefont {Aneesh}\ \bibnamefont
  {Manohar}}\ and\ \bibinfo {author} {\bibfnamefont {Howard}\ \bibnamefont
  {Georgi}},\ }\bibfield  {title} {\enquote {\bibinfo {title} {{Chiral Quarks
  and the Nonrelativistic Quark Model}},}\ }\href {\doibase
  10.1016/0550-3213(84)90231-1} {\bibfield  {journal} {\bibinfo  {journal}
  {Nucl. Phys.}\ }\textbf {\bibinfo {volume} {B234}},\ \bibinfo {pages}
  {189--212} (\bibinfo {year} {1984})}\BibitemShut {NoStop}%
\bibitem [{\citenamefont {Beane}\ \emph {et~al.}(2002)\citenamefont {Beane},
  \citenamefont {Bedaque}, \citenamefont {Savage},\ and\ \citenamefont {van
  Kolck}}]{Beane:2001bc}%
  \BibitemOpen
  \bibfield  {author} {\bibinfo {author} {\bibfnamefont {S.~R.}\ \bibnamefont
  {Beane}}, \bibinfo {author} {\bibfnamefont {Paulo~F.}\ \bibnamefont
  {Bedaque}}, \bibinfo {author} {\bibfnamefont {M.~J.}\ \bibnamefont {Savage}},
  \ and\ \bibinfo {author} {\bibfnamefont {U.}~\bibnamefont {van Kolck}},\
  }\bibfield  {title} {\enquote {\bibinfo {title} {{Towards a perturbative
  theory of nuclear forces}},}\ }\href {\doibase 10.1016/S0375-9474(01)01324-0}
  {\bibfield  {journal} {\bibinfo  {journal} {Nucl. Phys.}\ }\textbf {\bibinfo
  {volume} {A700}},\ \bibinfo {pages} {377--402} (\bibinfo {year} {2002})},\
  \Eprint {http://arxiv.org/abs/nucl-th/0104030} {arXiv:nucl-th/0104030
  [nucl-th]} \BibitemShut {NoStop}%
\bibitem [{\citenamefont {Pav\'on~Valderrama}\ and\ \citenamefont
  {Phillips}(2015)}]{Valderrama:2014vra}%
  \BibitemOpen
  \bibfield  {author} {\bibinfo {author} {\bibfnamefont {M.}~\bibnamefont
  {Pav\'on~Valderrama}}\ and\ \bibinfo {author} {\bibfnamefont {Daniel~R.}\
  \bibnamefont {Phillips}},\ }\bibfield  {title} {\enquote {\bibinfo {title}
  {{Power Counting of Contact-Range Currents in Effective Field Theory}},}\
  }\href {\doibase 10.1103/PhysRevLett.114.082502} {\bibfield  {journal}
  {\bibinfo  {journal} {Phys. Rev. Lett.}\ }\textbf {\bibinfo {volume} {114}},\
  \bibinfo {pages} {082502} (\bibinfo {year} {2015})},\ \Eprint
  {http://arxiv.org/abs/1407.0437} {arXiv:1407.0437 [nucl-th]} \BibitemShut
  {NoStop}%
\bibitem [{\citenamefont {Wang}\ \emph {et~al.}(2018)\citenamefont {Wang},
  \citenamefont {Engel},\ and\ \citenamefont {Yao}}]{Wang:2018htk}%
  \BibitemOpen
  \bibfield  {author} {\bibinfo {author} {\bibfnamefont {Long-Jun}\
  \bibnamefont {Wang}}, \bibinfo {author} {\bibfnamefont {Jonathan}\
  \bibnamefont {Engel}}, \ and\ \bibinfo {author} {\bibfnamefont {Jiang~Ming}\
  \bibnamefont {Yao}},\ }\bibfield  {title} {\enquote {\bibinfo {title}
  {{Quenching of nuclear matrix elements for $0\nu\beta\beta$ decay by chiral
  two-body currents}},}\ }\href {\doibase 10.1103/PhysRevC.98.031301}
  {\bibfield  {journal} {\bibinfo  {journal} {Phys. Rev.}\ }\textbf {\bibinfo
  {volume} {C98}},\ \bibinfo {pages} {031301} (\bibinfo {year} {2018})},\
  \Eprint {http://arxiv.org/abs/1805.10276} {arXiv:1805.10276 [nucl-th]}
  \BibitemShut {NoStop}%
\bibitem [{\citenamefont {Miller}\ and\ \citenamefont
  {Spencer}(1976)}]{Miller:1975hu}%
  \BibitemOpen
  \bibfield  {author} {\bibinfo {author} {\bibfnamefont {Gerald~A.}\
  \bibnamefont {Miller}}\ and\ \bibinfo {author} {\bibfnamefont {James~E.}\
  \bibnamefont {Spencer}},\ }\bibfield  {title} {\enquote {\bibinfo {title} {{A
  Survey of Pion Charge-Exchange Reactions with Nuclei}},}\ }\href {\doibase
  10.1016/0003-4916(76)90073-7} {\bibfield  {journal} {\bibinfo  {journal}
  {Annals Phys.}\ }\textbf {\bibinfo {volume} {100}},\ \bibinfo {pages} {562}
  (\bibinfo {year} {1976})}\BibitemShut {NoStop}%
\bibitem [{\citenamefont {Simkovic}\ \emph {et~al.}(2009)\citenamefont
  {Simkovic}, \citenamefont {Faessler}, \citenamefont {Muther}, \citenamefont
  {Rodin},\ and\ \citenamefont {Stauf}}]{Simkovic:2009pp}%
  \BibitemOpen
  \bibfield  {author} {\bibinfo {author} {\bibfnamefont {Fedor}\ \bibnamefont
  {Simkovic}}, \bibinfo {author} {\bibfnamefont {Amand}\ \bibnamefont
  {Faessler}}, \bibinfo {author} {\bibfnamefont {Herbert}\ \bibnamefont
  {Muther}}, \bibinfo {author} {\bibfnamefont {Vadim}\ \bibnamefont {Rodin}}, \
  and\ \bibinfo {author} {\bibfnamefont {Markus}\ \bibnamefont {Stauf}},\
  }\bibfield  {title} {\enquote {\bibinfo {title} {{The 0 nu bb-decay nuclear
  matrix elements with self-consistent short-range correlations}},}\ }\href
  {\doibase 10.1103/PhysRevC.79.055501} {\bibfield  {journal} {\bibinfo
  {journal} {Phys. Rev.}\ }\textbf {\bibinfo {volume} {C79}},\ \bibinfo {pages}
  {055501} (\bibinfo {year} {2009})},\ \Eprint {http://arxiv.org/abs/0902.0331}
  {arXiv:0902.0331 [nucl-th]} \BibitemShut {NoStop}%
\bibitem [{\citenamefont {Engel}\ \emph {et~al.}(2011)\citenamefont {Engel},
  \citenamefont {Carlson},\ and\ \citenamefont {Wiringa}}]{Engel:2011ss}%
  \BibitemOpen
  \bibfield  {author} {\bibinfo {author} {\bibfnamefont {J.}~\bibnamefont
  {Engel}}, \bibinfo {author} {\bibfnamefont {J.}~\bibnamefont {Carlson}}, \
  and\ \bibinfo {author} {\bibfnamefont {R.~B.}\ \bibnamefont {Wiringa}},\
  }\bibfield  {title} {\enquote {\bibinfo {title} {{Jastrow functions in
  double-beta decay}},}\ }\href {\doibase 10.1103/PhysRevC.83.034317}
  {\bibfield  {journal} {\bibinfo  {journal} {Phys. Rev.}\ }\textbf {\bibinfo
  {volume} {C83}},\ \bibinfo {pages} {034317} (\bibinfo {year} {2011})},\
  \Eprint {http://arxiv.org/abs/1101.0554} {arXiv:1101.0554 [nucl-th]}
  \BibitemShut {NoStop}%
\bibitem [{\citenamefont {Benhar}\ \emph {et~al.}(2014)\citenamefont {Benhar},
  \citenamefont {Biondi},\ and\ \citenamefont {Speranza}}]{Benhar:2014cka}%
  \BibitemOpen
  \bibfield  {author} {\bibinfo {author} {\bibfnamefont {Omar}\ \bibnamefont
  {Benhar}}, \bibinfo {author} {\bibfnamefont {Riccardo}\ \bibnamefont
  {Biondi}}, \ and\ \bibinfo {author} {\bibfnamefont {Enrico}\ \bibnamefont
  {Speranza}},\ }\bibfield  {title} {\enquote {\bibinfo {title} {{Short-range
  correlation effects on the nuclear matrix element of neutrinoless double-beta
  decay}},}\ }\href {\doibase 10.1103/PhysRevC.90.065504} {\bibfield  {journal}
  {\bibinfo  {journal} {Phys. Rev.}\ }\textbf {\bibinfo {volume} {C90}},\
  \bibinfo {pages} {065504} (\bibinfo {year} {2014})},\ \Eprint
  {http://arxiv.org/abs/1401.2030} {arXiv:1401.2030 [nucl-th]} \BibitemShut
  {NoStop}%
\bibitem [{\citenamefont {Pastore}\ \emph
  {et~al.}(2018{\natexlab{a}})\citenamefont {Pastore}, \citenamefont {Baroni},
  \citenamefont {Carlson}, \citenamefont {Gandolfi}, \citenamefont {Pieper},
  \citenamefont {Schiavilla},\ and\ \citenamefont {Wiringa}}]{Pastore:2017uwc}%
  \BibitemOpen
  \bibfield  {author} {\bibinfo {author} {\bibfnamefont {S.}~\bibnamefont
  {Pastore}}, \bibinfo {author} {\bibfnamefont {A.}~\bibnamefont {Baroni}},
  \bibinfo {author} {\bibfnamefont {J.}~\bibnamefont {Carlson}}, \bibinfo
  {author} {\bibfnamefont {S.}~\bibnamefont {Gandolfi}}, \bibinfo {author}
  {\bibfnamefont {Steven~C.}\ \bibnamefont {Pieper}}, \bibinfo {author}
  {\bibfnamefont {R.}~\bibnamefont {Schiavilla}}, \ and\ \bibinfo {author}
  {\bibfnamefont {R.~B.}\ \bibnamefont {Wiringa}},\ }\bibfield  {title}
  {\enquote {\bibinfo {title} {{Quantum Monte Carlo calculations of weak
  transitions in $A=6-10$ nuclei}},}\ }\href {\doibase
  10.1103/PhysRevC.97.022501} {\bibfield  {journal} {\bibinfo  {journal} {Phys.
  Rev.}\ }\textbf {\bibinfo {volume} {C97}},\ \bibinfo {pages} {022501}
  (\bibinfo {year} {2018}{\natexlab{a}})},\ \Eprint
  {http://arxiv.org/abs/1709.03592} {arXiv:1709.03592 [nucl-th]} \BibitemShut
  {NoStop}%
\bibitem [{\citenamefont {Gysbers}\ \emph {et~al.}(2019)\citenamefont {Gysbers}
  \emph {et~al.}}]{Gysbers:2019uyb}%
  \BibitemOpen
  \bibfield  {author} {\bibinfo {author} {\bibfnamefont {P.}~\bibnamefont
  {Gysbers}} \emph {et~al.},\ }\bibfield  {title} {\enquote {\bibinfo {title}
  {{Discrepancy between experimental and theoretical $\beta$-decay rates
  resolved from first principles}},}\ }\href {\doibase
  10.1038/s41567-019-0450-7} {\bibfield  {journal} {\bibinfo  {journal} {Nature
  Phys.}\ }\textbf {\bibinfo {volume} {15}},\ \bibinfo {pages} {428--431}
  (\bibinfo {year} {2019})},\ \Eprint {http://arxiv.org/abs/1903.00047}
  {arXiv:1903.00047 [nucl-th]} \BibitemShut {NoStop}%
\bibitem [{\citenamefont {Shanahan}\ \emph {et~al.}(2017)\citenamefont
  {Shanahan}, \citenamefont {Tiburzi}, \citenamefont {Wagman}, \citenamefont
  {Winter}, \citenamefont {Chang}, \citenamefont {Davoudi}, \citenamefont
  {Detmold}, \citenamefont {Orginos},\ and\ \citenamefont
  {Savage}}]{Shanahan:2017bgi}%
  \BibitemOpen
  \bibfield  {author} {\bibinfo {author} {\bibfnamefont {Phiala~E.}\
  \bibnamefont {Shanahan}}, \bibinfo {author} {\bibfnamefont {Brian~C.}\
  \bibnamefont {Tiburzi}}, \bibinfo {author} {\bibfnamefont {Michael~L.}\
  \bibnamefont {Wagman}}, \bibinfo {author} {\bibfnamefont {Frank}\
  \bibnamefont {Winter}}, \bibinfo {author} {\bibfnamefont {Emmanuel}\
  \bibnamefont {Chang}}, \bibinfo {author} {\bibfnamefont {Zohreh}\
  \bibnamefont {Davoudi}}, \bibinfo {author} {\bibfnamefont {William}\
  \bibnamefont {Detmold}}, \bibinfo {author} {\bibfnamefont {Kostas}\
  \bibnamefont {Orginos}}, \ and\ \bibinfo {author} {\bibfnamefont {Martin~J.}\
  \bibnamefont {Savage}},\ }\bibfield  {title} {\enquote {\bibinfo {title}
  {{The isotensor axial polarisability and lattice QCD input for nuclear
  double-$\beta$ decay phenomenology}},}\ }\href {\doibase
  10.1103/PhysRevLett.119.062003} {\bibfield  {journal} {\bibinfo  {journal}
  {Phys. Rev. Lett.}\ }\textbf {\bibinfo {volume} {119}},\ \bibinfo {pages}
  {062003} (\bibinfo {year} {2017})},\ \Eprint
  {http://arxiv.org/abs/1701.03456} {arXiv:1701.03456 [hep-lat]} \BibitemShut
  {NoStop}%
\bibitem [{\citenamefont {Tiburzi}\ \emph {et~al.}(2017)\citenamefont
  {Tiburzi}, \citenamefont {Wagman}, \citenamefont {Winter}, \citenamefont
  {Chang}, \citenamefont {Davoudi}, \citenamefont {Detmold}, \citenamefont
  {Orginos}, \citenamefont {Savage},\ and\ \citenamefont
  {Shanahan}}]{Tiburzi:2017iux}%
  \BibitemOpen
  \bibfield  {author} {\bibinfo {author} {\bibfnamefont {Brian~C.}\
  \bibnamefont {Tiburzi}}, \bibinfo {author} {\bibfnamefont {Michael~L.}\
  \bibnamefont {Wagman}}, \bibinfo {author} {\bibfnamefont {Frank}\
  \bibnamefont {Winter}}, \bibinfo {author} {\bibfnamefont {Emmanuel}\
  \bibnamefont {Chang}}, \bibinfo {author} {\bibfnamefont {Zohreh}\
  \bibnamefont {Davoudi}}, \bibinfo {author} {\bibfnamefont {William}\
  \bibnamefont {Detmold}}, \bibinfo {author} {\bibfnamefont {Kostas}\
  \bibnamefont {Orginos}}, \bibinfo {author} {\bibfnamefont {Martin~J.}\
  \bibnamefont {Savage}}, \ and\ \bibinfo {author} {\bibfnamefont {Phiala~E.}\
  \bibnamefont {Shanahan}},\ }\bibfield  {title} {\enquote {\bibinfo {title}
  {{Double-$\beta$ Decay Matrix Elements from Lattice Quantum
  Chromodynamics}},}\ }\href {\doibase 10.1103/PhysRevD.96.054505} {\bibfield
  {journal} {\bibinfo  {journal} {Phys. Rev.}\ }\textbf {\bibinfo {volume}
  {D96}},\ \bibinfo {pages} {054505} (\bibinfo {year} {2017})},\ \Eprint
  {http://arxiv.org/abs/1702.02929} {arXiv:1702.02929 [hep-lat]} \BibitemShut
  {NoStop}%
\bibitem [{\citenamefont {Feng}\ \emph {et~al.}(2019)\citenamefont {Feng},
  \citenamefont {Jin}, \citenamefont {Tuo},\ and\ \citenamefont
  {Xia}}]{Feng:2018pdq}%
  \BibitemOpen
  \bibfield  {author} {\bibinfo {author} {\bibfnamefont {Xu}~\bibnamefont
  {Feng}}, \bibinfo {author} {\bibfnamefont {Lu-Chang}\ \bibnamefont {Jin}},
  \bibinfo {author} {\bibfnamefont {Xin-Yu}\ \bibnamefont {Tuo}}, \ and\
  \bibinfo {author} {\bibfnamefont {Shi-Cheng}\ \bibnamefont {Xia}},\
  }\bibfield  {title} {\enquote {\bibinfo {title} {{Light-Neutrino Exchange and
  Long-Distance Contributions to $0\nu2\beta$ Decays: An Exploratory Study on
  $\pi\pi\to ee$}},}\ }\href {\doibase 10.1103/PhysRevLett.122.022001}
  {\bibfield  {journal} {\bibinfo  {journal} {Phys. Rev. Lett.}\ }\textbf
  {\bibinfo {volume} {122}},\ \bibinfo {pages} {022001} (\bibinfo {year}
  {2019})},\ \Eprint {http://arxiv.org/abs/1809.10511} {arXiv:1809.10511
  [hep-lat]} \BibitemShut {NoStop}%
\bibitem [{\citenamefont {Cirigliano}\ \emph {et~al.}(2019)\citenamefont
  {Cirigliano}, \citenamefont {Davoudi}, \citenamefont {Bhattacharya},
  \citenamefont {Izubuchi}, \citenamefont {Shanahan}, \citenamefont
  {Syritsyn},\ and\ \citenamefont {Wagman}}]{Cirigliano:2019jig}%
  \BibitemOpen
  \bibfield  {author} {\bibinfo {author} {\bibfnamefont {Vincenzo}\
  \bibnamefont {Cirigliano}}, \bibinfo {author} {\bibfnamefont {Zohreh}\
  \bibnamefont {Davoudi}}, \bibinfo {author} {\bibfnamefont {Tanmoy}\
  \bibnamefont {Bhattacharya}}, \bibinfo {author} {\bibfnamefont {Taku}\
  \bibnamefont {Izubuchi}}, \bibinfo {author} {\bibfnamefont {Phiala~E.}\
  \bibnamefont {Shanahan}}, \bibinfo {author} {\bibfnamefont {Sergey}\
  \bibnamefont {Syritsyn}}, \ and\ \bibinfo {author} {\bibfnamefont
  {Michael~L.}\ \bibnamefont {Wagman}},\ }\bibfield  {title} {\enquote
  {\bibinfo {title} {{The Role of Lattice QCD in Searches for Violations of
  Fundamental Symmetries and Signals for New Physics}},}\ }\href@noop {} {\
  (\bibinfo {year} {2019})},\ \Eprint {http://arxiv.org/abs/1904.09704}
  {arXiv:1904.09704 [hep-lat]} \BibitemShut {NoStop}%
\bibitem [{\citenamefont {Nicholson}\ \emph {et~al.}(2018)\citenamefont
  {Nicholson} \emph {et~al.}}]{Nicholson:2018mwc}%
  \BibitemOpen
  \bibfield  {author} {\bibinfo {author} {\bibfnamefont {A.}~\bibnamefont
  {Nicholson}} \emph {et~al.},\ }\bibfield  {title} {\enquote {\bibinfo {title}
  {{Heavy physics contributions to neutrinoless double beta decay from QCD}},}\
  }\href {\doibase 10.1103/PhysRevLett.121.172501} {\bibfield  {journal}
  {\bibinfo  {journal} {Phys. Rev. Lett.}\ }\textbf {\bibinfo {volume} {121}},\
  \bibinfo {pages} {172501} (\bibinfo {year} {2018})},\ \Eprint
  {http://arxiv.org/abs/1805.02634} {arXiv:1805.02634 [nucl-th]} \BibitemShut
  {NoStop}%
\bibitem [{\citenamefont {Monge-Camacho}\ \emph {et~al.}(2019)\citenamefont
  {Monge-Camacho} \emph {et~al.}}]{Monge-Camacho:2019nby}%
  \BibitemOpen
  \bibfield  {author} {\bibinfo {author} {\bibfnamefont {Henry}\ \bibnamefont
  {Monge-Camacho}} \emph {et~al.},\ }\bibfield  {title} {\enquote {\bibinfo
  {title} {{Short Range Operator Contributions to $0\nu\beta\beta$ decay from
  LQCD}},}\ }in\ \href@noop {} {\emph {\bibinfo {booktitle} {{36th
  International Symposium on Lattice Field Theory (Lattice 2018) East Lansing,
  MI, United States, July 22-28, 2018}}}}\ (\bibinfo {year} {2019})\ \Eprint
  {http://arxiv.org/abs/1904.12055} {arXiv:1904.12055 [hep-lat]} \BibitemShut
  {NoStop}%
\bibitem [{\citenamefont {Bedaque}\ and\ \citenamefont {van
  Kolck}(1998)}]{Bedaque:1997qi}%
  \BibitemOpen
  \bibfield  {author} {\bibinfo {author} {\bibfnamefont {Paulo~F.}\
  \bibnamefont {Bedaque}}\ and\ \bibinfo {author} {\bibfnamefont
  {U.}~\bibnamefont {van Kolck}},\ }\bibfield  {title} {\enquote {\bibinfo
  {title} {{Nucleon deuteron scattering from an effective field theory}},}\
  }\href {\doibase 10.1016/S0370-2693(98)00430-4} {\bibfield  {journal}
  {\bibinfo  {journal} {Phys. Lett.}\ }\textbf {\bibinfo {volume} {B428}},\
  \bibinfo {pages} {221--226} (\bibinfo {year} {1998})},\ \Eprint
  {http://arxiv.org/abs/nucl-th/9710073} {arXiv:nucl-th/9710073 [nucl-th]}
  \BibitemShut {NoStop}%
\bibitem [{\citenamefont {Bedaque}\ \emph {et~al.}(1998)\citenamefont
  {Bedaque}, \citenamefont {Hammer},\ and\ \citenamefont {van
  Kolck}}]{Bedaque:1998mb}%
  \BibitemOpen
  \bibfield  {author} {\bibinfo {author} {\bibfnamefont {Paulo~F.}\
  \bibnamefont {Bedaque}}, \bibinfo {author} {\bibfnamefont {H.~W.}\
  \bibnamefont {Hammer}}, \ and\ \bibinfo {author} {\bibfnamefont
  {U.}~\bibnamefont {van Kolck}},\ }\bibfield  {title} {\enquote {\bibinfo
  {title} {{Effective theory for neutron deuteron scattering: Energy
  dependence}},}\ }\href {\doibase 10.1103/PhysRevC.58.R641} {\bibfield
  {journal} {\bibinfo  {journal} {Phys. Rev.}\ }\textbf {\bibinfo {volume}
  {C58}},\ \bibinfo {pages} {R641--R644} (\bibinfo {year} {1998})},\ \Eprint
  {http://arxiv.org/abs/nucl-th/9802057} {arXiv:nucl-th/9802057 [nucl-th]}
  \BibitemShut {NoStop}%
\bibitem [{\citenamefont {Chen}\ \emph {et~al.}(1999)\citenamefont {Chen},
  \citenamefont {Rupak},\ and\ \citenamefont {Savage}}]{Chen:1999tn}%
  \BibitemOpen
  \bibfield  {author} {\bibinfo {author} {\bibfnamefont {Jiunn-Wei}\
  \bibnamefont {Chen}}, \bibinfo {author} {\bibfnamefont {Gautam}\ \bibnamefont
  {Rupak}}, \ and\ \bibinfo {author} {\bibfnamefont {Martin~J.}\ \bibnamefont
  {Savage}},\ }\bibfield  {title} {\enquote {\bibinfo {title} {{Nucleon-nucleon
  effective field theory without pions}},}\ }\href {\doibase
  10.1016/S0375-9474(99)00298-5} {\bibfield  {journal} {\bibinfo  {journal}
  {Nucl. Phys.}\ }\textbf {\bibinfo {volume} {A653}},\ \bibinfo {pages}
  {386--412} (\bibinfo {year} {1999})},\ \Eprint
  {http://arxiv.org/abs/nucl-th/9902056} {arXiv:nucl-th/9902056 [nucl-th]}
  \BibitemShut {NoStop}%
\bibitem [{\citenamefont {Kong}\ and\ \citenamefont
  {Ravndal}(2000)}]{Kong:1999sf}%
  \BibitemOpen
  \bibfield  {author} {\bibinfo {author} {\bibfnamefont {Xinwei}\ \bibnamefont
  {Kong}}\ and\ \bibinfo {author} {\bibfnamefont {Finn}\ \bibnamefont
  {Ravndal}},\ }\bibfield  {title} {\enquote {\bibinfo {title} {{Coulomb
  effects in low-energy proton proton scattering}},}\ }\href {\doibase
  10.1016/S0375-9474(99)00406-6} {\bibfield  {journal} {\bibinfo  {journal}
  {Nucl. Phys.}\ }\textbf {\bibinfo {volume} {A665}},\ \bibinfo {pages}
  {137--163} (\bibinfo {year} {2000})},\ \Eprint
  {http://arxiv.org/abs/hep-ph/9903523} {arXiv:hep-ph/9903523 [hep-ph]}
  \BibitemShut {NoStop}%
\bibitem [{\citenamefont {Bedaque}\ \emph {et~al.}(2000)\citenamefont
  {Bedaque}, \citenamefont {Hammer},\ and\ \citenamefont {van
  Kolck}}]{Bedaque:1999ve}%
  \BibitemOpen
  \bibfield  {author} {\bibinfo {author} {\bibfnamefont {Paulo~F.}\
  \bibnamefont {Bedaque}}, \bibinfo {author} {\bibfnamefont {H.~W.}\
  \bibnamefont {Hammer}}, \ and\ \bibinfo {author} {\bibfnamefont
  {U.}~\bibnamefont {van Kolck}},\ }\bibfield  {title} {\enquote {\bibinfo
  {title} {{Effective theory of the triton}},}\ }\href {\doibase
  10.1016/S0375-9474(00)00205-0} {\bibfield  {journal} {\bibinfo  {journal}
  {Nucl. Phys.}\ }\textbf {\bibinfo {volume} {A676}},\ \bibinfo {pages}
  {357--370} (\bibinfo {year} {2000})},\ \Eprint
  {http://arxiv.org/abs/nucl-th/9906032} {arXiv:nucl-th/9906032 [nucl-th]}
  \BibitemShut {NoStop}%
\bibitem [{\citenamefont {Bedaque}\ and\ \citenamefont {van
  Kolck}(2002)}]{Bedaque:2002mn}%
  \BibitemOpen
  \bibfield  {author} {\bibinfo {author} {\bibfnamefont {Paulo~F.}\
  \bibnamefont {Bedaque}}\ and\ \bibinfo {author} {\bibfnamefont {Ubirajara}\
  \bibnamefont {van Kolck}},\ }\bibfield  {title} {\enquote {\bibinfo {title}
  {{Effective field theory for few nucleon systems}},}\ }\href {\doibase
  10.1146/annurev.nucl.52.050102.090637} {\bibfield  {journal} {\bibinfo
  {journal} {Ann. Rev. Nucl. Part. Sci.}\ }\textbf {\bibinfo {volume} {52}},\
  \bibinfo {pages} {339--396} (\bibinfo {year} {2002})},\ \Eprint
  {http://arxiv.org/abs/nucl-th/0203055} {arXiv:nucl-th/0203055 [nucl-th]}
  \BibitemShut {NoStop}%
\bibitem [{\citenamefont {Vanasse}(2013)}]{Vanasse:2013sda}%
  \BibitemOpen
  \bibfield  {author} {\bibinfo {author} {\bibfnamefont {Jared}\ \bibnamefont
  {Vanasse}},\ }\bibfield  {title} {\enquote {\bibinfo {title} {{Fully
  Perturbative Calculation of $nd$ Scattering to
  Next-to-next-to-leading-order}},}\ }\href {\doibase
  10.1103/PhysRevC.88.044001} {\bibfield  {journal} {\bibinfo  {journal} {Phys.
  Rev.}\ }\textbf {\bibinfo {volume} {C88}},\ \bibinfo {pages} {044001}
  (\bibinfo {year} {2013})},\ \Eprint {http://arxiv.org/abs/1305.0283}
  {arXiv:1305.0283 [nucl-th]} \BibitemShut {NoStop}%
\bibitem [{\citenamefont {K{\"o}nig}\ \emph {et~al.}(2016)\citenamefont
  {K{\"o}nig}, \citenamefont {Grie{\ss}hammer}, \citenamefont {Hammer},\ and\
  \citenamefont {van Kolck}}]{Konig:2015aka}%
  \BibitemOpen
  \bibfield  {author} {\bibinfo {author} {\bibfnamefont {Sebastian}\
  \bibnamefont {K{\"o}nig}}, \bibinfo {author} {\bibfnamefont {Harald~W.}\
  \bibnamefont {Grie{\ss}hammer}}, \bibinfo {author} {\bibfnamefont {H.-W.}\
  \bibnamefont {Hammer}}, \ and\ \bibinfo {author} {\bibfnamefont
  {U.}~\bibnamefont {van Kolck}},\ }\bibfield  {title} {\enquote {\bibinfo
  {title} {{Effective theory of $^3$H and $^3$He}},}\ }\href {\doibase
  10.1088/0954-3899/43/5/055106} {\bibfield  {journal} {\bibinfo  {journal} {J.
  Phys.}\ }\textbf {\bibinfo {volume} {G43}},\ \bibinfo {pages} {055106}
  (\bibinfo {year} {2016})},\ \Eprint {http://arxiv.org/abs/1508.05085}
  {arXiv:1508.05085 [nucl-th]} \BibitemShut {NoStop}%
\bibitem [{\citenamefont {Platter}\ \emph {et~al.}(2005)\citenamefont
  {Platter}, \citenamefont {Hammer},\ and\ \citenamefont
  {Meissner}}]{Platter:2004zs}%
  \BibitemOpen
  \bibfield  {author} {\bibinfo {author} {\bibfnamefont {L.}~\bibnamefont
  {Platter}}, \bibinfo {author} {\bibfnamefont {H.~W.}\ \bibnamefont {Hammer}},
  \ and\ \bibinfo {author} {\bibfnamefont {Ulf-G.}\ \bibnamefont {Meissner}},\
  }\bibfield  {title} {\enquote {\bibinfo {title} {{On the correlation between
  the binding energies of the triton and the alpha-particle}},}\ }\href
  {\doibase 10.1016/j.physletb.2004.12.068} {\bibfield  {journal} {\bibinfo
  {journal} {Phys. Lett.}\ }\textbf {\bibinfo {volume} {B607}},\ \bibinfo
  {pages} {254--258} (\bibinfo {year} {2005})},\ \Eprint
  {http://arxiv.org/abs/nucl-th/0409040} {arXiv:nucl-th/0409040 [nucl-th]}
  \BibitemShut {NoStop}%
\bibitem [{\citenamefont {Stetcu}\ \emph {et~al.}(2007)\citenamefont {Stetcu},
  \citenamefont {Barrett},\ and\ \citenamefont {van Kolck}}]{Stetcu:2006ey}%
  \BibitemOpen
  \bibfield  {author} {\bibinfo {author} {\bibfnamefont {I.}~\bibnamefont
  {Stetcu}}, \bibinfo {author} {\bibfnamefont {B.~R.}\ \bibnamefont {Barrett}},
  \ and\ \bibinfo {author} {\bibfnamefont {U.}~\bibnamefont {van Kolck}},\
  }\bibfield  {title} {\enquote {\bibinfo {title} {{No-core shell model in an
  effective-field-theory framework}},}\ }\href {\doibase
  10.1016/j.physletb.2007.07.065} {\bibfield  {journal} {\bibinfo  {journal}
  {Phys. Lett.}\ }\textbf {\bibinfo {volume} {B653}},\ \bibinfo {pages}
  {358--362} (\bibinfo {year} {2007})},\ \Eprint
  {http://arxiv.org/abs/nucl-th/0609023} {arXiv:nucl-th/0609023 [nucl-th]}
  \BibitemShut {NoStop}%
\bibitem [{\citenamefont {Contessi}\ \emph {et~al.}(2017)\citenamefont
  {Contessi}, \citenamefont {Lovato}, \citenamefont {Pederiva}, \citenamefont
  {Roggero}, \citenamefont {Kirscher},\ and\ \citenamefont {van
  Kolck}}]{Contessi:2017rww}%
  \BibitemOpen
  \bibfield  {author} {\bibinfo {author} {\bibfnamefont {L.}~\bibnamefont
  {Contessi}}, \bibinfo {author} {\bibfnamefont {A.}~\bibnamefont {Lovato}},
  \bibinfo {author} {\bibfnamefont {F.}~\bibnamefont {Pederiva}}, \bibinfo
  {author} {\bibfnamefont {A.}~\bibnamefont {Roggero}}, \bibinfo {author}
  {\bibfnamefont {J.}~\bibnamefont {Kirscher}}, \ and\ \bibinfo {author}
  {\bibfnamefont {U.}~\bibnamefont {van Kolck}},\ }\bibfield  {title} {\enquote
  {\bibinfo {title} {{Ground-state properties of $^{4}$He and $^{16}$O
  extrapolated from lattice QCD with pionless EFT}},}\ }\href {\doibase
  10.1016/j.physletb.2017.07.048} {\bibfield  {journal} {\bibinfo  {journal}
  {Phys. Lett.}\ }\textbf {\bibinfo {volume} {B772}},\ \bibinfo {pages}
  {839--848} (\bibinfo {year} {2017})},\ \Eprint
  {http://arxiv.org/abs/1701.06516} {arXiv:1701.06516 [nucl-th]} \BibitemShut
  {NoStop}%
\bibitem [{\citenamefont {Bansal}\ \emph {et~al.}(2018)\citenamefont {Bansal},
  \citenamefont {Binder}, \citenamefont {Ekstr{\"o}m}, \citenamefont {Hagen},
  \citenamefont {Jansen},\ and\ \citenamefont {Papenbrock}}]{Bansal:2017pwn}%
  \BibitemOpen
  \bibfield  {author} {\bibinfo {author} {\bibfnamefont {A.}~\bibnamefont
  {Bansal}}, \bibinfo {author} {\bibfnamefont {S.}~\bibnamefont {Binder}},
  \bibinfo {author} {\bibfnamefont {A.}~\bibnamefont {Ekstr{\"o}m}}, \bibinfo
  {author} {\bibfnamefont {G.}~\bibnamefont {Hagen}}, \bibinfo {author}
  {\bibfnamefont {G.~R.}\ \bibnamefont {Jansen}}, \ and\ \bibinfo {author}
  {\bibfnamefont {T.}~\bibnamefont {Papenbrock}},\ }\bibfield  {title}
  {\enquote {\bibinfo {title} {{Pion-less effective field theory for atomic
  nuclei and lattice nuclei}},}\ }\href {\doibase 10.1103/PhysRevC.98.054301}
  {\bibfield  {journal} {\bibinfo  {journal} {Phys. Rev.}\ }\textbf {\bibinfo
  {volume} {C98}},\ \bibinfo {pages} {054301} (\bibinfo {year} {2018})},\
  \Eprint {http://arxiv.org/abs/1712.10246} {arXiv:1712.10246 [nucl-th]}
  \BibitemShut {NoStop}%
\bibitem [{\citenamefont {Barnea}\ \emph {et~al.}(2015)\citenamefont {Barnea},
  \citenamefont {Contessi}, \citenamefont {Gazit}, \citenamefont {Pederiva},\
  and\ \citenamefont {van Kolck}}]{Barnea:2013uqa}%
  \BibitemOpen
  \bibfield  {author} {\bibinfo {author} {\bibfnamefont {N.}~\bibnamefont
  {Barnea}}, \bibinfo {author} {\bibfnamefont {L.}~\bibnamefont {Contessi}},
  \bibinfo {author} {\bibfnamefont {D.}~\bibnamefont {Gazit}}, \bibinfo
  {author} {\bibfnamefont {F.}~\bibnamefont {Pederiva}}, \ and\ \bibinfo
  {author} {\bibfnamefont {U.}~\bibnamefont {van Kolck}},\ }\bibfield  {title}
  {\enquote {\bibinfo {title} {{Effective Field Theory for Lattice Nuclei}},}\
  }\href {\doibase 10.1103/PhysRevLett.114.052501} {\bibfield  {journal}
  {\bibinfo  {journal} {Phys. Rev. Lett.}\ }\textbf {\bibinfo {volume} {114}},\
  \bibinfo {pages} {052501} (\bibinfo {year} {2015})},\ \Eprint
  {http://arxiv.org/abs/1311.4966} {arXiv:1311.4966 [nucl-th]} \BibitemShut
  {NoStop}%
\bibitem [{\citenamefont {Beane}\ \emph {et~al.}(2015)\citenamefont {Beane},
  \citenamefont {Chang}, \citenamefont {Detmold}, \citenamefont {Orginos},
  \citenamefont {Parre{\~n}o}, \citenamefont {Savage},\ and\ \citenamefont
  {Tiburzi}}]{Beane:2015yha}%
  \BibitemOpen
  \bibfield  {author} {\bibinfo {author} {\bibfnamefont {Silas~R.}\
  \bibnamefont {Beane}}, \bibinfo {author} {\bibfnamefont {Emmanuel}\
  \bibnamefont {Chang}}, \bibinfo {author} {\bibfnamefont {William}\
  \bibnamefont {Detmold}}, \bibinfo {author} {\bibfnamefont {Kostas}\
  \bibnamefont {Orginos}}, \bibinfo {author} {\bibfnamefont {Assumpta}\
  \bibnamefont {Parre{\~n}o}}, \bibinfo {author} {\bibfnamefont {Martin~J.}\
  \bibnamefont {Savage}}, \ and\ \bibinfo {author} {\bibfnamefont {Brian~C.}\
  \bibnamefont {Tiburzi}} (\bibinfo {collaboration} {NPLQCD}),\ }\bibfield
  {title} {\enquote {\bibinfo {title} {{Ab initio Calculation of the $np\to d
  \gamma$ Radiative Capture Process}},}\ }\href {\doibase
  10.1103/PhysRevLett.115.132001} {\bibfield  {journal} {\bibinfo  {journal}
  {Phys. Rev. Lett.}\ }\textbf {\bibinfo {volume} {115}},\ \bibinfo {pages}
  {132001} (\bibinfo {year} {2015})},\ \Eprint
  {http://arxiv.org/abs/1505.02422} {arXiv:1505.02422 [hep-lat]} \BibitemShut
  {NoStop}%
\bibitem [{\citenamefont {Kirscher}\ \emph {et~al.}(2015)\citenamefont
  {Kirscher}, \citenamefont {Barnea}, \citenamefont {Gazit}, \citenamefont
  {Pederiva},\ and\ \citenamefont {van Kolck}}]{Kirscher:2015yda}%
  \BibitemOpen
  \bibfield  {author} {\bibinfo {author} {\bibfnamefont {Johannes}\
  \bibnamefont {Kirscher}}, \bibinfo {author} {\bibfnamefont {Nir}\
  \bibnamefont {Barnea}}, \bibinfo {author} {\bibfnamefont {Doron}\
  \bibnamefont {Gazit}}, \bibinfo {author} {\bibfnamefont {Francesco}\
  \bibnamefont {Pederiva}}, \ and\ \bibinfo {author} {\bibfnamefont
  {Ubirajara}\ \bibnamefont {van Kolck}},\ }\bibfield  {title} {\enquote
  {\bibinfo {title} {{Spectra and Scattering of Light Lattice Nuclei from
  Effective Field Theory}},}\ }\href {\doibase 10.1103/PhysRevC.92.054002}
  {\bibfield  {journal} {\bibinfo  {journal} {Phys. Rev.}\ }\textbf {\bibinfo
  {volume} {C92}},\ \bibinfo {pages} {054002} (\bibinfo {year} {2015})},\
  \Eprint {http://arxiv.org/abs/1506.09048} {arXiv:1506.09048 [nucl-th]}
  \BibitemShut {NoStop}%
\bibitem [{\citenamefont {Savage}\ \emph {et~al.}(2017)\citenamefont {Savage},
  \citenamefont {Shanahan}, \citenamefont {Tiburzi}, \citenamefont {Wagman},
  \citenamefont {Winter}, \citenamefont {Beane}, \citenamefont {Chang},
  \citenamefont {Davoudi}, \citenamefont {Detmold},\ and\ \citenamefont
  {Orginos}}]{Savage:2016kon}%
  \BibitemOpen
  \bibfield  {author} {\bibinfo {author} {\bibfnamefont {Martin~J.}\
  \bibnamefont {Savage}}, \bibinfo {author} {\bibfnamefont {Phiala~E.}\
  \bibnamefont {Shanahan}}, \bibinfo {author} {\bibfnamefont {Brian~C.}\
  \bibnamefont {Tiburzi}}, \bibinfo {author} {\bibfnamefont {Michael~L.}\
  \bibnamefont {Wagman}}, \bibinfo {author} {\bibfnamefont {Frank}\
  \bibnamefont {Winter}}, \bibinfo {author} {\bibfnamefont {Silas~R.}\
  \bibnamefont {Beane}}, \bibinfo {author} {\bibfnamefont {Emmanuel}\
  \bibnamefont {Chang}}, \bibinfo {author} {\bibfnamefont {Zohreh}\
  \bibnamefont {Davoudi}}, \bibinfo {author} {\bibfnamefont {William}\
  \bibnamefont {Detmold}}, \ and\ \bibinfo {author} {\bibfnamefont {Kostas}\
  \bibnamefont {Orginos}},\ }\bibfield  {title} {\enquote {\bibinfo {title}
  {{Proton-Proton Fusion and Tritium $\beta$ Decay from Lattice Quantum
  Chromodynamics}},}\ }\href {\doibase 10.1103/PhysRevLett.119.062002}
  {\bibfield  {journal} {\bibinfo  {journal} {Phys. Rev. Lett.}\ }\textbf
  {\bibinfo {volume} {119}},\ \bibinfo {pages} {062002} (\bibinfo {year}
  {2017})},\ \Eprint {http://arxiv.org/abs/1610.04545} {arXiv:1610.04545
  [hep-lat]} \BibitemShut {NoStop}%
\bibitem [{\citenamefont {Kaplan}\ \emph
  {et~al.}(1998{\natexlab{a}})\citenamefont {Kaplan}, \citenamefont {Savage},\
  and\ \citenamefont {Wise}}]{Kaplan:1998tg}%
  \BibitemOpen
  \bibfield  {author} {\bibinfo {author} {\bibfnamefont {David~B.}\
  \bibnamefont {Kaplan}}, \bibinfo {author} {\bibfnamefont {Martin~J.}\
  \bibnamefont {Savage}}, \ and\ \bibinfo {author} {\bibfnamefont {Mark~B.}\
  \bibnamefont {Wise}},\ }\bibfield  {title} {\enquote {\bibinfo {title} {{A
  New expansion for nucleon-nucleon interactions}},}\ }\href {\doibase
  10.1016/S0370-2693(98)00210-X} {\bibfield  {journal} {\bibinfo  {journal}
  {Phys. Lett.}\ }\textbf {\bibinfo {volume} {B424}},\ \bibinfo {pages}
  {390--396} (\bibinfo {year} {1998}{\natexlab{a}})},\ \Eprint
  {http://arxiv.org/abs/nucl-th/9801034} {arXiv:nucl-th/9801034 [nucl-th]}
  \BibitemShut {NoStop}%
\bibitem [{\citenamefont {Kaplan}\ \emph
  {et~al.}(1998{\natexlab{b}})\citenamefont {Kaplan}, \citenamefont {Savage},\
  and\ \citenamefont {Wise}}]{Kaplan:1998we}%
  \BibitemOpen
  \bibfield  {author} {\bibinfo {author} {\bibfnamefont {David~B.}\
  \bibnamefont {Kaplan}}, \bibinfo {author} {\bibfnamefont {Martin~J.}\
  \bibnamefont {Savage}}, \ and\ \bibinfo {author} {\bibfnamefont {Mark~B.}\
  \bibnamefont {Wise}},\ }\bibfield  {title} {\enquote {\bibinfo {title} {{Two
  nucleon systems from effective field theory}},}\ }\href {\doibase
  10.1016/S0550-3213(98)00440-4} {\bibfield  {journal} {\bibinfo  {journal}
  {Nucl. Phys.}\ }\textbf {\bibinfo {volume} {B534}},\ \bibinfo {pages}
  {329--355} (\bibinfo {year} {1998}{\natexlab{b}})},\ \Eprint
  {http://arxiv.org/abs/nucl-th/9802075} {arXiv:nucl-th/9802075 [nucl-th]}
  \BibitemShut {NoStop}%
\bibitem [{\citenamefont {Beane}\ \emph {et~al.}(1998)\citenamefont {Beane},
  \citenamefont {Cohen},\ and\ \citenamefont {Phillips}}]{Beane:1997pk}%
  \BibitemOpen
  \bibfield  {author} {\bibinfo {author} {\bibfnamefont {S.~R.}\ \bibnamefont
  {Beane}}, \bibinfo {author} {\bibfnamefont {T.~D.}\ \bibnamefont {Cohen}}, \
  and\ \bibinfo {author} {\bibfnamefont {Daniel~R.}\ \bibnamefont {Phillips}},\
  }\bibfield  {title} {\enquote {\bibinfo {title} {{The Potential of effective
  field theory in N N scattering}},}\ }\href {\doibase
  10.1016/S0375-9474(98)00007-4} {\bibfield  {journal} {\bibinfo  {journal}
  {Nucl. Phys.}\ }\textbf {\bibinfo {volume} {A632}},\ \bibinfo {pages}
  {445--469} (\bibinfo {year} {1998})},\ \Eprint
  {http://arxiv.org/abs/nucl-th/9709062} {arXiv:nucl-th/9709062 [nucl-th]}
  \BibitemShut {NoStop}%
\bibitem [{\citenamefont {van Kolck}(1999)}]{vanKolck:1998bw}%
  \BibitemOpen
  \bibfield  {author} {\bibinfo {author} {\bibfnamefont {U.}~\bibnamefont {van
  Kolck}},\ }\bibfield  {title} {\enquote {\bibinfo {title} {{Effective field
  theory of short range forces}},}\ }\href {\doibase
  10.1016/S0375-9474(98)00612-5} {\bibfield  {journal} {\bibinfo  {journal}
  {Nucl. Phys.}\ }\textbf {\bibinfo {volume} {A645}},\ \bibinfo {pages}
  {273--302} (\bibinfo {year} {1999})},\ \Eprint
  {http://arxiv.org/abs/nucl-th/9808007} {arXiv:nucl-th/9808007 [nucl-th]}
  \BibitemShut {NoStop}%
\bibitem [{\citenamefont {Bedaque}\ \emph
  {et~al.}(1999{\natexlab{a}})\citenamefont {Bedaque}, \citenamefont {Hammer},\
  and\ \citenamefont {van Kolck}}]{Bedaque:1998kg}%
  \BibitemOpen
  \bibfield  {author} {\bibinfo {author} {\bibfnamefont {Paulo~F.}\
  \bibnamefont {Bedaque}}, \bibinfo {author} {\bibfnamefont {H.~W.}\
  \bibnamefont {Hammer}}, \ and\ \bibinfo {author} {\bibfnamefont
  {U.}~\bibnamefont {van Kolck}},\ }\bibfield  {title} {\enquote {\bibinfo
  {title} {{Renormalization of the three-body system with short range
  interactions}},}\ }\href {\doibase 10.1103/PhysRevLett.82.463} {\bibfield
  {journal} {\bibinfo  {journal} {Phys. Rev. Lett.}\ }\textbf {\bibinfo
  {volume} {82}},\ \bibinfo {pages} {463--467} (\bibinfo {year}
  {1999}{\natexlab{a}})},\ \Eprint {http://arxiv.org/abs/nucl-th/9809025}
  {arXiv:nucl-th/9809025 [nucl-th]} \BibitemShut {NoStop}%
\bibitem [{\citenamefont {Bedaque}\ \emph
  {et~al.}(1999{\natexlab{b}})\citenamefont {Bedaque}, \citenamefont {Hammer},\
  and\ \citenamefont {van Kolck}}]{Bedaque:1998km}%
  \BibitemOpen
  \bibfield  {author} {\bibinfo {author} {\bibfnamefont {Paulo~F.}\
  \bibnamefont {Bedaque}}, \bibinfo {author} {\bibfnamefont {H.~W.}\
  \bibnamefont {Hammer}}, \ and\ \bibinfo {author} {\bibfnamefont
  {U.}~\bibnamefont {van Kolck}},\ }\bibfield  {title} {\enquote {\bibinfo
  {title} {{The Three boson system with short range interactions}},}\ }\href
  {\doibase 10.1016/S0375-9474(98)00650-2} {\bibfield  {journal} {\bibinfo
  {journal} {Nucl. Phys.}\ }\textbf {\bibinfo {volume} {A646}},\ \bibinfo
  {pages} {444--466} (\bibinfo {year} {1999}{\natexlab{b}})},\ \Eprint
  {http://arxiv.org/abs/nucl-th/9811046} {arXiv:nucl-th/9811046 [nucl-th]}
  \BibitemShut {NoStop}%
\bibitem [{\citenamefont {Pandharipande}\ \emph {et~al.}(2005)\citenamefont
  {Pandharipande}, \citenamefont {Phillips},\ and\ \citenamefont {van
  Kolck}}]{Pandharipande:2005sx}%
  \BibitemOpen
  \bibfield  {author} {\bibinfo {author} {\bibfnamefont {V.~R.}\ \bibnamefont
  {Pandharipande}}, \bibinfo {author} {\bibfnamefont {Daniel~R.}\ \bibnamefont
  {Phillips}}, \ and\ \bibinfo {author} {\bibfnamefont {U.}~\bibnamefont {van
  Kolck}},\ }\bibfield  {title} {\enquote {\bibinfo {title} {{Delta effects in
  pion-nucleon scattering and the strength of the two-pion-exchange
  three-nucleon interaction}},}\ }\href {\doibase 10.1103/PhysRevC.71.064002}
  {\bibfield  {journal} {\bibinfo  {journal} {Phys. Rev.}\ }\textbf {\bibinfo
  {volume} {C71}},\ \bibinfo {pages} {064002} (\bibinfo {year} {2005})},\
  \Eprint {http://arxiv.org/abs/nucl-th/0501061} {arXiv:nucl-th/0501061
  [nucl-th]} \BibitemShut {NoStop}%
\bibitem [{\citenamefont {Ordonez}\ \emph {et~al.}(1994)\citenamefont
  {Ordonez}, \citenamefont {Ray},\ and\ \citenamefont {van
  Kolck}}]{Ordonez:1993tn}%
  \BibitemOpen
  \bibfield  {author} {\bibinfo {author} {\bibfnamefont {C.}~\bibnamefont
  {Ordonez}}, \bibinfo {author} {\bibfnamefont {L.}~\bibnamefont {Ray}}, \ and\
  \bibinfo {author} {\bibfnamefont {U.}~\bibnamefont {van Kolck}},\ }\bibfield
  {title} {\enquote {\bibinfo {title} {{Nucleon-nucleon potential from an
  effective chiral Lagrangian}},}\ }\href {\doibase
  10.1103/PhysRevLett.72.1982} {\bibfield  {journal} {\bibinfo  {journal}
  {Phys. Rev. Lett.}\ }\textbf {\bibinfo {volume} {72}},\ \bibinfo {pages}
  {1982--1985} (\bibinfo {year} {1994})}\BibitemShut {NoStop}%
\bibitem [{\citenamefont {Ordonez}\ \emph {et~al.}(1996)\citenamefont
  {Ordonez}, \citenamefont {Ray},\ and\ \citenamefont {van
  Kolck}}]{Ordonez:1995rz}%
  \BibitemOpen
  \bibfield  {author} {\bibinfo {author} {\bibfnamefont {C.}~\bibnamefont
  {Ordonez}}, \bibinfo {author} {\bibfnamefont {L.}~\bibnamefont {Ray}}, \ and\
  \bibinfo {author} {\bibfnamefont {U.}~\bibnamefont {van Kolck}},\ }\bibfield
  {title} {\enquote {\bibinfo {title} {{The Two nucleon potential from chiral
  Lagrangians}},}\ }\href {\doibase 10.1103/PhysRevC.53.2086} {\bibfield
  {journal} {\bibinfo  {journal} {Phys. Rev.}\ }\textbf {\bibinfo {volume}
  {C53}},\ \bibinfo {pages} {2086--2105} (\bibinfo {year} {1996})},\ \Eprint
  {http://arxiv.org/abs/hep-ph/9511380} {arXiv:hep-ph/9511380 [hep-ph]}
  \BibitemShut {NoStop}%
\bibitem [{\citenamefont {Kaplan}\ \emph {et~al.}(1996)\citenamefont {Kaplan},
  \citenamefont {Savage},\ and\ \citenamefont {Wise}}]{Kaplan:1996xu}%
  \BibitemOpen
  \bibfield  {author} {\bibinfo {author} {\bibfnamefont {David~B.}\
  \bibnamefont {Kaplan}}, \bibinfo {author} {\bibfnamefont {Martin~J.}\
  \bibnamefont {Savage}}, \ and\ \bibinfo {author} {\bibfnamefont {Mark~B.}\
  \bibnamefont {Wise}},\ }\bibfield  {title} {\enquote {\bibinfo {title}
  {{Nucleon - nucleon scattering from effective field theory}},}\ }\href
  {\doibase 10.1016/0550-3213(96)00357-4} {\bibfield  {journal} {\bibinfo
  {journal} {Nucl. Phys.}\ }\textbf {\bibinfo {volume} {B478}},\ \bibinfo
  {pages} {629--659} (\bibinfo {year} {1996})},\ \Eprint
  {http://arxiv.org/abs/nucl-th/9605002} {arXiv:nucl-th/9605002 [nucl-th]}
  \BibitemShut {NoStop}%
\bibitem [{\citenamefont {Birse}(2006)}]{Birse:2005um}%
  \BibitemOpen
  \bibfield  {author} {\bibinfo {author} {\bibfnamefont {Michael~C.}\
  \bibnamefont {Birse}},\ }\bibfield  {title} {\enquote {\bibinfo {title}
  {{Power counting with one-pion exchange}},}\ }\href {\doibase
  10.1103/PhysRevC.74.014003} {\bibfield  {journal} {\bibinfo  {journal} {Phys.
  Rev.}\ }\textbf {\bibinfo {volume} {C74}},\ \bibinfo {pages} {014003}
  (\bibinfo {year} {2006})},\ \Eprint {http://arxiv.org/abs/nucl-th/0507077}
  {arXiv:nucl-th/0507077 [nucl-th]} \BibitemShut {NoStop}%
\bibitem [{\citenamefont {Wu}\ and\ \citenamefont {Long}(2019)}]{Wu:2018lai}%
  \BibitemOpen
  \bibfield  {author} {\bibinfo {author} {\bibfnamefont {Shaowei}\ \bibnamefont
  {Wu}}\ and\ \bibinfo {author} {\bibfnamefont {Bingwei}\ \bibnamefont
  {Long}},\ }\bibfield  {title} {\enquote {\bibinfo {title} {{Perturbative $NN$
  scattering in chiral effective field theory}},}\ }\href {\doibase
  10.1103/PhysRevC.99.024003} {\bibfield  {journal} {\bibinfo  {journal} {Phys.
  Rev.}\ }\textbf {\bibinfo {volume} {C99}},\ \bibinfo {pages} {024003}
  (\bibinfo {year} {2019})},\ \Eprint {http://arxiv.org/abs/1807.04407}
  {arXiv:1807.04407 [nucl-th]} \BibitemShut {NoStop}%
\bibitem [{\citenamefont {Kaplan}(2019)}]{Kaplan:2019znu}%
  \BibitemOpen
  \bibfield  {author} {\bibinfo {author} {\bibfnamefont {David~B.}\
  \bibnamefont {Kaplan}},\ }\bibfield  {title} {\enquote {\bibinfo {title} {{On
  the convergence of nuclear effective field theory with perturbative
  pions}},}\ }\href@noop {} {\  (\bibinfo {year} {2019})},\ \Eprint
  {http://arxiv.org/abs/1905.07485} {arXiv:1905.07485 [nucl-th]} \BibitemShut
  {NoStop}%
\bibitem [{\citenamefont {Fleming}\ \emph {et~al.}(2000)\citenamefont
  {Fleming}, \citenamefont {Mehen},\ and\ \citenamefont
  {Stewart}}]{Fleming:1999ee}%
  \BibitemOpen
  \bibfield  {author} {\bibinfo {author} {\bibfnamefont {Sean}\ \bibnamefont
  {Fleming}}, \bibinfo {author} {\bibfnamefont {Thomas}\ \bibnamefont {Mehen}},
  \ and\ \bibinfo {author} {\bibfnamefont {Iain~W.}\ \bibnamefont {Stewart}},\
  }\bibfield  {title} {\enquote {\bibinfo {title} {{NNLO corrections to
  nucleon-nucleon scattering and perturbative pions}},}\ }\href {\doibase
  10.1016/S0375-9474(00)00221-9} {\bibfield  {journal} {\bibinfo  {journal}
  {Nucl. Phys.}\ }\textbf {\bibinfo {volume} {A677}},\ \bibinfo {pages}
  {313--366} (\bibinfo {year} {2000})},\ \Eprint
  {http://arxiv.org/abs/nucl-th/9911001} {arXiv:nucl-th/9911001 [nucl-th]}
  \BibitemShut {NoStop}%
\bibitem [{\citenamefont {Epelbaum}\ and\ \citenamefont
  {Meissner}(2013)}]{Epelbaum:2006pt}%
  \BibitemOpen
  \bibfield  {author} {\bibinfo {author} {\bibfnamefont {E.}~\bibnamefont
  {Epelbaum}}\ and\ \bibinfo {author} {\bibfnamefont {U.~G.}\ \bibnamefont
  {Meissner}},\ }\bibfield  {title} {\enquote {\bibinfo {title} {{On the
  Renormalization of the One-Pion Exchange Potential and the Consistency of
  Weinberg`s Power Counting}},}\ }\href {\doibase 10.1007/s00601-012-0492-1}
  {\bibfield  {journal} {\bibinfo  {journal} {Few Body Syst.}\ }\textbf
  {\bibinfo {volume} {54}},\ \bibinfo {pages} {2175--2190} (\bibinfo {year}
  {2013})},\ \Eprint {http://arxiv.org/abs/nucl-th/0609037}
  {arXiv:nucl-th/0609037 [nucl-th]} \BibitemShut {NoStop}%
\bibitem [{\citenamefont {Epelbaum}\ \emph {et~al.}(2005)\citenamefont
  {Epelbaum}, \citenamefont {Glockle},\ and\ \citenamefont
  {Meissner}}]{Epelbaum:2004fk}%
  \BibitemOpen
  \bibfield  {author} {\bibinfo {author} {\bibfnamefont {E.}~\bibnamefont
  {Epelbaum}}, \bibinfo {author} {\bibfnamefont {W.}~\bibnamefont {Glockle}}, \
  and\ \bibinfo {author} {\bibfnamefont {Ulf-G.}\ \bibnamefont {Meissner}},\
  }\bibfield  {title} {\enquote {\bibinfo {title} {{The Two-nucleon system at
  next-to-next-to-next-to-leading order}},}\ }\href {\doibase
  10.1016/j.nuclphysa.2004.09.107} {\bibfield  {journal} {\bibinfo  {journal}
  {Nucl. Phys.}\ }\textbf {\bibinfo {volume} {A747}},\ \bibinfo {pages}
  {362--424} (\bibinfo {year} {2005})},\ \Eprint
  {http://arxiv.org/abs/nucl-th/0405048} {arXiv:nucl-th/0405048 [nucl-th]}
  \BibitemShut {NoStop}%
\bibitem [{\citenamefont {de~Vries}\ \emph {et~al.}(2013)\citenamefont
  {de~Vries}, \citenamefont {Mei{\ss}ner}, \citenamefont {Epelbaum},\ and\
  \citenamefont {Kaiser}}]{deVries:2013fxa}%
  \BibitemOpen
  \bibfield  {author} {\bibinfo {author} {\bibfnamefont {J.}~\bibnamefont
  {de~Vries}}, \bibinfo {author} {\bibfnamefont {Ulf-G.}\ \bibnamefont
  {Mei{\ss}ner}}, \bibinfo {author} {\bibfnamefont {E.}~\bibnamefont
  {Epelbaum}}, \ and\ \bibinfo {author} {\bibfnamefont {N.}~\bibnamefont
  {Kaiser}},\ }\bibfield  {title} {\enquote {\bibinfo {title} {{Parity
  violation in proton-proton scattering from chiral effective field theory}},}\
  }\href {\doibase 10.1140/epja/i2013-13149-9} {\bibfield  {journal} {\bibinfo
  {journal} {Eur. Phys. J.}\ }\textbf {\bibinfo {volume} {A49}},\ \bibinfo
  {pages} {149} (\bibinfo {year} {2013})},\ \Eprint
  {http://arxiv.org/abs/1309.4711} {arXiv:1309.4711 [nucl-th]} \BibitemShut
  {NoStop}%
\bibitem [{\citenamefont {Stoks}\ \emph {et~al.}(1993)\citenamefont {Stoks},
  \citenamefont {Klomp}, \citenamefont {Rentmeester},\ and\ \citenamefont
  {de~Swart}}]{Stoks:1993tb}%
  \BibitemOpen
  \bibfield  {author} {\bibinfo {author} {\bibfnamefont {V.~G.~J.}\
  \bibnamefont {Stoks}}, \bibinfo {author} {\bibfnamefont {R.~A.~M.}\
  \bibnamefont {Klomp}}, \bibinfo {author} {\bibfnamefont {M.~C.~M.}\
  \bibnamefont {Rentmeester}}, \ and\ \bibinfo {author} {\bibfnamefont {J.~J.}\
  \bibnamefont {de~Swart}},\ }\bibfield  {title} {\enquote {\bibinfo {title}
  {{Partial wave analaysis of all nucleon-nucleon scattering data below
  350-MeV}},}\ }\href {\doibase 10.1103/PhysRevC.48.792} {\bibfield  {journal}
  {\bibinfo  {journal} {Phys. Rev.}\ }\textbf {\bibinfo {volume} {C48}},\
  \bibinfo {pages} {792--815} (\bibinfo {year} {1993})}\BibitemShut {NoStop}%
\bibitem [{\citenamefont {Bernard}\ \emph {et~al.}(1995)\citenamefont
  {Bernard}, \citenamefont {Kaiser},\ and\ \citenamefont
  {Mei{\ss}ner}}]{Bernard:1995dp}%
  \BibitemOpen
  \bibfield  {author} {\bibinfo {author} {\bibfnamefont {V.}~\bibnamefont
  {Bernard}}, \bibinfo {author} {\bibfnamefont {Norbert}\ \bibnamefont
  {Kaiser}}, \ and\ \bibinfo {author} {\bibfnamefont {Ulf-G.}\ \bibnamefont
  {Mei{\ss}ner}},\ }\bibfield  {title} {\enquote {\bibinfo {title} {{Chiral
  dynamics in nucleons and nuclei}},}\ }\href {\doibase
  10.1142/S0218301395000092} {\bibfield  {journal} {\bibinfo  {journal} {Int.
  J. Mod. Phys.}\ }\textbf {\bibinfo {volume} {E4}},\ \bibinfo {pages}
  {193--346} (\bibinfo {year} {1995})},\ \Eprint
  {http://arxiv.org/abs/hep-ph/9501384} {arXiv:hep-ph/9501384 [hep-ph]}
  \BibitemShut {NoStop}%
\bibitem [{\citenamefont {van Kolck}(1993)}]{VanKolck:1993ee}%
  \BibitemOpen
  \bibfield  {author} {\bibinfo {author} {\bibfnamefont {Ubirajara~Lourencao}\
  \bibnamefont {van Kolck}},\ }\emph {\bibinfo {title} {{Soft Physics:
  Applications of Effective Chiral Lagrangians to Nuclear Physics and Quark
  Models}}},\ \href {http://wwwlib.umi.com/dissertations/fullcit?p9401021}
  {Ph.D. thesis},\ \bibinfo  {school} {Texas U.} (\bibinfo {year}
  {1993})\BibitemShut {NoStop}%
\bibitem [{\citenamefont {Gasser}\ \emph {et~al.}(2002)\citenamefont {Gasser},
  \citenamefont {Ivanov}, \citenamefont {Lipartia}, \citenamefont {Mojzis},\
  and\ \citenamefont {Rusetsky}}]{Gasser:2002am}%
  \BibitemOpen
  \bibfield  {author} {\bibinfo {author} {\bibfnamefont {J.}~\bibnamefont
  {Gasser}}, \bibinfo {author} {\bibfnamefont {M.~A.}\ \bibnamefont {Ivanov}},
  \bibinfo {author} {\bibfnamefont {E.}~\bibnamefont {Lipartia}}, \bibinfo
  {author} {\bibfnamefont {M.}~\bibnamefont {Mojzis}}, \ and\ \bibinfo {author}
  {\bibfnamefont {A.}~\bibnamefont {Rusetsky}},\ }\bibfield  {title} {\enquote
  {\bibinfo {title} {{Ground state energy of pionic hydrogen to one loop}},}\
  }\href {\doibase 10.1007/s10052-002-1013-z} {\bibfield  {journal} {\bibinfo
  {journal} {Eur. Phys. J.}\ }\textbf {\bibinfo {volume} {C26}},\ \bibinfo
  {pages} {13--34} (\bibinfo {year} {2002})},\ \Eprint
  {http://arxiv.org/abs/hep-ph/0206068} {arXiv:hep-ph/0206068 [hep-ph]}
  \BibitemShut {NoStop}%
\bibitem [{\citenamefont {Ananthanarayan}\ and\ \citenamefont
  {Moussallam}(2004)}]{Ananthanarayan:2004qk}%
  \BibitemOpen
  \bibfield  {author} {\bibinfo {author} {\bibfnamefont {B.}~\bibnamefont
  {Ananthanarayan}}\ and\ \bibinfo {author} {\bibfnamefont {B.}~\bibnamefont
  {Moussallam}},\ }\bibfield  {title} {\enquote {\bibinfo {title} {{Four-point
  correlator constraints on electromagnetic chiral parameters and resonance
  effective Lagrangians}},}\ }\href {\doibase 10.1088/1126-6708/2004/06/047}
  {\bibfield  {journal} {\bibinfo  {journal} {JHEP}\ }\textbf {\bibinfo
  {volume} {06}},\ \bibinfo {pages} {047} (\bibinfo {year} {2004})},\ \Eprint
  {http://arxiv.org/abs/hep-ph/0405206} {arXiv:hep-ph/0405206 [hep-ph]}
  \BibitemShut {NoStop}%
\bibitem [{\citenamefont {Feng}()}]{Feng:private}%
  \BibitemOpen
  \bibfield  {author} {\bibinfo {author} {\bibfnamefont {Xu}~\bibnamefont
  {Feng}},\ }\bibfield  {title} {\enquote {\bibinfo {title} {{private
  communication}},}\ }\href@noop {} {\ }\BibitemShut {NoStop}%
\bibitem [{\citenamefont {van Kolck}\ \emph {et~al.}(1996)\citenamefont {van
  Kolck}, \citenamefont {Friar},\ and\ \citenamefont
  {Goldman}}]{vanKolck:1996rm}%
  \BibitemOpen
  \bibfield  {author} {\bibinfo {author} {\bibfnamefont {U.}~\bibnamefont {van
  Kolck}}, \bibinfo {author} {\bibfnamefont {James~Lewis}\ \bibnamefont
  {Friar}}, \ and\ \bibinfo {author} {\bibfnamefont {J.~Terrance}\ \bibnamefont
  {Goldman}},\ }\bibfield  {title} {\enquote {\bibinfo {title}
  {{Phenomenological aspects of isospin violation in the nuclear force}},}\
  }\href {\doibase 10.1016/0370-2693(96)00009-3} {\bibfield  {journal}
  {\bibinfo  {journal} {Phys. Lett.}\ }\textbf {\bibinfo {volume} {B371}},\
  \bibinfo {pages} {169--174} (\bibinfo {year} {1996})},\ \Eprint
  {http://arxiv.org/abs/nucl-th/9601009} {arXiv:nucl-th/9601009 [nucl-th]}
  \BibitemShut {NoStop}%
\bibitem [{\citenamefont {van Kolck}\ \emph {et~al.}(1998)\citenamefont {van
  Kolck}, \citenamefont {Rentmeester}, \citenamefont {Friar}, \citenamefont
  {Goldman},\ and\ \citenamefont {de~Swart}}]{vanKolck:1997fu}%
  \BibitemOpen
  \bibfield  {author} {\bibinfo {author} {\bibfnamefont {U.}~\bibnamefont {van
  Kolck}}, \bibinfo {author} {\bibfnamefont {M.~C.~M.}\ \bibnamefont
  {Rentmeester}}, \bibinfo {author} {\bibfnamefont {James~Lewis}\ \bibnamefont
  {Friar}}, \bibinfo {author} {\bibfnamefont {J.~Terrance}\ \bibnamefont
  {Goldman}}, \ and\ \bibinfo {author} {\bibfnamefont {J.~J.}\ \bibnamefont
  {de~Swart}},\ }\bibfield  {title} {\enquote {\bibinfo {title}
  {{Electromagnetic corrections to the one pion exchange potential}},}\ }\href
  {\doibase 10.1103/PhysRevLett.80.4386} {\bibfield  {journal} {\bibinfo
  {journal} {Phys. Rev. Lett.}\ }\textbf {\bibinfo {volume} {80}},\ \bibinfo
  {pages} {4386--4389} (\bibinfo {year} {1998})},\ \Eprint
  {http://arxiv.org/abs/nucl-th/9710067} {arXiv:nucl-th/9710067 [nucl-th]}
  \BibitemShut {NoStop}%
\bibitem [{\citenamefont {Walzl}\ \emph {et~al.}(2001)\citenamefont {Walzl},
  \citenamefont {Mei{\ss}ner},\ and\ \citenamefont {Epelbaum}}]{Walzl:2000cx}%
  \BibitemOpen
  \bibfield  {author} {\bibinfo {author} {\bibfnamefont {Markus}\ \bibnamefont
  {Walzl}}, \bibinfo {author} {\bibfnamefont {Ulf-G.}\ \bibnamefont
  {Mei{\ss}ner}}, \ and\ \bibinfo {author} {\bibfnamefont {Evgeny}\
  \bibnamefont {Epelbaum}},\ }\bibfield  {title} {\enquote {\bibinfo {title}
  {{Charge dependent nucleon-nucleon potential from chiral effective field
  theory}},}\ }\href {\doibase 10.1016/S0375-9474(01)00969-1} {\bibfield
  {journal} {\bibinfo  {journal} {Nucl. Phys.}\ }\textbf {\bibinfo {volume}
  {A693}},\ \bibinfo {pages} {663--692} (\bibinfo {year} {2001})},\ \Eprint
  {http://arxiv.org/abs/nucl-th/0010019} {arXiv:nucl-th/0010019 [nucl-th]}
  \BibitemShut {NoStop}%
\bibitem [{\citenamefont {Jackson}\ and\ \citenamefont
  {Blatt}(1950)}]{Jackson:1950zz}%
  \BibitemOpen
  \bibfield  {author} {\bibinfo {author} {\bibfnamefont {John~David}\
  \bibnamefont {Jackson}}\ and\ \bibinfo {author} {\bibfnamefont {John~M.}\
  \bibnamefont {Blatt}},\ }\bibfield  {title} {\enquote {\bibinfo {title} {{The
  Interpretation of Low Energy Proton-Proton Scattering}},}\ }\href {\doibase
  10.1103/RevModPhys.22.77} {\bibfield  {journal} {\bibinfo  {journal} {Rev.
  Mod. Phys.}\ }\textbf {\bibinfo {volume} {22}},\ \bibinfo {pages} {77--118}
  (\bibinfo {year} {1950})}\BibitemShut {NoStop}%
\bibitem [{\citenamefont {Miller}\ \emph {et~al.}(1990)\citenamefont {Miller},
  \citenamefont {Nefkens},\ and\ \citenamefont {Slaus}}]{Miller:1990iz}%
  \BibitemOpen
  \bibfield  {author} {\bibinfo {author} {\bibfnamefont {G.~A.}\ \bibnamefont
  {Miller}}, \bibinfo {author} {\bibfnamefont {B.~M.~K.}\ \bibnamefont
  {Nefkens}}, \ and\ \bibinfo {author} {\bibfnamefont {I.}~\bibnamefont
  {Slaus}},\ }\bibfield  {title} {\enquote {\bibinfo {title} {{Charge symmetry,
  quarks and mesons}},}\ }\href {\doibase 10.1016/0370-1573(90)90102-8}
  {\bibfield  {journal} {\bibinfo  {journal} {Phys. Rept.}\ }\textbf {\bibinfo
  {volume} {194}},\ \bibinfo {pages} {1--116} (\bibinfo {year}
  {1990})}\BibitemShut {NoStop}%
\bibitem [{\citenamefont {Friar}\ and\ \citenamefont {van
  Kolck}(1999)}]{Friar:1999zr}%
  \BibitemOpen
  \bibfield  {author} {\bibinfo {author} {\bibfnamefont {James~Lewis}\
  \bibnamefont {Friar}}\ and\ \bibinfo {author} {\bibfnamefont
  {U.}~\bibnamefont {van Kolck}},\ }\bibfield  {title} {\enquote {\bibinfo
  {title} {{Charge independence breaking in the two pion exchange
  nucleon-nucleon force}},}\ }\href {\doibase 10.1103/PhysRevC.60.034006}
  {\bibfield  {journal} {\bibinfo  {journal} {Phys. Rev.}\ }\textbf {\bibinfo
  {volume} {C60}},\ \bibinfo {pages} {034006} (\bibinfo {year} {1999})},\
  \Eprint {http://arxiv.org/abs/nucl-th/9906048} {arXiv:nucl-th/9906048
  [nucl-th]} \BibitemShut {NoStop}%
\bibitem [{\citenamefont {Friar}\ \emph {et~al.}(2003)\citenamefont {Friar},
  \citenamefont {van Kolck}, \citenamefont {Payne},\ and\ \citenamefont
  {Coon}}]{Friar:2003yv}%
  \BibitemOpen
  \bibfield  {author} {\bibinfo {author} {\bibfnamefont {James~Lewis}\
  \bibnamefont {Friar}}, \bibinfo {author} {\bibfnamefont {U.}~\bibnamefont
  {van Kolck}}, \bibinfo {author} {\bibfnamefont {G.~L.}\ \bibnamefont
  {Payne}}, \ and\ \bibinfo {author} {\bibfnamefont {S.~A.}\ \bibnamefont
  {Coon}},\ }\bibfield  {title} {\enquote {\bibinfo {title} {{Charge symmetry
  breaking and the two pion exchange two nucleon interaction}},}\ }\href
  {\doibase 10.1103/PhysRevC.68.024003} {\bibfield  {journal} {\bibinfo
  {journal} {Phys. Rev.}\ }\textbf {\bibinfo {volume} {C68}},\ \bibinfo {pages}
  {024003} (\bibinfo {year} {2003})},\ \Eprint
  {http://arxiv.org/abs/nucl-th/0303058} {arXiv:nucl-th/0303058 [nucl-th]}
  \BibitemShut {NoStop}%
\bibitem [{\citenamefont {Kong}\ and\ \citenamefont
  {Ravndal}(1999)}]{Kong:1998sx}%
  \BibitemOpen
  \bibfield  {author} {\bibinfo {author} {\bibfnamefont {Xinwei}\ \bibnamefont
  {Kong}}\ and\ \bibinfo {author} {\bibfnamefont {Finn}\ \bibnamefont
  {Ravndal}},\ }\bibfield  {title} {\enquote {\bibinfo {title} {{Proton proton
  scattering lengths from effective field theory}},}\ }\href {\doibase
  10.1016/S0370-2693(99)00619-X, 10.1016/S0370-2693(99)00144-6} {\bibfield
  {journal} {\bibinfo  {journal} {Phys. Lett.}\ }\textbf {\bibinfo {volume}
  {B450}},\ \bibinfo {pages} {320--324} (\bibinfo {year} {1999})},\ \bibinfo
  {note} {[Erratum: Phys. Lett.B458,565(1999)]},\ \Eprint
  {http://arxiv.org/abs/nucl-th/9811076} {arXiv:nucl-th/9811076 [nucl-th]}
  \BibitemShut {NoStop}%
\bibitem [{\citenamefont {Baroni}\ \emph {et~al.}(2018)\citenamefont {Baroni}
  \emph {et~al.}}]{Baroni:2018fdn}%
  \BibitemOpen
  \bibfield  {author} {\bibinfo {author} {\bibfnamefont {A.}~\bibnamefont
  {Baroni}} \emph {et~al.},\ }\bibfield  {title} {\enquote {\bibinfo {title}
  {{Local chiral interactions, the tritium Gamow-Teller matrix element, and the
  three-nucleon contact term}},}\ }\href {\doibase 10.1103/PhysRevC.98.044003}
  {\bibfield  {journal} {\bibinfo  {journal} {Phys. Rev.}\ }\textbf {\bibinfo
  {volume} {C98}},\ \bibinfo {pages} {044003} (\bibinfo {year} {2018})},\
  \Eprint {http://arxiv.org/abs/1806.10245} {arXiv:1806.10245 [nucl-th]}
  \BibitemShut {NoStop}%
\bibitem [{\citenamefont {Pieper}(2008)}]{IL7}%
  \BibitemOpen
  \bibfield  {author} {\bibinfo {author} {\bibfnamefont {Steven~C.}\
  \bibnamefont {Pieper}},\ }\bibfield  {title} {\enquote {\bibinfo {title} {The
  illinois extension to the fujita-miyazawa three-nucleon force},}\ }\href
  {\doibase 10.1063/1.2932280} {\bibfield  {journal} {\bibinfo  {journal} {AIP
  Conference Proceedings}\ }\textbf {\bibinfo {volume} {1011}},\ \bibinfo
  {pages} {143} (\bibinfo {year} {2008})}\BibitemShut {NoStop}%
\bibitem [{\citenamefont {Pastore}\ \emph
  {et~al.}(2018{\natexlab{b}})\citenamefont {Pastore}, \citenamefont {Carlson},
  \citenamefont {Cirigliano}, \citenamefont {Dekens}, \citenamefont
  {Mereghetti},\ and\ \citenamefont {Wiringa}}]{Pastore:2017ofx}%
  \BibitemOpen
  \bibfield  {author} {\bibinfo {author} {\bibfnamefont {S.}~\bibnamefont
  {Pastore}}, \bibinfo {author} {\bibfnamefont {J.}~\bibnamefont {Carlson}},
  \bibinfo {author} {\bibfnamefont {V.}~\bibnamefont {Cirigliano}}, \bibinfo
  {author} {\bibfnamefont {W.}~\bibnamefont {Dekens}}, \bibinfo {author}
  {\bibfnamefont {E.}~\bibnamefont {Mereghetti}}, \ and\ \bibinfo {author}
  {\bibfnamefont {R.~B.}\ \bibnamefont {Wiringa}},\ }\bibfield  {title}
  {\enquote {\bibinfo {title} {{Neutrinoless double-$\beta$ decay matrix
  elements in light nuclei}},}\ }\href {\doibase 10.1103/PhysRevC.97.014606}
  {\bibfield  {journal} {\bibinfo  {journal} {Phys. Rev.}\ }\textbf {\bibinfo
  {volume} {C97}},\ \bibinfo {pages} {014606} (\bibinfo {year}
  {2018}{\natexlab{b}})},\ \Eprint {http://arxiv.org/abs/1710.05026}
  {arXiv:1710.05026 [nucl-th]} \BibitemShut {NoStop}%
\bibitem [{\citenamefont {Nollett}\ \emph {et~al.}(2001)\citenamefont
  {Nollett}, \citenamefont {Wiringa},\ and\ \citenamefont
  {Schiavilla}}]{Nollett:2000ch}%
  \BibitemOpen
  \bibfield  {author} {\bibinfo {author} {\bibfnamefont {K.~M.}\ \bibnamefont
  {Nollett}}, \bibinfo {author} {\bibfnamefont {R.~B.}\ \bibnamefont
  {Wiringa}}, \ and\ \bibinfo {author} {\bibfnamefont {R.}~\bibnamefont
  {Schiavilla}},\ }\bibfield  {title} {\enquote {\bibinfo {title} {{A Six body
  calculation of the alpha deuteron radiative capture cross-section}},}\ }\href
  {\doibase 10.1103/PhysRevC.63.024003} {\bibfield  {journal} {\bibinfo
  {journal} {Phys. Rev.}\ }\textbf {\bibinfo {volume} {C63}},\ \bibinfo {pages}
  {024003} (\bibinfo {year} {2001})},\ \Eprint
  {http://arxiv.org/abs/nucl-th/0006064} {arXiv:nucl-th/0006064 [nucl-th]}
  \BibitemShut {NoStop}%
\bibitem [{\citenamefont {\v{S}imkovic}\ \emph {et~al.}(2008)\citenamefont
  {\v{S}imkovic}, \citenamefont {Faessler}, \citenamefont {Rodin},
  \citenamefont {Vogel},\ and\ \citenamefont {Engel}}]{Simkovic:2007vu}%
  \BibitemOpen
  \bibfield  {author} {\bibinfo {author} {\bibfnamefont {Fedor}\ \bibnamefont
  {\v{S}imkovic}}, \bibinfo {author} {\bibfnamefont {Amand}\ \bibnamefont
  {Faessler}}, \bibinfo {author} {\bibfnamefont {Vadim}\ \bibnamefont {Rodin}},
  \bibinfo {author} {\bibfnamefont {Petr}\ \bibnamefont {Vogel}}, \ and\
  \bibinfo {author} {\bibfnamefont {Jonathan}\ \bibnamefont {Engel}},\
  }\bibfield  {title} {\enquote {\bibinfo {title} {{Anatomy of nuclear matrix
  elements for neutrinoless double-beta decay}},}\ }\href {\doibase
  10.1103/PhysRevC.77.045503} {\bibfield  {journal} {\bibinfo  {journal} {Phys.
  Rev.}\ }\textbf {\bibinfo {volume} {C77}},\ \bibinfo {pages} {045503}
  (\bibinfo {year} {2008})},\ \Eprint {http://arxiv.org/abs/0710.2055}
  {arXiv:0710.2055 [nucl-th]} \BibitemShut {NoStop}%
\bibitem [{\citenamefont {Men\'endez}\ \emph {et~al.}(2009)\citenamefont
  {Men\'endez}, \citenamefont {Poves}, \citenamefont {Caurier},\ and\
  \citenamefont {Nowacki}}]{Menendez:2008jp}%
  \BibitemOpen
  \bibfield  {author} {\bibinfo {author} {\bibfnamefont {J.}~\bibnamefont
  {Men\'endez}}, \bibinfo {author} {\bibfnamefont {A.}~\bibnamefont {Poves}},
  \bibinfo {author} {\bibfnamefont {E.}~\bibnamefont {Caurier}}, \ and\
  \bibinfo {author} {\bibfnamefont {F.}~\bibnamefont {Nowacki}},\ }\bibfield
  {title} {\enquote {\bibinfo {title} {{Disassembling the Nuclear Matrix
  Elements of the Neutrinoless beta beta Decay}},}\ }\href {\doibase
  10.1016/j.nuclphysa.2008.12.005} {\bibfield  {journal} {\bibinfo  {journal}
  {Nucl. Phys.}\ }\textbf {\bibinfo {volume} {A818}},\ \bibinfo {pages}
  {139--151} (\bibinfo {year} {2009})},\ \Eprint
  {http://arxiv.org/abs/0801.3760} {arXiv:0801.3760 [nucl-th]} \BibitemShut
  {NoStop}%
\bibitem [{\citenamefont {\v{S}imkovic}\ \emph {et~al.}(1999)\citenamefont
  {\v{S}imkovic}, \citenamefont {Pantis}, \citenamefont {Vergados},\ and\
  \citenamefont {Faessler}}]{Simkovic:1999re}%
  \BibitemOpen
  \bibfield  {author} {\bibinfo {author} {\bibfnamefont {F.}~\bibnamefont
  {\v{S}imkovic}}, \bibinfo {author} {\bibfnamefont {G.}~\bibnamefont
  {Pantis}}, \bibinfo {author} {\bibfnamefont {J.~D.}\ \bibnamefont
  {Vergados}}, \ and\ \bibinfo {author} {\bibfnamefont {Amand}\ \bibnamefont
  {Faessler}},\ }\bibfield  {title} {\enquote {\bibinfo {title} {{Additional
  nucleon current contributions to neutrinoless double beta decay}},}\ }\href
  {\doibase 10.1103/PhysRevC.60.055502} {\bibfield  {journal} {\bibinfo
  {journal} {Phys. Rev.}\ }\textbf {\bibinfo {volume} {C60}},\ \bibinfo {pages}
  {055502} (\bibinfo {year} {1999})},\ \Eprint
  {http://arxiv.org/abs/hep-ph/9905509} {arXiv:hep-ph/9905509 [hep-ph]}
  \BibitemShut {NoStop}%
\bibitem [{\citenamefont {Men\'endez}\ \emph {et~al.}(2011)\citenamefont
  {Men\'endez}, \citenamefont {Gazit},\ and\ \citenamefont
  {Schwenk}}]{Menendez:2011qq}%
  \BibitemOpen
  \bibfield  {author} {\bibinfo {author} {\bibfnamefont {J.}~\bibnamefont
  {Men\'endez}}, \bibinfo {author} {\bibfnamefont {D.}~\bibnamefont {Gazit}}, \
  and\ \bibinfo {author} {\bibfnamefont {A.}~\bibnamefont {Schwenk}},\
  }\bibfield  {title} {\enquote {\bibinfo {title} {{Chiral two-body currents in
  nuclei: Gamow-Teller transitions and neutrinoless double-beta decay}},}\
  }\href {\doibase 10.1103/PhysRevLett.107.062501} {\bibfield  {journal}
  {\bibinfo  {journal} {Phys. Rev. Lett.}\ }\textbf {\bibinfo {volume} {107}},\
  \bibinfo {pages} {062501} (\bibinfo {year} {2011})},\ \Eprint
  {http://arxiv.org/abs/1103.3622} {arXiv:1103.3622 [nucl-th]} \BibitemShut
  {NoStop}%
\bibitem [{\citenamefont {Graf}\ \emph {et~al.}(2018)\citenamefont {Graf},
  \citenamefont {Deppisch}, \citenamefont {Iachello},\ and\ \citenamefont
  {Kotila}}]{Graf:2018ozy}%
  \BibitemOpen
  \bibfield  {author} {\bibinfo {author} {\bibfnamefont {Lukas}\ \bibnamefont
  {Graf}}, \bibinfo {author} {\bibfnamefont {Frank~F.}\ \bibnamefont
  {Deppisch}}, \bibinfo {author} {\bibfnamefont {Francesco}\ \bibnamefont
  {Iachello}}, \ and\ \bibinfo {author} {\bibfnamefont {Jenni}\ \bibnamefont
  {Kotila}},\ }\bibfield  {title} {\enquote {\bibinfo {title} {{Short-Range
  Neutrinoless Double Beta Decay Mechanisms}},}\ }\href {\doibase
  10.1103/PhysRevD.98.095023} {\bibfield  {journal} {\bibinfo  {journal} {Phys.
  Rev.}\ }\textbf {\bibinfo {volume} {D98}},\ \bibinfo {pages} {095023}
  (\bibinfo {year} {2018})},\ \Eprint {http://arxiv.org/abs/1806.06058}
  {arXiv:1806.06058 [hep-ph]} \BibitemShut {NoStop}%
\end{thebibliography}%

\end{document}